\RequirePackage{fix-cm} 
\documentclass[oneside,phd,4x6]{snuthesis}

\usepackage{tgpagella} 

\usepackage{biolinum} 
\usepackage[T1]{fontenc}

\usepackage{etoolbox}
\usepackage{caption}
\usepackage{booktabs} 
\usepackage{relsize} 
\usepackage{subcaption}
\usepackage{afterpage}
\usepackage{enumitem}
\usepackage{makecell}
\usepackage{multirow}
\usepackage{array}
\usepackage{amssymb}
\usepackage{tabularx}
\usepackage{ifthen}
\usepackage[english]{babel}
\usepackage{varwidth}
\usepackage[hidelinks, breaklinks]{hyperref}
\usepackage{rotating} 
\usepackage[font=sf]{floatrow}
\usepackage[raggedright]{titlesec}
\usepackage{xcolor,colortbl}
\usepackage{longtable}
\usepackage{threeparttable}
\usepackage[all]{nowidow}
\usepackage{setspace}
\usepackage{sectsty}
\usepackage[bottom, flushmargin]{footmisc}
\usepackage{xspace}
\usepackage{amsmath}
\usepackage[dvipsnames]{xcolor}
\usepackage{tikz} 
\usepackage{changepage}

\DeclareMathOperator*{\argmax}{arg\,max}

\definecolor{mymygreen}{RGB}{77,175,74}
\definecolor{mymyorange}{RGB}{255,127,0}
\definecolor{mymypurple}{RGB}{152,78,164}

\setlist{itemsep=2pt, parsep=0pt} 

\allsectionsfont{\raggedright\setstretch{1.2}\normalfont\sffamily\bfseries}
\subsubsectionfont{\raggedright\setstretch{1.2}\sffamily\fontseries{sb}\itshape\large}

\addto\extrasenglish{%
}
\addto\extrasenglish{%
}
\addto\extrasenglish{%
}

\newlist{todolist}{itemize}{10} 
\setlist[todolist]{label=$\square$}
\newlist{radiolist}{itemize}{10}
\setlist[radiolist]{label=$\circ$}

\usepackage[sortcites, backend=biber, natbib=true, style=trad-plain, sorting=nty, maxbibnames=99, url=false, isbn=false, doi=false]{biblatex}
\definecolor{tableheader}{HTML}{EFEFEF}

\definecolor{todo}{RGB}{255,0,205}

\definecolor{needrevise}{RGB}{255,70,200}

\usepackage[titles]{tocloft} 
\makeatletter
\if@snu@ko
	\renewcommand\cftchappresnum{제~}
	\renewcommand\cftchapaftersnum{~장}
	\renewcommand\cftfigpresnum{그림~}
	\renewcommand\cfttabpresnum{표~}
\else
	\renewcommand\cftchappresnum{Chapter~}
	\renewcommand\cftfigpresnum{Figure~}
	\renewcommand\cfttabpresnum{Table~}
\fi
\makeatother
\newlength{\mytmplen}
\makeatletter
\g@addto@macro\appendix{%
	\addtocontents{toc}{%
		\settowidth{\mytmplen}{\bfseries\protect\cftchappresnum\protect\cftchapaftersnum}%
		\addtolength{\cftchapnumwidth}{-\mytmplen}%
		\protect\renewcommand{\protect\cftchappresnum}{\appendixname~}%
		\protect\renewcommand{\protect\cftchapaftersnum}{}%
		\setlength\cftchapnumwidth{0em} 
	}%
}
\makeatother

\title{Dimensionality Reduction Considered Harmful \\ (Some of the Time)}

\title*{차원축소는 위험하다 (생각보다)}

\academicko{공학}
\schoolen{College of Engineering}
\departmenten{Dept. of Computer Science and Engineering}
\departmentko{컴퓨터공학부}

\author{Hyeon Jeon}
\author*{전~현} 

\studentnumber{2020-24899}

\advisor{Jinwook Seo}
\advisor*{서~진~욱}

\graddate{February 2026}

\submissiondate{December 2025}

\approvaldate{December 2025}

\committeemembers%
{Jaesik Park}%
{\advisor}%
{Ghulam Jilani Quadri}%
{Micha\"el Aupetit}
{Kwan-Liu Ma}%

\renewcommand{\paragraph}[1]{\vspace{4pt}\noindent\textbf{#1.}}
\newcommand{\paragraphit}[1]{\vspace{4pt}\noindent\textit{#1.}}

\begin{document}

\pagenumbering{Roman}
\makefrontcover

\makeapproval

\cleardoublepage
\pagenumbering{roman}

\keywordalt{시각적 분석; 차원축소; 최적화; 평가; 상호작용; 신뢰성}

\begin{abstract}

\vspace{-10mm}
Dimensionality reduction (DR) is one of the most commonly used yet most easily misinterpreted tools in visual analytics. 
Visual analytics using DR can thus easily be unreliable:
insights derived from analysis may not accurately reflect the underlying data, potentially leading to flawed knowledge and decision-making. This problem occurs because DR projections inherently cannot capture all characteristics of the original data, yet these limitations are often not adequately accounted for during analysis.

In this dissertation, we enhance the reliability of visual analytics with DR.
At the beginning, we understand the reliability challenges practitioners encounter when using DR for visual analytics. Following a human-centric approach,
we detail how practitioners leverage DR in practice by combining interview studies and literature review. 
We then address three challenges by designing technical solutions. At first, we \textbf{mitigate the common misuse of famous DR techniques}: $t$-SNE and UMAP. 
These techniques are misused for unsuitable analytical tasks as they exaggerate cluster and class separability, which practitioners perceive as ``aesthetically pleasing''. 
We find that existing DR evaluation metrics that leverage class labels amplify this bias as they favor projections that well separate the classes. 
This is because these existing metrics assume classes as ground truth clusters. 
We propose new metrics that escape from this assumption, provoking that \textit{classes are not clusters}, enabling practitioners to identify projections that more reliably support cluster analysis.
Second, we \textbf{address the prevalent cherry-picking of hyperparameters}. 
Even with proper evaluation metrics, selecting appropriate DR techniques and optimizing hyperparameters to maximize metric scores requires extensive trial and error, leading practitioners to rely on default settings or cherry-pick hyperparameters. 
We introduce a \textit{dataset-adaptive optimization workflow} that significantly reduces the computational cost of optimizing DR projections, motivating practitioners to avoid cherry-picking hyperparameters and instead systematically optimize projections.
Third, we \textbf{make interactions in DR projections less erroneous.}
High-dimensional space has a significantly higher degree of freedom compared to a low-dimensional space. Thus, even properly optimized DR projections cannot fully escape from distortions in representing the original structure. Such distortions cause interactions on DR projections, such as brushing, to erroneously reflect users’ intentions. We address this problem by proposing a new brushing technique called \textit{Distortion-aware brushing}, which corrects distortions as users investigate clusters, thereby helping them precisely capture the high-dimensional clusters they target. 

Building on the insights from these studies, we outline future directions that can fundamentally enhance the reliability of DR-based visual analytics---spanning from facilitating relevant discourse to fully automating the selection of optimal DR projections. We conclude the thesis by discussing how our contributions lay the foundation for achieving more reliable visual analytics practices.

\end{abstract}

\setlength{\cftfigindent}{0pt}  
\setlength{\cfttabindent}{0pt}

\setlength\bibitemsep{2\itemsep} 


\setlength{\cftsecindent}{0cm}
\setlength{\cftsubsecindent}{0.9cm}

\renewcommand\cftfigpresnum{\bfseries\sffamily{Figure~}}
\renewcommand\cfttabpresnum{\bfseries\sffamily{Table~}}
\renewcommand\cftfigpagefont{\sffamily}
\renewcommand\cfttabpagefont{\sffamily}
\renewcommand\cftfigfont{\sffamily\small}
\renewcommand\cfttabfont{\sffamily\small}

\renewcommand\cftchappresnum{\small{CHAPTER~}}
\renewcommand\appendixname{\small{APPENDIX~}}

\renewcommand{\cftchapleader}{\cftdotfill{\cftdotsep}}

\renewcommand{\cftchapfont}{\sffamily\bfseries\Large}
\renewcommand{\cftsecfont}{\sffamily\bfseries}
\renewcommand{\cftsubsecfont}{\sffamily}

\renewcommand{\cftchapaftersnumb}{\\\hskip0em}
\setlength\cftchapnumwidth{0em}

\renewcommand{\cftchappagefont}{\sffamily\bfseries}
\renewcommand{\cftsecpagefont}{\sffamily\bfseries}
\renewcommand{\cftsubsecpagefont}{\sffamily}

\renewcommand\cftchapafterpnum{\vskip0.5em}
\setlength\cftbeforechapskip{2em} 

\def\perplexity{\texttt{perplexity}\xspace}
\def\nneighbors{\texttt{n\_neighbors}\xspace}

\definecolor{appledarkblue}{RGB}{30, 110, 244}
\definecolor{appledarkred}{RGB}{233,21,45}

\definecolor{applebluedark}{RGB}{30, 110, 244}

\newcommand{\blue}[1]{\textcolor{appledarkblue}{#1}}
\newcommand{\red}[1]{\textcolor{appledarkred}{#1}}

\def\LS{Label-Trustworthiness\xspace}
\def\LC{Label-Continuity\xspace}
\def\lc{Label-C\xspace}
\def\ls{Label-T\xspace}
\def\lsc{Label-T\&C\xspace}

\def\LT{Label-Trustworthiness\xspace}
\def\LC{Label-Continuity\xspace}
\def\lc{Label-C\xspace}
\def\lt{Label-T\xspace}
\def\ltc{Label-T\&C\xspace}
\def\CHb{$CH_{A}$\xspace}

\newcommand{\revise}[1]{#1}

\def\MNC{\textsc{Mnc}\xspace}
\def\PDS{\textsc{Pds}\xspace}
\def\LogPDS{\textsc{LogPDS}\xspace}
\def\MNCPDS{\textsc{Pds+Mnc}\xspace}
\def\PDSMNC{\textsc{Pds+Mnc}\xspace}
\def\MNCLogPDS{\textsc{MNC+LogPDS}\xspace}

\definecolor{applereddark}{RGB}{215, 0, 21}

\newcommand{\redd}[1]{\textcolor{applereddark}{#1}}

\def\AB{A\&B\xspace} 

\def\brush{Distortion-aware brushing\xspace}

\newcommand{\close}[2]{\texttt{close}$_{\kappa}$\(({#1}, {#2})\)}
\newcommand{\simil}[2]{\texttt{sim}$_k$\(({#1}, {#2})\)}
\newcommand{\dens}[1]{\texttt{dens}\(({#1})\)}

\def\tsne{\textit{t}-SNE\xspace}
\def\umap{UMAP\xspace}

\definecolor{appleredlight}{RGB}{255, 105, 97}
\definecolor{appleorangelight}{RGB}{255, 179, 64}
\definecolor{appleyellowlight}{RGB}{255, 212, 38}
\definecolor{applegreenlight}{RGB}{48, 219, 91}
\definecolor{applemintlight}{RGB}{102, 212, 207}
\definecolor{appleteallight}{RGB}{93, 230, 255}
\definecolor{applecyanlight}{RGB}{112, 215, 255}
\definecolor{applebluelight}{RGB}{64, 156, 255}
\definecolor{appleindigolight}{RGB}{125, 122, 255}
\definecolor{applepurplelight}{RGB}{218, 143, 255}
\definecolor{applepinklight}{RGB}{255, 100, 130}
\definecolor{applebrownlight}{RGB}{181, 148, 105}

\definecolor{applerednormal}{RGB}{255, 69, 58}
\definecolor{appleorangenormal}{RGB}{255, 159, 10}
\definecolor{appleyellownormal}{RGB}{255, 214, 10}
\definecolor{applegreennormal}{RGB}{48, 209, 88}
\definecolor{applemintnormal}{RGB}{99, 230, 226}
\definecolor{appletealnormal}{RGB}{64, 200, 224}
\definecolor{applecyannormal}{RGB}{100, 210, 255}
\definecolor{applebluenormal}{RGB}{10, 132, 255}
\definecolor{appleindigonormal}{RGB}{94, 92, 230}
\definecolor{applepurplenormal}{RGB}{191, 90, 242}
\definecolor{applepinknormal}{RGB}{255, 55, 95}
\definecolor{applebrownnormal}{RGB}{172, 142, 104}

\newcommand{\redt}[1]{\textcolor{appleredlight}{#1}}
\newcommand{\bluet}[1]{\textcolor{applebluelight}{#1}}

\newcommand{\redn}[1]{\textcolor{applerednormal}{#1}}
\newcommand{\bluen}[1]{\textcolor{applebluenormal}{#1}}


\newcommand{\oone}{O1}

\newcommand{\otwo}{O2}

\newcommand{\othree}{O3}

\newcommand{\ofour}{O4}

\newcommand{\stepone}{Step 1}

\newcommand{\steptwo}{Step 2}

\newcommand{\stepthree}{Step 3}

\newcommand{\stepfour}{Step 4}

\newcommand{\finding}[2]{
\begin{adjustwidth}{5.5mm}{5.5mm}
\vspace{2.5mm}
\noindent
\textbf{\textit{Finding #1. }} \textit{
#2
}
\vspace{2.5mm}
\end{adjustwidth}
}


{\hypersetup{hidelinks}\tableofcontents
\listoffigures
\listoftables}

\cleardoublepage
\pagenumbering{arabic}

\chapter{Introduction}

Visual analytics, the process of analytical reasoning supported by interactive visual interfaces, plays a central role in informed decision-making across domains, from scientific discovery \cite{tu24pvis} to public health (e.g., COVID-19) \cite{leung21iv}. As data grows in scale and complexity, the importance and impact of visual analytics continue to grow. However, visual analytics often becomes unreliable: decisions or knowledge from the analysis may not precisely reflect the underlying data, e.g., due to algorithmic errors or human bias. Such unreliability can have serious real-world consequences, such as weakening public safety (e.g., failures to detect terrorism \cite{cook2005illuminating}) or undermining the trustworthiness of the scientific process.

In this dissertation, we improve the reliability of visual analytics with a focus on the use of dimensionality reduction (DR).
DR has become one of the most widely used techniques for analyzing high-dimensional data across scientific disciplines \cite{cashman25tvcg}.
However, DR-based visual analytics often becomes unreliable due to errors introduced by DR techniques and human bias.
This dissertation enhances the reliability of DR-based visual analytics by identifying the prevalent challenges that undermine it and by presenting technical solutions that directly address these challenges.


\section{Backgrounds and Motivation}

\label{sec:introback}

DR denotes a set of techniques, including $t$-SNE \cite{maaten08jmlr, maaten14jmlr}, UMAP \cite{mcinnes2020arxiv}, and PCA \cite{pearson01pmjs}, that receive high-dimensional data as input and output its low-dimensional representations \cite{nonato19tvcg, jeon21tvcg}. The output dimension is typically two or three when used for visual analytics. Using DR, any high-dimensional data can be visualized and analyzed as a scatterplot.
Such intuitiveness makes DR commonly adopted in many disciplines, including machine learning \cite{jeon25tpami}, bioinformatics \cite{cheng23tvcg, becht19nature}, and human-computer interaction \cite{hamalainen23chi, ha24chi}.

However, despite its usefulness, practitioners should account for the following weaknesses when using DR:

\paragraph{DR projections cannot preserve entire structural characteristics of the original data}
As high-dimensional spaces have a significantly higher degree of freedom compared to low-dimensional spaces, DR projections cannot show all aspects of the original data \cite{nonato19tvcg, faust19tvcg, sedlmair13tvcg, jiazhi21tvcg, etemadpour15tvcg}. DR techniques thus focus on specific structural characteristics to preserve \cite{etemadpour15tvcg, choo09vast, }. For example, $t$-SNE and UMAP focus on preserving the local neighborhood structures, while PCA focuses on preserving global structures like the density of clusters or the distances between points. It is thus important to select appropriate DR techniques and hyperparameters that produce projections that faithfully represent the structure of interest. 

\paragraph{DR projections inherently suffer from distortions}
Even though DR projections are properly configured or optimized to preserve the structure of interest, they cannot fully escape from distortions in representing the original structure of the data \cite{nonato19tvcg, lespinats07tnn, lespinats11cgf, jeon21tvcg}. 
Therefore, to ensure reliable visual analytics, it is important to evaluate DR projections in advance to know how much and how they are distorted \cite{jeon21tvcg, martins14cg, colange19vis}.

However, practitioners are often unaware of these weaknesses, leading them to use DR in ways that undermine the reliability of visual analytics. For example, practitioners select inappropriate DR techniques that do not match target tasks or use biased metrics to evaluate DR projections. Given the widespread use of DR across many disciplines, addressing these inappropriate uses is crucial not only for enhancing the reliability of individual analyses but also for ensuring the rigor of scientific discoveries and communication.
This dissertation achieves this goal by (1) investigating prevalent reliability challenges that practitioners encounter, and (2) providing technical solutions to remedy these challenges.

\section{Thesis Contribution}

\label{sec:thesiscontribution}

\textbf{Thesis statement:} \textit{
An in-depth understanding of practitioners’ use of DR, together with carefully designed technical solutions, can jointly advance the reliability of visual analytics using DR.}

\vspace{4pt}
\noindent
In response to this statement, we first contribute a detailed investigation of how practitioners---visual analytics researchers and domain experts from fields outside of visualization---employ DR in their analytical workflows.
This is done by (1) reviewing papers that incorporate visual analytics with DR projections, and (2) conducting semi-structured interviews with practitioners.
As a result, we identify three prevalent reliability challenges that practitioners encounter in their analysis, then subsequently develop corresponding technical solutions for each. Detailed descriptions of these challenges and their associated solutions are provided below.

\paragraph{Challenge 1: Misuse of commonly used DR techniques}
We identify that the two most widely used DR techniques---$t$-SNE and UMAP---are also commonly misused for tasks for which they are not suitable. 
A prevalent reason for its misuse is that it produces aesthetically pleasing or interpretable projections by exaggerating separability between classes or clusters. 
Although evaluation metrics should serve as objective gatekeepers against such biases, we show that a widely used set of DR evaluation metrics ironically reinforces the bias. These metrics typically assume class labels as ground-truth clusters, using class separability in the low-dimensional projection as a proxy for faithfulness. Consequently, they tend to reward projections that exaggerate class separation even when the underlying class structure is not well separated, thereby encouraging practitioners to rely more heavily on $t$-SNE and UMAP.

\begin{sidewaysfigure}
    \centering
    \includegraphics[width=\linewidth]{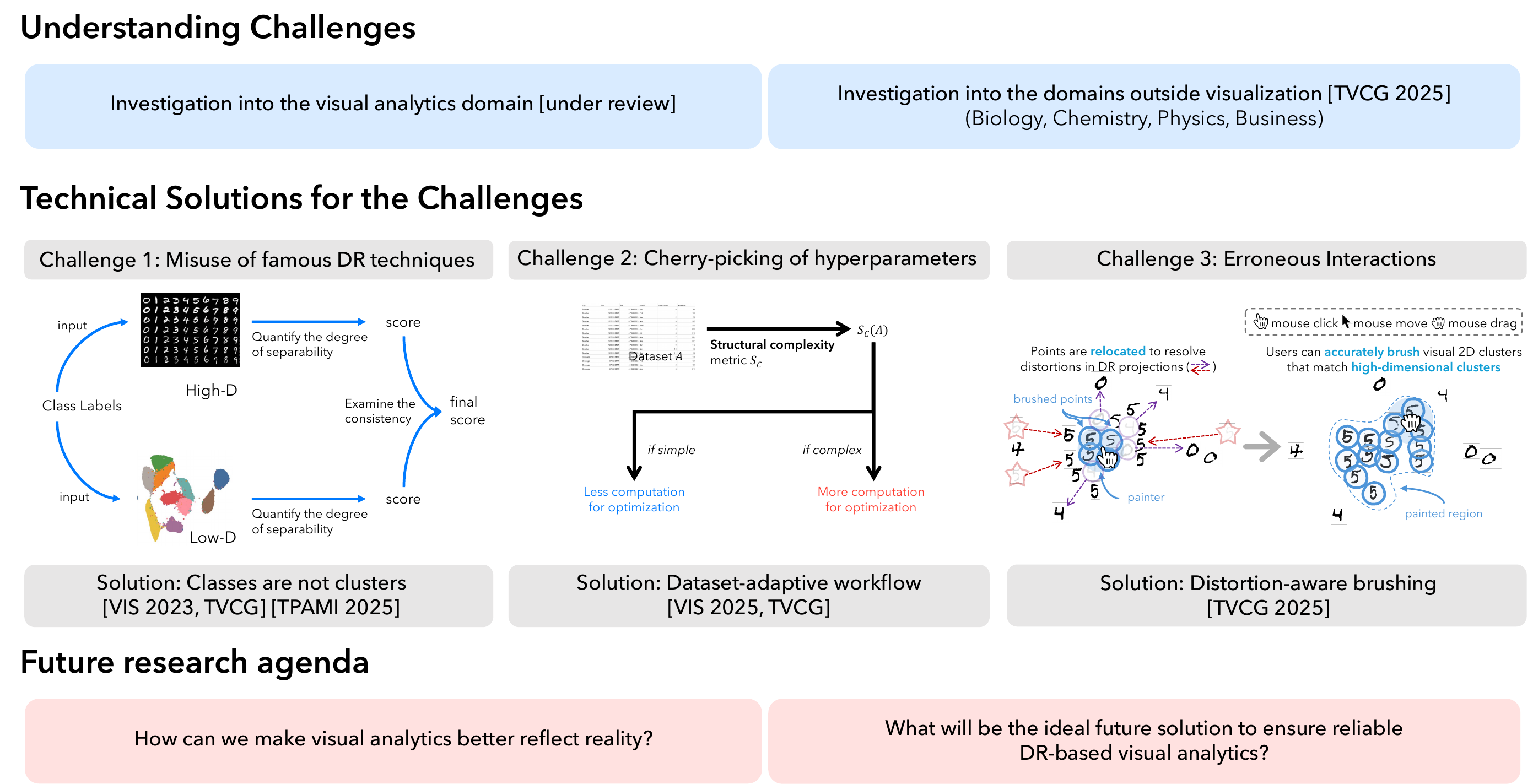}
    \caption{The overview of the contributions provided by this thesis (\autoref{sec:thesiscontribution}). In summary, the thesis identifies challenges that undermine the reliability of visual analytics and presents technical solutions to address them. }
    \label{fig:intro}
\end{sidewaysfigure}

To address this problem, we propose an improved way of evaluating DR projections using class labels, urging that \textit{classes are not clusters} (\autoref{fig:intro}a).
We rely on a simple idea: instead of measuring class separability only in the low-dimensional space, we measure it in both high- and low-dimensional spaces and quantify their consistency as faithfulness. 
To implement this idea, we first develop clustering metrics that produce class separability scores that are comparable across different dimensions and data patterns by adjusting existing clustering metrics such as Silhouette \cite{rousseuw87silhouette}. Then, using these adjusted clustering metrics, we design and propose two new DR evaluation metrics called \textit{Label-Trustworthiness and Continuity (Label-T\&C)}.
Controlled experiments verify that these metrics detect distortions overlooked by conventional metrics and mitigate bias introduced by previous metrics.




\paragraph{Challenge 2: Cherry-picking of hyperparameters}
Selecting appropriate DR techniques and optimizing hyperparameters contributes to reliable visual analytics by directly minimizing the amount of distortions in the projections. 
However, this optimization process is computationally expensive, leading practitioners to occasionally rely on famous techniques and, moreover, cherry-pick hyperparameters to produce projections with desirable visual patterns \cite{doh25arxiv}. 
This is because the process requires extensive trial-and-error that repeatedly tests various sets of techniques and hyperparameter values, e.g., by leveraging random search \cite{bergstra12jmlr} or Bayesian optimization \cite{snoek12nips}.


To motivate practitioners to systematically optimize DR projections, we accelerate the optimization process by introducing a \textit{dataset-adaptive workflow} (\autoref{fig:intro}b).
In this workflow, the optimization process is guided based on the characteristics of the input dataset. 
In detail, we first measure the \textit{structural complexity} of datasets and predict the maximum accuracy achievable by DR techniques using the scores. The output scores enhance the efficiency of DR optimization by (1) guiding the selection of an appropriate DR technique for a given dataset and (2) enabling early termination of optimization by detecting whether optimization has reached near-optimum, avoiding unnecessary computations.

\paragraph{Challenge 3: Erroneous interactions}
Brushing is a fundamental interaction technique in visual analytics for scatterplots, which are commonly used to represent DR projections. This interaction typically involves selecting clusters by drawing a rectangular box or lasso around them \cite{heer08chi}, with the selected regions then explored through auxiliary visualizations. 
However, even highly optimized projections cannot completely escape from distortions \cite{lespinats11cgf, lespinats07tnn, jeon21tvcg}. 
Consequently, brushing a contiguous region in the projection space does not guarantee that the selected points form a true cluster in the original high-dimensional space that users intended to brush. This issue undermines not only visual analytics workflows but also other applications that rely on brushing, e.g., interactive data labeling \cite{meng24tvcg}.

We mitigate this problem by proposing a new brushing technique called \textit{Distortion-aware brushing} (\autoref{fig:intro}c). 
This technique resolves distortion by dynamically relocating points: as users perform brushing, the system automatically pulls points that are close in the high-dimensional space toward the brushed 2D region and pushes those that are distant away. By doing so, Distortion-aware brushing helps users accurately brush a set of points that form a cluster in high-dimensional space, providing benefits in both accuracy and user confidence.

\section{Prior Publications and Authorship}

I (Hyeon Jeon) am a primary author of this dissertation, but much work has been executed with my advisor, Jinwook Seo, and many other collaborators (see \autoref{tab:priorpubs} for the list of prior publications). At first, the workflow model that explains visual analytics with DR is published as part of the CHI 2025 paper \cite{jeon25chi}, which I co-authored with Hyunwook Lee and Sungahn Ko from UNIST, Yun-Hsin Kuo and Kwan-Liu Ma from UC Davis, Taehyun Yang from SNU, Daniel Archambault from Newcastle University, and Takanori Fujiwara from Link\"oping University. 
Our investigation of DR use within visual analytics practice was conducted in collaboration with Jeongin Park from SNU and Sungbok Shin from INRIA \cite{jeon25arxiv}, while our examination of DR use outside the visualization domain \cite{cashman25tvcg} was carried out jointly with Dylan Cashman from Brandeis University, Mark Keller from Harvard Medical School, Bum Chul Kwon from IBM Research, and Qianwen Wang from the University of Minnesota Twin Cities and published in TVCG in 2025.
\textit{Classes are not clusters} project is published in VIS 2023 and TVCG \cite{jeon24tvcg} with the help of Yun-Hsin Kuo, Kwan-Liu Ma, and Micha\"el Aupetit from QCRI, Hamad Bin Khalifa University. 
Adjusted internal clustering metrics used in this paper are introduced in our TPAMI paper \cite{jeon25tpami}, where I work with Micha\"el Aupetit, along with Seokhyeon Park and Aeri Cho from SNU and DongHwa Shin from Kwangwoon University.
\textit{Dataset-adaptive workflow}, which is presented in VIS 2025 and also published in TVCG \cite{jeon25tvcg3}, is designed jointly with Jeongin Park and Soohyun Lee at SNU, Dae Hyun Kim from Yonsei University, and Sungbok Shin. 
Finally, I appreciate Micha\"el Aupetit, Soohyun Lee, Youngtaek Kim from Samsung Electronics, Kwon Ko from Stanford University, and Ghulam Jilani Quadri from the University of Oklahoma for helping me design \textit{Distortion-aware brushing} \cite{jeon25tvcg2} and publish the results at TVCG in 2025. 



\begin{table}[t]
    \centering
    \scalebox{0.97}{
    \begin{tabular}{ll}
    \toprule
       \textbf{Contents and Associated Publications}  &   \textbf{Chapter}\\
       \midrule
       Workflow model for visual analytics with dimensionality reduction  & 
       \multirow{3.8}{*}{\autoref{sec:vadr}} \\ 
       \textit{\makecell[l]{\footnotesize \textbf{H. Jeon}, H. Lee, Y.-H. Kuo, T. Yang, D. Archambault, S. Ko, T. Fujiwara, K.-L. Ma, and J. Seo \\ \footnotesize Unveiling high-dimensional backstage: a survey for reliable visual analytics with dimensionality reduction \\ \footnotesize 2025 CHI Conference on Human Factors in Computing Systems (CHI 2025)}}
       \\ 
              \midrule 
       Identifying reliability challenges & \multirow{6.8}{*}{\autoref{sec:investigation}} \\ 
       \textit{\makecell[l]{\footnotesize \textbf{H. Jeon}, J. Park, S. Shin, and J. Seo \\ \footnotesize Stop misusing t-SNE and UMAP for visual analytics \\ \footnotesize under review}} \\  
       \textit{\makecell[l]{\footnotesize D. Cashman, M. Keller, \textbf{H. Jeon}, B.C. Kwon, and Q. Wang \\ \footnotesize A critical analysis of the usage of dimensionality reduction in four domains \\ \footnotesize IEEE Transactions on Visualization and Computer Graphics (TVCG)}} \\

       \midrule
       Addressing the misuse of famous dimensionality reduction techniques & \multirow{6.8}{*}{\autoref{sec:clcl}} \\ 
       \textit{\makecell[l]{\footnotesize \textbf{H. Jeon}, Y.-H. Kuo, M. Aupetit, K.-L. Ma, and J. Seo \\ \footnotesize Classes are not clusters: Improving label-based evaluation of dimensionality reduction \\ \footnotesize IEEE Transactions on Visualization and Computer Graphics (TVCG, Proc. VIS 2023)}} \\ 
       \textit{\makecell[l]{\footnotesize \textbf{H. Jeon}, M. Aupetit, D. Shin, A. Cho, S. Park, and J. Seo \\ \footnotesize Measuring the validity of clustering validation datasets \\ \footnotesize IEEE Transactions on Pattern Analysis and Machine Intelligence (TPAMI)}} \\
       \midrule
       Preventing the prevalent cherry-picking of hyperparameters & \multirow{3.8}{*}{\autoref{sec:dawadr}} \\
       \textit{\makecell[l]{\footnotesize \textbf{H. Jeon}, J. Park, S. Lee, D.H. Kim, S. Shin, and J. Seo \\ \footnotesize Dataset-adaptive dimensionality reduction \\ \footnotesize IEEE Transactions on Visualization and Computer Graphics (TVCG, Proc. VIS 2025)}} \\ 
       \midrule 
       Making interactions less erroneous & \multirow{3.8}{*}{\autoref{sec:dabrca}} \\ 
       \textit{\makecell[l]{\footnotesize \textbf{H. Jeon}, M. Aupetit, S. Lee, K. Ko, Y. Kim, G.J. Quadri, and J. Seo \\ \footnotesize Distortion-aware brushing for reliable cluster analysis in multidimensional projections \\ \footnotesize IEEE Transactions on Visualization and Computer Graphics (TVCG)}} \\

       \bottomrule
    \end{tabular}
    }
    \caption{Associated prior publications to the contents of this dissertation. Although I am the primary author of this thesis, it would not have been possible to complete it without the help of my collaborators and advisors. This thesis also contains texts from my other prior publications in IEEE VIS and TVCG \cite{jeon23vis, jeon21tvcg, jeon24tvcg, jeon25eurovis, jeon22vis, jeon25tvcg}. }
    \label{tab:priorpubs}
\end{table}

\chapter{Background: Dimensionality Reduction}

We provide brief introductions on dimensionality reduction (DR). 
We first define DR and discuss representative techniques.
We then detail the major characteristics of DR that are highly relevant to this dissertation: inherent distortions in DR projections and the variability introduced by hyperparameter selection.


\section{Overview}

DR techniques denote functions that produce a low-dimensional projection preserving the structure of the input high-dimensional dataset \cite{nonato19tvcg, lee07springer}.
Formally, a DR technique gets $\mathbf{X} = \{x_i \in \mathbb{R}^D, i = 1, 2, \cdots, N \}$ as input and produces $\mathbf{Y} = \{y_i \in \mathbb{R}^d, i = 1, 2, \cdots, N \}$, where $d<D$ and $d$ is typically two or three when used for visual analytics. $Y$ is optimized to minimize the structural discrepancies between $X$ and $Y$, where it is conventionally visualized by scatterplots.

As the degree of freedom is substantially larger in the high-dimensional space compared to the low-dimensional space, the low-dimensional representations cannot preserve entire structural characteristics of the input data \cite{jeon21tvcg, nonato19tvcg, venna06nn}. This constraint leads to the design of diverse DR techniques that focus on different structural characteristics.  
For example, local techniques like $t$-SNE, UMAP, or LLE \cite{roweis00science} focus on minimizing discrepancies between high- and low-dimensional data in local neighborhood structure. This is typically done by constructing $k$-nearest neighbors graph in the high-dimensional space and focusing only on point pairs that are connected by graph edges \cite{mcinnes2020arxiv}.
The projections produced by these techniques are thus suitable for local techniques like neighborhood identification. 
Meanwhile, global techniques like PCA or MDS \cite{ramsay77pm} aim to directly preserve the pairwise distances, making it appropriate to investigate global structure like distances between points or cluster density.
Recently, hybrid techniques that aim to preserve both global and local techniques like UMATO \cite{jeon25tvcg, jeon22vis} or At-SNE \cite{fu19kdd} have also emerged.

\section{Distortions in DR Projections}

\label{sec:distortionsproj}

We discuss how distortions of DR projections are defined in the literature, then describe metrics for evaluating the amount of distortions.

\begin{figure}
    \centering
    \includegraphics[width=0.7\linewidth]{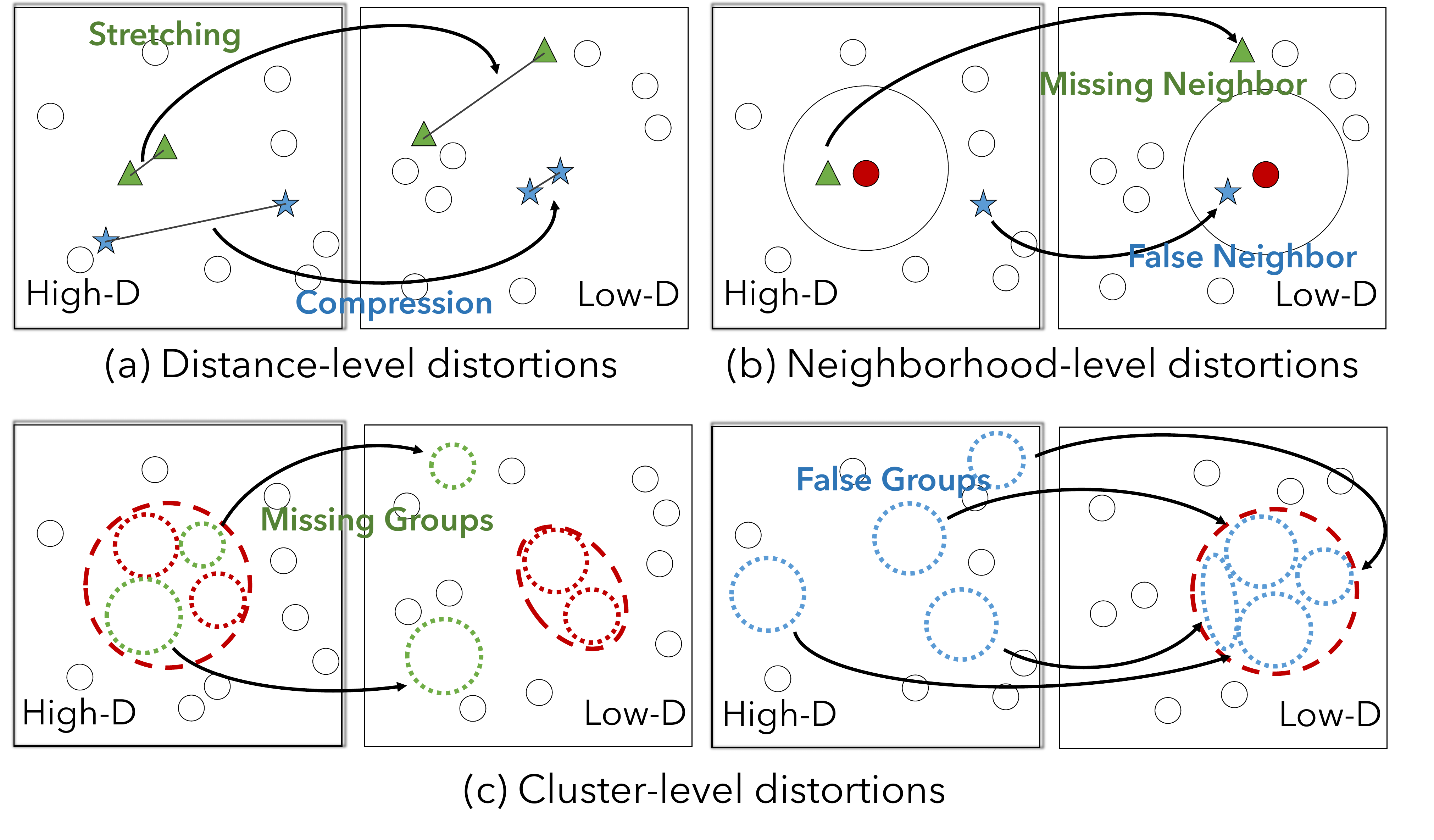}
    \caption{Three different perspectives in defining distortions in DR projections (\autoref{sec:distortionsproj}). Each type of distortion is measured with different sets of DR evaluation metrics.}
    \label{fig:distortions}
\end{figure}

\paragraph{Definition of distortions}
DR projections inherently suffer from distortions \cite{lespinats07tnn, nonato19tvcg, aupetit07neurocomputing, bae25arxiv}. Even for the target structure that the techniques aim to preserve (e.g., the neighborhood structure of UMAP), completely escaping from distortion is not possible. 
These fundamental limits prompt researchers to invest effort in formally defining distortions.
In his seminar work, Aupetit \cite{aupetit07neurocomputing} defines \textit{stretching} and \textit{compression} (\autoref{fig:distortions}a). Stretching describes the situation in which the pairwise distances in the projected space are increased compared to those of the original space; conversely, compression indicates the case in which the pairwise distances decrease.
\textit{Missing Neighbors} and \textit{False Neighbors} \cite{lespinats07tnn, lespinats11cgf, venna10jmlr, lee07springer} are then introduced as an interpretation of stretching and compression in terms of the neighborhood structure (\autoref{fig:distortions}b). 
Given a high-dimensional point $x$ and its corresponding low-dimensional point $y$, Missing Neighbors are defined as the $k$-nearest neighbors of $x$ that are not among the ones of $y$. Conversely, False Neighbors are defined as the $k$-nearest neighbors of $y$ that are not among the ones of $x$. 
However, Missing and False Neighbors are insufficient to explain the distortions of complex, intertwined cluster structures. 
For example,  the relative increase of cluster density in the projection does not incur Missing and False Neighbors distortions, because it does not alter the $k$-nearest neighbor structure for small $k$ values.
As alternatives, \textit{cluster-level} distortions, named \textit{Missing Groups} and \textit{False Groups} (\autoref{fig:distortions}c), are proposed by Jeon et al. \cite{jeon21tvcg}. Missing Groups occur when a cluster in the original space splits into multiple separated clusters in the embedding, and False Groups occur when a cluster in the embedding consists of multiple separated clusters in the original space.

\paragraph{Evaluation metrics}
Aligned with the diversity in how we define distortions, metrics that focus on different levels of structural granularity have been proposed. 
These metrics are divided into \textit{global}, \textit{local}, and \textit{cluster-level} metrics. 
\textit{Global metrics}, such as Kullback-Liebler divergence (KL Divergence) \cite{hinton02nips} and Distance to Measure (DTM) \cite{chazal11fcm, chazal17jmlr}, quantify how well the projections preserve the global structure of the original data against stretching and compression.
Meanwhile, \textit{local measures} focus on neighborhood preservation.
Trustworthiness and Continuity (T\&C) \cite{venna06nn} measure how Missing and False Neighbors affected the distance-based ranking of the nearest neighbor for every data point in both spaces. Mean Relative Rank Errors (MRREs) \cite{lee07springer} extends T\&C by additionally regarding the ranking of True Neighbors: the points that are neighbors in both the original and projection spaces. 
Finally, \textit{cluster-level metrics} capture distortions on cluster structures. Steadiness and Cohesiveness (S\&C) \cite{jeon21tvcg}  assess how much Missing and False Groups  distortions have occurred by  (1) extracting clusters from one space and (2) evaluating their dispersion in the other space. Clustering metrics like Silhouette \cite{rousseuw87silhouette} or Distance consistency \cite{sips09cgf} are also widely used as cluster-level metrics. Here, the evaluation is done by preparing ground truth high-dimensional clusters and investigating how well these clusters are well separated in the low-dimensional projection.

\section{Hyperparameters of DR Techniques}

\label{sec:relhyperparameter}

Hyperparameters denote user-specified constant values that control the behavior of DR techniques but are not learned during execution.
While various hyperparameters exist, those with the most substantial impact on the visual patterns of output projections are the ones that control the locality, i.e., the degree to which a DR technique defines the neighborhood of each point---whether narrowly or broadly.
Representative hyperparameters include \perplexity in $t$-SNE and \nneighbors in techniques like UMAP, LLE \cite{roweis00science}, and Isomap \cite{tenenbaum00science}.
These hyperparameters directly control the number of nearest neighbors for each data point that should be preserved as neighbors in the low-dimensional space \cite{wattenberg2016tsnetuning, coenen19fiar, mcinnes2020arxiv}. Therefore, small \nneighbors or \perplexity makes DR techniques
more focus on local neighborhoods, and large values make them emphasize global structures. 
Meanwhile, hyperparameters like learning rate (or step size) in gradient-based techniques like $t$-SNE and UMAP control the trade-off between efficiency and optimality; higher learning rates can lead to faster convergence but may cause instability or result in suboptimal projections \cite{kobak19nc, belkina19nc}. 

\begin{figure*}
    \centering
    \includegraphics[width=\linewidth]{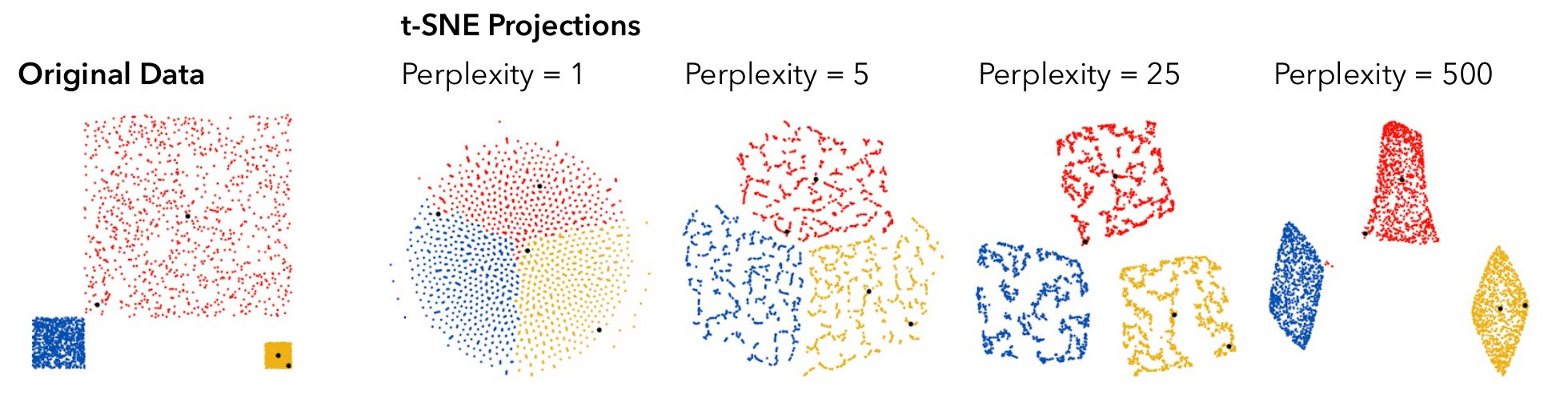}
    \caption{$t$-SNE projections (right) of a 2D dataset (left) with different perplexity values. The resulting projections fail to faithfully depict the structure of the original data and also show varying patterns by hyperparameter.}
    \label{fig:hp}
\end{figure*}

As hyperparameters substantially impact the visual patterns and faithfulness of DR projections (\autoref{fig:hp}) \cite{jeon21tvcg, wattenberg2016tsnetuning, coenen19fiar}, setting appropriate hyperparameters that produce projections with faithful representation of the structure of interest is crucial for reliable visual analytics. A systematic way to do so is to optimize hyperparameters while using evaluation metric scores as the target function \cite{jeon25tvcg3} using optimization techniques like Bayesian optimization \cite{snoek12nips}.


\section*{\textit{Our contribution}}
The fact that DR projections cannot preserve the entire structural characteristics of the high-dimensional dataset can cause significant reliability concerns in visual analytics. For example, using highly distorted projections due to improper hyperparameter choice may lead to erroneous conclusions about the underlying data. 
To address this problem, this dissertation thoroughly investigates reliability challenges in DR-based visual analytics, then provides technical solutions for these challenges. 


\chapter{Visual Analytics with Dimensionality Reduction}

\label{sec:vadr}


We want to detail how this dissertation impacts visual analytics. 
To this end, we introduce a workflow model that provides overview of how DR is used for visual analytics. 
This workflow model is established based on an extensive survey of 133 papers that improve the reliability of DR-based visual analytics, where the detailed procedure is provided in our reference paper \cite{jeon25chi}.
In the following sections, we first discuss the main components and stages in our workflow model (\autoref{sec:vaworkflow}). We then detail how our contributions improve visual analytics with DR using the model (\autoref{sec:vaimpact}).

\section{Workflow Model}

\label{sec:vaworkflow}

\def\Preprocessing{\textsc{Preprocessing}\xspace}
\def\DR{\textsc{Dr}\xspace}
\def\DimRed{\textsc{Dimensionality Reduction}\xspace}
\def\Evaluation{\textsc{Evaluation}\xspace}
\def\Visualization{\textsc{Visualization}\xspace}
\def\Sensemaking{\textsc{Sensemaking}\xspace}
\def\Interaction{\textsc{Interaction}\xspace}

\begin{figure*}[t]
    \centering
    \includegraphics[width=\linewidth]{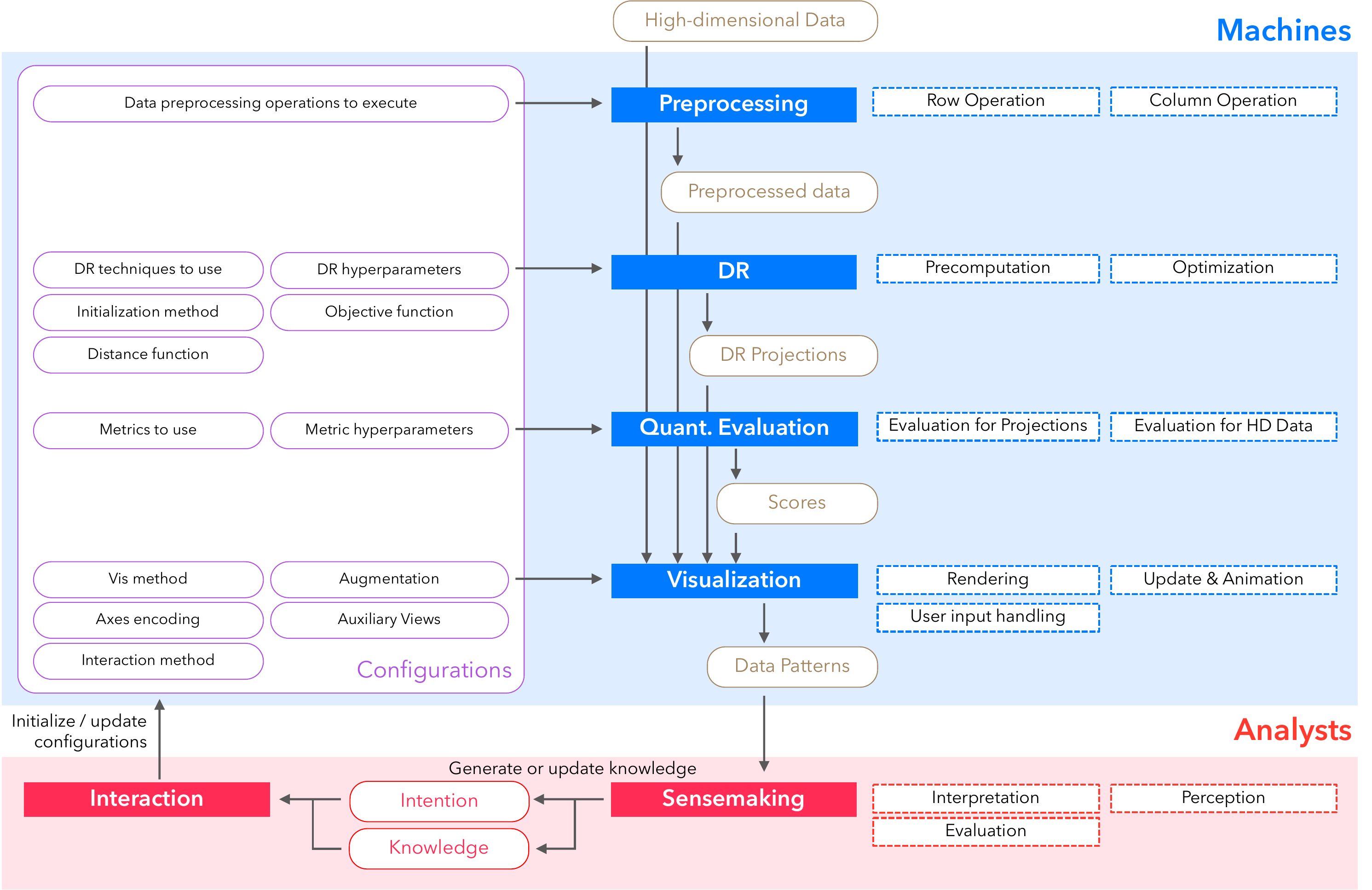}
    \caption{The illustration of the workflow model. The model explains how an analyst and a machine interact while conducting visual analytics using DR. 
    Each stage of visual analytics executed by analysts and machines is represented by \red{red} and \blue{blue} rectangles, respectively, and the input and output of each stage are designated by arrows. }
    \label{fig:workflow}
\end{figure*}

The workflow model (\autoref{fig:workflow}) explains the interaction between two main actors (\textit{Analysts} and \textit{Machines}) in six workflow stages (\Preprocessing, \DR or \DimRed, \Evaluation, \Visualization, \Sensemaking, and \Interaction).
The first four stages are carried out by machines, and the last two stages are executed by analysts. 
Inspired by van Wijk's visualization model \cite{wijk06tvcg}, we describe visual analytics using DR as an iterative, looped process where analysts continually adjust configurations to update visualizations based on newly acquired knowledge.
The model also reflects a common workflow of using machine learning techniques through iterative monitoring and refinement of specifications \cite{kapoor10chi, endert17cgf}.
We further suggest that high-dimensional data cascades from previous stages to subsequent stages executed by machines, with each stage enriching the data, for example, by generating DR projections (\DR) or evaluating their faithfulness (\Evaluation). 


\subsection{Analysts and Machines}

Visual analytics cannot be established without both data analysts and machines.
Analysts initialize and update configurations that capture both their intentions (e.g., task, requirements, hypothesis) and knowledge, which is subsequently fed into machines to guide their behavior.

\paragraph{Analysts}
As demonstrated in the knowledge generation model proposed by Sacha et al. \cite{sacha14tvcg}, analysts' main goal in our workflow is to perform \Sensemaking, thereby generating knowledge that is useful and also reliably reflects the original data. Detailed descriptions of analysts' actions are as follows.

\textit{At first, analysts initialize configurations.}
This is done by interacting with machines, e.g., by manipulating visualizations or writing a program (\Interaction stage). 
The setup is done based on analysts' knowledge (e.g., domain knowledge about the target data) and intention.
For example, analysts may configure machines to execute DR techniques that focus on local structure (e.g., $t$-SNE or UMAP) to perform local neighborhood investigation.
Operational knowledge, encompassing the analysts' expertise in configurations, serves as a ``mental toolbox'' to support this process. Without prior knowledge of the configurations that fit their needs, analysts will have difficulty establishing appropriate configurations. 
For example, analysts may hardly use $t$-SNE or UMAP for tasks related to the local structure if they do not know that such techniques fit the task well. 

\textit{Analysts then generate or update knowledge based on output visualizations.}
After analysts deliver the configurations, the machines process the input high-dimensional data and visually convey data patterns to the analysts. Then, analysts perform \Sensemaking, e.g., perceiving and interpreting data patterns, thereby updating knowledge and intention. 
Please refer to Sacha et al. \cite{sacha14tvcg} or Kim et al. \cite{kim21tvcg, kim19chi} for a general explanation of the sensemaking procedure in visual analytics.

\textit{Analysts again interact with machines to update configurations based on the updated knowledge and intention}. This happens when analysts think the current configurations and resulting data patterns can hardly validate new hypotheses. For example, analysts may newly want to explore the detailed attribute values of clusters they discovered in the current DR projection. In such a case, the analysts need to reconfigure the \Visualization, e.g., by adding auxiliary views like parallel coordinate plots that can better depict the detailed attribute values of the clusters.

\paragraph{Machines}
In our workflow model, machines process the input high-dimensional data based on the given configurations. 
It comprises four stages: \Preprocessing, \DimRed, \Evaluation, and \Visualization.
At each stage, the machine enriches the data. The enriched data are summarized as data patterns and delivered by visualizations. 
We detail the labors of machines while we detail each stage in \autoref{sec:stages}.

\subsection{Workflow Stages}

\label{sec:stages}

We outline six workflow stages (text boxes in \autoref{fig:workflow})of the workflow model. The stages' composition draws inspiration from the KDD process model \cite{fayyad96aaai}, which describes how raw data passes through preprocessing, transformation, and data mining to yield final visualizations. Here, we substitute the transformation and data mining steps with \DR and \Evaluation stages to better match the visual analytics workflow using DR. 
We thus describe the stages in the sequence that are aligned with the KDD process. 
It is still important to note that the nature of interactive visual analytics causes the execution of these stages to be nonlinear: analysts can interact with machines to update configurations, which can cycle the analysis back to the machine-side stages.

\paragraph{\Preprocessing}
The preprocessing stage involves a variety of data manipulations that are executed before applying the DR techniques. The step receives raw \textit{high-dimensional data} as input and outputs the \textit{preprocessed data} based on the configurations specifying preprocessing operations to execute.
It can be largely divided into row operations and column operations. Data imputation, subsampling, or outlier removal are representative row operations. Regarding column operations, attribute selection that samples out or less weighs non-significant attributes is widely adopted. Attribute selection can also be conducted automatically, e.g., to reveal hidden patterns \cite{fujiwara23pacificvis} or maximize specific patterns like class separation \cite{wang18tvcg, sips09cgf}. Another typical column operation is to introduce new columns that represent the original structure's feature, e.g., by executing clustering techniques and adding the resulting clusters as labels to the datasets \cite{wenskovitch20iui, wenskovitch18tvcg}.

\paragraph{\DimRed or \DR}
This stage gets the \textit{preprocessed data} as input and executes DR algorithms, producing \textit{DR projections} as outputs. The stage is controlled by five configurations: DR techniques, hyperparameters, initialization method, objective function (i.e., loss function), and distance function. The first two configurations should be necessarily set to execute the stage. The last three configurations are widely set for nonlinear DR techniques (e.g., UMAP, $t$-SNE, and Isomap \cite{tenenbaum00science}), which work by optimizing the 2D positions of data points to preserve the original distances between them.
The initialization method defines how the points will be positioned before starting optimization, and the objective function defines how the preservation of distances will be quantified. 
These configurations substantially impact the resulting DR projections. 
For example, Kobak et al. \cite{kobak21nb} show that initializing points using PCA before the optimization leads to more accurate DR results than using random initialization. Lee and Verleysen \cite{lee11pcs} show that the design of the distance function affects the projection accuracy.

\paragraph{\textsc{Quantitative} \Evaluation}
This stage gets \textit{high-dimensional data}, \textit{preprocessed data}, and \textit{DR projections} as inputs and produces evaluation results as numerical \textit{scores}. These scores (1) explain how well the projections support the analytic tasks of investigating high-dimensional data or (2) the validation of hypotheses.
For example, when analysts want to conduct cluster analysis, it is important to evaluate the projections in advance using the evaluation metrics assessing the preservation of cluster structure \cite{sips09cgf, jeon21tvcg}.
This stage is affected by two configurations: evaluation metrics and metric hyperparameters. For example, Trustworthiness \& Continuity \cite{venna06nn}, which examine how well DR projections preserve the local neighborhood structure of original data, require a hyperparameter that designates the number of nearest neighbors to be considered.
Note that these DR evaluation metrics differ from ``perceptual metrics'' (e.g., Scagnostics \cite{dang14pvis} or the class separation measure \cite{aupetit15pvis}), which focus on the perceived patterns of 2D projections without accounting for the original high-dimensional (HD) data.

\paragraph{\Visualization}
Machines get all data cascaded from the previous stages (raw \textit{high-dimensional data}, \textit{preprocessed data}, \textit{projections}, and \textit{scores}) as input, generating a visualization or a set of visualizations delivering \textit{data patterns}. The main configuration that affects the patterns is the visualization methods for depicting DR projection. This incorporates the selection of not only the type and number of visual idioms (e.g., 2D or 3D scatterplots \cite{sedlmair13tvcg}) but also displays (e.g., 2D screen or VR headsets \cite{in24chi}) or the use of animations \cite{asimov85siam}. Configuring how monochrome scatterplots can be augmented also substantially affects the resulting data patterns. Color-encoding the classes, for example, can cause analysts to overlook the separation between intrinsic clusters within the data \cite{sedlmair12cgf}. Also, how and how much projections suffer from distortions can be augmented by encoding the scores from the evaluation on the projections \cite{aupetit07neurocomputing, lespinats07tnn, jeon21tvcg}, making analysts more aware of distortions. Finally, auxiliary visualizations can be configured to make the original high-dimensional data more interpretable \cite{chatzimparmpas20tvcg, yan24pvis, kwon17tvcg}.

\paragraph{\Sensemaking}
Analysts make sense of data by investigating \textit{data patterns} delivered by the \Visualization. The procedure starts by perceiving the patterns, e.g., cluster \cite{jeon24tvcg2, abbas19cgf} or local neighborhood \cite{lespinats11cgf, aupetit05neurocomputing} patterns, aligned with the target task. The perception usually relies on the Gestalt principle of proximity and similarity \cite{palmer99gestalt}, where analysts perceive data points that have high proximity or similar color to be closely located in the original high-dimensional space.

\paragraph{\Interaction}
Analysts can also interact with data, meaning they ``signal'' machines to update configurations, thereby updating data patterns to align with their intentions \cite{sacha17tvcg}. 
To do so, analysts first evaluate whether the data patterns are suitable for validating hypotheses. If not suitable, they can interact with machines to correct the visualizations. For instance, when analysts notice that the evaluation scores of DR projections indicate insufficient faithfulness to properly support the target task, they can reconfigure the DR techniques or adjust hyperparameters to correct the projections. The machines then again process the data based on the updated configuration, which leads to a new cycle of sensemaking. 

\section{Impact of Our Contributions}

\label{sec:vaimpact}

We discuss how our main contributions impact visual analytics using DR by referring to the workflow model.


\paragraph{We avoid human bias in \Sensemaking by improving \Evaluation stage}
We find that analysts are biased to favor projections with high class or cluster separability, which often leads them to misuse techniques such as $t$-SNE and UMAP that exaggerate these properties.
As a result, analysts may adopt visually appealing projections that do not faithfully represent the underlying data, leading to erroneous \Sensemaking.
However, existing evaluation metrics that emphasize cluster structure cannot serve as reliable gatekeepers against this bias, as they implicitly assume that class labels form ground-truth clusters.
To address this issue, we introduce Label-T\&C, a new family of evaluation metrics that avoids this pitfall and instead rewards projections that better support reliable \Sensemaking in visual analytics.

\paragraph{We accelerate the feedback loop between \DR and \Evaluation stages}
Optimizing DR hyperparameters can be viewed as a feedback loop between these two stages. When \DR stage produces a projection, \Evaluation stage outputs its faithfulness scores, then \DR is re-run using updated configuration based on this information. Although this process is crucial in obtaining projections that faithfully represent the original data, it requires extensive trial-and-error. We address this gap by proposing a \textit{dataset-adaptive workflow} that accelerates the optimization process based on the dataset properties, making it easier for practitioners to obtain faithful projections. 

\paragraph{We support practitioners to execute \Interaction with less error}
Even if analysts appropriately configure machines and thus obtain projections well-suited to their analysis, the projections cannot completely avoid distortions. 
This problem can make the way analysts interact with projections erroneous, resulting in insights that thus not well reflect the underlying data. 
To mitigate this reliability concern, we propose \textit{Distortion-aware brushing}, a new brushing technique that resolves distortions while users interact with DR projections.

\chapter{Identifying Reliability Challenges}

\label{sec:investigation}

We aim to identify the prevalent challenges that may degrade the reliability of visual analytics using DR. 
To this end, we investigate how practitioners in the visual analytics field and domains outside of visualizations use DR for their analysis by conducting a literature review and an interview study. 
As a result, we reveal three challenges: (1) misuse of famous DR techniques, (2) prevalent cherry-picking of hyperparameters, and (3) erroneous interactions. 



\section{Introduction}

Recent years have witnessed the ubiquitous application of DR techniques across various domains \cite{cashman25tvcg}.
For example, now DR is one of the most widely used components in modern visual analytics systems \cite{li23tvcg, kwon18tvcg, shilpika22tvcg, ye25tvcg, he24tvcg}. DR is also commonly used to deliver data insights in diverse domains, including bioinformatics \cite{cheng23tvcg, lause24plos, kobak19nc}, business \cite{feuillet21esmq}, machine learning \cite{bau21ijcai}, and HCI \cite{hamalainen23chi, ha24chi, lim23chi}.

On the other hand, DR is also a tool that can be easily misinterpreted and misused, leading to unreliable visual analytics.
DR cannot preserve all structural characteristics of high-dimensional data and also intrinsically suffers from distortions (\autoref{sec:introback}).
Without a clear understanding of how to properly apply DR techniques, analysts may easily draw incorrect insights and conclusions from their data.
However, it is unclear whether practitioners are aware of these weaknesses and use DR appropriately.
Indeed, prior reports across various domains suggest that practitioners frequently misuse DR, e.g., by misinterpreting visual representations of DR projections \cite{wattenberg2016tsnetuning} or selecting techniques that do not align with target tasks \cite{lause24plos, chari23plos}. 
Yet, we still lack a systematic investigation that identifies the specific challenges that undermine the reliability and explains their underlying causes. 
This gap makes it difficult to design effective approaches for improving the reliability of visual analytics with DR.

In this chapter, we identify three challenges that harm the reliability of DR-based visual analytics and their underlying causes.
To do so, we conduct a literature review that investigates the use of DR in (1) the visual analytics domain, and (2) four domains outside of visualizations, which are biology, chemistry, business, and physics. 
Then, we execute an interview study with researchers in visual analytics and in other domains like bioinformatics and machine learning to complement the literature review.

Our investigation identifies three prevailing challenges and their causes. First, practitioners frequently \textbf{misuse two famous techniques}---$t$-SNE and UAMP---even for analytical tasks for which they are not suitable.
Second, practitioners frequently \textbf{cherry-pick hyperparameters} rather than systematically optimize them, leading to projections that offer little assurance of faithfully representing the original data.
Finally, practitioners often \textbf{interact with DR projections in erroneous ways}, for example, by brushing or highlighting apparent 2D clusters that may not correspond to true clusters in the high-dimensional space.
We discuss our technical solutions to address these three challenges in the subsequent chapters (\autoref{sec:clcl}, \ref{sec:dawadr}, and \ref{sec:dabrca}).







\section{Related Work}

We review related studies that investigate the usage of DR within visualization, our four selected domains (biology, chemistry, physics, and business), and science at large.

\subsection{Investigation from Visualization Researchers}

Visualizing and interacting with DR techniques have become important topics in the visualization community, sparking visualization researchers to conduct various surveys. We categorize previous literature into the target of their analysis.

\paragraph{Literature review on DR techniques}
The most common type of DR-related literature review is, not surprisingly, the ones about DR techniques. These surveys aim to clarify the advantages and disadvantages of DR techniques, thereby supporting practitioners in selecting proper DR techniques in their analysis \cite{engel2012survey, nonato19tvcg, espadoto21tvcg, etemadpour15tvcg}. 
For example, Espadoto et al. \cite{espadoto21tvcg} survey 44 DR techniques and quantitatively examine their performance using five quality metrics. Nonato and Aupetit \cite{nonato19tvcg} also compared 28 different techniques, providing guidelines to select DR techniques by the analytic task. Etemadpour et al. \cite{etemadpour15tvcg}, Xia et al. \cite{jiazhi21tvcg}, and Sedlmair et al. \cite{sedlmair13tvcg} also provide similar guidelines, where empirical user studies ground their guides. 

\paragraph{Literature review on tasks}
This family of surveys taxonomizes analytic tasks in DR projections. They aim to gain a further understanding of how practitioners use and interact with DR projections. 
For example, Sacha et al. surveyed visualization papers that included interaction techniques with dimensionality reduction algorithms, revealing the procedure in which analysts interact with DR~\cite{sacha17tvcg}.  
Nonato and Aupetit \cite{nonato19tvcg} systematically survey and taxonomize the type of analytic tasks. It is not a literature survey, but Brehmer et al. \cite{brehmer14beliv} also report on the task sequence using DR projections for HD data analysis, based on interviews with analysts. 

\noindent \textbf{Literature review on DR evaluation metrics}
Evaluation metrics for DR assess the extent to which DR projections are distorted \cite{jeon23vis}. As different evaluation metrics focus on distinct structural characteristics (e.g., local neighborhood structure \cite{venna10jmlr} or cluster structure \cite{jeon21tvcg}), selecting appropriate DR metrics that align with target analytic tasks is important for reliable data analysis. Surveys regarding evaluation metrics, therefore, aim to guide analysts in choosing appropriate metrics. 
Thurn et al. \cite{thrun2023analyzing} and Bertini et al. \cite{bertini2011quality}, for example, organize DR quality metrics regarding which structural characteristics they focus on. Lee and Verleysen \cite{lee09neurocomp} share a similar goal but concentrate on neighborhood preservation-based methods.

\paragraphit{Our contribution}
While prior studies focus on organizing techniques practitioners can use for DR-based visual analytics, our work examines how people actually use these tools in research and analysis and the problems that arise in the process.
By uncovering these real-world practices and challenges, we provide insights into what additional technical solutions are needed.

\subsection{Investigation from Domain Researchers}

Within individual subject areas, surveys, meta-analyses, or guidelines, papers serve as implicit or explicit references for using various DR techniques in the analysis of that domain area's data.  
Within biology, there exist several works comparing the usage of different DR techniques on various types of biological data~\cite{becht19nature, ujas2023guide, clarke2021tutorial, kimball2018beginner, liechti2021updated}.  Similar works can be also found in physics and astronomy~\cite{traven2017galah}, epidemiology~\cite{sakaue2020dimensionality}, and chemistry~\cite{anders2018dissecting}.  These works provide a review of practices within the silo of a single field and present recommendations on the use of DR within a particular high-dimensional data analysis pipeline.  

\paragraphit{Our contribution}
In contrast to these previous works, we examine practices both within and across a wide range of domains, spanning visual analytics as well as subject areas that employ DR.
By doing so, we provide a more comprehensive overview of how practitioners use DR in practice and the challenges they encounter when applying it to visual analytics.



\section{Literature Review on the Visual Analytics Field}

\label{sec:litreviewva}

We want to understand the practice of using DR in the visual analytics fields. 
We thus execute a literature review on visual analytics papers leveraging DR. 
We focus on examining whether DR techniques are properly selected to align with the intended analytic tasks, as a mismatch between task and technique is the most frequently reported form of misuse in the literature \cite{lause24plos, chari23plos}.


\subsection{Protocol}

The review consists of four steps: paper search, categorization, task suitability review, and quantitative analysis.

\paragraph{Paper search}
We aim to find \textit{papers that propose a visual analytics approach, framework, or systems that incorporate the existing DR techniques}.
We query papers that satisfy two conditions: (1) the term \texttt{``visual analytics''} or \texttt{``visual analysis''} appears in the title or abstract, and (2) one or more of the terms between \texttt{``dimensionality reduction''}, \texttt{``dimension reduction''}, \texttt{``multidimensional projection''}, and \texttt{``multidimensional scaling''} are mentioned in the full text. 
We use IEEE Xplore, Wiley online library, and Sagepub to search papers published in major data visualization journals and conferences (e.g., IEEE VIS, TVCG, CG\&A, PacificVis, EuroVis, CGF, IVIS).
We filter out papers published before 2008, the year \tsne is announced.
Then, we inspect each paper and remove the papers that do not stay within our search scope from our list. The following are the example papers that we excluded:
\begin{itemize}
    \item Papers that propose novel DR technique or pipeline \cite{fujiwara20tvcg, kwon17tvcg, doraiswamy21tvcg}.
    \item Papers that propose novel interaction techniques or visual encodings that can be applied to any DR technique \cite{jeon25tvcg2, aupetit14vast, stahnke16tvcg, lespinats11cgf}.
    \item Papers that focus on other kinds of data (e.g., network data \cite{shi22tvcg, col18tvcg}) or visualization idioms (e.g., parallel coordinates \cite{tyagi23tvcg, junpeng17tvcg})
    \item Papers that study human perception on scatterplots \cite{jeon24tvcg2, quadri21tvcg}.
    \item Papers that provide survey or theory contributions \cite{wenskovitch18tvcg, nonato19tvcg}.
\end{itemize}
Two coders engaged in this process, where we exclude a paper only when both coders agree to exclude it to minimize true negatives. The eligibility of the papers is determined based on the title, abstract, introduction, and method sections of the paper.

\paragraph{Categorization}
We categorize the identified papers according to the following criteria:

\paragraphit{DR techniques}
We mark the paper to use a certain DR technique if the paper explicitly mentions that they use the technique to visualize high-dimensional data. That is, we do not consider the case in which DR is used for data preprocessing or compression \cite{jeon25chi, fujiwara20tvcg}. This is because we focus on identifying whether DR techniques, including \tsne and \umap, are used for unsuitable visual analytics tasks or not.

\paragraphit{Target analytic tasks}
We categorize each paper as performing a specific analytic task if it explicitly states that it targets to support the task. For example, some papers mention the intended analytic tasks and explain that they selected their DR techniques accordingly.
Meanwhile, several papers do not explicitly specify their target analytic tasks. In such cases, we first examine the user studies described in the paper and infer the target tasks from those studies. If no user studies involve analytic tasks related to DR projections, we then successively examine the case studies or use cases. For instance, if a case study includes analyses that ``lasso'' a cluster and conduct a detailed investigation of it, we classify the paper as executing a cluster identification task.

\paragraphit{Reasonings}
We examine the reasoning behind the selection of DR techniques by analyzing the system design section of each paper. A paper is marked as having a specific reasoning only if the reasoning is explicitly stated. If no explicit reasoning is provided, we label the paper as ``No reason''.

\vspace{4pt}
\noindent
We select the first two criteria to investigate the extent to which DR techniques are misused for tasks that are not suitable for them.
We select reasonings as an additional criterion to check whether they justify the use of \tsne, \umap, and other DR techniques in an inappropriate way.
We follow the thematic coding process for the categorization.
To begin with, two coders interdependently categorize the papers. 
The initial agreement between two coders measured by Cohen's $\kappa$ is 0.62. 
Then, they merge and revise their categorizations through three iterative discussions.

\paragraph{Task suitability review}
We identify the suitability of major DR techniques for analytic tasks.
This is done by examining research that verifies the weaknesses and strengths of different DR techniques \cite{jiazhi21tvcg, jeon22vis, jeon24tvcg, narayan21nature}. 
Here, we define that a DR technique is suitable for an analytic task if it preserves the structural characteristics corresponding to that task. This means that analytic tasks can be reliably conducted when suitable DR techniques are used.

Two coders conduct the investigation through the following procedure.
At first, the coders review the papers and record the task suitability of the DR techniques by quoting the texts from the papers. We only consider claims supported by evidence such as computational benchmarks, user studies, or case studies. The coders then unified their findings to derive a final conclusion regarding the task suitability of the DR techniques. If conflicts arise among papers, we prioritize claims supported by stronger evidence (e.g., we place greater weight on findings from user studies than those from case studies) and by a larger number of papers.
We depict the results in \autoref{sec:suitability}.

\paragraph{Quantitative analysis}
We quantitatively analyze the categorized papers. Detailed statistical results are presented in \autoref{sec:quantianal}.

\subsection{Paper Search and Categorization Results}

\label{sec:litreviewresults}

We retrieve 312 papers published after 2008. 
After screening and filtering, we retain 136 papers. 
The categories we identify from them are described below.

\subsubsection{DR Techniques}
We identify 18 DR techniques in total. 
Among them, we find four commonly used techniques (\tsne, \umap, PCA, and MDS) for which each is used more than 20 times. 
The other 14 techniques are used fewer than five times each, and we categorize them all under ``others.''

\begin{figure*}[t]
    \centering
    \includegraphics[width=0.7\linewidth]{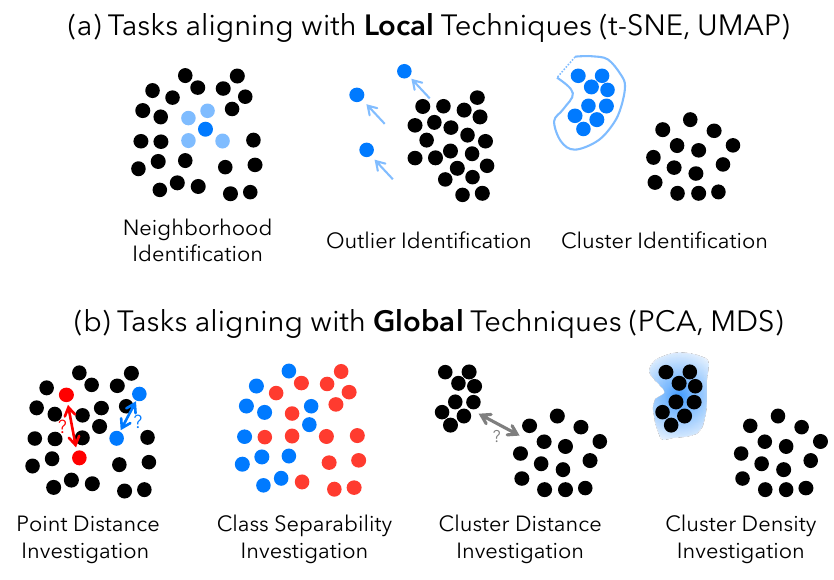}
    \caption{Illustrations of the analytic tasks using DR and their alignment to local and global DR techniques. Our literature review identifies seven types of analytic tasks using DR. }
    \label{fig:tasks}
\end{figure*}

\subsubsection{Analytic Tasks}
Our review reveals that analytic tasks using DR can be divided into seven categories (\autoref{fig:tasks}). 
The first three of them are \textit{identification} tasks: the tasks that find or extract certain data points or their set. 
The other four are \textit{investigation} tasks: the tasks that measure or quantify certain characteristics of the data.
Detailed descriptions of each task are as follows.


\paragraph{Neighborhood identification}
This task aims to find data points similar to a target point based on the proximity within the projection. 
Since neighborhood identification supports many other analytic tasks, such as cluster identification, preserving local neighborhood structure is often considered the most important criterion when evaluating DR projections \cite{nonato19tvcg, jeon22vis, moor20icml}.

\paragraph{Outlier identification}
This task is about identifying outliers within projections. 
Analysts often count the number of outliers in the data \cite{etemadpour15tvcg} or determine whether a point is a cluster member or outlier \cite{jiazhi21tvcg}. 
The task is typically leveraged for examining the quality of class labelings by identifying points that have high uncertainty about their class membership \cite{hong23pvis}.

\paragraph{Cluster identification}
This task involves identifying clusters within DR projections. 
Analysts typically count the number of clusters \cite{etemadpour15tvcg} or label clusters interactively using selection tools such as lasso or box-shaped brushes \cite{jiazhi21tvcg, choo10tvcg, meng24tvcg}. 
This task often includes investigating subclusters within existing clusters \cite{etemadpour15tvcg}. 
Visual analytics systems often provide auxiliary visualizations to show detailed information about the identified clusters \cite{chatzimparmpas20tvcg, li23tvcg}.

\paragraph{Point distance investigation}
Similar to the cluster distance investigation task, this task investigates the distance between data points as a proxy for their high-dimensional dissimilarity. 
It can be interpreted as a ``continuous'' version of the neighborhood identification task. 

\paragraph{Class separability investigation}
This task investigates how distinctly different classes are separated or distinct in the projections, where the classes are color-coded. 
It involves analyzing both the degree to which classes are mixed with each other and the relative distances among classes within the projection.
The task is commonly performed when DR techniques are employed to explain the behavior of a supervised machine learning model, particularly to illustrate how the model distinguishes between different classes \cite{solunke24tvcg, ploshchik23pvis}.

\paragraph{Cluster distance investigation}
This task uses the distance between well-separated clusters as a proxy for their similarity in the original high-dimensional space.
The clusters can be explicitly labeled (i.e., color-coded) or implicitly represented by data distribution \cite{sedlmair12report}. 

\paragraph{Cluster density investigation}
This task identifies and compares the density of clusters using cluster density as a proxy for the variability of data points within each cluster \cite{narayan21nature}.


\vspace{4pt}
\noindent
\textit{Task coverage validation.}
To validate the comprehensiveness of our categorization, we examine whether the analytic tasks in our list are included in prior task taxonomies within the visualization field. 
We review Etemadpour et al. \cite{etemadpour15tvcg}, Xia et al. \cite{jiazhi21tvcg}, Brehmer et al. \cite{brehmer14beliv}, Nonato and Aupetit \cite{nonato19tvcg}, Sedlmair et al. \cite{sedlmair12report}, and Cashman et al. \cite{cashman25tvcg}. 
We find that all tasks are covered by at least two previous studies (\autoref{tab:tasks}), confirming that our categorization covers all important tasks in DR.

\begin{table}[t]
    \centering

    \caption{\textit{The coverage of analytic tasks using DR (rows) by references (columns).} We represent the references using the last name of the first author. (Xia: Xia et al. \cite{jiazhi21tvcg}, Ete.: Etemadpour et al. \cite{etemadpour15tvcg}, Bre.: Brehmer et al. \cite{brehmer14beliv}, Non.: Nonato and Aupetit \cite{nonato19tvcg}, Sed.: Sedlmair et al. \cite{sedlmair12report}), Cas. : Cashman et al. \cite{cashman25tvcg}}
    \scalebox{1}{
    \begin{tabular}{r|cccccc}
    \toprule
    \textbf{Task} & \textbf{Xia} & \textbf{Ete.}  & \textbf{Bre.} & \textbf{Non.}  & \textbf{Sed.} & \textbf{Cas.}\\
    \midrule
         Neighborhood Identification &  & $\bullet$ & & $\bullet$ & & $\bullet$\\
         Outlier Identification &  $\bullet$  & $\bullet$ & & $\bullet$ & & $\bullet$\\
         Cluster Identification &  $\bullet$& $\bullet$ & $\bullet$ & $\bullet$ & $\bullet$ & $\bullet$\\
         Point Distance Investigation &  & & & $\bullet$ & & $\bullet$\\
         Class Separability Investigation & & & $\bullet$ & $\bullet$ & $\bullet$ & $\bullet$ \\
         Cluster Distance Investigation & $\bullet$ & $\bullet$ & $\bullet$ & $\bullet$ & $\bullet$ & $\bullet$\\
         Cluster Density Investigation&  $\bullet$ & $\bullet$& & $\bullet$\\
    \bottomrule
    \end{tabular}
    }
    \label{tab:tasks}
\end{table}

\subsubsection{Reasonings}

\begin{table*}[t]
    \centering
    
    \caption{The definition of the reasonings that we identify from the literature review. }
    \scalebox{0.61}{
    \begin{tabular}{ll}
    \toprule 
    \textbf{Reasoning} & \textbf{Definition} \\
    \midrule
     \textbf{Faithfulness} & The degree to which DR techniques accurately represent the original structure of the high-dimensional data without distortions. \\
        \textbf{Popularity} &  The degree to which DR techniques are widely known and used by practitioners in the visual analytics field. \\
        \textbf{Scalability} & Computational efficiency in executing DR techniques. \\
        \textbf{Interpretability} & The degree to which DR techniques yield visually distinct, analyzable clusters, enabling clear explanation of the data. \\
        \textbf{Stability} & The degree to which DR techniques produce projections that are stable against hyperparameter change or the stochastic nature of DR. \\ 
        \textbf{Extensibility} &  The degree to which DR techniques can be adapted or expanded to accommodate diverse data conditions or input formats. \\
        \textbf{Simplicity} & The degree to which practitioners can readily understand and apply DR techniques. \\
        \bottomrule 
    \end{tabular}
    }
    \label{tab:defreasonings}
\end{table*}

We identify seven large categories of reasoning used to justify the selection of DR techniques. 
We define each reasoning in \autoref{tab:defreasonings}.
It is worth noting that a substantial amount of papers (44\%; \autoref{fig:reasonings}) do not mention specific reasoning, i.e., marked as ``No reason''.

\paragraph{Faithfulness} 
Researchers justify the usage of DR techniques based on their faithfulness, or the capability to accurately represent the original structure of the high-dimensional data without distortions.
This reasoning mostly relies on references to benchmark studies that examine the faithfulness of DR techniques~\cite{espadoto21tvcg, jiazhi21tvcg}. 
One notable finding is that researchers often cite the original papers introducing DR techniques to support claims about their capability to preserve global structure, e.g., the original UMAP paper \cite{mcinnes2020arxiv}, which is not always correct \cite{coenen19fiar, jeon24tvcg2, jeon22vis, jiazhi21tvcg, narayan21nature}. We also find that several papers claim the faithfulness of UMAP without references. 

\paragraph{Popularity} 
Researchers justify the use of DR techniques based on their popularity, which indicates the degree to which the techniques are widely acquainted and used by practitioners in the visual analytics field.
For example, papers mention employing \tsne because it is a ``default option'' in visualizing high-dimensional data or is ``widely recognized'' by the research community. 
These papers also highlight specific research communities, such as biology, computer vision, and document clustering, where these techniques are commonly used \cite{cashman25tvcg}. 

\paragraph{Scalability}
The use of DR techniques is also justified by their
computational efficiency, which enhances the responsiveness of visual analytics systems. 
For example, papers state that efficient GPU implementations \cite{nolet22biorxiv, pezzotti20tvcg} facilitate the effectiveness of DR techniques.

\paragraph{Interpretability}
Researchers use DR techniques because they enable a clear explanation of the data with projections that contain visually distinct and, thus, analyzable clusters.
This finding aligns with the work by Morariu et al. \cite{morariu23tvcg} and Doh et al. \cite{doh25arxiv}, in that they also identify clear cluster separation as a key factor influencing the preference of DR projections.

\paragraph{Stability}
Researchers justify the selection of DR techniques by highlighting their stability (i.e., the degree to which DR techniques produce projections that are stable against hyperparameter change or the stochastic nature of DR) as a means to improve the reproducibility of their research. 
For example, one paper argues that \tsne is stable due to its non-convex optimization \cite{arora18clt}. 

\paragraph{Extensibility}
We find that researchers justify the use of DR techniques based on their extensibility or their ability to adapt or expand to accommodate diverse data conditions or input formats.
For example, some papers use DR techniques because they are parametric, i.e., support new data points to be dynamically projected after initial projection \cite{sainburg21nc}, particularly for analyzing streaming, online data. 

\paragraph{Simplicity}
A few papers mention that they select DR techniques that are simple and thus can be easily understood and applied by practitioners. 
Among the four major techniques, only PCA has been explicitly justified by this reasoning.

\subsection{Suitability of DR Techniques to Analytic Tasks}

\label{sec:suitability}

We assess the suitability of four major DR techniques (\tsne, \umap, PCA, and MDS) to the analytic tasks identified in \autoref{sec:litreviewresults}. 
This is done by revisiting previous studies that evaluate DR techniques and analyze the alignment between the DR techniques and analytic tasks.
This analysis helps determine whether researchers are applying DR to suitable tasks. 

Note that we adopt a coarse-grained approach that determines suitability primarily based on the selection of DR techniques for two reasons.
First, we identify that nearly all studies we investigated did not report their hyperparameter settings or other configurations, like seed selection. 
Second, the choice of DR technique is the most influential factor affecting the task suitability of DR projections \cite{espadoto21tvcg, nonato19tvcg}.

\subsubsection{Tasks Suitable for \tsne and \umap}
\tsne and \umap focus on preserving local neighborhoods by positioning neighboring points close together in the projection while separating non-neighboring points. They are thus commonly referred to as local techniques. Several studies demonstrate that they show state-of-the-art performance in preserving local structures, both empirically \cite{jeon22vis, espadoto21tvcg, moor20icml} and theoretically \cite{lee11pcs}. 
We categorize the following tasks that are suitable for \tsne and \umap.

\paragraph{Neighborhood identification task is more suitable for \tsne and \umap}
As aforementioned, \tsne and \umap directly aim to preserve local neighborhood structure. This makes them better suited for neighborhood identification tasks than alternative techniques by design \cite{venna10jmlr, amid22arxiv, jeon24tvcg, zhou22nc}, which also have been empirically validated \cite{jeon22vis, moor20icml}.

\paragraph{Outlier identification task is more suitable for \tsne and \umap}
Since projections generated by local techniques clearly distinguish neighboring and non-neighboring points, they can effectively separate outliers from clusters. 
Xia et al. \cite{jiazhi21tvcg} empirically validate that \tsne and \umap are the most effective DR techniques in supporting the outlier identification task, outperforming alternative techniques like PCA.

\paragraph{Cluster identification task is more suitable for \tsne and \umap}
As \tsne and \umap locate neighboring points close and non-neighboring points far away \cite{jeon22vis, moor20icml, venna10jmlr}, they clearly represent individual high-dimensional clusters as 2D clusters, thus suitable for the cluster identification task. 
The recent user study by Xia et al. \cite{jiazhi21tvcg} shows that the participants perform best when identifying clusters with \tsne and \umap projections.

\subsubsection{Tasks Suitable for PCA and MDS}
PCA and MDS are DR techniques that are more effective in preserving global pairwise distances between data points compared to local techniques like \tsne and \umap \cite{nonato19tvcg, jeon22vis, jiazhi21tvcg, sorzano14arxiv, van09jmlr}. They are usually referred to as global techniques. The following tasks are suitable for these DR techniques.

\paragraph{Point distance investigation task is more suitable for PCA and MDS}
These techniques are designed to directly preserve the pairwise distance structures more effectively compared to local techniques. They are thus more suitable than \tsne and \umap in investigating distances between data points. 
Several studies \cite{jeon22vis, moor20icml, amid22arxiv} propose techniques that improve UMAP in preserving global pairwise distances between points, such as UMATO \cite{jeon22vis} and TriMap \cite{amid22arxiv}.

\paragraph{Class separability investigation task is more suitable for PCA and MDS}
The superiority of PCA and MDS in preserving distances between data points makes them more precisely exhibit the separability between class labels \cite{wattenberg2016tsnetuning, jeon24tvcg, jiazhi21tvcg}.
Meanwhile, local techniques like \tsne and \umap can well depict the extent to which classes are ``mixed'' in the original space \cite{jeon24tvcg}. However, they are widely reported to exaggerate the distance between classes compared to other techniques \cite{bernard21cng, atzberger25tvcg, benato24cvicg, jeon24tvcg}. Wattenberg et al. \cite{wattenberg2016tsnetuning} show that hyperparameter choices can significantly distort the global relationship between classes in \tsne, including their distances.

\paragraph{Cluster distance investigation task is more suitable for PCA and MDS}
PCA and MDS better preserve pairwise distances between points within each cluster compared to alternatives, making the inter-cluster distances in the resulting projections meaningful. 
These techniques are thus suitable for tasks involving cluster distance analysis.
Many computational benchmarks have validated the superiority of PCA and MDS in supporting the cluster distance investigation task \cite{jeon22vis, moor20icml, wang21jmlr, bentocilla20ivs, jeon24tvcg}.
This implies the appropriateness of these techniques for supporting cluster distance investigation.  
Xia et al. \cite{jiazhi21tvcg} empirically show that global techniques like PCA enable users to perform this task more accurately than local techniques. 
In contrast, Wattenberg et al. \cite{wattenberg2016tsnetuning} and Coenen et al. \cite{coenen19fiar} also inform that the distance between clusters lacks meaning in \tsne and \umap projections, respectively.

\paragraph{Cluster density investigation task is more suitable for PCA and MDS}
These techniques depict the similarity between data points as low-dimensional proximity and thus can more sensitively depict the differences in cluster densities. 
In contrast, local techniques like \tsne and \umap poorly reflect their true similarity in high-dimensional space \cite{jeon22vis, amid22arxiv, narayan21nature} as they only focus on neighboring points. 
As a result, \tsne and \umap projections poorly represent cluster density, which motivates the development of improved techniques such as den-SNE and densMAP \cite{narayan21nature}. 
The superiority of global techniques in supporting the density investigation task is also validated by Xia et al. \cite{jiazhi21tvcg} through user studies and Jeon et al. \cite{jeon24tvcg} via case studies.

\subsubsection{Validity of the Suitability Analysis} 
One notable finding in our suitability analysis is that \tsne and \umap perform better on all ``identification'' tasks while PCA and MDS excel at ``investigation'' tasks. 
This result aligns with the fundamental differences in how these methods interpret distances. \tsne and \umap prioritize preserving local neighborhood structures by effectively treating similarity as a binary function (neighbors or non-neighbors), making them well-suited for tasks that require identifying distinct clusters or groups. In contrast, PCA and MDS interpret distances as continuous values, enabling more accurate interpretation of relative distances between points. This finding supports the validity of our task categorization in capturing the alignment between DR techniques and the tasks that are suitable for.

\subsection{Findings}

\label{sec:quantianal}

We present the findings from the quantitative analysis of the identified papers.
We reveal that \tsne and UMAP are the most commonly adopted DR techniques. 
Yet, researchers often utilize them across any tasks, making them simultaneously the most commonly misused DR techniques.
We also observe that many papers leverage \tsne and \umap without justifications or with improper reasoning. 

\begin{figure}
    \centering
    \includegraphics[width=0.8\linewidth]{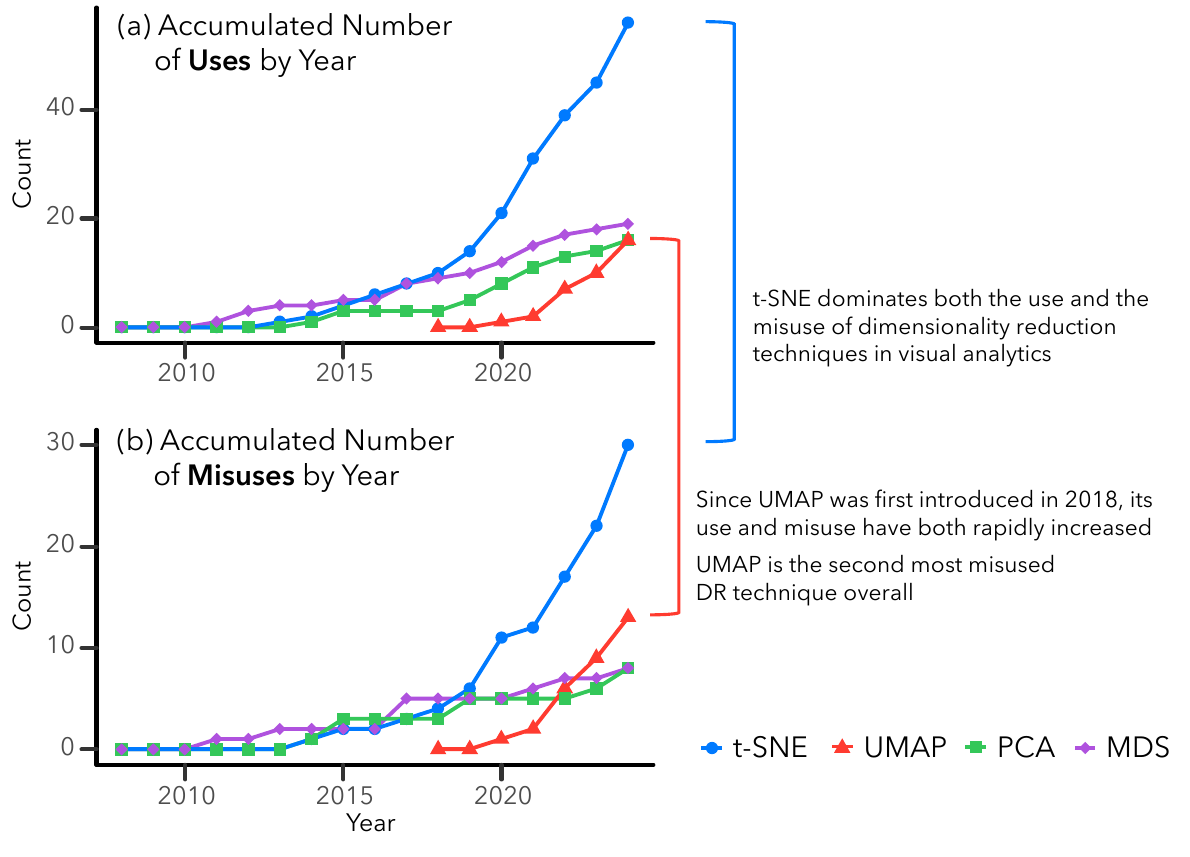}
    \caption{The trend of the accumulated number of papers that use (a) or misuse (b) four major DR techniques. We collect papers published from 2008, the year \tsne is introduced. Note that UMAP's data also starts from the year it is released (2018). }
    \label{fig:trend}
\end{figure}

\paragraph{(Finding 1) \tsne and \umap dominate the use of DR}
While the number of visual analytics papers using DR has increased over the years, this growth is largely driven by \tsne and \umap (\autoref{fig:trend}).
\tsne, for example, appears in more than half of the identified papers (75 out of 136), more than twice as often as the runner-up.
\umap is used in 31 papers. However, UMAP's adoption is increasing at a much steeper rate compared to PCA and MDS, enabling it to achieve parity with these established techniques within only six years.

\begin{figure}[t]
    \centering
    \includegraphics[width=0.65\linewidth]{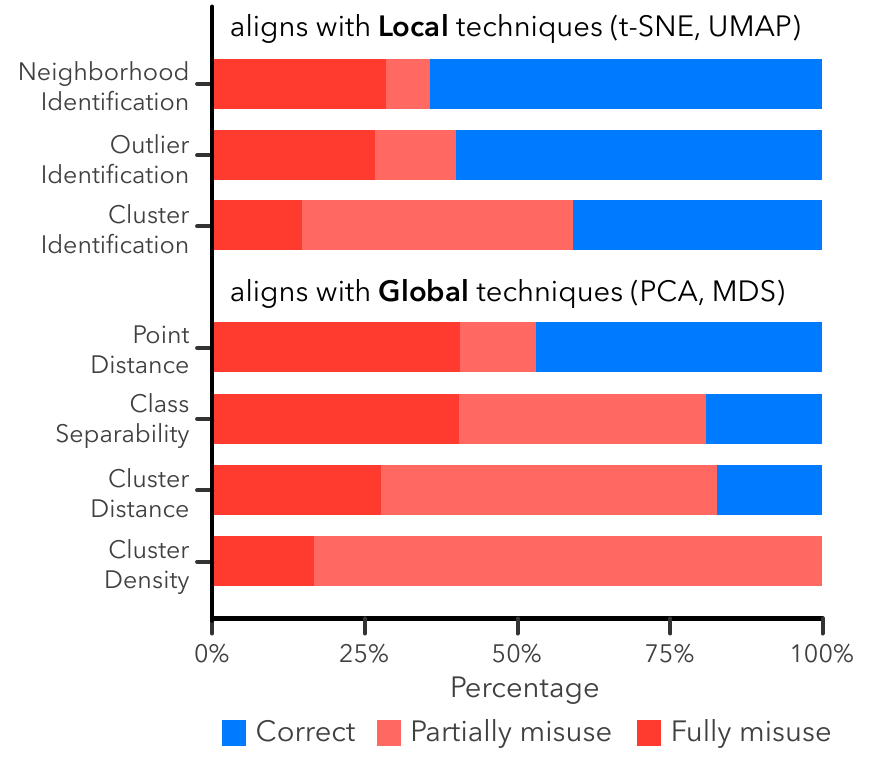}
    \caption{The ratio of appropriate use and misuse of DR techniques by analytic tasks. DR is appropriately used for tasks that align with local techniques (top 3) but not for those that align with global techniques (bottom 4). This result indicates that local techniques (\tsne and \umap) are overtrusted even for tasks that are not suitable. Papers are marked as ``fully misused'' if all tasks targeted by the paper are not supported by the employed DR technique. Papers with partial support are marked as ``partially misused.''}
    \label{fig:error-tasks}
\end{figure}

\paragraph{(Finding 2) \tsne and \umap are used for \textit{any} tasks}
We identify that \tsne and UMAP are commonly used for both suitable and unsuitable tasks.
To do so, we automatically determine whether the paper properly uses DR techniques for their suitable tasks or not. Based on our task suitability analysis results (\autoref{sec:suitability}), we mark the paper to be \bluen{``correct''} if all DR techniques used in the paper aligns with the tasks. We mark the paper to \redn{``fully misuse''} or \redt{``partially misuse''} DR when entire or part of DR techniques are used for unsuitable tasks, respectively. Then,
for each analytic task, we compute the proportion of papers that employ suitable DR techniques relative to all papers addressing the task.
As a result (\autoref{fig:error-tasks}), we find that tasks supported by local techniques have a high proportion of proper usage. However, tasks requiring global techniques have a substantially lower rate of proper usage.
This indicates that researchers appropriately employ local techniques (\tsne and \umap) when required, while also correctly avoiding global techniques for these tasks.
However, tasks requiring global techniques have a substantially lower rate of proper usage, indicating that \tsne and \umap are misused even for unsuitable tasks.

\begin{figure}[t]
    \centering
    \includegraphics[width=0.9\linewidth]{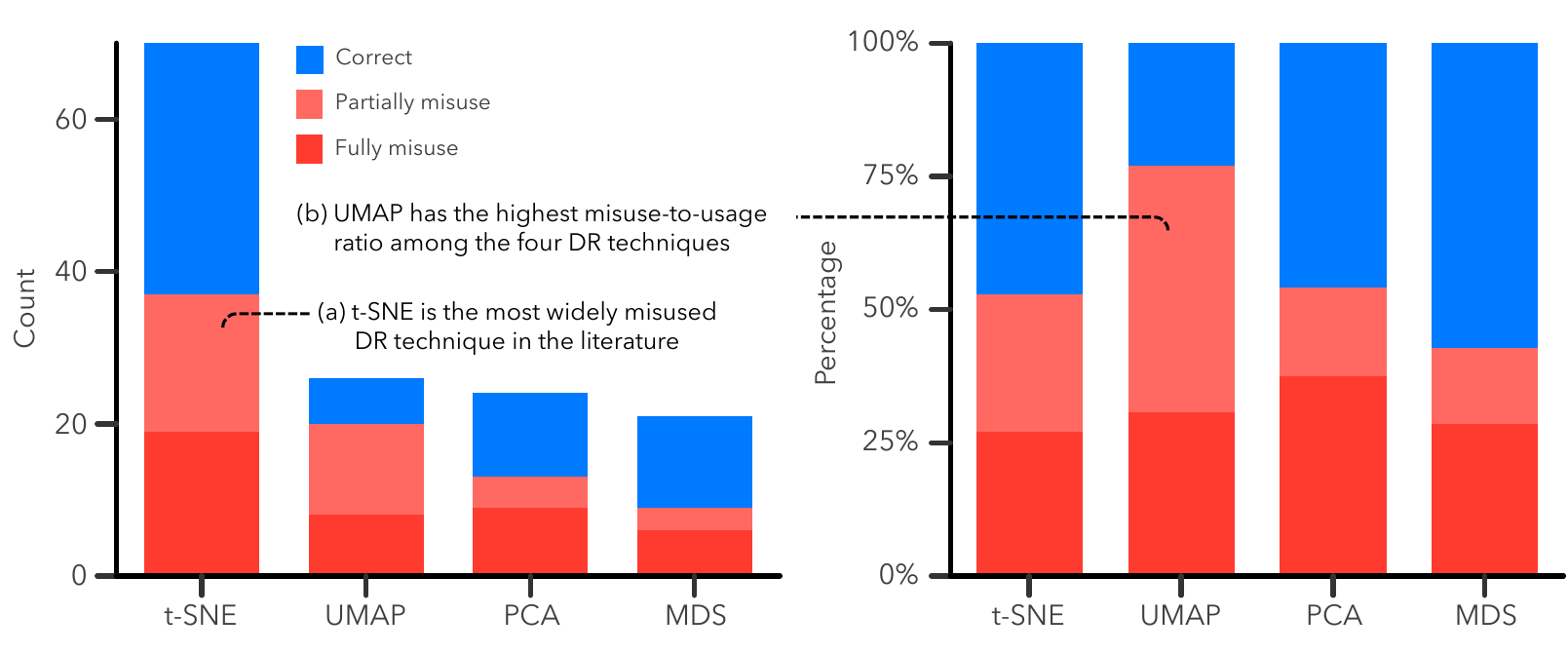}
    \caption{The number of appropriate use and misuse of DR by techniques (left) and their ratio (right). As with \autoref{fig:error-tasks}, papers are marked as ``fully misused'' and ``partially misused'' if all tasks targeted by the paper are entirely or partially not supported by the used DR techniques. The analysis reveals that \tsne and \umap dominate the misuse of DR.}
    \label{fig:misuse_summary}
\end{figure}

\begin{figure*}
    \centering
    \includegraphics[width=\textwidth]{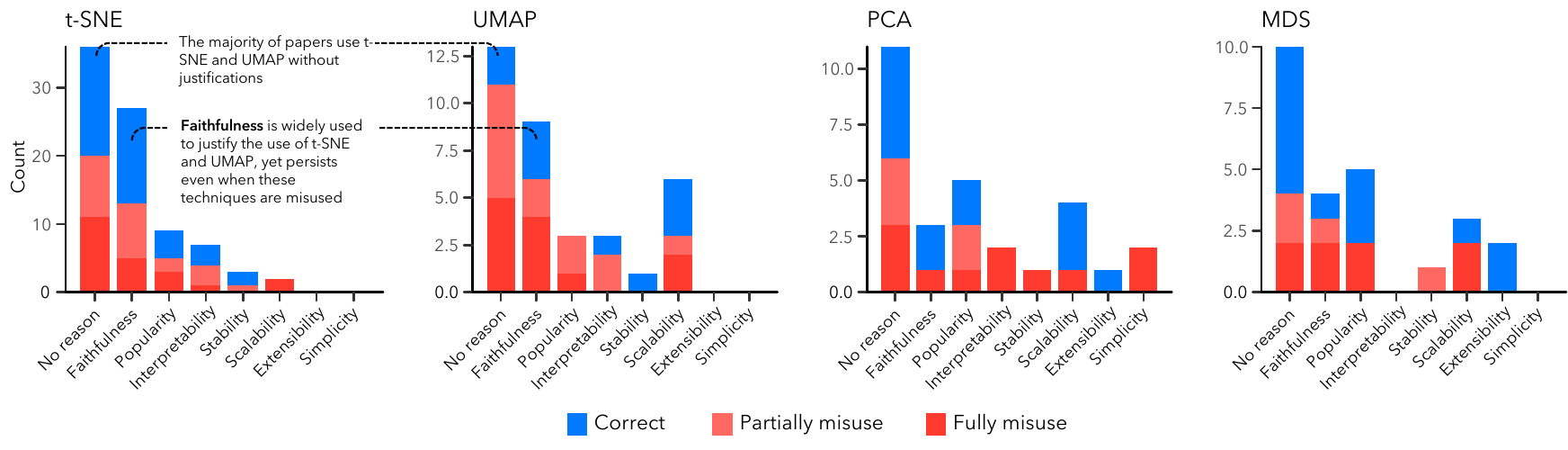}
    \caption{The number of appropriate use and misuse of DR techniques by reasonings.The reasonings (x-axis) are sorted in descending order based on the number they are referenced to justify the use of \tsne (leftmost chart). 
    }
    \label{fig:reasonings}
\end{figure*}

This tendency makes \tsne and \umap to dominate the misuse of DR techniques in practice. We find that \tsne is the most widely misused DR technique in our list of papers (\autoref{fig:misuse_summary}a). Regarding \umap, we observe that its misuse has recently increased rapidly, making it the runner-up (\autoref{fig:trend}b). We also find that \umap has the highest misuse-to-usage ratio among the four major DR techniques (\autoref{fig:misuse_summary}b). In summary, \tsne and \umap are misapplied more frequently than other techniques, underscoring the need for increased efforts to address these misuses.

\paragraph{(Finding 3) \tsne and \umap are used without reasonings or with improper reasonings}
We find that more than 40\% of papers do not explicitly justify their choice of DR techniques (\autoref{fig:reasonings}), and this trend persists for \tsne and \umap. 
These papers often discuss the general need for DR or describe the techniques’ characteristics rather than explaining why specific techniques are chosen.
This result implies that practitioners may perceive DR technique selection---including the case of \tsne and \umap---as requiring less critical evaluation, suggesting a lack of clear understanding of the appropriate way of using DR. Our subsequent interviews (\autoref{sec:interstudy}) further reinforce this observation in the context of \tsne and \umap usage.

We also identify that faithfulness is more widely used to justify the use of \tsne and \umap compared to PCA and MDS, but is referred to even when these techniques are misused (\autoref{fig:reasonings}). This indicates a bias among practitioners to assume that \tsne and \umap are inherently more faithful.
In reality, however, these techniques well preserve local structure but do not faithfully capture global structure (\autoref{sec:suitability}).

\section{Literature Review Beyond the Visualization Field}

\label{sec:litreviewothers}

We complement our examination of reliability challenges in DR-based visual analytics by investigating disciplines beyond data visualization. 
We specifically focus on four domains: Biology, Chemistry, Physics, and Business. 
We select for two reasons. First, these are core domains that commonly use DR for visual analytics. Second, we (the author of this thesis and collaborators) have experience working with domain scientists in these four fields.

\subsection{Paper Search Protocol}

The detailed paper search and filtering procedure is as follows:

\paragraph{Initial paper search}
We want to identify papers that \textit{(1) use a dimensionality reduction algorithm as part of their data analysis and 2) present a visualization of the dimensionally reduced data} in four domains. We query Scopus considering its reproducibility and the wide coverage of the subject areas we select.
We search for keywords (\texttt{``Dimensionality Reduction''}, \texttt{``UMAP''}, \texttt{``t-SNE''}, \texttt{``PCA''}, and \texttt{``Projection''}) within Scopus subject areas of chemistry, biology, physics, and business from 2019 to 2024. 
Subject areas are determined based on the Scopus database labels. 
Our initial query on Scopus returns 52141 matching papers.

\paragraph{Automatic filtering}
We filter out papers with fewer than 5 citations according to Scopus to filter out papers with low impact.
This results in 21083 papers. 
Next, we conducted a stratified sample from this set down to 2000 publications, with 500 from each of our 4 subject areas.  
From this set of 2000 publications, we used the Zotero Reference Manager's \textit{Find Available PDF} function from our academic library's connection to identify publications for which we can find the PDF for review. 
This sampling is allowable because we were not aiming to completely survey all works using DR in these fields; rather, we aim to merely sample them into a small number, in which we can manually screen and review.
This automatic filtering results in 930 papers with PDF available.

\paragraph{Manual filtering}
Finally, we divide these 930 publications to four reviewers (i.e., authors of the reference paper \cite{cashman25tvcg}) and scan them to filter out any publications that do not have any visualization of DR projections, which remove 62\% of the publications.
From the resulting set of 347 papers, each author randomly selected five publications from each of the four subject areas. 
As a result, we obtain \textbf{71 papers}: 20 from Biology, 20 from Chemistry, 17 from Business, and 14 from Physics; note that the total number is smaller than 80 as three Business papers and six Physics papers are incorrectly classified by Scopus and instead came from other fields, i.e. industrial design and engineering, thus are excluded from our analysis.

We discuss the reliability problems identified from these papers in \autoref{sec:domainfinding}.
Note that the relatively small sample of 71 papers included in our survey are not completely representative of the more than fifty thousand papers returned in our initial search in Scopus. 
However, we also believe that 71 papers is a large enough sample to provide valuable insights into the usage of DR outside of computer science. 
 In addition, we investigate the author list of these 71 papers to ensure that there is not an overrepresentation of any particular authors.  Across these 71 papers, there were 705 unique authors with only a single author appearing on more than one paper (which is two).
 We believe this sample represented a diversity of authors across these domains.  In addition, we analyzed the listed affiliations of these papers and found that only six out of 705 unique authors listed an affiliation in a department of computer science or information science \cite{cheng2018monitoring, kuchroo2022multiscale, lee2019dynamic, wei2021novel, xu2022pattern, xu2020t}.

\subsection{Findings}

\label{sec:domainfinding}

The following are our findings on practices in four domains that harm the reliability of DR-based visual analytics.

\paragraph{(Finding 1) Inappropriate selection of DR techniques}
Aligned with our findings in the visual analytics field (\autoref{sec:quantianal}),  we identify that research works in four domains often use DR techniques that do not match with their task. 

\begin{figure}[t]
    \centering
    \includegraphics[width=0.8\linewidth]{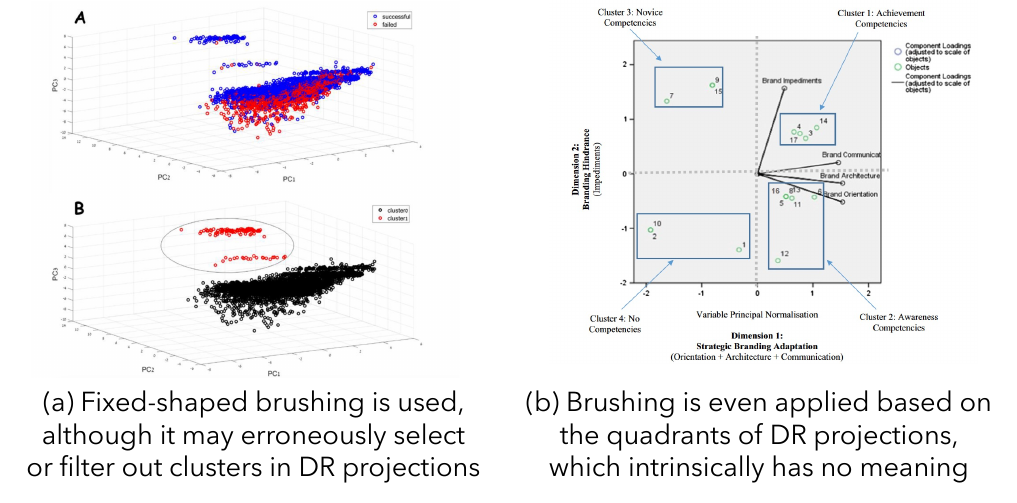}
    \caption{The examples of our findings on inappropriate usage of DR techniques (\autoref{sec:domainfinding}) \cite{song2019mining, lin2020exploring}. These improper usages lead data analysis to have limited reliability, casting doubt on the conclusions that research works are based on. }
    \label{fig:incorrect_usages}
\end{figure}

\paragraph{(Finding 2) Inappropriate use of brushing interaction}
We observe that many papers select and highlight clusters using fixed-shape brushes (e.g., rectangular, elliptical, or spherical) (\autoref{fig:incorrect_usages}a) \cite{song2019mining}.
However, due to distortions in the projection, these 2D clusters often do not accurately correspond to their original high-dimensional counterparts, making subsequent analyses on such brushed clusters unreliable.
We also identify cases in which brushing is performed not based on point-wise proximity but instead by selecting entire quadrants of a DR projection (\autoref{fig:incorrect_usages}b) \cite{lin2020exploring}---an approach that is misleading because the absolute values and signs of DR coordinates have no intrinsic meaning.

\paragraph{(Finding 3) Lack of reproducibility}
We find that, in most cases, papers do not specify how optional hyperparameters (e.g., \texttt{perplexity} in \tsne) are set, thereby degrading the reproducibility of their analyses. 
There are two possibilities: (1) practitioners used the default hyperparameter settings of the library. (2) hyperparameters are cherry-picked to generate DR projections that best align with the papers' hypotheses.
The absence of a hyperparameter report also raises concerns about whether the hyperparameter values have been properly optimized, negatively impacting not only reproducibility but also the credibility of the data analysis.
Our interview study provides evidence that this problem may primarily stem from the widespread cherry-picking of hyperparameters (\autoref{sec:interstudy}).




\section{Interview Study}

\label{sec:interstudy}

We aim to uncover the reliability challenges that lie beneath the surface of research papers and to understand why practitioners fall into such pitfalls.
To this end, we conduct interviews with practitioners who regularly use DR techniques in their research or visual analytics workflows.
In the following sections, we describe our study design (\autoref{sec:interstudydesign}) and analysis procedure (\autoref{sec:interviewanalysis}).
We then present our findings in \autoref{sec:addfindings}.

\begin{table*}[t]
    \centering
    
    \caption{The demographics of the participants in our interview study with practitioners. Our aim in recruiting participants is to maximize diversity in research fields and experience (exp.). VA stands for visual analytics. We order the participants chronologically by interview date.}
    \scalebox{0.83}{
    \begin{tabular}{rlllllll}
    \toprule
     & \textbf{Occupation}  & \textbf{Age} &\textbf{Gender} &\textbf{Type} & \textbf{exp. in VA}  & \textbf{exp. in DR} & \textbf{Domain}\\
    \midrule
        P1 & Professor &  35& Male & VA Researcher & 10 years & 7 years & $\cdot$\\ 
        P2 & Undergraduate & 22 & Female & VA Researcher & 2 years & 1 year & $\cdot$\\ 
        P3 & Research Scientist & 30 & Male & Domain Researcher & $\cdot$ & 6 years & Computer Vision \\ 
        P4 & Research Scientist & 30 & Male &Domain Researcher & $\cdot$ & 4 years & Signal processing\\ 
        P5 & Ph.D. Student & 28 & Female & VA Researcher & 6 years & 5 years & $\cdot$\\ 
        P6 & Ph.D. Student & 28 & Male & Domain Researcher & $\cdot$ & 4 years & HCI  \\ 
        P7 & Ph.D. Student & 29 & Male & Domain Researcher& $\cdot$ & 4 years  & Chemistry\\ 
        P8 & Ph.D. Student & 24& Female & VA Researcher &3 years& 2 years & $\cdot$\\ 
        P9 & Research Scientist & 30 & Male & Domain Researcher &  $\cdot$ & 7 years & NLP \\ 
        P10 & MS Student &  22 & Male & VA Researcher & 2 years & 1 year & $\cdot$\\ 
        P11 & Staff Engineer & 34 & Male & VA Researcher & 2 years & 2 years & $\cdot$\\ 
        P12 & Postdoc. & 34 & Male & Domain Researcher & $\cdot$ & 4 years & Bioinformatics \\ 
    \bottomrule
    \end{tabular}
    }
    \label{tab:demo}
\end{table*}

\subsection{Study Design}

\label{sec:interstudydesign}

\noindent
\textbf{Participants and recruitment.}
We want to diversify our participants in terms of their experience on DR. We first aim to achieve diversity in the domains in which participants work.
To do so, we recruit both visual analytics researchers and domain researchers who have experience in visually analyzing and presenting their data using DR.
For visual analytics researchers, we randomly select papers from our literature review on visual analytics (\autoref{sec:litreviewva}), prioritizing diversity in target data and problem domains. We then contact either the first or the corresponding author via email to increase diversity in participants' expertise and visualization literacy.
For domain researchers, we ensure that they are from distinct disciplines without overlap.
We recruit participants from a local university through an internal web community. We also employ snowball sampling \cite{goodman61ams} to expand our participant pool.
In total, we interview 12 participants (six visual analytics researchers and six domain researchers) with diverse occupations and research experience (\autoref{tab:demo}).

\paragraph{Interview protocol}
We interview participants in a semi-structured manner.
We first ask the participants to give consent for their participation, then ask participants a series of questions.
The questions mainly ask the participants (1) their expertise in DR,  (2) their experience and justifications in using DR techniques, specifically \tsne and \umap, and (3) the difficulties that occur while using DR techniques. 
The interviews are conducted via a recorded Zoom call, where we transcribe the interview using a commercial speech-to-text service. We compensate participants with the equivalent of 15 USD. 
All interviews are finished within 40 minutes. 

\subsection{Analysis Procedure}

\label{sec:interviewanalysis}

One coder initially analyzes the interview transcripts using thematic analysis \cite{joffe11qrm}. Each code focuses on describing either the practitioners’ behaviors in analyzing high-dimensional data using DR. The coder first extracts participant quotes as initial codes and then hierarchically clusters them into two levels based on semantic similarity. Each lower-level cluster is grouped into a subtheme, and these subthemes are used to derive the final set of themes, where each represents individual higher-level clusters. 
Note that we group codes as subthemes only when we have quotes from more than two different participants. 
Finally, two additional authors newly engage in the process to review and refine the results of the thematic analysis.

\subsection{Findings}

\label{sec:addfindings}

We discuss the main challenges we identified from our interview study.

\paragraph{(Finding 1) Practitioners often cherry-pick hyperparameters}
We observe that participants routinely cherry-pick hyperparameters, which aligns with our findings in the literature review that most papers do not report their parameter settings. Eight out of 12 participants report experience manually tuning the hyperparameters of \tsne and \umap. 
Four participants explicitly mention that they tune hyperparameters without understanding their effect on projection results.


\paragraph{(Finding 2) Practitioners prefer projections with high class and cluster separability}
Participants report that they wanted to generate either an interpretable or aesthetically pleasing projection, i.e., projections that show high class or cluster separability (\autoref{tab:defreasonings}). 
For example, participants report cherry-picking hyperparameters to achieve aesthetically pleasing projections with well-separated classes or clusters. 
This finding resonates with the common misuse of \tsne and \umap, as both techniques exaggerate class or cluster separability by design.

\section{Summary of Identified Challenges}

The series of literature reviews (\autoref{sec:litreviewva}, \ref{sec:litreviewothers}) and an interview study (\autoref{sec:interstudy}) reveal three prevalent challenges practitioners encounter that undermine the reliability of DR-based visual analytics.
We address these challenges in following chapters (\autoref{sec:clcl}, \ref{sec:dawadr}, and \ref{sec:dabrca}).

\paragraph{Challenge 1: Misuse of commonly used DR techniques}
Our studies reveal that two popular DR techniques, \tsne and \umap, are also widely misused for analytic tasks for which they are unsuitable. 
As a result, visual analytics becomes unreliable due to errors introduced while executing these tasks. 
Our literature review on visual analytics papers reveals that \tsne is the most widely misused DR technique in practice, and \umap has the highest misuse-to-usage ratio (Finding 2 in \autoref{sec:quantianal}). Our interviews indicate that this tendency arises, in part, from practitioners’ preference for projections with high class or cluster separability---properties that \tsne and \umap intentionally exaggerate (Finding 2 in \autoref{sec:addfindings}). The literature review also indicates that practitioners often assume that \tsne and \umap generally provide a more faithful representation than alternative techniques (Finding 3 in \autoref{sec:quantianal}).

\paragraph{Challenge 2: Cherry-picking of hyperparameters}
We find that practitioners routinely cherry-pick hyperparameters to generate projections with interpretable or aesthetically pleasing projections (Finding 1 in \autoref{sec:addfindings}). 
This means the resulting projections no longer guarantee a faithful representation of the original high-dimensional data, ultimately leading to unreliable visual analytics.
Such a finding aligns with the observation in literature reviews that specific parameter settings used are rarely reported in published papers (Finding 3 in \autoref{sec:domainfinding}).


\paragraph{Challenge 3: Erroneous interactions}
One finding from the literature review with domain papers is that brushing interactions with fixed-shaped brushes (E.g., rectangular brushes) are widely adopted (Finding 2 in \autoref{sec:domainfinding}). 
However, distortions in DR projections make the 2D clusters identified by these brushes not faithfully represent the original high-dimensional clusters, making successive analysis on these clusters unreliable.
\section{Conclusion}


In this chapter, we systematically investigated how practitioners across visualization and various domain areas employ DR in their analytical workflows. Through literature reviews and interview studies, we uncovered three prevalent reliability challenges: 
(1) the misuse of popular DR techniques, notably t-SNE and UMAP, 
(2) the cherry-picking of hyperparameters, and 
(3) erroneous interaction due to distortions in DR projections. 
Our findings reveal that reliability issues originate not only from algorithmic limitations but also from practitioners’ misconceptions or biases. 
In the subsequent chapters, we address each of these challenges with tailored technical solutions.

\chapter{Classes are Not Clusters: Improving Label-based Evaluation of Dimensionality Reduction}

\label{sec:clcl}

Practitioners commonly misuse the popular techniques \tsne and \umap, assuming that their projections with exaggerated class and cluster separability faithfully represent the underlying data.
DR evaluation metrics should serve as gatekeepers against such bias, but several commonly used metrics instead reinforce it, favoring projections with high class separability.
This occurs because these metrics assume that class labels correspond to well-separated ground-truth clusters, even though classes may overlap or split into multiple clusters in reality \cite{aupetit14beliv}.

In this chapter, we discuss the design and evaluation of two new DR evaluation metrics---\textit{Label-Trustworthiness and Label-Continuity (L-T\&C)}---that address this problem and thus reduce bias in the evaluation of DR projections.



\section{Introduction}

\label{sec:cncintro}

Using DR for cluster analysis relies on the assumption that the cluster structure of the original data is accurately represented in the low-dimensional DR projections. However, as DR inherently generates distortions, i.e., the original cluster structure is imprecisely represented in the projections \cite{aupetit07neurocomputing, aupetit14vast, heulot13vamp, lespinats07tnn, lespinats11cgf}, cluster analysis with DR can often become unreliable \cite{jeon21tvcg, martins15cgvc}.
It is thus important to evaluate how well the original cluster structure is preserved in the DR projections \cite{jiazhi21tvcg, jeon21tvcg, martins14cg, joia11lamp}, prior to the analysis. Various ways to evaluate the faithfulness of cluster structures in DR projections, in either a perceptual \cite{etemadpour15ivapp, jiazhi21tvcg, sedlmair13tvcg} or computational \cite{jeon21tvcg, venna06nn, lee07springer, motta15neurocomputing} way, have been thus proposed.  


However, widely used evaluation metrics for cluster structure tend to erroneously favor projections with high class separability, reinforcing a bias toward DR techniques that exaggerate this separability like \tsne and \umap.
These metrics evaluate DR projections by assessing \textit{cluster-label matching} (CLM), that is, the extent to which class labels form clusters in the embedded space \cite{joia11lamp, fadel15neurocomp, jiazhi21tvcg, becht19nature, xiang21fig, yang21cellreports}, using \textit{clustering quality metrics} \cite{liu10icdm, wu09kdd} like Silhouette \cite{rousseuw87silhouette}.
These clustering quality metrics inform how well the class labels form clear position-based clusters in the data space. 
The labels that contain mutually separated and individually condensed groups are preferred. 
Projections with good CLM are considered of high quality, assuming the original data also have good CLM. 
Yet this assumption is hardly guaranteed \cite{aupetit14beliv, jiazhi21tvcg, farber10multiclust, jeon25tpami, vanderhoorn25arxiv}.
Labels can come from an external source (e.g., human annotation), possibly unrelated to the features of the data space. Labels can also result from clustering techniques, which may not align with the actual clusters. 
As a result, a single class may contain multiple distinct clusters, and different classes may lie near one another or overlap \cite{aupetit14beliv}.
These possibilities cast doubt on the validity of label-based DR evaluation: a projection that faithfully preserves overlapping classes in the original space may receive a low score simply because it exhibits poor CLM.

To address this problem, we revisit the process of evaluating DR projections using class labels. 
In contrast to the previous label-based evaluation process, we propose to use clustering quality metrics to quantify the difference between the CLM estimated in both original and projection spaces. 
We first design clustering quality metrics that produce comparable scores across different dimensions and data patterns, making it available to calculate the difference of CLM across the data and projection. 
Then, using these new clustering quality metrics, we design two DR evaluation metrics---\textit{\LT} (\lt) and \textit{\LC} (\lc)---that implement our new evaluation process. 
\lt quantifies the distortion due to the degradation of CLM: the score is lower when the points of two different classes get closer in the projection than in the original space. 
Conversely, \lc evaluates the distortion regarding the exaggeration of CLM: the score is lower when the points of two different classes get farther apart in the projection than in the original space.
The rationale behind our metrics is that in visual analytics using DR, it is important to investigate how class labels span the original cluster structure as seen through the projection \cite{brehmer14beliv, aupetit14beliv, aupetit22arxiv, aupetit22topoinvis, wenskovitch18tvcg} (e.g., examine the individual density of a class or the pairwise proximity between classes). Since distortions in CLM reduce the faithfulness of cluster structures represented by projections, \ltc scores can be interpreted as proxies for the reliability of DR-based visual analytics that targets the investigation of cluster structure.

We conduct a series of quantitative experiments to validate the effectiveness of \ltc. 
The results show that \ltc better capture the distortions of cluster structures than the existing evaluation metrics (e.g., Steadiness \& Cohesiveness \cite{jeon21tvcg} and Trustworthiness  \& Continuity \cite{venna06nn}) and resolve the bias in existing label-based metrics (i.e., naive application of clustering quality metrics).
From the scalability analysis, we validate that the runtime of using \ltc is competitive with that of the existing methods.
Finally, we present two case studies demonstrating that \ltc can reveal how different DR techniques or hyperparameter settings affect embedding results. 

\section{Related Works}

We discuss three relevant areas: (1) evaluation of DR projections with class labels, (2) evaluation of clustering quality, and (3) evaluation of CLM.

\subsection{Evaluating DR Projections Using Class Labels}


\label{sec:cnclabelsmetric}

Exploiting labels is a common scheme in evaluating DR projections \cite{colange20neurips, joia11lamp, fadel15neurocomp, jiazhi21tvcg, becht19nature, xiang21fig, yang21cellreports}.
Historically, this strategy has started to evaluate the ``visual interestingness'' \cite{seo06tvcg, huber85ims, dang14pvis} of projections when exploring a large set of projections representing different subspaces \cite{wang18tvcg, huber85ims}. Here, the aim is to find projections that represent a subspace of data maximizing class separability, thereby revealing which data attributes compose these subspaces, under the assumption that such subspaces are worthwhile to investigate. This approach has advanced beyond finding projections with high class separability to also identifying other patterns (e.g., overlap between specific classes or class density) \cite{fujiwara22tvcg, fujiwara19tvcg}.

Recently, this label-based evaluation has been exploited to evaluate not only the interestingness but also the faithfulness of DR projections \cite{colange20neurips, joia11lamp, fadel15neurocomp, jiazhi21tvcg, becht19nature, xiang21fig, yang21cellreports}. 
A general process is to use clustering quality metrics to measure the CLM of projections as a proxy of faithfulness \cite{joia11lamp, fadel15neurocomp, becht19nature, yang21cellreports}, assuming that class labels form ground truth clusters (i.e., CLM in the high-dimensional space is good).
However, the approach is prone to producing errors while examining the faithfulness of DR projections. 
For example, if the CLM of the original data is bad (e.g., some classes overlap), projections that have good CLM for bad reasons (e.g., DR artificially separates each class into a distinct cluster) will be considered to be highly faithful.
Such an error makes the DR evaluation to praise DR projections with exaggerated class separability, notably the ones produced by \tsne and \umap.


To the best of our knowledge, a sole pair of metrics that relies on class labels but is independent of clustering quality metrics is Class-Aware Trustworthiness and Continuity (CA-T\&C) \cite{colange20neurips}.
CA-T\&C is a variant of T\&C that assess the degradation of CLM (i.e., False Groups distortions) by estimating 
the extent to which Missing and False Neighbors occurred within and between classes, respectively. 
However, CA-T\&C hardly captures the Missing Groups distortions as they do not consider the increase of CLM as distortions. 
These metrics also mainly focus on local structures and thus cannot comprehensively examine CLM distortions.

\paragraphit{Our contribution}
We propose \ltc as novel metrics utilizing class labels to evaluate DR projections.
As with the previous process of label-based DR evaluation (i.e., the process of naively applying clustering metrics in the projection space), our new metrics utilize clustering quality metrics to evaluate CLM; however, by applying clustering metrics to both the original and projection spaces and assessing their difference, our measures precisely capture distortions at the cluster level, mitigating bias towards favoring projections with high class separability.



\subsection{Clustering Quality Metrics} 

Clustering quality metrics aim to evaluate how well-clustered the given partition (i.e., clustering) is in the given data. 
We commonly use clustering metrics to find the optimal clustering technique or hyperparameter setting that produces the partition of the data that best matches its cluster structure.
In our case, clustering metrics are used to examine the CLM of datasets by receiving class labels as partitions.
These metrics are largely divided into two types: \textit{external} and  \cite{liu10icdm, liu13tsmcb} and \textit{internal} clustering quality metrics \cite{wu09kdd}. 
External metrics quantify how much the resulting clustering matches a ground truth partition of a given dataset. For example, adjusted mutual information \cite{vinh10jmlr} measures the agreement of two assignments (clustering and ground truth partitions) in terms of information gain corrected for chance effects. 
In contrast, internal metrics examine cluster structure based on the data distribution without ground truth clusters. They focus on two criteria, namely \textit{compactness} (i.e., the pairwise proximity of data points within a cluster) and \textit{separability} (i.e., the degree to which clusters lie apart from one another) \cite{liu10icdm, tan05idm}. For example, famous metrics like Silhouette \cite{rousseuw87silhouette} or Calinski-Harabasz index \cite{calinski74cis} produce a quality score by dividing separability by compactness.


\paragraphit{Our contribution}
Cluster quality metrics examine whether a given partition of the dataset forms clusters, thus have the capability to measure CLM. However, existing clustering metrics cannot be directly used to evaluate the comparison of the CLM of high-dimensional data and low-dimensional projections. To overcome this problem, we systematically adjust existing internal clustering quality metrics (\autoref{sec:adjusted}) and apply these new metrics to \ltc.

\subsection{Evaluation of Cluster-Label Matching}

\label{sec:cncclm}

We discuss possible approaches to evaluate the CLM of different data and their limitations. 

\paragraph{Classifiers}
A natural approach is to use classification scores as a proxy for CLM \cite{abul03smc, rodriques18kbs}. This approach is based on the assumption that the classes of a labeled dataset with good classification scores would provide well-separated ground truth clusters. Still, classifiers can hardly distinguish between two ``adjacent'' classes forming a single cluster and two ``separated'' classes forming distant clusters, because both pairs are easily distinguishable by classifiers. It also cannot distinguish different within-class structures, such as
a class forming a single cluster and one made of several distant clusters. In addition, classifiers require expensive training time (\autoref{sec:cnceval}).

\paragraph{Clustering techniques and external clustering metrics}
A more direct approach to evaluate CLM is to examine how well the clustering techniques capture class labels, as well-separated classes will be easily captured by clustering techniques. 
This can be done by evaluating the difference of clustering results and class labels using external clustering quality metrics. 
However, this approach is also computationally expensive (\autoref{sec:cnceval}).
Moreover, this approach is likely to be biased with respect to a certain type of cluster that the clustering techniques used aim to capture. 
Still, we can approximate ground truth CLM by running multiple and diverse clustering techniques \cite{vega11jprai} and aggregating results.  For lack of a better option, we use this ensemble approach to obtain an approximate ground truth in our experiments to validate our axiom-based solution.

\paragraph{Internal clustering quality metrics}
These metrics examine how well given partitions form a cluster both in terms of mutual separability and individual compactness, thus can detail CLM of a given dataset or projection. 
Furthermore, using internal metrics are relatively inexpensive compared to using classifiers or clustering techniques. 
However, these metrics are originally designed to compare and rank different partitions of the \textit{same} dataset.
Therefore, they are not only affected by cluster structure but also dependent on the characteristics of the datasets, such as the number of points, classes, and dimensions, which means that they are cannot properly compare CLM \textit{across} different datasets. 
This means that they cannot also used to compare CLM between high-dimensional data and resulting projections. 

\paragraphit{Our contribution}
We want to use internal clustering quality metrics to evaluate CLM of high-dimensional data and its projections due to its capability to detail cluster structure and efficiency. 
We thus propose four axioms that internal clustering metrics should satisfy to compare CLM across datasets.
We then derive new \textit{adjusted} internal clustering metrics satisfying the axioms, and use these metrics to design \ltc.

\section{Proposed Label-Based Evaluation Process}

\label{sec:proposedevaluation}

The previous label-based DR evaluation process promotes projections with good CLM (i.e., high class separability) regardless of the original data's CLM (\autoref{sec:cncintro}).
In other words, the process examines the extent to which CLM is harmed in projections while assuming that the original data has good CLM.
If the assumption is broken, the process will frame projections that correctly represent overlapped classes to have False group distortions. 
As the process considers good CLM projections as high-quality ones, it is also incapable of detecting Missing Groups distortions that may arise from CLM amplification. These pitfalls were identified for the first time by Aupetit \cite{aupetit14beliv}. The previous label-based evaluation thus erroneously prefers DR techniques that maximize class separation like \tsne and \umap, rather than those that aim to preserve the original data structure when datasets have poor CLM.

To mitigate such a bias, we propose to quantify CLM not only in low-dimensional projection but also in high-dimensional data. The central idea here is to examine how well the CLM in the original data is consistently maintained in the projection. If CLM becomes worse in the projection, it means that False Groups distortions have occurred. In contrast, the increment of CLM can be interpreted as Missing Groups distortions.

\section{Adjusted Clustering Quality Metrics}

\label{sec:adjusted}

We propose \textit{adjusted} internal clustering quality metrics that can compare CLM across datasets with different dimensionality and patterns, thus can be used for our new proposed DR evaluation process. We first discuss the axioms (\autoref{sec:cncreq}) for the adjusted metrics and protocols to adjust existing internal clustering metrics (\autoref{sec:cncprotocol}) to make them satisfy axioms. We then detail how we convert an existing internal metric to be adjusted (\autoref{sec:cncadjust}). Finally, we evaluate the proposed adjusted metrics in accuracy in quantifying CLM and runtime (\autoref{sec:cnceval}).

\subsection{Axioms}

\label{sec:cncreq}

Ackerman and Ben-David (\AB) introduce \textit{within-dataset} axioms \cite{bendavid08nips} that specify the requirements for internal clustering metrics to properly evaluate clustering partitions: \textbf{W1: Scale Invariance} requires metrics to be invariant to distance scaling; \textbf{W2: Consistency} is satisfied by a metric that increases when within-cluster compactness or between-cluster separability increases; \textbf{W3: Richness} requires metrics possible to give any fixed cluster partition the best score over the domain by only modifying the distance function; and \textbf{W4: Isomorphism Invariance} ensures that an internal clustering metrics does not depend on the external identity of points (e.g., class labels). 
However, these within-dataset axioms do not consider the case of comparing scores across datasets; rather, they assume that the dataset is invariant. 
We propose four additional \textit{across-dataset} axioms that a function should satisfy to fairly compare cluster partitions across datasets with different dimensionality and data distributions, thus can be used to compare CLM for evaluating DR projections.

\paragraph{Notations}
We begin by defining four fundamental building blocks of our axioms, using notation identical to that of \AB:
\vspace{3pt}
\begin{itemize}[leftmargin=1.2em]
    \item A \textbf{finite domain set}, i.e., dataset, $X \subset \mathcal{D}$ of dimension $\Delta_X$, where $\mathcal{D}$ denotes the data space.
    \item A \textbf{clustering partition} of $X$ as $C =\{C_1, C_2, \cdots, C_{|C|}\}$, where  $\forall i\neq j, C_i \cap C_j = \emptyset$ and $\cup_{i=1}^{|C|} C_i = X$.
    \item A \textbf{distance function} $\delta : \mathcal{D} \times \mathcal{D} \rightarrow \mathbb{R}$, satisfying $\delta(x, y) \geq 0$, $\delta(x, y) = \delta(y, x)$ and $\delta(x, y) = 0$ if $x=y$ $\forall x, y \in \mathcal{D}$. We do not require the triangle inequality.
    \item A \textbf{metric} $f$ as a function that takes $C, X, \delta$ as input and returns a real number. Higher $f$ implies a better clustering. 
\end{itemize}
\vspace{3pt}
We extend these notations with:
\vspace{3pt}
\begin{itemize}[leftmargin=1.2em]
    \item $\overline{X'}$ the \textbf{centroid} of $X'$, where $X' \subset X$. 
    \item $\underline{W}_\alpha$ a \textbf{random subsample}  of the set $W$ ($\underline{W}_\alpha\overset{D}{=}W$) such that $|\underline{W}_\alpha|/|W|=\alpha$, and the corresponding clustering partition is noted $\underline{C}_\alpha=\{\underline{C_i}_\alpha\}_{i=1, \dots, |C|}.$
\end{itemize}

\paragraph{Goals and factors at play}
\AB's within-dataset axioms are based on the assumption that the clustering metrics that satisfy these axioms properly evaluate the quality of a clustering partition.
However, the axioms do not consider that the metrics $f$ could operate on varying $C$, $\delta$, and $X$. For example, isomorphism invariance (W4) assumes fixed $X$ and $\delta$; consistency (W2) and richness (W3) define how functions $f$ should react to the change of $\delta$, but do not consider how $\delta$ changes in real terms, affected by various aspects of $X$ (e.g., dimensionality); scale invariance (W1) considers such variations, but only in terms of the global scaling. Thus, the satisfaction of \AB's axioms is a way to ensure internal clustering quality metrics focus on measuring clustering quality within a single dataset but not across different datasets.

In contrast, our new adjusted internal clustering metrics shall operate on varying $C$, $\delta$, and $X$. 
Thus, several aspects of the varying datasets now come into play, and their influence on adjusted metrics shall be minimized. The sample size $|X|$ is one of them (\textbf{Axiom A1}), and the dimension $\Delta_X$ of the data is another one (\textbf{Axiom A2}). 
Moreover, what matters is the \textit{matching} between natural clusters and data labels more than the number of clusters or labels; therefore, 
we shall reduce the influence of the number of labels $|C|$ (\textbf{Axiom A3}). Lastly, we need to align adjusted metrics to a comparable range of values (\textbf{Axiom A4}) across datasets, in essence capturing all remaining hard-to-control factors unrelated to clustering quality. We now explain the new axioms in detail:

\subsubsection*{Axiom A1: Data-Cardinality Invariance}
Invariance of the sample size $|X|$ is ensured if subsampling all clusters in the same proportion does not affect the metric score. This leads to the first axiom:

\paragraph{A1: Data-Cardinality Invariance} \textit{A metric $f$ satisfies data-cardinality invariance if
$\forall X,\forall \delta$ and $\forall C$ over $(X,\delta)$, $f(C,X,\delta) = f(\underline{C}_\alpha,X_\alpha,\delta)$ with $X_\alpha=\cup_{i=1}^{|C|} \underline{C_i}_\alpha$  $\forall \alpha\in ]0,1]$.}

\subsubsection*{Axiom A2: Shift Invariance}
We shall consider that data dimension $\Delta_X$ varies across datasets. An important aspect of the dimension called the concentration of distance phenomenon, which is related to the curse of dimensionality, affects the distance measures involved in adjusted metrics. As the dimension grows, the variance of the pairwise distance for any data tends to be constant, while its mean value increases with the dimension \cite{beyer99icdt, francois07tkde, lee11pcs}. Therefore, in high-dimensional spaces, $\delta$ will act as a constant function for any data $X$, and thus an clustering quality metrics $f$ will generate similar scores for all datasets. 
To mitigate this phenomenon, and as a way to reduce the influence of the dimension, we require that the metric $f$ be shift invariant \cite{lee11pcs, lee14cidm} so that the shift of the distances (i.e., growth of the mean) can be canceled out.

\paragraph{A2: Shift Invariance} \textit{A metric $f$ satisfies the shift invariance if $\forall X, \forall \delta$, and  $\forall C$ over $(X, \delta)$, $f(C, X, \delta) = f(C, X, \delta+ \beta)$ $ \forall \beta > 0$, where $\delta + \beta$ is a distance function satisfying $(\delta + \beta)(x,y) = \delta(x,y) + \beta$, $\forall x, y \in X$.}

\subsubsection*{Axiom A3: Class-Cardinality Invariance}

The number of classes should not affect the results of adjusted metrics; for example, two well-clustered classes should get an score similar to 10 well-clustered classes. \AB proposed that the minimum, maximum, and average class-pairwise aggregations of clustering quality metrics form yet other valid metrics. We follow this principle as an axiom for adjusted metrics.

\paragraph{A3: Class-Cardinality Invariance}
\textit{A metric $f$ satisfies class-cardinality invariance if $\forall X,  \forall \delta$ and $\forall C$ over $(X, \delta)$, $f(C, X, \delta) = \textrm{agg}_{S \subseteq C, |S| = 2} f'(S, X, \delta)$ \textit{where function}  $\textrm{agg}_S\in\{\textrm{avg}_S,\min_S,\max_S\}$ and $f'$ is an IVM.}


\subsubsection*{Axiom A4: Range Invariance} Lastly, we need to ensure that an adjusted metrics  takes a common range of values across datasets. In detail, we want their minimum and maximum values to correspond to the datasets with the worst and the best CLM, respectively, and that these extrema are aligned across datasets (we set them arbitrarily to 0 and 1), as follows:

\sloppy
\paragraph{A4: Range Invariance} \textit{A metric $f$ satisfies range invariance if $\forall X, \forall \delta$, and $\forall C$ over $(X,\delta)$, $\min_{C} f(C,X,\delta)=0$ and $\max_{C} f(C,X,\delta)=1$.}

\fussy

\subsection{Protocols}

\label{sec:cncprotocol}

We introduce four technical protocols (T1-T4), designed to generate adjusted internal clustering metrics that satisfy the corresponding axioms A1-A4, respectively.

\subsubsection*{T1: Approaching Data-Cardinality Invariance (A1)}

We cannot guarantee the invariance of an adjusted metric for any subsampling of the data (e.g., very small sample size). However, we can obtain robustness to random subsampling if we use consistent estimators of population statistics \cite{VaartAsymptoticStat1998} as building blocks of the metric. For example, we can use the mean, the median, or the standard deviation of the points within a class or the whole dataset or quantities derived from them, such as the average distance between all points of two classes.

\subsubsection*{T2: Achieving Shift Invariance (A2)}

\paragraph{T2-a,b: Exponential protocol} Considering a vector of distances $u=(u_1\dots u_n)$, we can define a shift-invariant function by using a ratio of exponential functions: 
\begin{equation}
    g_j(u)=\frac{e^{u_j}}{\sum_k e^{u_k}}.
\end{equation}
We observe that $\forall \beta \in \mathbb{R}$,
\begin{equation}
g_j(u+\beta) = \frac{e^{u_j+\beta}}{\sum_k e^{u_k+\beta}}=\frac{e^{u_j}}{\sum_k e^{u_k}}\frac{e^{\beta}}{e^{\beta}}=g_j(u), 
\end{equation}
hence $g_j$ is shift invariant. Thus, the metric $f$ is shift invariant if it consists of ratios of the exponential distances (\textbf{T2-a}). Note that this protocol is at the core of the $t$-SNE loss function \cite{lee11pcs}.
If a building block is a sum or average of distances, the exponential should be applied to the average of distances rather than individuals (\textbf{T2-b}), as the shift occurs to the average distances \cite{francois07tkde}.

\paragraph{T2-c: Equalizing shifting}
The exponential protocol can be safely applied only if the measure incorporates the distance between data points within $X$ (Type-1 distance).
We do not know, in general, how the shift of type-1 distances affects the distances between data points and their centroid (Type-2),
nor do we know how the shift affects the distance between two centroids (Type-3),
even though they are common building blocks in internal clustering metrics \cite{liu10icdm}. Fortunately, if $\delta$ is the square of Euclidean distances (i.e., $\delta =d^2$ where $d(x,y)$ denotes the Euclidean distance between points $x$ and $y$), we can prove that the shift of type-1 distances by $\beta$ results in the shift of type-2 distances by $\beta/2$, and in no shift of type-3 distances, which is stated by the following theorems (proof in the Appendix of the reference paper \cite{jeon25tpami}).

\paragraph{Theorem 1 (Type-2 Shift)} \textit{$\forall X' \subset X$,  $\forall \beta > 0$, and for any Euclidean distance functions $d_L$ and $d_H$ satisfying $d_H^2 = d_L^2 + \beta$, 
    $\sum_{x \in X'}d_H^2(x, c) = \sum_{x \in X'}d^2_L(x,c) + \beta/2$, where $c = \overline{X'}$.}
    
\paragraph{Theorem 2 (Type-3 Shift).} \textit{$\forall X', X'' \subset X$, $\forall \beta > 0$, and for any Euclidean distance functions $d_L$ and $d_H$ satisfying $d_H^2 = d_L^2 + \beta$, $d^2_H(c', c'') = d^2_L(c', c'')$, where $c' = \overline{X'}$ and $c''= \overline{X''}$.}
\vspace{4pt}

\noindent
Therefore, if an internal clustering metrics consists of different types of distances, we should use $\delta = d^2$ and apply the exponential protocol with the same type of distances for both its numerator and denominator (\textbf{T2-c}).

\paragraph{T2-d: Recovering Scale Invariance}
After applying the exponential protocol, $g_j$ is no more scale-invariant:
\begin{equation}
    \forall \lambda\in \mathbb{R},g_j(\lambda u) = \frac{e^{\lambda u_j}}{\sum_k e^{\lambda u_k}}\neq g_j(u),
\end{equation}
and so it will not satisfy axiom W1. We can recover scale-invariance by normalizing each distance $u_i$ by a term that scales with all of the $u_k$ together, such as their standard deviation, $\sigma(u)$. Now, 
\begin{equation}
    g_j(\lambda u/\sigma(\lambda u))=g_j(\lambda u/\lambda \sigma(u))=g_j(u/\sigma(u)),
\end{equation}
is both shift and scale invariant (\textbf{T2-d}).

\subsubsection*{T3: Achieving Class-Cardinality Invariance (A3)}

\sloppy
Class-cardinality invariance can be achieved by following the definition of Axiom A3; thas is, by defining the global metric as the aggregation of class-pairwise local metrics, $f_c(C, X, d) = \textrm{agg}_{S \subseteq C, |S| = 2} f(S, X, \delta)$, where $\textrm{agg} = \{\textrm{avg}, \min, \max\}$.

\fussy

\subsubsection*{T4: Achieving Range Invariance (A4)}

\label{sec:trick4}

\paragraph{T4-a,b: Scaling} 
A common approach to get a unit range for $f$ is to use min-max scaling $f_u=(f-f_{\min})/(f_{\max}-f_{\min})$.
However, determining the minimum and maximum values of $f$ for any data $X$ is nontrivial. Theoretical extrema are usually computed for edge cases far from realistic $X$ and $C$. Wu et al. \cite{wu09kdd}  proposed estimating the worst score over a given dataset $X$ by the expectation $\hat{f}_{\min}=E_{\pi}(f(C^{\pi},X,\delta))$ of $f$ computed over random partitions $C^{\pi}$ of $(X,\delta)$ preserving class proportions $|C^{\pi}_i|=|C_i|{\forall i}$ (\textbf{T4-a}), which are arguably the worst possible clustering partitions of $X$. In contrast, it is hard to estimate the maximum achievable score over $X$, as this is the very objective of clustering techniques. If the theoretical maximum $f_{\max}$ is known and finite, we use it by default; otherwise, if $f_{\max}\rightarrow+\infty$ then the scaled metric $f_u \rightarrow 0, \forall f$. We propose to use a logistic function $f'= 1 / (1 + e^{-k\cdot f})$ (\textbf{T4-b}) before applying the normalization so $f'_{\max}=1$ and $f'_{\min}=\hat{f'}_{\min}$.

\paragraph{T4-c: Calibrating logistic growth rate \textit{k}} 
We can arbitrarily make a logistic function to pull or push all scores toward the minimum or maximum value by tuning the growth rate $k$. 
We thus propose calibrating $k$ with datasets with ground truth CLM scores. 
 Assume a set of labeled datasets $\mathcal{X} = \{X^{1}, \cdots, X^{n}\}$ with class labels
 $\mathcal{C} = \{C^{1}, \cdots, C^{n}\}$ and the corresponding ground truth CLM scores
 $\mathcal{S} = \{s^{1}, \cdots, s^{n}\}$ where $\min \mathcal{S} =0$ (worst) and 
 $\max \mathcal{S} = 1$ (best). Here, we can optimize $k$ to make
 $\mathcal{S}' = \{s'^{1}, \cdots, s'^{n}\}$ best matches with $S$, where 
 $s'^{i} = f(C^{i}, X^{i}, \delta)$. In practice, we use Bayesian optimization \cite{snoek12nips} targeting the $R^2$ score. 

 We propose using human-driven separability scores as a proxy for CLM, building upon available human-labeled datasets acquired from a user study \cite{abbas19cgf} and used in several works on visual perception of cluster patterns \cite{AupetitVIS19},\cite{Abbas2024},\cite{jeon24tvcg}. Each dataset consists of a pair of Gaussian clusters (classes) with diverse hyperparameters (e.g., covariance, position), graphically represented as a monochrome scatterplot. The perceived separability score of each pair of clusters was obtained by aggregating the judgments of 34 participants of whether they could see one or more than one cluster in these plots. The separability score of each dataset is defined as the proportion of participants who detected more than one cluster. 
We used these datasets $\mathcal{X}$ and separability scores $\mathcal{S}$ for calibration because they are not biased by a certain clustering technique or validation measure; they are based on human perception following a recent research trend in clustering \cite{AupetitVIS19},\cite{BlasilliVisualCQM2024}, and the probabilistic scores naturally range from 0 to 1.
 However, as the scores are not uniformly distributed, we bin them and weigh each dataset in proportion to the inverse size of the bin they belong to (see the Appendix of the reference paper \cite{jeon25tpami}).

\subsection{Adjusting Internal Clustering Quality Metrics}

\label{sec:cncadjust}

We use the proposed protocols (\autoref{sec:cncprotocol}) to adjust six internal clustering quality metrics: Calinski-Harabasz index ($CH$) \cite{calinski74cis}, Dunn index ($DI$) \cite{dunn74joc}, I index ($II$) \cite{maulik02tpami}, Xie-Beni index ($XB$) \cite{xie91tpami}, Davies-Bouldin index ($DB$) \cite{davies79tpami}, and Silhouette coefficient ($SC$) \cite{rousseuw87silhouette}, into adjusted internal clustering metrics that satisfy both within- and across-dataset axioms. 
We select the clustering metrics from the survey by Liu et al. 
\cite{liu10icdm}. 
We select every metric except those optimized by the elbow rule (e.g., the modified Hubert $\Gamma$ statistic \cite{hubert85classification}) and those that require multiple clustering results (e.g., the S\_Dbw index \cite{halkidi01icdm}). 
Our choice covers the most widely used internal clustering metrics, which offer a range of approaches for examining cluster structure.

Here, we explain the adjustment of $CH$. We select $CH$ because it does not satisfy any of the across-dataset axioms, thereby allowing us to demonstrate the application of all protocols (\autoref{sec:cncprotocol}). The adjusted $CH$ ($CH_A$) also turned out to be the best adjusted metric in our evaluations (\autoref{sec:cnceval}).

\subsection{Adjusting the Calinski-Harabasz Index}

$CH$ \cite{calinski74cis} is defined as:
\begin{equation}
 CH(C, X, d^2) = \displaystyle\frac{|X| - |C|}{|C| - 1}\cdot\frac{\sum_{i=1}^{|C|}|C_i|d^2(c_i, c)}{\sum_{i=1}^{|C|} \sum_{x \in C_i} d^2(x, c_i)}, 
\end{equation}
where $c_i=\overline{C_i}$ and $c = \overline{X}$. A higher value implies a better CLM. The denominator and numerator measure compactness and separability, respectively. The adjustment procedure is as follows:

\paragraph{Applying T1 (Data-cardinality invariance)}
Both the denominator and numerator of $CH$ are already robust estimators of population statistics (T1). However, as the term $(|X|-|C|)$ makes the score grow proportional to the size of the datasets, we remove the term to eliminate the influence of data-cardinality, resulting in: 
\begin{equation}
 CH_{1}(C, X, d^2) = \displaystyle\frac{\sum_{i=1}^{|C|}|C_i|d^2(c_i, c)}{(|C| - 1)\sum_{i=1}^{|C|} \sum_{x \in C_i} d^2(x, c_i)}.
\end{equation}

\paragraph{Applying T2 (Shift invariance)}
$CH_1$'s numerator and denominator consists of \mbox{type-3}  and type-2 distances, respectively. To equalize the shift before applying exponential (T2-c), we add the sum of the squared distances of the data points to their centroid as a factor to the numerator, which does not affect separability or compactness. This leads to:\
\begin{equation}
\displaystyle
CH_{2}(C, X, d^2) = \displaystyle\frac{\displaystyle\sum_{x \in X} d^2(x, c)}{\displaystyle\sum_{i=1}^{|C|} \displaystyle\sum_{x \in C_i} d^2(x, c_i)} \cdot \frac{\displaystyle\sum_{i=1}^{|C|}|C_i|d^2(c_i, c)}{|C| - 1}.
\end{equation}
As the left term is a fraction of the sum of type-2 distances, we get shift invariance by dividing both the numerator and the denominator by $|X|$ (i.e., the sum becomes an average; T2-b), then by applying the exponential normalized by the standard deviation $\sigma_{d^2}$ of the square distances of the data points to their centroid (T2-a); i.e., $\sigma_{d^2} = \text{std}(\{d^2(x, c) | x \in X\})$.
The right term does not need an exponential protocol as type-3 distances do not shift as the dimension grows. We still divide the term with $|X|$ and $\sigma_{d^2}$ to ensure data-cardinality and scale invariance, respectively. This leads to:
\begin{equation}
\begin{split}
CH_{3}(C, X, d^2) &= \frac{e^{\sum_{x \in X} \frac{d^2(x, c)}{\sigma_{d^2} \cdot |X|}}}{e^{\sum_{i=1}^{|C|} \frac{\sum_{x \in C_i} d^2(x, c_i)}{\sigma_{d^2} \cdot|X|}}} \cdot  \frac{\sum_{i=1}^{|C|}|C_i|d^2(c_i, c)}{\sigma_{d^2} \cdot|X| \cdot (|C| - 1)}.
\end{split}
\end{equation}



\paragraph{Applying T4 (Range invariance)}
We apply min-max scaling to make the measure range invariant. 
As $\max(CH_3) \rightarrow +\infty$, we transform it through a logistic function (T4-b), resulting in:
\begin{equation}
    CH_4 = 1 / (1 +e^{-k \cdot CH_3}), \text{ }\therefore CH_{4\max} \rightarrow 1.
\end{equation}
We then estimate the worst score $CH_{4\min}$ (T4-a) as the average $CH_4$ score computed over $T$ Monte-Carlo simulations with random clustering partitions $C^{\pi}$: 
\begin{equation} \label{eq:montecarlo}
    CH_{4\min}=\frac{1}{T}\sum_{t=1}^T CH_4(C^{\pi_t}, X, d^2),
\end{equation}.

We then get:
\begin{equation}
    CH_5= (CH_4 - CH_{4\min}) / (CH_{4\max} - CH_{4\min}),
\end{equation}
where we set the logistic growth rate $k$ by calibrating the $CH_5$ scores ($\mathcal{S'}$) with human-judgment scores ($\mathcal{S}$) (T4-c).

\paragraph{Applying T3 (Class-cardinality invariance)}
Lastly, we make our measure satisfy class-cardinality invariance (Axiom A3) by averaging class-pairwise scores (T3), which finally determines the adjusted Calinski-Harabasz index:
\begin{equation}
CH_A(C, X, d^2) = \frac{1}{{|C| \choose 2}}\sum_{S \subseteq C, |S| = 2} CH_5(S, X, d^2). 
\end{equation}
Unlike $CH$, which misses all across-dataset axioms, $CH_A$ satisfies all of them (Refer to the Appendix of the original paper \cite{jeon25tpami} for the proofs).

\paragraph{Removing Monte-Carlo simulations}
We can reduce the computing time of $CH_A$ by removing Monte-Carlo Simulations for estimating $CH_{4\min}$. Indeed, as randomly permuting class labels make all $C_i \in C$ satisfy $C_i \overset{D}{=} X$, we can assume $c \simeq c_i$ $\forall c_i$. Therefore, $CH_3(C^{\pi}, X, d^2) \simeq 0$ as it contains $d^2(c_i, c) \simeq 0$ in the second term, which leads to:
\begin{equation} 
\begin{split}
    CH_{4\min} &= E_\pi(CH_4(C^{\pi}, X, d^2)) \\ &= E_\pi(1/2) = 1/2.
\end{split}
\end{equation}
This approximation also makes $CH_A$ deterministic.

\paragraph{Computational complexity}
\autoref{tab:comp} presents the time complexities of the IVMs. 
Since the complexity of $CH_A$ is linear with respect to all parameters, where the only additional parameter compared to $CH$ is $|C|$, the measure scales efficiently to datasets with large sizes and high dimensionality (\autoref{sec:cnceval}).

\begin{table}[t]
\centering
\caption{Time complexity of $CH$, $CH_A$ and their variants generated by applying our protocols (\autoref{sec:cncadjust}). The simulation refers to $T$ runs of Monte-Carlo experiments used to compute $CH_{4\min}$ (\autoref{eq:montecarlo}). IVMs in the first two rows do not depend on Monte-Carlo simulations. }
\label{tab:comp}
\begin{tabular}{lll}
\toprule
\textbf{IVMs} & \textbf{w/ simulation} & \textbf{w/o simulation} \\
\midrule
$CH, CH_1, CH_2, CH_3, CH_4$ & \multicolumn{2}{c}{$O(|X|\Delta_X)$} \\
$CH_{4\max}$ & \multicolumn{2}{c}{$O(1)$} \\
$CH_{4\min}$ & $O(T|X|\Delta_X)$ & $O(|X|\Delta_X)$ \\
$CH_5$ & $O(T|X|\Delta_X)$ & $O(|X|\Delta_X)$ \\
$CH_A$ & $O(T|X|\Delta_X\,|C|)$ & $O(|X|\Delta_X \,|C|)$ \\
\bottomrule
\end{tabular}
\end{table}

\subsection{Adjusting the Remaining IVMs}

\label{sec:remaining}

The adjustment processes of the remaining clustering quality metrics are not significantly different from those of $CH$.
As $DI$ misses all across-dataset axioms, it goes through all protocols like $CH$. $II$, $XB$, and $DB$ require the shift (T2), range (T4), and class-cardinality (T3) invariance protocols as they are already data-cardinality invariant (A1). After passing through these protocols, $II$ and $XB$ become identical (i.e., $II_A = XB_A$). For $SC$, only shift and class-cardinality invariance protocols are required, as data-cardinality and range invariance (A4) are already satisfied. We thus obtain five adjusted internal clustering quality metrics: $CH_A$, $DI_A$, $\{II, XB\}_A$, $DB_A$, and $SC_A$. Please refer to the Appendix of the original reference paper \cite{jeon25tpami} for detailed adjustment processes.

\subsection{Quantitative Evaluation}

\label{sec:cnceval}

We conduct three experiments to evaluate our protocols (\autoref{sec:cncprotocol}) and adjusted metrics. The first and second experiments investigate how well the metrics correlate with the ground truth CLM---across and within datasets---comparing their performance to competitors like standard internal clustering metrics and supervised classifiers (\autoref{sec:cncclm}).
We then investigate the runtime of adjusted metrics and competitors in quantifying the CLM of datasets. 

\subsubsection{Across-Dataset Rank Correlation Analysis}

\paragraph{Objectives and design}
We evaluate five adjusted metrics ($CH_A$, $DI_A$, $\{II, XB\}_A$, $DB_A$, and $SC_A$) against competitors (IVMs and classifiers) for estimating the CLM ranking of labeled datasets. We approximate a ground truth CLM ranking of labeled datasets using multiple clustering techniques. We then compare the rankings made by all competitors to the GT using Spearman's rank correlation.

\paragraph{Datasets} We collect 96 publicly available benchmark labeled datasets from various sources (e.g., UCI ML repository \cite{asuncion07uci} and Kaggle \cite{kaggle2022}), with diverse numbers of data points, class labels, cluster patterns (presumably), and dimensionality\footnote{\href{https://hyeonword.com/clm-datasets/}{hyeonword.com/clm-datasets/}}.

\paragraph{Approximating the ground truth CLM} For the lack of definite ground truth clusters in multidimensional real datasets, we use the maximum external clustering quality score achieved by nine various clustering techniques (see below) on a labeled dataset as an approximation of the ground truth CLM score for that dataset. These ground truth scores are used to obtain the ranking of all data sets based on CLM. 
This scheme relies on the fact that a high external clustering quality implies good CLM (\autoref{sec:cncclm}).
We use Bayesian optimization \cite{snoek12nips} to find the best hyperparameter setting for each clustering technique. 
We obtain the ground truth CLM ranking based on the following four external metrics: adjusted rand index (\texttt{arand}) \cite{santos09icann}, adjusted mutual information (\texttt{ami}) \cite{vinh10jmlr}, v-measure (\texttt{vm}) \cite{rosenberg07emnlp}, and normalized mutual information (\texttt{nmi}) \cite{strehl02jmlr} with geometric mean. 
We select these metrics because they are ``normalized'' or ``adjusted'' so that their scores can be compared across datasets \cite{wu09kdd}, and also widely used in literature \cite{liu18tpami, xiong17tpami, zhang22tpami, chakraborty22tpami}. 
For clustering techniques, we use HDBSCAN \cite{campello13akddm}, DBSCAN \cite{schubert17tds},  $K$-Means \cite{hartigan79jstor, likas03pr}, $K$-Medoids  \cite{park09esa}, $X$-Means \cite{pelleg00icml}, Birch \cite{zhang96sigmod}, and single, average, and complete variants of Agglomerative clustering \cite{mullner11arxiv}.

\paragraph{Competitors} We compare adjusted and standard internal clustering metrics and classifiers, which are natural competitors in measuring CLM (\autoref{sec:cncclm}), to the ground truth ranking. For classifiers, we use Support Vector Machine (SVM), $k$-Nearest Neighbors ($k$NN), Multilayer Perceptron (MLP), Naive Bayesian Networks (NB), Random Forest (RF), Logistic Regression (LR), Linear Discriminant Analysis (LDA), following  Rodr\'iguez et al. \cite{rodriques18kbs}. 
We also use XGBoost (XGB), an advanced classifier based on tree boosting \cite{chen16kdd}. 
We use XGBoost as it adapts well regardless of the datasets' format \cite{chen16kdd, bohacek24trustnlp}, thus being suitable to all the 96 datasets composed of tabular, image, and text datasets.
XGBoost also outperforms recent deep-learning-based models in classifying tabular datasets \cite{grin22nips}, a preponderant type among our datasets.
Finally, we test the ensemble of classifiers.
We measure the classification score of a given labeled dataset using five-fold cross-validation and Bayesian optimization \cite{snoek12nips} to ensure the fairness of the evaluation. The accuracy in predicting class labels is averaged over the five validation sets to get a proxy of the CLM score for that dataset. For the ensemble, we get the proxy as the highest accuracy score among the eight classifiers for each dataset independently \cite{rodriques18kbs}.

\definecolor{lightred}{RGB}{247, 163, 180}
\definecolor{lightlightred}{RGB}{245, 208, 216}

\newcommand{\lred}{\cellcolor{lightred}}
\newcommand{\llred}{\cellcolor{lightlightred}}

\begin{table}[t]
    \centering
    \caption{
    \label{tab:between_rank}
    The results of the across-dataset rank correlation analysis. The numbers are the rank correlations between the approximated ground truth CLM ranking based on nine clustering techniques and the estimated CLM ranking obtained by adjusted metrics, standard metrics, and classifiers. }  
    \scalebox{1}{
    \begin{tabular}{crrrrr}
    \toprule 
   %
    & &  \multicolumn{4}{c}{\makebox[0pt]{Ground truth CLM ranking}} \\     
    & & \texttt{ami} & \texttt{arand} & \texttt{vm} & \texttt{nmi} \\ 
    \midrule
    \multirow{8}{*}{\rotatebox[origin=c]{90}{Classifiers}}                                  & NB & $.4126$ & $.5276$ & $.3157$  & $.3130$\\  
                                & MLP & $.4405$ & $.5386$ & $.3600$  & $.3761$\\ 
                                 & LR & $.4456$ & $.5382$ & $.3666$  & $.3873$\\ 
                                 & XGB & $.4543$ &  $.5247$  &  $.3373$     & $.3377$ \\
& $k$NN & $.4876$ & $.5810$ & $.3974$ & $.4094$\\ 
                                 & RF & $.4893$ & $.5741$ & $.3991$ & $.3889$\\ 
                                     & LDA & $.4999$ &  $.5726$ & $.3945$ & $.3606$\\ 
                                 & SVM & $.5427$ & \llred $.6235$ & $.4625$ & $.4827$\\  
                                 & Ensemble & $.5536$ &  $.6162$ \llred & $.4486$   & $.4531$ \\ 
    \midrule
         \multirow{6}{*}{\rotatebox[origin=c]{90}{IVM}}  
          & $CH$             & $.5923$& \llred $.6222$ & $.4487$ & $.3810$\\
          & $DI$            & $.4026$    & $.3534$ & $.5366$ & $.5979$\\
          & $II$          & $.5668$  & $.5957$ & \llred $.6086$ & \llred $.6454$\\
          & $XB$   & \llred $.6201$ & \llred $.7019$ & $.4934$ & $.4446$\\
          & $DB$   & \llred $.7091$ & \llred $.7513$ &  $.5719$ & $.5015$\\ 
          & $SC$ & $.5648$ & \llred $.6800$ & $.4549$ &   $.4208$\\
    \midrule
           \multirow{5}{*}{\rotatebox[origin=c]{90}{IVM$_{A}$}} 
           & $CH_{A}$   & \lred $^{**}$$.8714$& \lred $^{**}$$.8472$ & \lred $^{***}$$.8300$  & \llred $^{***}$$.7836$ \\
           & $DI_{A} $& \llred $.7293$ & \llred $.7177$& \llred $.7504$& \llred $.7427$ \\
           & $\{II, XB\}_{A}$& \lred $^{*}$$.8463$& \lred $^{*}$$.8442$& \lred $^{*}$$.8060$ & \llred $^{**}$$.7818$ \\
           & $DB_{A}$ & \lred $.8315$ & \lred $.8111$ & \llred $.7856$ & \llred $.7436$ \\ 
           & $SC_{A}$ & \lred$^{***}$$.8955$ & \lred $^{***}$$.8769$ & \lred $^{**}$$.8217$& \llred $^{*}$$.7733$ \\
 
    \bottomrule
    \addlinespace[0.115cm]
    \multicolumn{6}{l}{
        \footnotesize 
        \makecell[l]{ 
            (1) Every result was validated to be statistically significant  \\
            \hspace{3.3mm} ($p < .001)$ by Spearman's rank correlation test.
        }
    } \\
    \multicolumn{6}{l}{
        \footnotesize
        (2) $^{***}$ / $^{**}$ / $^{*}$: 1st- / 2nd- / 3rd-highest scores for each external clustering metrics
    } \\ 
    \multicolumn{6}{l}{\footnotesize
        (3) \textcolor{lightred}{$\blacksquare$} / \textcolor{lightlightred}{$\blacksquare$}: very strong ($>0.8)$ / strong ($>0.6)$ correlation \cite{prion14csn}
    } 
    \end{tabular}
    }
    
\end{table}

\paragraph{Results and discussions} 
\autoref{tab:between_rank} shows that for every external clustering metric used to compute ground truth CLM, adjusted internal metrics outperform the competitors; first (***), second (**), and third (*) places are all part of the adjusted metric category. These metrics
achieve about 17\% ($DB$) to 81\% ($DI$) of performance improvement compared to the standard form (average: 48\%), and have \textit{strong} (light-red cells) or \textit{very strong} (red cells) correlation with ground truth CLM ranking according to Prion et al.'s criteria \cite{prion14csn}. These results show that the adjustment procedure (T1-T4) relying on the new axioms (A1-A4), is beneficial to all internal clustering metrics, systematically increasing their correlation with the ground truth ranking. Hence, adjusted metrics are the most suitable ones to compare and rank datasets based on their CLM. Within the adjusted metrics, $CH_A$ and $SC_A$ show the best performances, with a slight advantage for $CH_A$ (first place for both \texttt{vm} and \texttt{nmi}, and runner-up for both \texttt{ami} and \texttt{arand}).

In contrast, as expected, supervised classifiers fall behind the standard and adjusted metrics, indicating that they should not be relied upon for predicting CLMs. 
A notable finding is the most advanced model, XGB, shows relatively poor performance in estimating CLM compared to classical models such as SVM, $k$NN, and LDA; even an ensemble of classifiers falls behind SVM in terms of \texttt{arand}, \texttt{vm}, and \texttt{nmi} (\autoref{tab:between_rank}). 
This is because XGB and ensemble classifiers effectively discriminate classes regardless of whether they are well-separated by a large margin or not in the data space, leading them to classify most datasets as having similarly good CLM. 
This finding indicates that improving classification accuracy does not necessarily help achieve better CLM measurement, further emphasizing the significance of our contribution.

\subsubsection{Within-Dataset Rank Correlation Analysis}

\begin{table}
    \centering
        \caption{The results of within-dataset rank correlation analysis.
    We compared the pairs of rankings obtained by standard (St.) and adjusted (Adj.) internal metrics, with the ground truth noisy labels ranking (NR) on the 96 collected datasets. 
    }
    \begin{tabular}{r|c|cc}
    \toprule
    IVM    & \textit{St.} vs. \textit{Adj.}  &  NR vs. \textit{St.}  & NR vs. \textit{Adj.} \\
    \midrule
    $CH$    & \lred $.848\pm.263$ & \lred $.876\pm.296$ & \lred $.879\pm.277$ \\
    $DI$ &        $.253\pm.429$ &        $.451\pm.335$ &        $.381\pm.727$ \\
    $II$   & \lred $.825\pm.296$ & \lred $.857\pm.284$ &  \lred $.881\pm.307$ \\
    $XB$   &\lred $.820\pm.308$  & \lred $.832\pm.341$ & \lred $.881\pm.307$ \\
    $DB$    & \lred $.855\pm.257$ & \lred $.884\pm.268$ &  \lred $.876\pm.294$ \\
    $SC$   &        $.515\pm.574$ &        $.530\pm.601$ &\lred $.878\pm.322$ \\
    \bottomrule
    \addlinespace[0.115cm]
    \multicolumn{4}{l}{\footnotesize
    \makecell[l]{ 
            \textcolor{lightred}{$\blacksquare$} / \textcolor{lightlightred}{$\blacksquare$} : very strong ($>0.8)$ / strong ($>0.6)$ \\ 
            \hspace{8mm} correlation \cite{prion14csn}
        }
        
    } 
    \end{tabular}

    \label{tab:within_rank}
\end{table}

\noindent
\textbf{Objectives and design.}
We want to evaluate the adjusted internal clustering metrics' ability to evaluate and compare CLM within a dataset, which is the original purpose of internal metrics.
For this purpose, we generate several noisy label variants of each dataset and compare how the scores achieved by IVMs and their adjusted counterparts are correlated with the ground truth noisy label ranking (NR).
Assume a set of datasets $\mathcal{X} = \{X_i | i = 1, \cdots, n\}$ and their corresponding labels $\{C_j | j = 1, \cdots, n\}$. 
For each dataset $(X_k, C_k)$, we run the following process.
First, we generate 11 noisy label variants of each dataset $\{C_{k,l}| l = 0, \cdots, 10\}$ by randomly shuffling $l \cdot 10$\% of their labels.
The $l$-th noisy label dataset is authoritatively ranked at the $(11-l)$-th place of the NR (i.e., the larger the proportion $l$ of shuffled labels is, the lower the expected CLM). 
Then, for each standard internal metrics $Z \in \{CH, DI, II, XB, DB, SC\}$ and its corresponding adjusted metrics (i.e., $Z_A$), we compute the CLM ranking of these noisy label datasets based on 
$Z(C_{k, l}, X_k, \delta)$ and $Z_A(C_{k, l}, X_k, \delta)$, respectively.
We examine how the ranking generated by standard and adjusted metrics
are similar to NR using Spearman's rank correlation. 
We also check the rank correlation between the rankings from standard and adjusted metrics.

\paragraph{Datasets}
For $\mathcal{X}$, we use the 96 labeled datasets from the between-dataset rank correlation analysis.

\paragraph{Results and discussions}
As shown in \autoref{tab:within_rank},
every adjusted metric has a very strong rank correlation ($> 0.8$) with both NR and IVM for every case except for $DI_A$. 
The adjusted metrics showed equal ($CH$, $DI$, $II$, $XB$, $DB$) or better ($SC$) performance in estimating the CLM within a dataset.
We also see that the discrepancy between the rankings made by standard and adjusted metrics follows the one between standard metrics and GT noisy labels ranking.
Such results verify the effectiveness of our protocols and adjusted metrics in precisely measuring CLM within a dataset.

\subsubsection{Runtime Analysis}

\begin{figure}[t]
    \centering
    \includegraphics[width=0.7\linewidth]{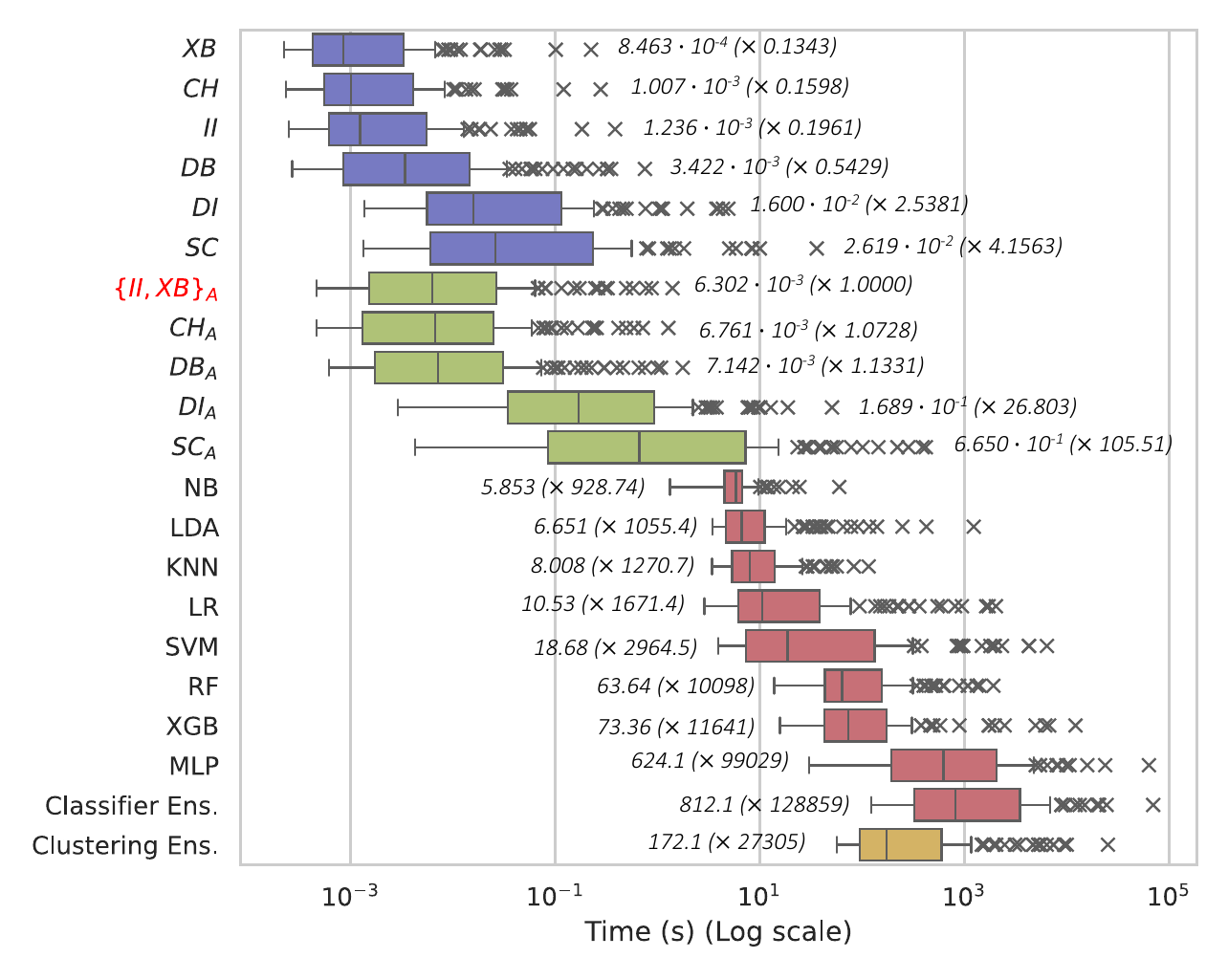}
    \caption{The runtime of the standard metrics (blue), adjusted metrics (green), classifiers (red), and the clustering ensemble (Ens.; yellow) in computing the CLM of 96 datasets. The numbers next to each box depict the median runtime of the corresponding measure (left) and the relative time compared to $\{II, XB\}_{A}$, the fastest adjusted metrics  (e.g., $CH_A$ is  1.07 times slower than $\{II, XB\}_{A}$).
    }
    \label{fig:cncruntime}
\end{figure}

\paragraph{Objectives and design}
We compare the runtime of the approaches explored in previous experiments to estimate the CLM of 96 labeled datasets.
For classifiers and the clustering ensemble, we measure the total optimization time. 
We run all the experiments in a Linux server machine equipped with 40-core Intel Xeon Silver 4210 CPUs, TITAN RTX, and 224GB RAM.

\paragraph{Results and discussion}
As a result (\autoref{fig:cncruntime}), adjusted metrics are up to one order of magnitude slower than $XB$, the fastest standard metrics
However, they are up to \textit{four orders of magnitude} ($\times 10,000$) faster than the competitors, like clustering ensembles used to estimate ground truth CLM. This verifies that most adjusted metrics, among which is $CH_A$, show an excellent tradeoff between accuracy and speed. Despite $SC_A$ being as accurate as $CH_A$, it is \textit{two orders of magnitude} slower ($\times 0.01$), making $CH_A$ the best adjusted metric to use in practice. This finding leads us to use $CH_A$ for our new metrics for DR in \autoref{sec:cncltlc}.

\section{Label-Trustworthiness and Label-Continuity}

\label{sec:cncltlc}

We introduce \LT and \LC (Label-T\&C) as metrics that implement our new process of evaluating DR projections using class labels (\autoref{sec:proposedevaluation}). 
\lt and \lc capture the False and Missing Groups distortions, respectively. 
It is worth noting that these metrics are named after Trustworthiness and Continuity, two local distortion measures that focus on capturing False and Missing Neighbors \cite{venna06nn}.

\subsection{Metric Design}

\label{sec:cncdesign}

We discuss the design of \ltc.

\paragraph{Notations}
We define a high-dimensional data $\mathbf{X} = \{\mathbf{x}_i \in \mathbb{R}^D, i = 1, 2, \cdots, N\}$.
We denote the low-dimensional embedding of $\mathbf{X}$ as $\mathbf{Z} = \{\mathbf{z}_i \in \mathbb{R}^d \mid i = 1, 2, \cdots, N\}$, where $D > d$. 
For any set $\mathbf{S}\in\{\mathbf{X},\mathbf{Z}\}$, the distance function $\delta$ satisfies $\delta(x,y) \geq 0$, $\delta(x, y) = \delta(y,x)$ and $\delta(x,y)=0$ if $x=y$  $\forall x, y \in \mathbf{S}$.
A partition of $\mathbf{S}$ is defined as $\mathbf{P}=\{P_1, P_2, \cdots, P_k\}$ satisfying $P_i\subseteq \mathbf{S}$, $P_i \cap P_j = \emptyset$ and $\cup^{k}_{i=1}P_i = \mathbf{S}$.
If a partition is defined by class labels, we denote the partition as $\mathbf{P}_L$.

\paragraph{Inputs, output, and hyperparameters} \ltc take (1) the high-dimensional data $\mathbf{X}$, (2) its DR projection $\mathbf{Z}$, and (3) class labels $\mathbf{\mathbf{P}_L} = \{P_{L,1}, P_{L,2}, \cdots P_{L,k} \}$ as inputs. Both \lt and \lc output a number between 0 and 1; a higher value indicates lower distortions and a better projection. For hyperparameters, an internal clustering quality metric $f$ is provided. The $f$ should assign higher scores to better clusterings and range from 0 to 1 (refer to \autoref{sec:cncltncreq} for a detailed explanation about this requirement). Note that we do not consider external clustering quality metrics as they reuqires clustering techniques as additional hyperparameters.


\noindent
\textbf{Step 1. Measuring CLM in the original data and projection}
We apply an internal clustering metric to both the original and projection spaces to examine CLM. Here, unlike the general process of label-based DR evaluation that applies clustering metric to all classes at once, we apply CVM to every \textit{pair} of classes, so that we can take account of the relationships of classes in more detail.
Formally, we construct the class-pairwise CLM matrices $M(\mathbf{X})$ and $M(\mathbf{Z})$, where the $(i, j)$-th cell of the matrices $M(\mathbf{S})_{i,j}$ ($\mathbf{S} \in \{\mathbf{X}, \mathbf{Z}\}$) is defined as:
\[
\left\{
\begin{array}{rcl} 
f(\{P_{L, i}, P_{L, j}\}, \mathbf{S}, \delta) & \text{if} & i \neq j \text{ and $m$ is an IVM} \\ 
f(C(P_{L, i} \cup P_{L, j}, \delta), \{P_{L, i}, P_{L, j}\}) & \text{if} & i \neq j \text{ and $m$ is an EVM} \\
0 & \text{if} & i=j 
\end{array}
\right..
\]

\noindent
\textbf{Step 2. Computing distortion matrices}
We construct a matrix $M^{*} = M(\mathbf{X}) - M(\mathbf{Z})$ representing CLM distortions. We then compute $M^{FG}$ and $M^{MG}$, where $M^{FG}_{i,j} = (M^{*}_{i,j}$ if $M^{*}_{i,j} >0$, else $0)$, and $M^{MG}_{i,j} = (-M^{*}_{i,j}$ if $M^{*}_{i,j} <0$, else $0)$. 
$M^{FG}$ and $M^{MG}$ abstract the CLM decrement (False Groups) and increment (Missing Groups), respectively.

\noindent
\textbf{Step 3. Aggregating distortions}
Finally, we average the upper-diagonal elements of $M^{FG}$ and $M^{MG}$ into final scores:

\begin{itemize}
    \item[] \hspace{30mm} \textsc{\LT}: $1 - \mbox{avg}_{i, j} M^{FG}_{i > j}$ \vspace{-1mm}
    \item[] \hspace{30mm} \textsc{\LC}: $1 - \mbox{avg}_{i, j} M^{MG}_{i > j}$.
\end{itemize}
Note that we subtract the average from 1 to make projections with fewer distortions receive higher quality scores.

\subsection{Selecting Internal Clustering Quality Metrics for Label-T\&C}

\label{sec:cncltncreq}


We select clustering quality metrics based on the following proposition: to be used for \ltc, a proper metric $f$ should be comparable across $\mathbf{X}$ and $\mathbf{Z}$.
In other words, $f$ shall consider only \textit{how well the given partition is clustered in the given data} and be invariant to the characteristics that differ between $\mathbf{X}$ and $\mathbf{Z}$ but are not related to the cluster structure. 
For example, the scaling of the pairwise distances should not alter the score. Otherwise, the evaluation will be unreliable; for example, we can simply manipulate \ltc scores by scaling the original data or projection while there is no change in the cluster structure.

We first consider our adjusted internal clustering metrics (\autoref{sec:adjusted}) as a natural candidate for $f$.
We consider adjusted metrics because they satisfy axioms that ensure comparable scores across datasets with different visual patterns, dimensionality, and numbers of classes. 
We select the \textbf{adjusted Calinski-Harabasz Index} ($CH_A$) as our final candidate metric because it demonstrates the best accuracy in estimating CLM and is also scalable (\autoref{sec:cnceval}).

Additionally, we select \textbf{distance consistency (DSC)} \cite{sips09cgf} as a candidate for $f$, as it also produces scores that are comparable across datasets.
DSC is defined as the number of data points closer to the centroid of another class than their own in the data, normalized by the total number of data points. As DSC ranges from $0.5$ to $1$ if the number of classes is two and assigns a lower score for a better CLM, we use the value $2(1 - \text{DSC})$. We discuss how DSC satisfies our requirements in Appendix X.

\subsection{Guidelines to Interpret \ltc}

\label{sec:cncguideline}

\begin{figure}[t]
    \centering
    \includegraphics[width=0.7\linewidth]{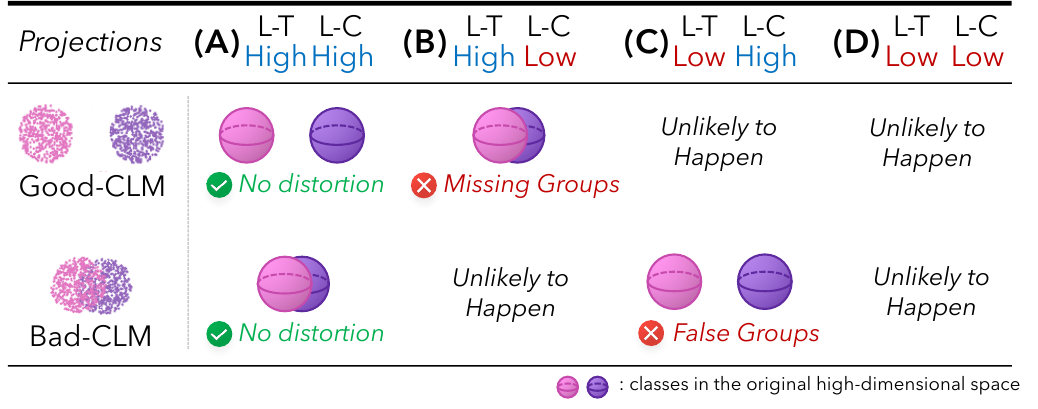}
    \caption{Guidelines to infer the CLM of the high-dimensional data based on the CLM of the projection (left column) and the scores given by \lt (L-T) and \lc (L-C) (first row) (see \autoref{sec:cncguideline} for details).
    }
    \label{fig:cncguide}
\end{figure}

We present guidelines for interpreting the faithfulness of DR projections based on \ltc.
If \lt and \lc are both high, the CLM of the projection accurately depicts the CLM in the original space (\autoref{fig:cncguide}A). 
High \lt and low \lc (\autoref{fig:cncguide}B) mean that Missing Groups distortions have occurred, i.e., the CLM of the original data is worse than it appears in the projection (first row); some pairs of classes appear more separated than they actually are in the data space. 
When the CLM of the projection is already low (\textit{e.g.} due to overlapping classes), Missing Groups distortions are more unlikely to happen as the CLM in the data would have to be even worse (second row).
In contrast, low \lt and high \lc (\autoref{fig:cncguide}C) inform that False Groups distortions have occurred; the CLM in the original data is better than in the projection (second row).
As False Groups distortions deteriorate the CLM of the projection, the situation is unlikely to occur if the projection has a good CLM, and thus can hardly become better (first row).
Due to such a tradeoff between False and Missing Groups (i.e., more Missing Groups lead to fewer False Groups, and vice versa), 
it is unlikely to get low \lt and \lc at the same time (\autoref{fig:cncguide}D). Our evaluation (\autoref{sec:cncltnceval}) confirms the existence of the tradeoff. 

\subsection{Sensitivity Analysis}

\label{sec:cncltnceval}

We conduct quantitative experiments to evaluate \ltc with DSC and \CHb, i.e., \ltc [DSC] and \ltc [\CHb{}], respectively. We do so by checking the accuracy of \ltc and competitors in quantifying distortions.

\subsubsection{Competitors}
We first consider DR evaluation metrics without labels as competitors. 
For global metrics, we use KL divergence and DTM. T\&C and MRRE are used as representative local metrics. MRRE [Missing] and MRRE [False] target Missing and False Neighbors, respectively.
We select S\&C as the sole pair of metrics targeting cluster-level distortions. 
For the metrics using labels (\autoref{sec:cnclabelsmetric}), we first add CA-T\&C. 
We then select Silhouette and DSC as representative clustering quality metrics used in the general label-based evaluation. 
For T\&C, MRRE, and CA-T\&C, we average their score across $k$-nearest neighbor values: $k = [5, 10, 15, 20, 25]$, following Jeon et al. \cite{jeon21tvcg}. For KL divergence and DTM, we average the scores across different standard deviation values of Gaussian kernels $\sigma$: $[0.01, 0.1, 1]$, following Moor et al. \cite{moor20icml}. For S\&C, we use the default hyperparameter setting \cite{jeon21tvcg}.

\subsubsection{Objectives and Design}

\begin{figure*}[]
    \centering
    \includegraphics[width=\textwidth]{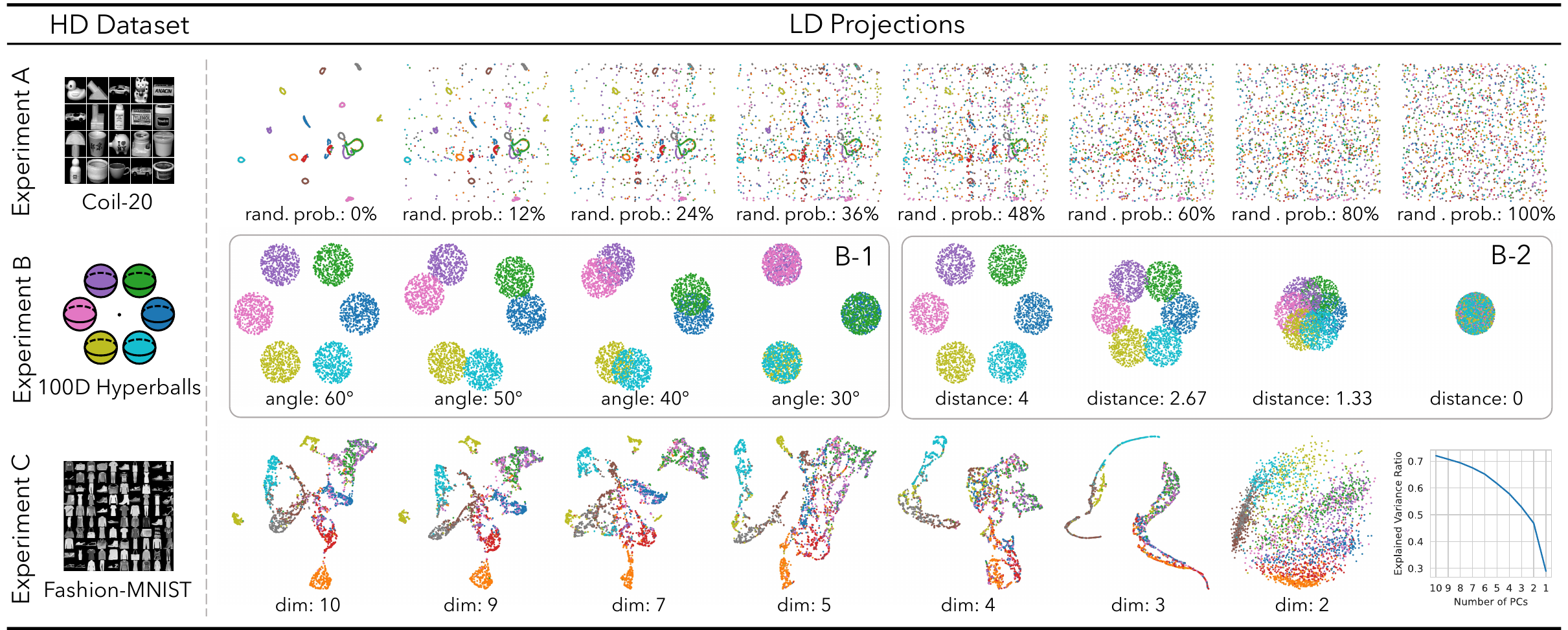} 
    \caption{The high-dimensional (HD) datasets and low-dimensional (LD) projections used in experiments A, B, and C of sensitivity analysis (\autoref{sec:cncltnceval}). The experiments aim to check the DR evaluation metrics'  ability to capture False Groups distortions. Class labels are mapped to colors. (A) The Coil-20 \cite{nene96tech} dataset and the projections generated by randomizing the positions of the embedded points with a certain probability. 
    (B) An HD dataset consists of six well-separated hyperballs (left) and their synthetic projections (right) made by initializing the embedding with six well-separated discs and gradually overlapping the discs in two different manners (B-1, 2). (C) The Fashion-MNIST \cite{xiao2017arxiv} dataset and the PCA projections with different numbers of principal components (PC); here we depict the UMAP projection of PCA embeddings if it has more than two PCs (i.e., dimensionality is higher than two). We depict the relation between the explained variance ratio and the number of PC in the line chart next to the projections.}
    \label{fig:sendataabc}
\end{figure*}

\begin{figure*}[]
    \centering
    \includegraphics[width=\textwidth]{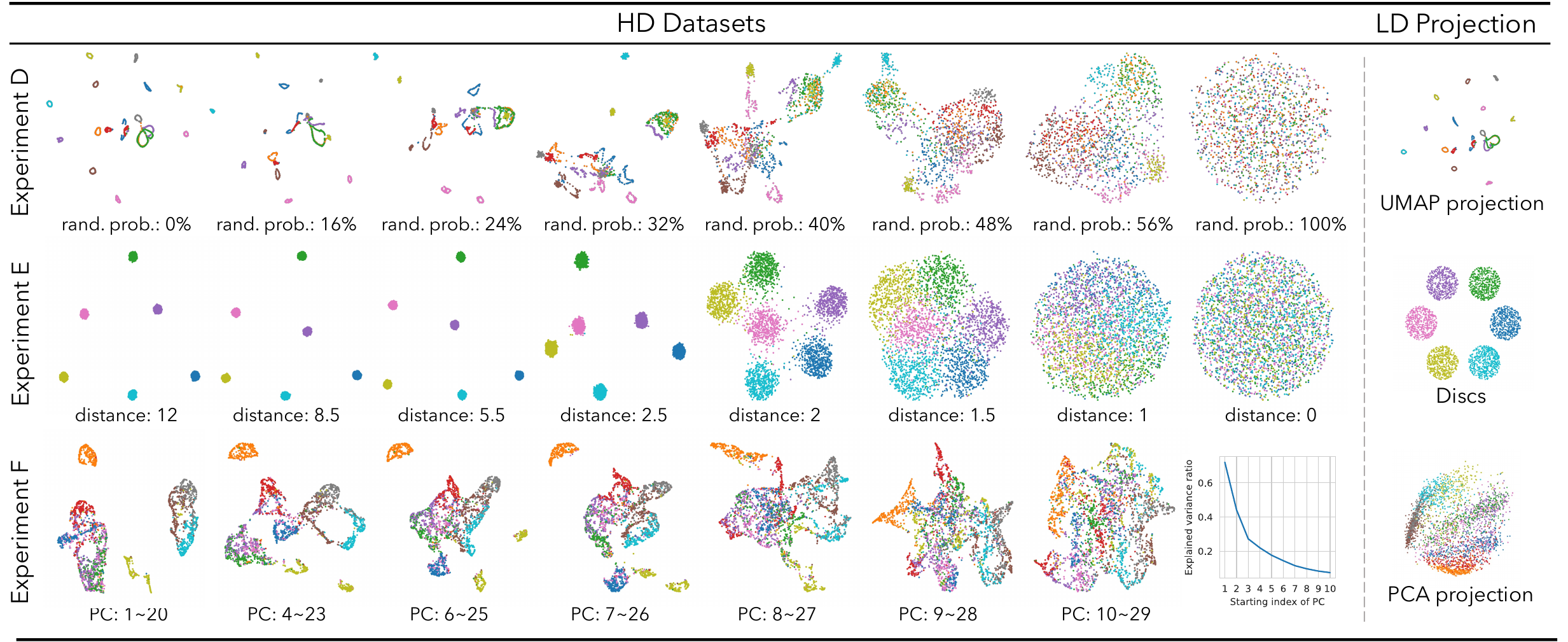} 
    \caption{The low-dimensional (LD) projections and corresponding high-dimensional (HD) datasets represented as UMAP projections, used in experiments D, E, and F of sensitivity analysis to examine DR evaluation metrics' ability to capture Missing Groups distortions.
    (D) An UMAP embedding of the Coil-20 \cite{nene96tech} dataset (right), and the variants of the Coil-20 dataset made by randomizing the coordinates of data points in HD space with a certain probability. 
    (E) A 2D projection with six well-separated discs and synthetic HD datasets. We create the datasets by generating six 100D hyperballs and gradually overlapping them. (F) A 2D PCA projection of the Fashion-MNIST dataset and corresponding HD datasets variants, created by slicing 20 principal components (PC) with different rankings. The line chart shows their corresponding explained variance ratio. }
    \label{fig:sendatadef}
\end{figure*}

We conduct six experiments (A-E) to examine \ltc{}'s sensitivity in quantifying False Groups (Fixed data and variable projections in experiments A, B, and C) or Missing Groups (Variable data and fixed projections in experiments D, E, and F) distortions.
By doing so, we want to check whether (1) \ltc precisely captures the distortions and (2) successfully mitigates the bias in previous label-based metrics towards good CLM. 
The labeled data and projections used in the experiments are shown in \autoref{fig:sendataabc} (A, B, and C) and \autoref{fig:sendatadef} (D, E, and F).
In all of them, we run \ltc and competitors to evaluate the projections.

\paragraph{Experiment A: Randomizing projections}
We examine whether \ltc and competitors can accurately quantify False Groups distortions. 
We generate a 2D UMAP projection of the Coil-20 \cite{nene96tech}  dataset. We then create variants of the projection with different levels of False Groups distortion by randomly permuting the point locations.
We create 21 variants, ranging the replacement probability from 0\% (same as the original projection) to 100\% (totally randomized) with an interval of 5\%. 
The original class assignments of Coil-20 are used as labels. 
We hypothesize that \lt will decrease as the replacement probability increases, properly capturing False Groups distortions, whereas \lc will ignore the distortions.


\paragraph{Experiment B: Overlapping discs}
We aim to check the DR evaluation metrics' ability to precisely capture False Groups distortions, as with experiment A.
We create a high-dimensional dataset consisting of six hyperballs with a radius of 5 lying in 100 dimensions. 
We set the hyperballs to be equidistant ($=10$) from the origin. We then create an artificial 2D projection consisting of six discs (radius of 1.5) evenly and equidistantly ($=4$) distributed around the origin $O$.
Data points and labels within each disc correspond to those of each hyperball.
The positions of each point within the disc and hyperball are determined randomly.
The label is also set based on the disc each point belongs to.
We gradually overlap the discs to artificially generate distortions. Here, we use two overlapping schemes to evaluate the sensitivity of \ltc in detail, resulting in two separate subexperiments (B-1, B-2). In B-1, three independent pairs of adjacent discs are overlapped; for each pair of discs $(A, B)$ with centers $C_A$, $C_B$, we adjusted $\angle C_A O C_B$ from $60^{\circ}$ to $0^{\circ}$ with an interval of $2.4^{\circ}$ (25 projection variants in total). In B-2, we overlap all discs at once by moving them toward the origin; for each disc $A$, we gradually decrease $C_A O$ from 4 to 0 with an interval of 0.16 (25 embedding variants). 
We hypothesize that the \lt score will go down as False Groups distortions increase due to the overlap of the discs, while \lc will stay still. 
We also hypothesize that \lt will decrease more in B-2 than in B-1, as the overlap is larger.

\paragraph{Experiment C: Decreasing the dimension of the projection space}
We generate False Groups distortions by decreasing the dimensionality of projections and check whether the evaluation metrics can detect the distortions. 
We prepare the Fashion-MNIST \cite{xiao2017arxiv} as a high-dimensional dataset. We generate PCA projections with a decreasing number of top principal components (10 to 1 with an interval of 1; 10 embeddings in total). 
We expect projections with fewer principal components (i.e., embeddings lying in lower-dimensional spaces) to exhibit more False Groups distortions, as they have a lower explained variance ratio (see the line chart in \autoref{fig:sendataabc}). 
We use the class assignments of the Fashion-MNIST dataset as labels.
Our hypothesis is that \lt will decrease as the dimensionality decreases, while \lc will stay still.

\paragraph{Experiment D: Randomizing the original data}
We want to evaluate \ltc and competitors' capability in accurately quantifying Missing Groups distortions. 
We first generate a fixed 2D UMAP projection of the Coil-20 \cite{nene96tech} dataset. 
We then generate the variants of the original data by mixing the points in the high-dimensional space with a fixed probability, producing Missing Groups distortions. We control the replacement probability from 0\% to 100\% with an interval of 5\%, resulting in 21 variants. 
The class assignments of the original data are used as labels.
We hypothesize that \lc will decrease as Missing Groups distortions increase (i.e., replacement probability increases), and that \lt will ignore the distortions. 
Accepting this hypothesis implies that \ltc mitigates the bias toward favoring projections with good CLM, i.e., projections that may understate Missing Groups distortions.

\paragraph{Experiment E: Overlapping hyperballs}
We want to evaluate whether \ltc and competitors can precisely capture Missing Groups distortions. 
We prepare variants of high-dimensional data and a fixed low-dimensional projection consisting of six 100D hyperballs and corresponding 2D discs, respectively. 
The points within the same disc have the same label. All discs are well separated from each other.
We artificially overlap hyperballs to generate Missing Groups distortions. 
For each hyperball $A_H$, we gradually decrease $C_{A_H}O$ from 4 to 0 with an interval of 0.16 (25 variants in total). 
As with experiment D, we hypothesize that \lc will decrease as hyperballs overlap, while \lt will stay still.

\paragraph{Experiment F: Decreasing the dimension of the original data space}
We examine whether the evaluation metrics can detect the Missing Groups distortions made by the decrease in the dimensionality of the original data. We prepare a 2D PCA projection of the Fashion-MNIST dataset. 
We then select ten 20D PCA projections with different sets of principal components as high-dimensional datasets; the $i$-th dataset variant consists of the $(i)$-th to $(i+19)$-th principal components, where $1 \leq i \leq 10$.
We expect the dataset with a higher order to exhibit more Missing Groups distortions in the projection, as it has a lower explained variance ratio (line chart in \autoref{fig:sendatadef}). 
We used the class labels from the Fashion-MNIST dataset.
Again, we hypothesize that \lc will decrease as the starting index of principal components increases, while \lt will remain the same.

\subsubsection{Results}

\begin{sidewaysfigure}
    \centering
    \includegraphics[width=\textwidth]{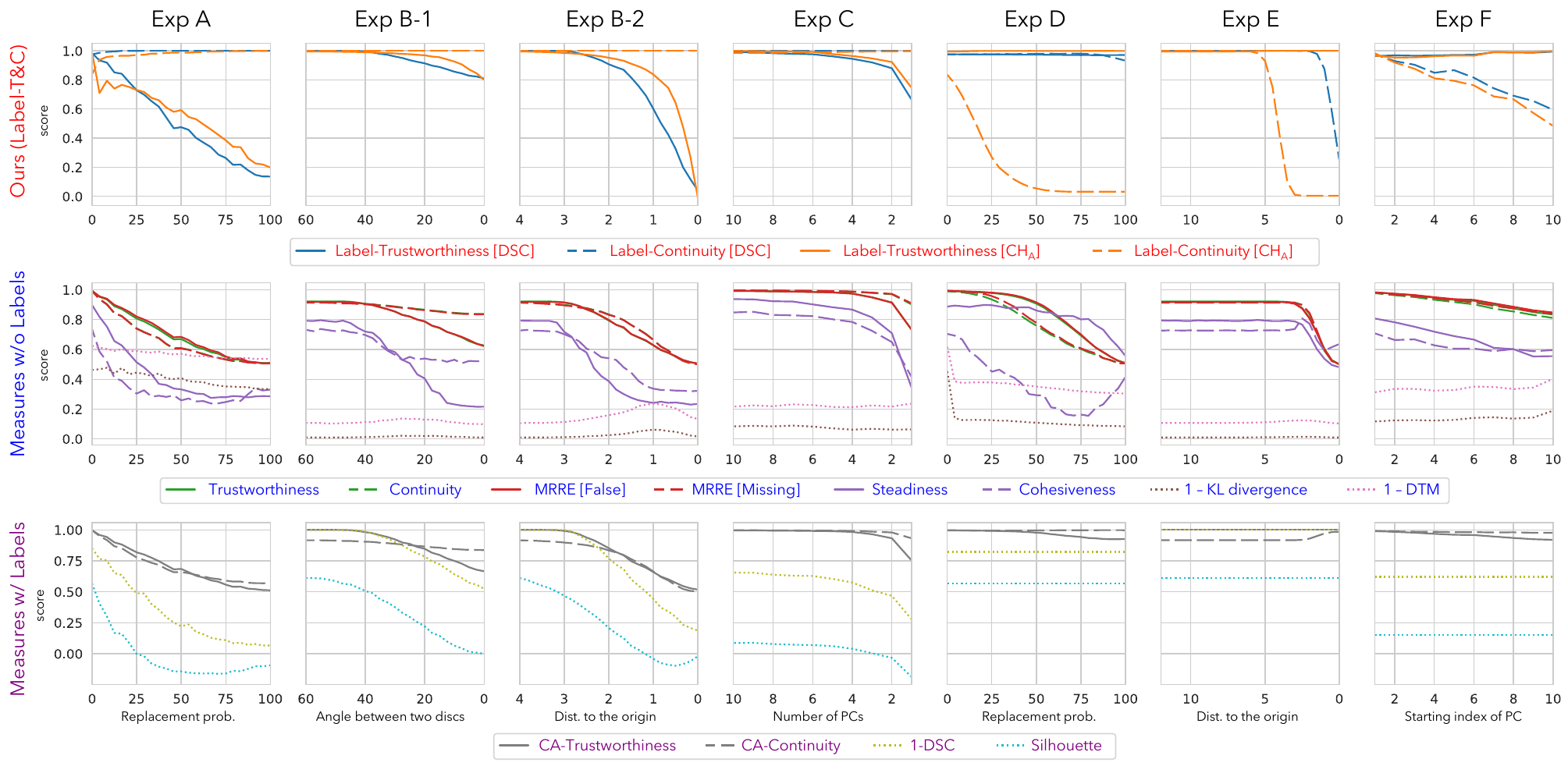} 
    \caption{The results of the sensitivity analysis (\autoref{sec:cncltnceval}; experiments A-F). Solid lines and dashed lines represent the evaluation metrics that focus on compression (e.g., False Groups, False Neighbors) and stretching (e.g., Missing Groups, Missing Neighbors), respectively. Dotted lines represent global metrics and clustering quality metrics. A pair of compression and stretching measures is represented with the same line color. 
    Measure names in red, blue, and purple correspond to our approach, the metrics without labels, and the metrics with labels, respectively.
    In summary, \lt (blue and orange bold line) and \lc (blue and orange dotted line) accurately detect Missing and False Groups distortions, respectively. Meanwhile, all other metrics, including the metrics representing the previous label-based DR evaluation process (i.e., DSC and Silhouette), fail to capture these distortions.
    These results also confirm that \ltc addresses the bias in previous label-based DR evaluation metrics.
    }
    \label{fig:cncsenexp}

\end{sidewaysfigure}

\autoref{fig:cncsenexp} shows the results of our experiments that we comment on below. 

\paragraph{Experiment A} As the randomization probability grows, both \lt [DSC] and \lt [\CHb{}] similarly decrease linearly while \lc [DSC] and \lc [\CHb{}] slightly increase, confirming our hypothesis. 
Meanwhile, S\&C and local metrics decrease regardless of the distortion type, while global metrics slightly increase.
In the case of label-based metrics, both CA-T\&C and the clustering quality metrics (DSC and Silhouette) show mainly decreasing scores.

\paragraph{Experiment B} In B-1, as the overlap between the discs grows, both \lt [DSC] and \lt [\CHb{}] decrease in a similar manner, while \lc{}s stay still. 
Such results validate our hypothesis, confirming \ltc's capability in properly detecting False Groups distortions. 
Meanwhile, S\&C,  T\&C, and MRREs all decrease, while Steadiness, Trustworthiness, and MRRE [False] decrease more than Cohesiveness, CA-Continuity, and MRRE [Missing], respectively. 
Global metrics stay still.
CA-T\&C partially succeed in properly detecting False Groups distortions; both CA-Continuity and CA-Trustworthiness decrease, but CA-Continuity's decrement was subtle compared to that of CA-Trustworthiness. 
Clustering quality metrics show a decreasing trend.
In B-2, the amount of decrement becomes bigger than in B-1 for \lt [DSC] and \lt [\CHb{}] while \lc{}s again stay still, confirming our second hypothesis.
The amount of decrement also becomes bigger than in B-1 for  T\&C, MRREs, and Cohesiveness, while Steadiness showed a similar drop as in B-1. 
In the case of KL divergence, DTM, and Silhouette, the patterns are almost identical to B-1 except that the scores rebound when the discs are nearly overlapped. 
The decrement also becomes larger for CA-T\&C and DSC.

\paragraph{Experiment C} As the number of PCs decreases, \lt{}s decrease while \lc{}s stay still, validating our hypothesis. 
Global metrics (KL divergence, DTM) stay still while all other measures decrease.

\paragraph{Experiment D} As we increase the randomization probability, both \lc [DSC] and \lc [\CHb{}] decrease, while \lt{}s stay still, verifying our hypothesis. However, while \lc [DSC] decreases right before the data are perfectly mixed, \lc [\CHb{}] decreases from the start. 
For local metrics, both T\&C and MRREs decrease. 
Steadiness decreases, while Cohesiveness suddenly goes up after decreasing for a while.
Global metrics (KL divergence, DTM) increase in general.
CA-Trustworthiness goes down while CA-Continuity and clustering quality metrics (DSC and Silhouette) remain unchanged. 

\paragraph{Experiment E} When the overlap between hyperballs increases, both \lc [DSC] and \lc [\CHb{}] decrease, while \lt{}s stay still, verifying our hypothesis. However, as in experiment D, \lc [DSC] and \lc [\CHb{}] decrease differently; while \lc [DSC] decreases right before the hyperballs perfectly overlap, \lc [\CHb{}] decreases before \lc [DSC] does. 
Meanwhile, local metrics (T\&C, MRRE) decrease, while global metrics (KL divergence, DTM) remain unchanged.
Steadiness decreases, while Cohesiveness temporarily rises as Steadiness declines. 
CA-Trustworthiness maintains a maximum score while the CA-Continuity score increases before the perfect overlap of the hyperballs. Clustering metrics stay still.

\paragraph{Experiment F} The results confirm our hypothesis;  as the starting index of the PCs that we slice increases, both \lc [DSC] and \lc [\CHb{}] decrease while \lt{}s stay still. 
Local metrics (T\&C, MRRE) decrease, and global metrics (KL divergence, DTM) stay still.
S\&C decrease, while Steadiness decreases more than Cohesiveness.
CA-T\&C show a similar trend; CA-Trustworthiness decreases, while CA-Continuity decreases to a smaller extent. Clustering quality metrics stay still.

\subsubsection{Discussions}

\noindent
\textbf{\ltc and competitors' capability in detecting cluster-level distortions.}
The results from experiments A-C confirm that \lt is sensitive to False Groups distortions, while \lc is not, as we intended. 
Moreover, the difference between the B-1 and B-2 results validates \lt{}'s accuracy at measuring the amount of False Groups distortions.
The results from experiments D-F, on the other hand, confirm that \lc accurately captures Missing Groups distortions, while \lt ignores them.

The general label-based DR evaluation using clustering quality metrics (DSC and Silhouette) successfully detects False Group distortions in experiments A-C.
However, in experiments D-F, the process fails to detect the Missing Groups distortions. Moreover, the process does not have a specific focus on distortion type and thus cannot explain whether the False or Missing Groups distortions occurred.
Such results confirm the bias of these evaluation metrics towards favoring projections that intentionally reduce Missing Groups distortions (i.e., exaggerate CLM), where \ltc successfully mitigate such bias.

Meanwhile, experiment results also validate that previous metrics fail to accurately detect the distortions or to distinguish specific distortion types.
Global metrics (KL Divergence, DTM) hardly discover distortions for all six experiments. 
Local metrics (T\&C, MRRE) fail to pinpoint specific distortion types; all measures decrease regardless of the type of distortion they aim to measure.
Cluster-level measures (S\&C) fail to distinguish False Groups distortions in experiments A-C. 
For experiments D-F, the situation is even worse; Steadiness reacts more sensitively to Missing Groups distortions, although it was originally designed to aim at False Groups distortions.
CA-T\&C succeed in pinpointing False Groups distortions for B-1, but fails to do so for the remaining experiments.


\noindent
\textbf{Effect of the choice of clustering quality metrics on \ltc.}
\ltc{}s with two different internal clustering quality metrics (DSC or \CHb{}) show a consistent pattern in experiments A-C. However, they behave differently in experiments D and E; 
\lc [\CHb{}] starts decreasing for the lower level of generated CLM distortions than \lc [DSC]. This observation may be specific to clustering quality metrics, as DSC and \CHb use different schemes in examining how the classes are clustered. 
In DSC, the score only drops when classes overlap. 
Therefore, \lc [DSC]  is sensitive to Missing Groups distortions only if the \textit{overlapped} classes in the original space are more separated in the embedding. 
In contrast, \CHb{} decreases as the proximity between classes increases, whether the classes overlap or not. Thus, when proximity increases, \lc [\CHb{}]  is more sensitive to Missing Groups distortions than \lc [DSC].
The results indicate that \CHb shows greater variation, is more sensitive to CLM than DSC, but is less sensitive to class overlap. 
Designing clustering quality metrics that are both sensitive to CLM and class overlap while fulfilling our requirements (\autoref{sec:cncltncreq}) constitutes an interesting area for future work. 

\subsection{Runtime Analysis}

We evaluate the scalability of \ltc and competitors by measuring their runtime.

\begin{figure}[t]
    \centering
    \includegraphics[width=0.65\linewidth]{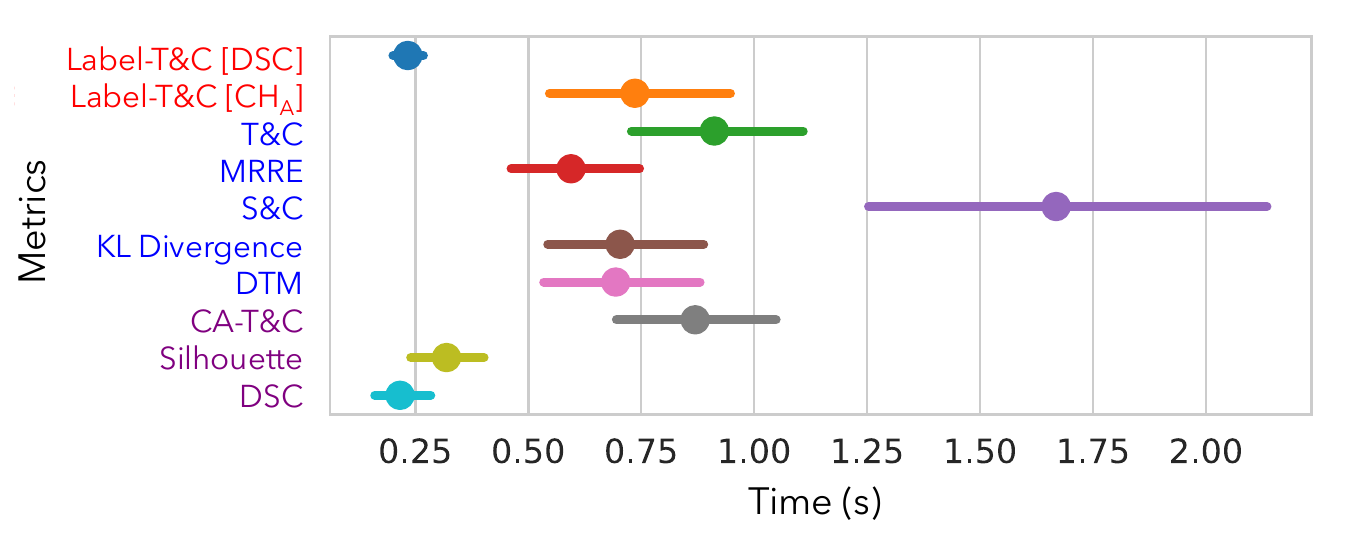}
    \caption{Results of the runtime analysis. Name and line colors match with \autoref{fig:cncsenexp}. \ltc [DSC] (dark blue) is on par with CVMs (Silhouette, DSC), while \ltc [\CHb] is similar to most of the other measures. S\&C is the slowest.}
    \label{fig:cncscal}
\end{figure}

\subsubsection{Objectives and Design}

We evaluate the runtime of \ltc against the competitors. 
We again use 96 labeled datasets that we used to evaluate the adjusted internal clustering metrics (\autoref{sec:cnceval}).
We generate projections using $t$-SNE, UMAP, PCA, and random projection for all 94 datasets.
We check the overall execution time by applying all measures to the projections, adding up the running times of the measures run in pairs (\ltc, T\&C, MRRE, S\&C, and CA-T\&C). We use the provided implementation for S\&C and scikit-learn \cite{pedregosa11jmlr} for the Silhouette.  We implement the remaining measures in Python using Numba's parallel computing \cite{lam15llvm} to maximize scalability.
We run the experiments on a Linux server with 40-core Intel Xeon Silver 4210 CPUs.

\subsubsection{Results and Discussion}

\autoref{fig:cncscal} shows that the running time of \ltc highly depends on the CVM. Among all measures, DSC is the fastest, followed by \ltc [DSC]. If $CH_{btwn}$ is used as the CVM, \ltc becomes less scalable. Still, \ltc [$CH_A$] has scalability similar to local (T\&C, MRRE) and global (KL Divergence, DTM) measures and to CA-T\&C,  all being more than twice faster than S\&C. The results confirm that \ltc can be scalably used in practice.
\section{Case Studies}

We report two case studies demonstrating the usefulness of \ltc for characterizing DR techniques and their hyperparameters, thereby contributing to reliable visual analytics with DR.

\subsection{Examining the Effect of \textbf{\textit{t}}-SNE Perplexity}

\label{sec:apptsne}

We first demonstrate that \ltc can be used to better understand the effects of hyperparameters in DR techniques and to leverage this information to perform visual analysis more reliably.

\subsubsection{Objectives and Design}

\begin{figure}[t]
    \centering
    \includegraphics[width=\linewidth]{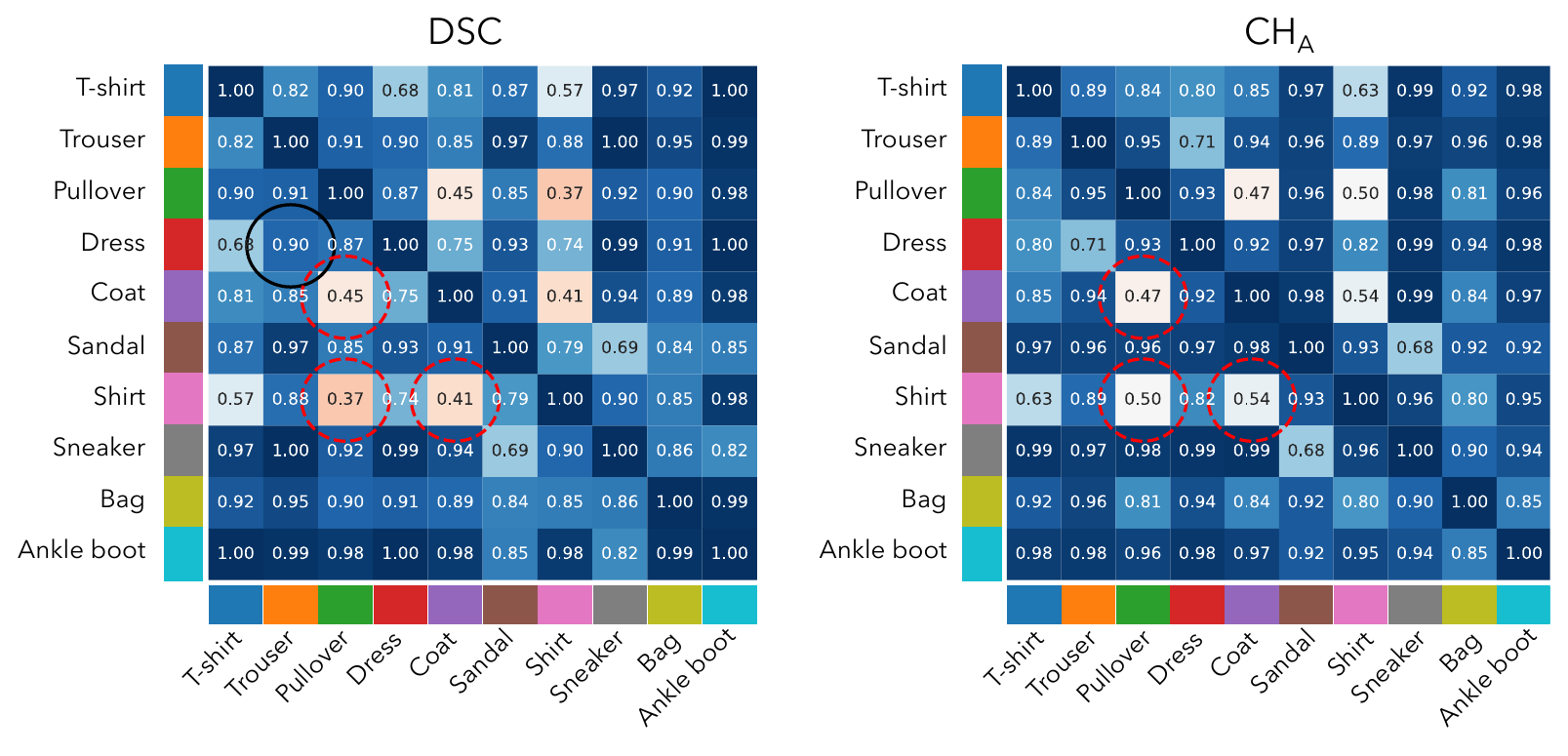}
    \caption{Heatmaps detailing the CLM matrix of the Fashion-MNIST dataset ($M(X)$ in \autoref{sec:cncdesign}). The color of each cell depicts the clustering quality metric (DSC, \CHb{}) score measured for each pair of classes corresponding to rows and columns.}
    \label{fig:app_pp_heatmap}
\end{figure}

\begin{figure*}[t]
    \centering
    \includegraphics[width=\textwidth]{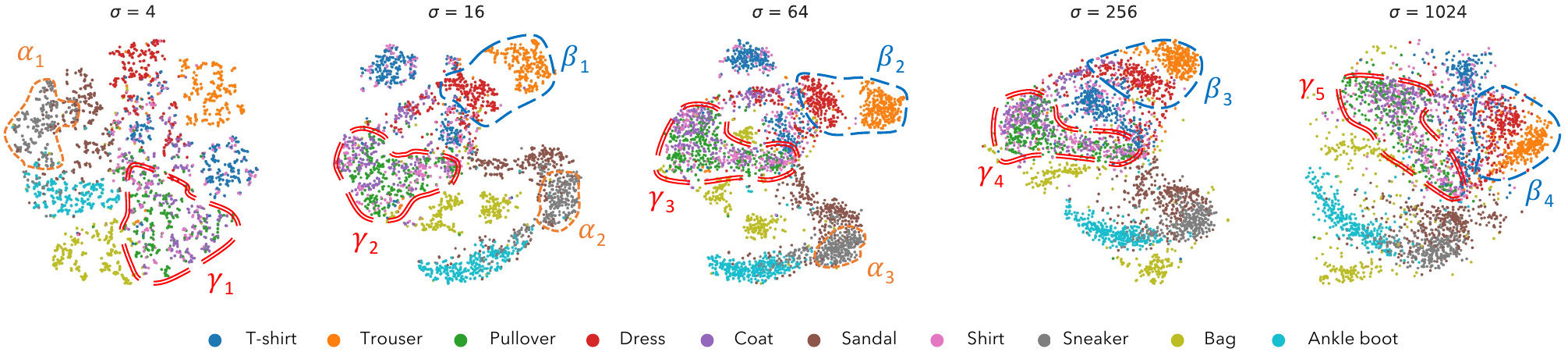} 
    \caption{$t$-SNE projections of Fashion-MNIST \cite{xiao2017arxiv} data with diverse perplexity ($\sigma$) values. Combined with the class-pairwise CLM of the original dataset (\autoref{fig:app_pp_heatmap}), the patterns in the projections qualitatively support the findings about the effect of $\sigma$ revealed by \ltc (\autoref{fig:app_hp}; \autoref{sec:apptsne}).}
    \label{fig:app_pp}
\end{figure*}

We want to use \ltc to evaluate the faithfulness of the cluster structures from $t$-SNE projection depending on its perplexity hyperparameter $\sigma$.
$\sigma$ adjusts the balance between local and global cluster structures \cite{wattenberg2016tsnetuning, cao17arxiv}. 
We generate the $t$-SNE projection of the 96 labeled datasets using different $\sigma$ values ($\sigma \in \{2^i \mid i=0,\cdots, 10 \}$) and evaluate them using \ltc [\CHb{}] and \ltc [DSC].
We also inspect the $t$-SNE projections of the Fashion-MNIST \cite{xiao2017arxiv} dataset with various perplexity values ($\sigma \in \{4, 16, 64, 256, 1024\}$; \autoref{fig:app_pp}) to gain more qualitative insights.
Moreover, we compute the ``ground-truth'' CLM matrix of the Fashion-MNIST dataset (\autoref{fig:app_pp_heatmap}), where the $(i, j)$-th cell represents the clustering metric score (\CHb or DSC) of the $i$-th and $j$-th classes. Note that this CLM matrix is identical to $M(\mathbf{X})$ in \autoref{sec:cncdesign}.

\begin{figure}[t]
    \centering
    \includegraphics[width=0.65\linewidth]{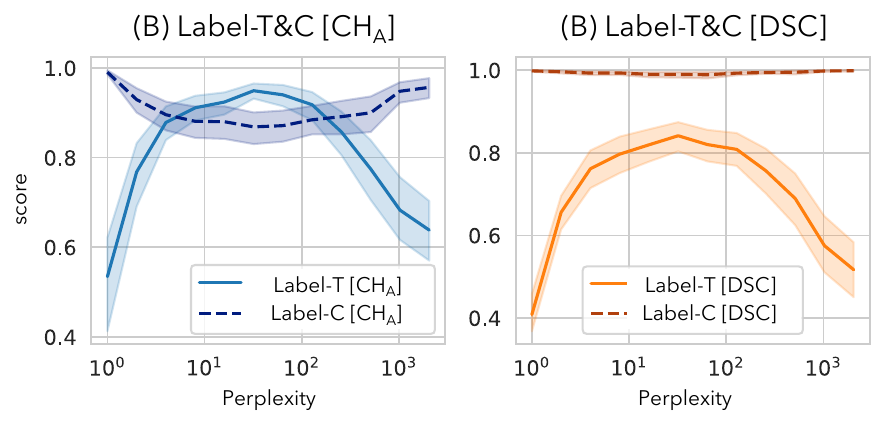} 
    \caption{Overall faithfulness of $t$-SNE projections according to the $\sigma$ value quantified by \ltc [DSC] and \ltc [\CHb{}]. For each $\sigma$ value, we average the score of the projections generated from 96 labeled datasets (95\% confidence interval shaded). }
    \label{fig:app_hp}
\end{figure}

\subsubsection{Results and Discussions}

\newcommand{\greytext}[1]{\textcolor{gray}{#1}}
\newcommand{\orangetext}[1]{\textcolor{orange}{#1}}
\newcommand{\redtext}[1]{\textcolor{Red}{#1}}
\newcommand{\greentext}[1]{\textcolor{ForestGreen}{#1}}
\newcommand{\purpletext}[1]{\textcolor{Plum}{#1}}
\newcommand{\pinktext}[1]{\textcolor{VioletRed}{#1}}


In the case of \ltc [\CHb{}], we found a clear tradeoff between \lt and \lc (\autoref{fig:app_hp}A).
When $\sigma$ is low or high, \lt [\CHb{}] gives low scores to $t$-SNE projections, indicating more False Groups distortions, while \lc [\CHb{}] gives high scores, meaning fewer Missing Groups distortions. 
This means that $t$-SNE underrepresents the extent to which classes are clustered. 
In contrast, when $\sigma$ has an intermediate value, \ltc [\CHb{}] indicates more Missing Groups and fewer False Groups distortions; hence, $t$-SNE exaggerates the degree to which classes are clustered. 

These results align well with the intent of $\sigma$. With low $\sigma$, $t$-SNE focuses more on a small number of neighbors, likely fewer than the clusters' sizes, interpreting each cluster as made of loosely-connected components in the data space. Thus, the projection is more likely to split classes into several clusters.
This phenomenon occurs in the Fashion-MNIST projection (\autoref{fig:app_pp}); the \greytext{\textit{Sneaker}} class is less dense if $\sigma$ is low (region $\alpha_1$) and relatively condensed when $\sigma$ has intermediate values ($\alpha_2$ and $\alpha_3$).
For the latter, the number of neighbors that $t$-SNE focuses on will likely match the size of natural clusters within the original data. Therefore, $t$-SNE projections will tend to dismiss the inter-cluster connections, exaggerating the between-cluster distances. The number of neighbors that $t$-SNE focuses on with high $\sigma$ values will likely be bigger than the clusters' sizes. Thus, $t$-SNE will detect all data clusters as one densely-packed component and generate projections with smaller inter-cluster distances.

The relation between the \orangetext{\textit{Trouser}} and \redtext{\textit{Dress}} classes of the Fashion-MNIST projections (\autoref{fig:app_pp}) qualitatively verifies these hypotheses. Their DSC scores are almost maximum (the black circle in \autoref{fig:app_pp_heatmap}), meaning they slightly overlap in the data space. However, their distance in the projection is exaggerated with intermediate $\sigma$ ($\beta_1$ and $\beta_2$) compared to high $\sigma$ ($\beta_3$ and $\beta_4$).
The same effect was observed qualitatively by Jeon et al. \cite{jeon21tvcg} while \ltc does so quantitatively.

Meanwhile, \lc [DSC] decreases slightly for intermediate values of $\sigma$ (\autoref{fig:app_hp}B dotted line). 
As \ltc [DSC] focuses more on class overlaps and less on between-class distances compared to \ltc [\CHb{}] (see D and E in \autoref{sec:cncltnceval}), it indicates that $t$-SNE preserves well the extent to which classes overlap regardless of $\sigma$.
To quantitatively validate these findings, we searched for the overlapped classes within the CLM matrices, assuming that $t$-SNE faithfully depicts class overlap for all $\sigma$ values. We observed that the \greentext{\textit{Pullover}}, \purpletext{\textit{Coat}}, and \pinktext{\textit{Shirt}} classes overlap in the high-dimensional space (red circles in \autoref{fig:app_pp_heatmap}; both their DSC and \CHb class-pairwise scores are low). 
We found that these classes overlap in all projections in \autoref{fig:app_pp} ($\gamma_1$ to $\gamma_5$), confirming our assumption.

In summary, we can conclude that for non-overlapping classes in $t$-SNE projections, the amount of proximity between them depends essentially on $\sigma$ and is not indicative of the proximity of these classes in the data space: $t$-SNE is not trustworthy regarding the original distance between visually separated classes.  However, classes with strong overlaps in the data are depicted as overlapping in the projection too: $t$-SNE is more trustworthy for overlapping classes. 
Such results align with the qualitative findings of Wattenberg et al. \cite{wattenberg2016tsnetuning}.

Overall, these findings demonstrate the effectiveness of \ltc to enhance our understanding of the effect of $\sigma$ on $t$-SNE results, which contributes to reliable visual analytics with DR.

\subsection{Analyzing DR Techniques' Performance in Detail}

\label{sec:app_hier}

In the second case study, we showcase that \ltc can be used to compare how different DR techniques behave in detail.

\subsubsection{Objectives and Design}

We use \ltc to analyze the quality of unsupervised DR techniques across fine-grained to coarse-grained cluster structures. 
We project each of the previous 96 datasets using six DR techniques: $t$-SNE, PCA, UMAP, Isomap, LLE, and Densmap \cite{narayan21nature}. 
We also apply hierarchical clustering, yielding 20 partitions with varying granularity for each dataset. 
The levels of granularity are obtained by thresholding the pairwise distances computed by Ward linkage \cite{ward64taylor} into 20 equal ranges.   
We use \ltc [\CHb{}] and \ltc [DSC] to evaluate the embeddings using each of the 20 clusterings as class labels.

We also want to check whether the results obtained by \ltc align with the ones made by previous measures. 
We thus evaluate the projections using T\&C and KL divergence as representative local and global measures, respectively. We use the same hyperparameter setting as the sensitivity analysis (\autoref{sec:cncltnceval}).

\subsubsection{Results and Discussions}

\begin{figure*}[t]
    \centering
    \includegraphics[width=\textwidth]{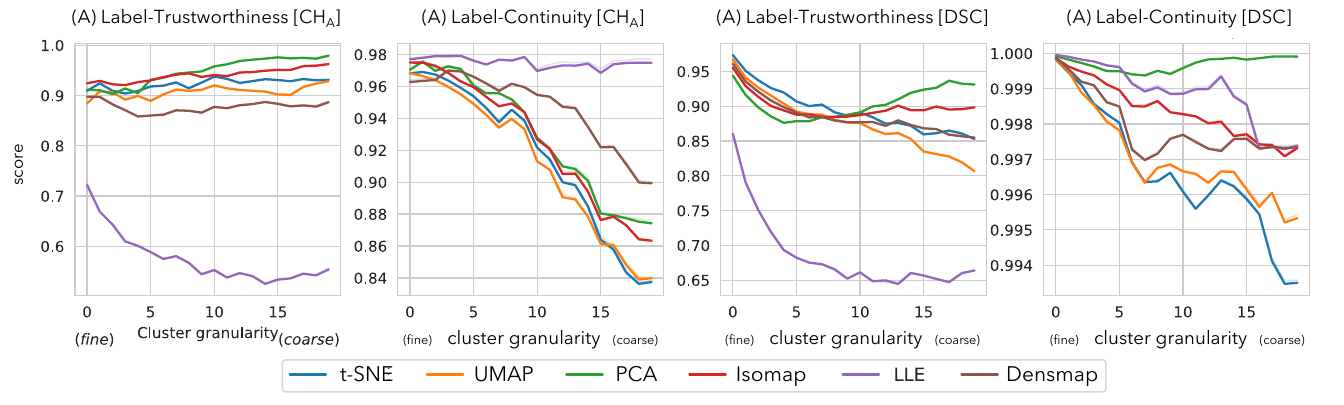}
    \caption{
    CLM distortion evaluation of a linear (PCA) and five nonlinear (t-SNE, UMAP, Isomap, LLE, and Densmap) unsupervised DR techniques with \ltc [\CHb{}/DSC] where class labels are obtained from the hierarchical clustering of the original data at multiple granularity levels (x-axis). \ltc evaluates more coarse-grained (global) clusterings for higher levels. 
    (E-F) } 
    
    \label{fig:app_hier}
\end{figure*}

\begin{figure*}[t]
    \centering
    \includegraphics[width=0.65\textwidth]{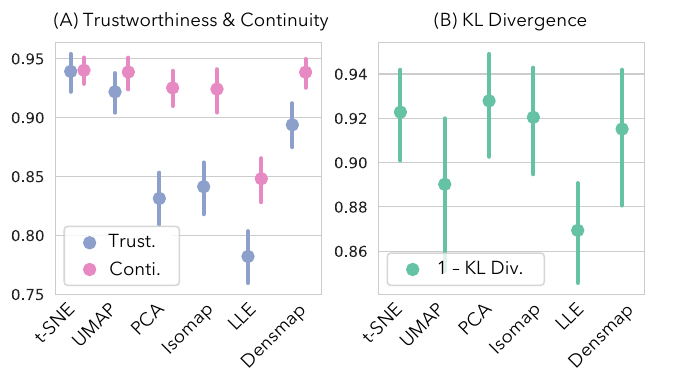}
    \caption{
    Evaluation results of DR techniques with T\&C (A) and KL Divergence (B). Higher scores indicate better projections
    }
    \label{fig:app_hier_others}
\end{figure*}

\autoref{fig:app_hier} and \ref{fig:app_hier_others} depicts the results. 
LLE generates few Missing Groups distortions (highest \lc score; \autoref{fig:app_hier}B, D) at any level, but more False Groups distortions as the granularity level increases (\lt decreases; \autoref{fig:app_hier}A, C). This finding aligns with the fact that LLE obtains the worst KL divergence score among all techniques (\autoref{fig:app_hier_others}B).
Such results are coherent with how LLE works, trying to reconstruct the ``local patches'' consisting of each point and its nearest neighbors while neglecting the overlap between the patches. 

There is a \lc downward trend across all other techniques as the level increases,
while \lc [DSC] shows higher scores than \lc [\CHb{}] (\autoref{fig:app_hier}B, D). 
This implies that Missing Groups distortions generally occur more for coarse-grained structures than for fine-grained ones; DR techniques exaggerate the separation between clusters at a global level.
$t$-SNE and UMAP especially give the worst  \lc scores because they focus on the preservation of local neighborhoods, casting doubts on their reliability in identifying global clusters. T\&C and KL divergence score provide strong evidence to the reliability of that claim. $t$-SNE and UMAP are in the top-2 highest ranks for T\&C but fail to do so for KL divergence (\autoref{fig:app_hier_others}A, B). 

For \ltc except \lc [\CHb{}], PCA gets the best score at higher  granularity, suggesting that PCA is more reliable to conduct global tasks such as the density and similarity identification of clusters.
These results align with the fact that PCA earns the best score for KL divergence.
The phenomenon confirms the experimental observation made by Xia et al. \cite{jiazhi21tvcg}. This is also coherent with the fact that PCA projects the data along the top two principal axes that preserve most of their variance, better representing coarse-grained structures than fine-grained ones.

We also find that Densmap, which is a variant of UMAP better preserving cluster density \cite{narayan21nature}, gets worse \lt [\CHb{}] scores than UMAP (\autoref{fig:app_hier}A) but better \lc [\CHb{}] scores (\autoref{fig:app_hier}B), at all levels.
This means that Densmap generates fewer Missing Groups but more False Groups distortions than UMAP.
As Densmap approximately maintains the cluster locations of UMAP \cite{narayan21nature}, such difference indicates that the clusters generally become bigger in Densmap compared to UMAP, hence the cluster density is relatively lower. 
Meanwhile, Densmap gets better \ltc [DSC] scores than UMAP for high granularity levels, confirming Densmap's advantage in investigating the overlap of clusters. The result is consistent with the KL divergence scores, indicating Densmap's advantage in preserving global structures when compared to UMAP (\autoref{fig:app_hier_others}B).

These findings confirm the ability of \ltc to reveal the characteristics of DR methods over a wide range of clustering granularities.
Although typical evaluation approaches of DR faithfulness using both local and global metrics (\autoref{fig:app_hier}E, F) \cite{jeon22vis, moor20icml, espadoto21tvcg} show consistent results, they cannot reveal how the quality changes across granularity levels, as different metrics are incomparable.

\section{Conclusion}

The general process of label-based DR evaluation relies on the assumption that the original data has good CLM, which can lead practitioners to erroneously favor projections with good CLM, such as those generated using \tsne and \umap.
To address this problem, we first develop adjusted internal clustering metrics that can evaluate and compare CLM across datasets, then leverage these metrics to design two new DR evaluation metrics: \LT and \LC  (\ltc).
These new DR evaluation metrics use class labels without the need to check that classes form clusters. Our quantitative experiments show that \ltc outperforms previous DR metrics in terms of precision and sensitivity in detecting Missing and False Groups distortions, thereby mitigating the bias in previous label-based metrics.
Use cases show that \ltc can be used to characterize DR techniques and their hyperparameters, contributing to reliable visual analytics with DR.

In summary, we address critical bias towards well-separated classes in DR evaluation. Using our metrics, analysts can select projections that faithfully represent the original data, and thus can conduct visual analytics in a more reliable manner. 


\chapter{Dataset-Adaptive Workflow for Accelerating Dimensionality Reduction Optimization}

\label{sec:dawadr}

We identify that practitioners prevalently cherry-pick hyperparameters to produce projections with desirable patterns. 
For reliable visual analytics, hyperparameters should be systematically optimized to obtain projections that faithfully represent the original high-dimensional data (\autoref{sec:relhyperparameter}).
However, this optimization process requires extensive iterative execution of DR techniques and is therefore computationally expensive.
As a result, practitioners rarely perform such optimization in practice.


In this chapter, we discuss \textit{dataset-adaptive workflow} that leverages dataset characteristics to accelerate DR optimization, motivating practitioners to optimize parameters rather than cherry-picking them.

\section{Introduction}

DR projections inherently suffer from distortions, leading to inaccurate representations of the original structure of high-dimensional data \cite{lespinats11cgf, lespinats07tnn}.
As distorted projections may occasionally lead analysts to inaccurate findings or cause them to overlook important structural characteristics \cite{aupetit07neurocomputing, jeon25arxiv}, analysts should \textit{optimize} hyperparameters of DR techniques to minimize distortions of key structural characteristics \cite{jeon25chi}.

However, optimizing hyperparameters of DR techniques to minimize distortions is typically computationally expensive.
Different hyperparameter settings must be tested iteratively, but the optimal number of iterations is rarely clear. 
Practitioners often run more iterations than necessary, observing minimal or no reduction in distortion after a certain point. 
This is because, unlike typical machine learning optimization methods such as gradient descent, DR hyperparameter optimization lacks explicit convergence criteria.
In visual analytics, a desirable DR hyperparameter optimization involves more than just minimizing a loss function---additional constraints, such as the preservation of neighborhood~\cite{venna06nn, colange19vis} and cluster structures~\cite{jeon21tvcg, martins14cg}, must also be considered. 
Representing these diverse criteria and DR techniques together within a single mathematical formulation is challenging, making the validation of convergence also challenging.
Practitioners thus more often cherry-pick hyperparameters or rely on default hyperparameters rather than executing optimization.

We propose a \textit{dataset-adaptive} workflow that improves the efficiency of DR optimization. 
Building upon previous findings \cite{lee11pcs, lee14cidm} that certain patterns are more prominent in HD data, our approach quantifies the prominence of these patterns to estimate the difficulty of accurately projecting the data into lower-dimensional spaces.
We introduce \textit{structural complexity metrics} to measure these patterns, and use these scores to predict the maximum accuracy achievable by DR techniques.
The metrics thus enhance the efficiency of DR optimization by (1) guiding the selection of an appropriate DR technique for a given dataset and (2) enabling early termination of optimization once near-optimal hyperparameters have been reached, avoiding unnecessary computations.

While existing metrics, such as intrinsic dimensionality metrics (\autoref{sec:intdim}), can potentially serve as structural complexity metrics, they lack the desired characteristics necessary for effective integration into our dataset-adaptive workflow (\autoref{sec:definition}). We thus introduce two novel structural complexity metrics---\textit{Pairwise Distance Shift} (\PDS) and \textit{Mutual Neighbor Consistency} (\MNC)---tailored to the dataset-adaptive workflow.
\PDS characterizes the complexity of a dataset's global structure by quantifying the shift in pairwise distances \cite{lee07springer, lee14cidm}, a well-established indicator associated with the curse of dimensionality. 
\MNC, in contrast, captures the complexity of a dataset's local structure by measuring the inconsistency between two neighborhood-based similarity functions: $k$-Nearest Neighbors ($k$NN) and Shared Nearest Neighbors (SNN) \cite{ertoz02siam}.
By jointly characterizing global and local structural complexity, these metrics effectively guide the optimization of DR techniques.
We theoretically and empirically verify that our metrics capture prominently observable patterns in higher-dimensional space, ensuring reliable guidance when identifying optimal DR techniques and hyperparameter settings.

A series of experiments with real-world datasets confirms the effectiveness of our structural complexity metrics and the dataset-adaptive workflow.
First, we verify that \PDS, \MNC, and their ensemble 
produce scores that highly correlate with ground truth structural complexity approximated by an ensemble of multiple state-of-the-art DR techniques, significantly outperforming baselines such as intrinsic dimensionality metrics. 
Second, we verify our metrics' utility in supporting the dataset-adaptive workflow of finding optimal DR projections. 
Finally, we show that the dataset-adaptive workflow significantly reduces the computational time required for DR optimization without compromising projection accuracy. 
\section{Related Work}

We first discuss previous approaches that leverage dataset characteristics to aid machine learning. We then describe intrinsic dimensionality metrics, which are natural candidates for structural complexity metrics.

\subsection{Dataset-Adaptive Machine Learning}

\label{sec:relprev}

Making machine learning models dataset-adaptive, i.e., measuring dataset properties to guide their use, has been considered an effective strategy to improve their reliability and efficiency.
We discuss how the fields of clustering, natural language processing, and computer vision benefit from dataset-adaptive approaches. 


\paragraph{Clustering} 
In the clustering field, \textit{Clusterability} metrics \cite{ackerman09icais, adolfsson19pr, mccarthy16cl, kalogeratos12nips} are proposed to quantify the extent to which clusters are clearly structured in datasets \cite{adolfsson19pr, ackerman09icais}. 
Adolfsson et al. \cite{adolfsson19pr} show that analysts can enhance the efficiency of data analysis by using clusterability to decide whether they should apply clustering techniques or not.
CLM (\autoref{sec:cncintro}) is also studied to guide the clustering benchmark, where Jeon et al. \cite{jeon25tpami} show that using high-CLM datasets yields more generalizable results when evaluating clustering techniques.


\paragraph{Natural language processing} 
The natural language processing field proposes metrics to predict the difficulty of datasets to be learned by machine learning models, i.e., \textit{dataset difficulty}, highlighting their importance in building the models with more reliable outputs \cite{ethayarajh22icml, wang22arxiv, byrd22acl}.
The literature shows that dataset difficulty can be used to predict model accuracy and overfitting before the training. Dataset difficulty is also verified to be effective in improving dataset quality, especially by identifying mislabeled \cite{ethayarajh22icml} or ambiguously-labeled \cite{wang22arxiv} data points. 

\paragraph{Computer vision} 
Measuring dataset difficulty is also studied in computer vision \cite{liu11mldmpr, scheidegger21vc, mayo23nips}. 
Liu et al. \cite{liu11mldmpr} develop a dataset difficulty metric that balances the model complexity and accuracy of image segmentation models.
Meanwhile, Mayo et al. \cite{mayo23nips} propose to use human viewing time as a proxy for dataset difficulty. 
These methods support scaling up the model architecture search or reducing the gap between benchmark results and real-world performance \cite{scheidegger21vc}. 

\paragraphit{Our contribution} All these works validate the effectiveness of the dataset-adaptive approach.
However, we lack discussion and validation on how DR---one of the most widely studied unsupervised machine learning domains---can be aligned with dataset properties.
We bridge this gap (1) by demonstrating the importance of measuring the property (which is structural complexity) of high-dimensional datasets before optimizing DR projections and (2) by introducing accurate and fast structural complexity metrics.

\subsection{Intrinsic Dimensionality Metrics}

\label{sec:intdim}

To the best of our knowledge, intrinsic dimensionality metrics \cite{fukunga71toc, kak20sr, bac21entropy} are currently the only available options for measuring the intricacy of a dataset's structure. 
Projection-based intrinsic dimensionality metrics identify the dimensionality in which further increments of dimensionality may slightly enhance the accuracy of a particular DR technique. PCA is widely used for this purpose \cite{espadoto21tvcg, fukunga71toc}.
On the other hand, geometric metrics (e.g., fractal dimension \cite{karbauskaite16ifm, theiler90josaa, falconer04wiley}) evaluate how detailed the geometric structure of datasets is.
Both align well with our definition of structural complexity (\autoref{sec:definition}); intuitively, if a dataset needs more dimensionality to be accurately projected, the dataset can be considered to be more complex and tends to be less accurately projected in low-dimensional spaces.
Recent works show that intrinsic dimensionality provides a grounded basis in selecting the appropriate machine learning model to train on the datasets \cite{bac21entropy} and improving their efficiency \cite{gong19cvpr, hasanlou12grsl}.



\paragraphit{Our contribution} 
Intrinsic dimensionality metrics may correlate with structural complexity by intuition.
However, they do not satisfy the desired properties of structural complexity metrics to support the dataset-adaptive workflow.
For example, projection-based metrics suffer from low generalizability across multiple DR techniques as they depend on a specific DR technique (\autoref{sec:definition} P1).
Also, geometric metrics produce scores that vary with the global scaling of the dataset, compromising their applicability to real-world datasets with diverse scales (\autoref{sec:definition} P2).
We thus propose novel structural complexity metrics (\autoref{sec:metrics}) that satisfy these desired properties.
Our metrics outperform intrinsic dimensionality metrics in terms of accurately predicting the ground truth structural complexity (\autoref{sec:acceval}) and in properly guiding the dataset-adaptive workflow (\autoref{sec:suitability}).

\section{Conventional Workflow for Finding Optimal DR Projections}

\label{sec:conventional}

We detail the conventional workflow (\autoref{fig:dadrworkflow} CW) to find optimal DR projections with high accuracy. 
We derive the workflow by reviewing prior work that optimizes hyperparameters of DR techniques to minimize distortions \cite{jeon22vis, moor20icml, jeon24tvcg2, jiazhi21tvcg, espadoto21tvcg}. 

The conventional workflow aims to find a DR technique $T \in \mathbf{T}$ and hyperparameter setting $H \in \mathbf{H}_T$ that maximizes $C(X, T_H(X))$, where $C$ is a DR evaluation metric. $T_H(X)$ denotes the projection generated using $T$ and $H$, and $\mathbf{H}_T$ indicates the hyperparameter domain of $T$. 
Here, hyperparameter optimization often yields only marginal improvements over well-chosen default parameters \cite{espadoto21tvcg}. This means that users can naively use default hyperparameters when efficiency is crucial (e.g., when interactivity is critical). Still, optimization remains essential, as it consistently improves reliability in visual analytics using DR and cannot be entirely avoided.

\subsection{Workflow}

\noindent
\textbf{(Step 1) Optimizing hyperparameters of DR technique.}
In the conventional workflow, we first optimize hyperparameters $H$ for individual DR techniques $T \in \mathbf{T}$ while using $C(X, T_H(X))$ as the target function (\autoref{fig:dadrworkflow} CW1).
We do so by repeatedly testing various hyperparameter settings for a fixed number of iterations.
Here, detecting convergence in DR hyperparameter optimization is nontrivial, given the difficulty of unifying the DR technique and execution metric into one differentiable expression.
Therefore, hyperparameter search methods for non-differentiable functions, such as grid search, random search \cite{bergstra12jmlr}, or Bayesian optimization \cite{snoek12nips}, are commonly used.

\paragraph{(Step 2) Selecting the projection with the best accuracy}
We compare the accuracy of the optimal projection achieved for each DR technique. We select the one with the highest accuracy as the optimal projection (\autoref{fig:dadrworkflow} CW2).

\subsection{Problems}
\begin{sidewaysfigure}
    \centering
    \includegraphics[width=\linewidth]{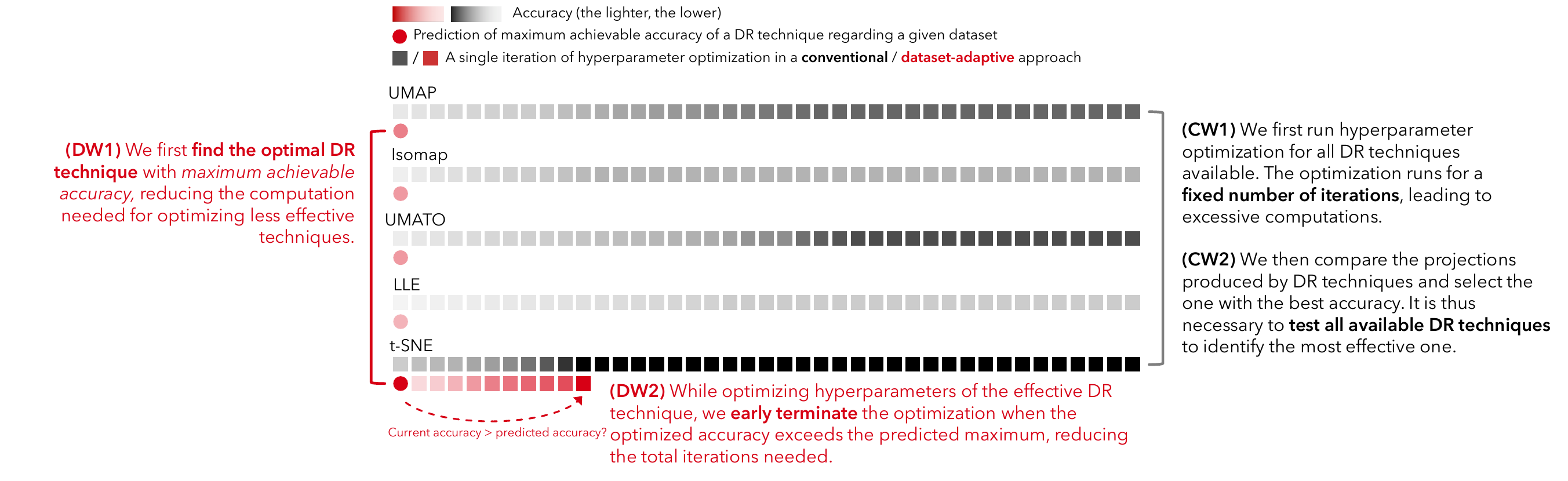}
    \caption{Illustrations of our \redd{\textit{dataset-adaptive}} workflow (DW1, DW2) and conventional workflow (CW1, CW2) of finding an optimal DR projection. 
  Each square depicts individual iterations of optimizing DR hyperparameters, where the opacity represents the maximum accuracy achieved by the current and previous iterations.  
  The dataset-adaptive workflow reduces the number of iterations required to discover the optimal projections (red squares) compared to the conventional approach (gray squares).}
    \label{fig:dadrworkflow}
\end{sidewaysfigure}

This conventional workflow is time-consuming for two reasons. 
First, the workflow should test all DR techniques available (\autoref{fig:dadrworkflow} CW2).
Second, as detecting convergence in DR hyperparameter optimization is nontrivial, the workflow relies on a fixed number of iterations.
If the number of iterations is set too low, the optimization may fail to reach an optimum; if it is set too high, computation is wasted after an optimum has been reached (\autoref{fig:dadrworkflow} CW1). 
Such inefficiency leads practitioners to avoid executing optimization and instead rely on cherry-picking. 
We detail how our dataset-adaptive workflow prevents these redundant computations in \autoref{sec:datasetadaptive}.

\section{Dataset-Adaptive Workflow}

\label{sec:datasetadaptive}

We introduce the dataset-adaptive workflow that reduces the redundant computations in the conventional workflow (\autoref{sec:conventional}) while still achieving an accuracy close to what is achievable in the conventional workflow. 
The central idea is to quantify how structurally complex a given dataset is, i.e., structural complexity, and then use this value to predict the maximum achievable accuracy of DR techniques in representing the original HD dataset. 
Here, we can reduce the redundant computation in the conventional workflow by (1) focusing on DR techniques with high maximum predicted accuracy (\autoref{fig:dadrworkflow} DW1), and (2) early terminating hyperparameter optimization when the predicted maximum accuracy is reached (\autoref{fig:dadrworkflow} DW2). 

In the following sections, we first define structural complexity and structural complexity metrics. 
We then detail our dataset-adaptive workflow for optimizing DR projection.

\subsection{Structural Complexity and Complexity Metrics}

\label{sec:definition}

If HD data is structurally complex, DR techniques will inevitably distort these structures, leading to less accurate low-dimensional representations.
Accordingly, we define the \textbf{\textit{structural complexity}} of a dataset $X$ as the degree to which even the best possible 2D representation having the same cardinality as $X$ falls short in accurately capturing its structure.
Formally, for $X \in \mathbb{R}^{N \times D}$ consisting of $N$ points in $D$-dimensional space, structural complexity of $X$ is defined as $S_C(X) = - \max_{Y\in \mathbb{R}^{N\times 2}}C(X,Y)$, where $C$ is a DR evaluation metric that measures a projection $Y$'s accuracy in representing $X$'s structural characteristics. Note that we use an additive inverse to align the formula with our definition.
We set the output dimension to two since DR is commonly visualized in 2D space in visual analytics.

The structural complexity of $X$ can be explained in various aspects, depending on the target structural characteristics of DR evaluation metrics (e.g., local neighborhood structure). 
This feature is crucial in leveraging structural complexity scores to guide the dataset-adaptive workflow that comprises multiple DR techniques emphasizing distinct structural characteristics.

A \textbf{\textit{structural complexity metric}} is a function $f: \mathbb{X} \rightarrow \mathbb{R}$ that is desired to have high predictive power with the ground truth structural complexity, where for any HD data $X$, $X \in \mathbb{X}$.
Formally, given a set of datasets $\{X_1, X_2, \cdots, X_N\}$, the predictive power of $f$ with respective to $S_C$ is defined as the degree to which $\{S_C(X_1), S_C(X_2),\cdots, S_C(X_N)\}$ can be accurately predicted by $\{f(X_1), f(X_2), \cdots, f(X_N)\}$.
To properly support the dataset-adaptive workflow, we want our metrics to be \textbf{(P1) independent of any DR technique}. This is because depending on a certain DR technique may make metrics work properly in predicting maximum achievable accuracy for some DR techniques while working poorly for others. 
Our metrics should also be \textbf{(P2) scale-invariant}, i.e., be independent of the global scale of datasets. 
Here, global scaling denotes the operation that multiplies an arbitrary positive real number $\alpha > 0$ to all values in the dataset.
Global scaling is a characteristic unrelated to data pattern or distribution, thus also independent of the accuracy of projections in representing HD data. 
Failing to be scale-invariant thus makes the metrics hardly support the dataset-adaptive workflow, as scores can be artificially manipulated through global scaling.
Finally, we want our metrics to be \textbf{(P3) computationally beneficial}, meaning that computing the metrics should be faster than a single run of the DR technique applied in the dataset-adaptive workflow. This requirement ensures that the workflow can achieve substantial efficiency gains when optimizing DR hyperparameters.

Here, it is worth noting that these requirements may be incomplete. Although satisfying these requirements results in effective structural complexity metrics, it remains uncertain whether they are the only requirements needed to characterize the metrics. Identifying new requirements that contribute to designing improved structural complexity metrics would be an interesting avenue to explore.

\paragraphit{Validity of our desired properties in the lens of existing dataset property metrics.}
The fact that P1--P3 are established as the requirements for existing dataset property metrics (\autoref{sec:relprev}) further confirms their validity \cite{adolfsson19pr, jeon22arxiv2, ackerman09icais}. Adolfsson et al. \cite{adolfsson19pr} state that P1, P2, and P3 as requirements for clusterability metrics. Jeon et al. \cite{jeon22arxiv2} state that P2 and P3 is an important requirement for CLM metrics.

\paragraphit{Violations of the desired properties by widely used intrinsic dimensionality metrics.}
Intrinsic dimensionality metrics are natural candidates for being structural complexity metrics (\autoref{sec:intdim}). However, commonly used intrinsic dimensionality metrics do not meet our desired properties (P1 and P2) and thus have limited capability in supporting the dataset-adaptive workflow, justifying the need to design new metrics.
For example, two existing methods to compute geometric metrics (correlation method \cite{grassberger83pnp} and box-counting method \cite{liebovitch89pla}) are not scale-invariant (P2). In terms of projection-based metrics, they are dependent on DR techniques by design (P1), making them weakly correlate with the ground truth and less applicable to DR optimization involving multiple techniques. 
For example, if DR techniques that focus on global structures like PCA are used, projection-based metrics may inaccurately predict the maximum achievable accuracy of local techniques like $t$-SNE or UMAP. 
We empirically show that these metrics fall behind our proposed metrics (\autoref{sec:metrics}) in correlating with ground truth structural complexity (\autoref{sec:acceval}) and also show that they are not suitable for the dataset-adaptive workflow (\autoref{sec:suitability}).

\subsection{Workflow}

\label{sec:daworkflow}

We detail our dataset-adaptive workflow.

\paragraph{Pretraining}
The workflow requires a pretraining of regression models that predict the maximum achievable accuracy of DR techniques from structural complexity scores. 
We first prepare a set of HD datasets $\mathbf{X}$, a set of structural complexity metrics $\mathbf{f}$, a set of DR techniques $\mathbf{T}$, and a DR evaluation metric $C$. 
Then, for each dataset $X \in \mathbf{X}$, we compute $f(X)$ for all structural complexity metrics $f \in \mathbf{f}$, and also compute the optimal accuracy achievable by each technique $T \in \mathbf{T}$ by optimizing hyperparameters $H$ while using $C(X, T_H(X))$ as target function, where we denote this value as $\hat S_{C, T}(X)$.
Finally, for each $T \in \mathbf{T}$, we train a regression model that predicts $\hat S_{C, T}(X)$ from $\{f(X) | f \in \mathbf{f}\}$.

\paragraph{(Step 1) Finding effective DR techniques}
Given an unseen dataset $X'$, we start by predicting $\hat S_{C, T}(X')$ for each 
technique $T \in \mathbf{T}$ from $\{f(X') | f \in \mathbf{f}\}$ (\autoref{fig:workflow} DW1). We then optimize the hyperparameters 
of only the techniques that have high predicted $\hat S_{C, T}(X')$, eliminating the 
redundant computations in running hyperparameter optimization on potentially less effective DR techniques. 

\paragraph{(Step 2) Early terminating hyperparameter optimization}
We then optimize the hyperparameters for a selected method $T$, where
we halt the iteration early if the accuracy reaches $\hat S_{C, T}(X)$.
This early termination prevents unnecessary optimization iterations that would likely result in a minimal gain in improving accuracy (\autoref{fig:workflow} DW2). 
Note that we set the threshold exactly at  $\hat S_{C, T}(X)$ to minimize the risk of sacrificing accuracy; choosing a lower threshold (e.g., 95\% of  $\hat S_{C, T}(X)$) will increase the efficiency but at the cost of further accuracy loss. 

\section{Structural Complexity Metrics for Dataset-Adaptive Workflow}

\label{sec:metrics}

We introduce two structural complexity metrics---\textit{Pairwise Distance Shift} (\PDS), \textit{Mutual Neighbor Consistency} (\MNC)---that are tailored for the dataset-adaptive workflow. 
These metrics quantify phenomena that become more pronounced in higher-dimensional spaces as a proxy for structural complexity. 
\PDS estimates the complexity of global structure, such as pairwise distances between data points, and \MNC focuses on local neighborhood structures.
\PDS and \MNC thus complement each other in estimating ground truth structural complexity (\autoref{sec:acceval}) and guiding dataset-adaptive workflow (\autoref{sec:suitability}), which motivates us to propose their ensemble (\PDSMNC) as an additional metric. 

Note that these two metrics depend solely on the distance matrix of the input data. This means that any high-dimensional data, regardless of its type (e.g., image, tabular), can be applied as long as the distances between pairs of points can be defined.

\subsection{Pairwise Distance Shift (\PDS)}

\label{sec:pds}

\PDS estimates the degree of shift made by pairwise distances between data points \cite{lee07springer, lee14cidm} (also known as distance concentration \cite{francois07tkde}) as a proxy for complexity regarding global structure.
The shift refers to a common phenomenon of HD space in which pairwise distances tend to have a large average and small deviation, shifting its distribution to the positive side \cite{francois07tkde, lee11pcs}. 
Since it is naturally easier to create complex, irreducible patterns with more dimensions, we design \PDS to use the degree of shift as a target measurand.

\paragraph{Theoretical verification} 
The following theorem confirms that distance shifts are more pronounced in HD spaces when the dataset is independent and identically distributed (i.i.d.); thus, measuring shifts in pairwise distances provides an effective proxy for structural complexity.

\paragraph{Theorem 1} \textit{ Let $X, Y \subset \mathbb{R}^{d}$ ($d \in$ $\mathbb{N}$) be random variables such that $\mathbf{x}\in X$ and $\mathbf{y}\in Y$ follow i.i.d. (independent and identically distributed) density functions. 
$\sigma(X)$ and }$\textsf{E}(X)$ \textit{refer to the standard deviation and expectation value of $X$.
We also define $\| X \| = \{\|\mathbf{x}\| \mid \mathbf{x} \in X  \}$, where $\|\mathbf{x}\|$ is the Euclidean norm of $\mathbf{x}$.
Suppose there exists $Z \subset \mathbb{R}^{d}$, such that $Z=\{ \mathbf{x} - \mathbf{y} \mid \mathbf{x} \in X, \mathbf{y} \in Y\}$. Then,
}
$$\frac{\sigma(\| Z\|)}{\textsf{E}(\| Z \|)} = \Theta(1/\sqrt{d}).$$

\vspace{3pt}

\paragraphit{proof} According to Demartines \cite{demartines94dis}, 
\[
\mathsf{E}(\|Z\|) = \sqrt{ad-b} + \Theta(1/d) \text{ and } \sigma(\|Z\|) = b + \Theta(1/\sqrt{d}),
\]
where $a, b \in \mathbb{R}$ and $a > 0$.\\
This implies that $\mathsf{E}(\|Z\|) = \Theta(\sqrt{d}) \text{ and } \sigma(\|Z\|) = \Theta(1)$ since $1/d$ is negligible compared to $\sqrt{d}$. Hence, 
\[
\frac{\sigma(\|Z \|)}{\textsf{E}(\|Z\|)} = \Theta(1/\sqrt{d}).
\]
\hfill $\Box$
\vspace{3pt}

\begin{figure*}
    \centering
    \includegraphics[width=0.8\linewidth]{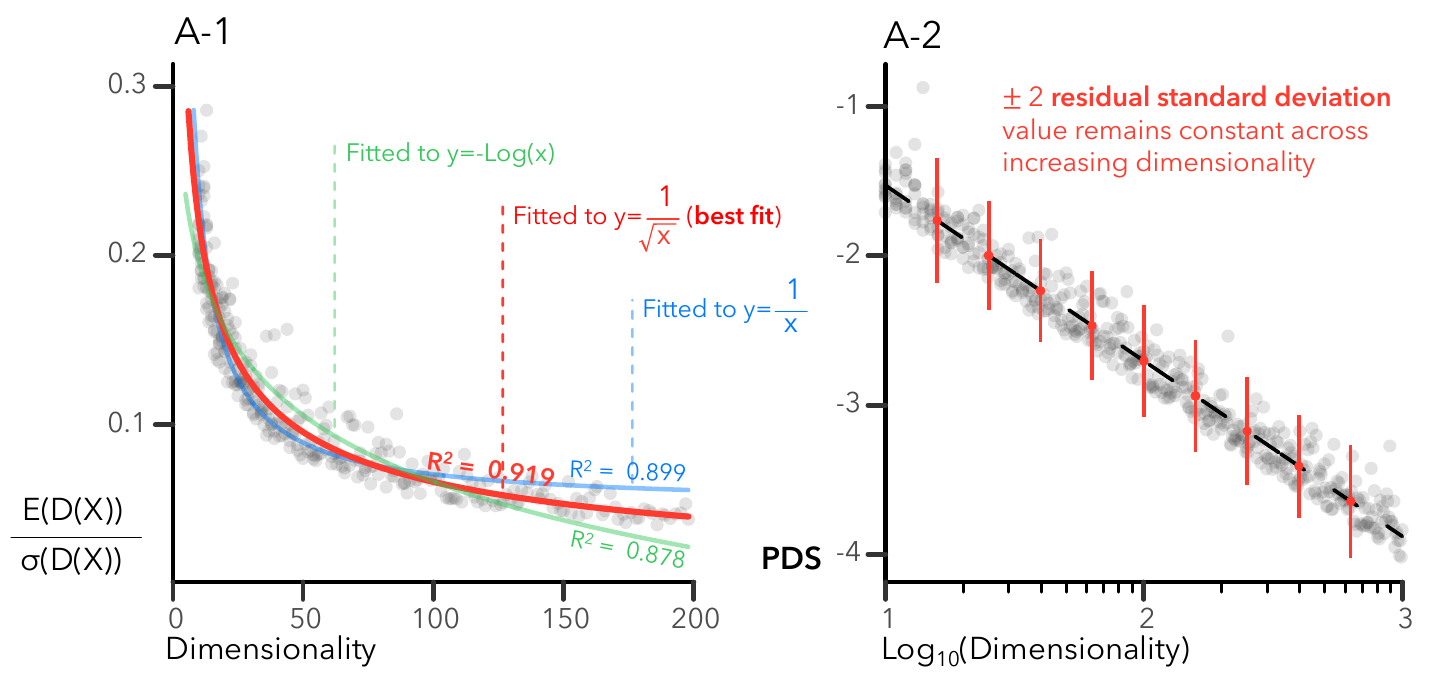}
    \caption{
    Empirical verifications for grounding our design of \PDS.
    We generate HD random distributions with diverse dimensionality following i.i.d. density functions randomly determined by the mixture of Gaussian distributions, denoted as $X$, and investigate how (A-1) $\frac{\mathbf{\sigma}(D(X))}{\mathbf{E}(D(X))}$ and (A-2) \PDS change over dimensionality. 
    }
    \label{fig:theoretical_pds}
\end{figure*}

\paragraph{Empirical verification} 
We generate a high-dimensional random Gaussian distribution with diverse dimensionality following an i.i.d. density function, denoted as $X$. We then calculate all pairwise Euclidean distances between distinct data points in $X$, represented as $D(X) = \{\delta(p_i, p_j) \mid p_i, p_j \in X,  p_i \neq p_j\}$, where $\delta(p_i, p_j)$ is the Euclidean distance between points $p_i$ and $p_j$.
We investigate how the ratio $\sigma(D(X))/E(D(X))$---degree of distance shift---changes with dimensionality. Our experimental results (\autoref{fig:theoretical_pds}) show that this ratio decreases proportionally to $1/\sqrt{d}$, which aligns with our theoretical prediction. Furthermore, we observe a linear decrease when we plot the logarithm of this ratio against the logarithm of the dimensionality. This linear relationship reaffirms our theoretical findings as $\log (1/\sqrt{d}) = -0.5 \log d$.

Note that, because we assume the data follow i.i.d. distributions, the phenomenon may not generalize to real-world datasets. We thus conduct experiments with real-world datasets (\autoref{sec:acceval}, \ref{sec:suitability}) to verify the practical applicability of our metrics. 


\subsubsection{Algorithm}
The detailed steps to compute \PDS are as follows.

\paragraph{(Step 1) Compute pairwise distances}
For a given HD dataset $X$, we first compute the set of pairwise distances $D(X)$.

\begin{figure}
    \centering
    \includegraphics[width=0.65\linewidth]{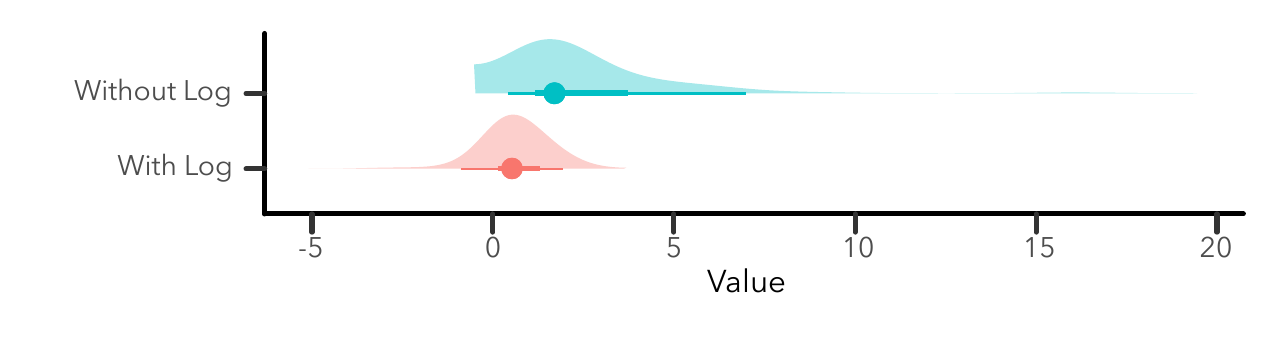}
    \caption{The distribution of \PDS scores across 96 datasets with and without log transformation. \PDS is highly skewed without log transformation.}
    \label{fig:skewness}
\end{figure}

\paragraph{(Step 2) Compute pairwise distance shift}
We then compute the final score representing the degree of shift as:
\[
\PDS = \log \mathbf{\sigma}(D(X))/\mathbf{E}(D(X)),
\]
where $\mathbf{\sigma}(D(X))$ and $\mathbf{E}(D(X))$ represent the standard deviation and the average of $D(X)$, respectively. 
Without log transformation, the distribution of \PDS scores is highly skewed (\autoref{fig:skewness}), which can harm its applicability \cite{fernandez98jasa}. 
As $0 < \mathbf{\sigma}(D(X)) < \infty$ and $0 <\mathbf{E}(D(X)) < \infty$, $-\infty < \PDS < \infty$. However, as $\mathbf{\sigma}(D(X))$ is typically less than $\mathbf{E}(D(X))$, \PDS mostly ranges from $-\infty$ to $0$. Lower scores indicate more distance shift, i.e., higher structural complexity.




\subsubsection{Compliance with the Desired Properties}

\label{sec:pdscompile}

We confirm that \PDS complies with our desired properties for structural complexity metrics (P1--P3).

\paragraph{(P1) Independence to any DR techniques}
The metric only relies on Euclidean distance, thus independent of a specific DR technique.

\paragraph{(P2) Scale-invariance}
For any scaling factor $\alpha > 0$ and dataset $X$, we have $D(\alpha X) = \alpha D(X)$. Here, we obtain:
\begin{align*}
\text{\textsc{Pds}}(\alpha X) &= \frac{\sigma(D(\alpha X))}{E(D(\alpha X))} = \frac{\sigma(\alpha D(X))}{E(\alpha D(X))} \\&=  \frac{\alpha \sigma(D(X))}{\alpha E(D(X))} = \frac{\sigma(D(X))}{E(D(X))} = \text{\textsc{Pds}}(X).
\end{align*}
Therefore, PDS is scale-invariant.

\paragraph{(P3) Computational benefit}
The computations of $D(X)$, $E(D(X))$, and $\sigma(D(X))$ are $O(N^2)$.
However, each step can be fully parallelized, making \PDS highly scalable (see \autoref{sec:imp}, \ref{sec:acceval}). We empirically verify that \PDS is computationally beneficial, i.e., faster than a single run of typical DR techniques like $t$-SNE and UMAP, in Appendix XX.

\subsection{Mutual Neighbor Consistency (\MNC)}

\MNC computes the consistency between $k$-Nearest Neighbors ($k$NN) similarity \cite{dong11www} and Shared Nearest Neighbors (SNN) similarity \cite{jarvis73toc, liu18infosci} as a proxy for the structural complexity of local structure. 
While $k$NN similarity considers a point and its $k$NN to be similar, SNN similarity regards the pair of points that share more $k$NN to be more similar. 

\MNC relies on the phenomenon that higher-dimensional spaces tend to make datasets exhibit greater inconsistency between $k$NN and SNN. 
This phenomenon originates from the pairwise distance shift. For any point $p$ in a HD dataset, the distance shift makes all other points equidistantly far from $p$, distributed on a thin hypersphere $C_p$ centered on $p$. Consequently, all other points are equally likely to be a $k$NN of $p$ with probability $k/(N-1)$, where $N$ denotes the number of points in the dataset. Now, consider $p$ and its $k$NN $q$. To make $p$ and $q$ to become SNN, a point $r$ that (1) stays within the intersection of $C_p$ and $C_q$, and (2) is a $k$NN of both $p$ and $q$ should exist. 
The first condition has a low probability of occurring as $C_p$ and $C_q$ are very thin. The second condition is also unlikely to be satisfied since the probability is $k^2/(N-1)^2$, where $N \gg k$. 
Therefore, lower $k$NN-SNN consistency likely indicates high dimensionality, which increases the likelihood of complex or irreducible patterns. We thus design \MNC to target this consistency as a key measurand.

\paragraph{Theoretical verification}
Theorem 1 leads to the following corollary:

\paragraph{Corollary 1}
\textit{
Given a random variable $X \subset \mathbb{R}^d$ such that $\mathbf{x}\in X$ follows an i.i.d. (independent and identically distributed) density function. Then, }
\[
\frac{\sigma(\|X \|)}{\textsf{E}(\|X\|)} \rightarrow 0  \text{ as } d \rightarrow \infty.
\].

\paragraphit{Proof}
For any $\epsilon > 0$, we should show that $\exists \delta$ such that $\forall d > \delta$, 
\[
\frac{\sigma(\|X \|)}{\textsf{E}(\|X\|)} < \epsilon.
\]
By Theorem 1, there exists $C > 0$ and $d_0 > 0$ satisfying 
\[
\frac{\sigma(\|X \|)}{\textsf{E}(\|X\|)} \leq \frac{C}{\sqrt{d}}
\]
$\forall d > d_0$. \\
Let
\[
\delta = \max\left(\left(\frac{C}{\epsilon}\right)^2, d_0\right).
\]
Then, $\forall d > \delta$,
\[
\frac{\sigma(\|X \|)}{\textsf{E}(\|X\|)} < \frac{C}{\sqrt{\delta}} = C \cdot \frac{\epsilon}{C} = \epsilon.
\]
\hfill $\Box$
\vspace{3pt}

Then, the following theorem, based on this corollary, shows that the inconsistency between $k$NN and SNN increases in high-dimensional spaces when the dataset is i.i.d., implying that measuring this inconsistency provides a proxy for structural complexity.

\paragraph{Theorem 2} \textit{
Let $\overline X_N$ be a set of N data samples from a multidimensional dataset $X \subset \mathbb{R}^d$. 
Let $i, j, k, \in N$. Two points $p_i, p_j \in \overline X_N$ are said to be in a $k$-nearest neighbor ($k$-NN) relationship if one is among the k nearest neighbors of the other, and are in an SNN relationship if the SNN similarity between $p_i$ and $p_j$ is above 0.
Suppose there exists a constant integer $k > 0$ where $k \ll N$. 
Then, $\forall p_i, p_j \in \overline X_N$, the probability in which $p_i$ and $p_j$ are both in $k$NN and SNN relationship approximates to $k^3/N^2$ as $d \rightarrow \infty$.}
\vspace{3pt}

\paragraphit{Proof}
According to Corollary 1, 
\[
\frac{\sigma(\|X \|)}{\textsf{E}(\|X\|)} \rightarrow 0  \text{ as } d \rightarrow \infty.
\]
Under such equidistance, the probability of $p_j$ to become $k$NN of $p_i$ approximates to:
\begin{equation}\label{eq:pknn}
\mathbb{P}\Bigl[p_j \in \text{\(k\)NN}(p_i)\Bigr] \approx \frac{k}{N-1} \sim \frac{k}{N}.
\end{equation}
and vice versa. \\
Let
\[
A_i = \{a_i | a_i \in \text{\(k\)NN of } p_i\}, \quad A_j = \{a_j | a_j \in\text{\(k\)NN of } p_j\}.
\]
An estimate for the probability that \(A_i\) and \(A_j\) have a nonempty intersection is obtained as follows. 
As mentioned above,
$\forall p \in \overline X $ where $p \neq p_i, p \neq p_j$, the probability of being in \(A_i\) and \(A_j\) is approximately \(k/N\). Hence, the probability of $p_j$ and $p_j$ to be in SNN relationship approaches:
\[
\mathbb{P}\Bigl[A_i\cap A_j\neq\varnothing\Bigr] \approx (N-2)\left(\frac{k}{N}\right)^2 
\]
Using Taylor expansion:
\[\left(1-\frac{k}{N}\right)^k = 1 - \frac{k^2}{N} + O\left(\frac{k^4}{N^2}\right),\]
we get
\[
\mathbb{P}\Bigl[A_i\cap A_j\neq\varnothing\Bigr] \approx 1 - \left(1-\frac{k}{N}\right)^k \sim \frac{k^2}{N}.
\]
Thus, the probability in which $p_i$ and $p_j$ is both $k$NN and SNN approximates to $k/N \cdot k^2/N = k^3/N^2$. \hfill $\Box$

\begin{figure*}
    \centering
    \includegraphics[width=0.6\linewidth]{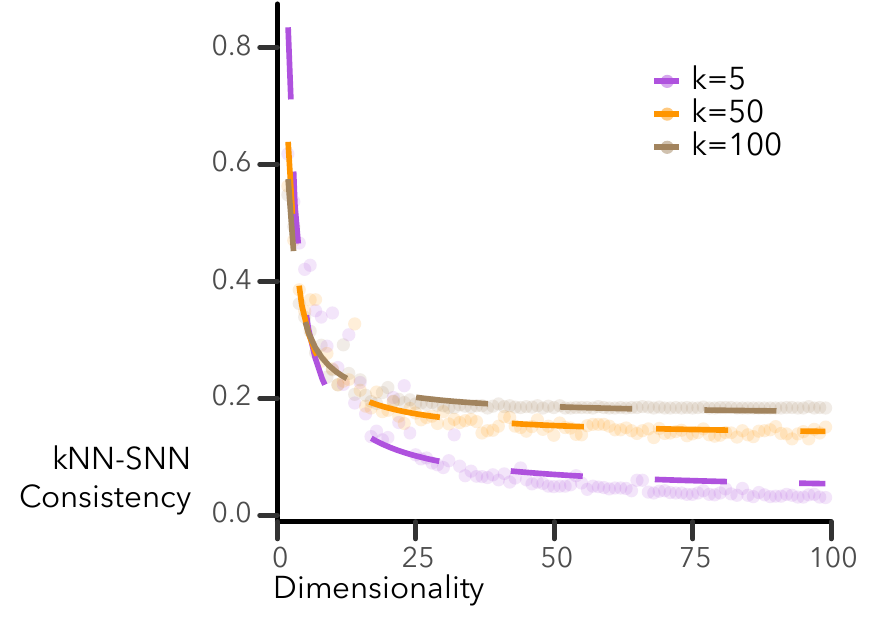}
    \caption{
    Empirical verifications for grounding our design of \MNC.
    We generate HD random distributions with diverse dimensionality following i.i.d. density functions randomly determined by the mixture of Gaussian distributions, denoted as $X$, and investigate how the consistency between $k$NN and SNN changes over dimensionality.
    }
    \label{fig:theoretical_mnc}
\end{figure*}

\paragraph{Empirical verification} 
We prepare high-dimensional random Gaussian distribution $X$ and examine how the average cosine similarity between the rows of $k$NN and SNN matrices (i.e., $k$NN-SNN consistency) changes over dimensionality, where the $(i, j)$-th cell of $k$NN and SNN matrices is 1 if the $i$-th and $j$-th points are in $k$NN or SNN relationship, respectively, and otherwise 0.
The results (\autoref{fig:theoretical_mnc}) show that consistency converges to a certain value as $d$ increases, supporting the theorem. Additionally, the fact that the converged value decreases as $k$ gets small supports our theoretical finding that the consistency converges to $k^3/N^2$.

\subsubsection{Algorithm} 

We detail the procedure of computing \MNC with a hyperparameter $k$.

\paragraph{(Step 1) Compute the $k$NN similarity matrix}
For a given dataset $X = \{p_1,p_2, \cdots, p_{N}\}$, we first compute $k$NN matrix $M^{kNN}$, which is an adjacency matrix of $X$ that regarding where two points are connected if they are in $k$NN relationship. Formally,
\[
M^{kNN}_{i,j} = \max(0, k-r+1) \text{ if } i\neq j \text{ else } 0,
\]
where $p_j$ is $p_i$'s $r$-th NN determined by Euclidean distance.

\paragraph{(Step 2) Compute the SNN similarity matrix}
We compute SNN similarity matrix $M^{SNN}$ by setting diagonal elements as 0 and every $(i, j)$-th non-diagonal elements as:
\[
M^{SNN}_{i,j} = \sum_{(m,n) \in N_{p_i,p_j}} (k+1-m) \cdot (k+1-n),
\]
where $(m,n) \in N_{p_i,p_j}$ if $m$-th NN of $p_i$ is identical to $n$-th NN of $p_j$.

\paragraph{(Step 3) Compute the discrepancy between $k$NN and SNN matrices}
At last, the final \MNC score of $X$ is computed as:
\[
\MNC(X) = \sum_{i \in \{1, \cdots, N\}} \cos(M^{kNN}_{i, *}, M^{SNN}_{i, *}) / N,
\]
where $M_{i,*}$ denotes the $i$-th row of $M$, and $\cos$ designates cosine similarity.
We use cosine similarity as it is invariant to the scaling of similarity matrices (P2) and works robustly regardless of $N$. As cosine similarity ranges from 0 to 1, \MNC also ranges from 0 to 1 ($0 \leq \MNC \leq 1$). Lower scores indicate more inconsistency, i.e., higher structural complexity.

\subsubsection{Compliance with the Desired Properties}

\label{sec:mnccompile}

We discuss how \MNC meets our desired properties for structural complexity metrics (\autoref{sec:definition}).

\paragraph{(P1) Independence to any DR techniques}
\MNC relies only on distance functions ($k$NN and SNN) and is independent of any DR technique. 

\paragraph{(P2) Scale-invariance}
\MNC scores do not change due to the global scaling of datasets, as both $k$NN and SNN are scale-invariant. 
$k$NN is scale-invariant as the ranking of NNs is not affected by scaling.
 SNN is also scale-invariant as it only relies on $k$NN similarity. 

\paragraph{(P3) Computational benefit}
$k$NN and SNN matrix construction requires $O(kN\log N)$ and $O(kN^2)$, respectively. 
However, $k$NN can be accelerated up to $O(\log^2|k|)$ \cite{johnson21tbd}, and for the SNN matrix, we can compute every matrix cell in parallel. 
Computing cosine similarity over rows can also be fully parallelized, as every pair of rows can be treated individually. Such parallelization makes the run of \MNC significantly faster than typical DR techniques (empirical validation in Appendix D). We empirically demonstrate the practical efficiency of \MNC in our quantitative analysis (\autoref{sec:acceval}).

\subsection{\PDSMNC}

We propose to leverage the ensemble of \PDS and \MNC, which we call \PDSMNC. This can be done by using both \PDS and \MNC scores as independent variables of regression models in predicting structural complexity or the maximum achievable accuracy of DR techniques. 
As \PDS and \MNC focus on global and local structure, respectively, we can expect \PDSMNC to generally work well in practice.
We recommend using multiple \MNC with different $k$ values as regression models theoretically work better when there are more variables.

\subsection{Implementation}

\label{sec:imp}

We develop our metrics using Python. 
We use \texttt{CUDA} provided by \texttt{numba} \cite{lam15llvm} to optimize and parallelize the algorithms. We also exploit \texttt{faiss} \cite{johnson21tbd} to parallelize $k$NN computation. 
While using \texttt{faiss}, we use \texttt{IndexFlatL2} option to compute $k$NN without approximation. 
One limitation here is that we store distance matrices within the memory of a single GPU; thus, the implementation may result in out-of-memory error for large datasets. Improving the implementation to handle larger data will be a critical future work to increase the practical applicability of the metrics.

\section{Experiment 1: Validity of \PDSMNC}

\label{sec:acceval}

We evaluate how accurately our structural complexity metrics can predict the ground truth structural complexity. 

\subsection{Objectives and Study Design}

We want to verify that \PDS, \MNC, and \PDSMNC have high predictive power towards ground truth structural complexity, comparing them against baselines (intrinsic dimensionality metrics).
We also aim to provide practical guidelines for selecting structural complexity metrics for different DR evaluation metrics $C$.
We first approximate the ground truth structural complexity of HD datasets following its definition (\autoref{sec:definition}).
We then evaluate the accuracy of our structural complexity metrics and baselines in predicting the approximated ground truth.

\paragraph{Approximating ground truth structural complexity}
The procedure of approximating ground truth structural complexity aligns with its definition (\autoref{sec:definition}).
For a given evaluation metric $C$, we approximate ground truth $S_C(X)$ by identifying a projection $Y$ that has maximal accuracy $C(X, Y)$, and using this accuracy as the ground truth structural complexity. 
Exhaustively testing every possible 2D projection guarantees finding the global optimum but requires testing an infinite number of projections. Instead, we test projections with a high probability of being local optima to approximate $Y$.
We do so by leveraging the ensemble of multiple DR techniques. We prepare DR techniques and identify the optimal accuracy of each technique using Bayesian optimization \cite{snoek12nips}.
Then, the maximum optimal accuracy obtained by DR techniques is set as an approximated ground truth structural complexity.

For DR techniques, we first use $t$-SNE, UMAP, LLE, and Isomap, which are currently (November 2025) the most widely referenced DR techniques according to their citation numbers in Google Scholar. 
We also use PCA as a representative global DR technique. 
Finally, we use UMATO \cite{jeon22vis} to add more diversity, as this technique aims to balance the preservation of local and global structures. Refer to Appendix XX for the technical details, e.g., hyperparameter settings. 

For $C$, we use five representative DR evaluation metrics \cite{jeon23vis} that quantify the distortions focusing on diverse structural characteristics.
We pick two local metrics (T\&C, MRRE), two global metrics (Spearman's $\rho$ and Pearson's $r$), and one cluster-level metric (Label-T\&C). 
For T\&C, MRREs, and Label-T\&C, we use the F1 score of the two scores produced by the metric. 
Note that we constrain $C$ to satisfy P2 to make their scores comparable across datasets. For example, Steadiness \& Cohesiveness \cite{jeon21tvcg}, KL divergence \cite{hinton02nips},  and DTM \cite{chazal11fcm} are excluded as they are not independent of global scaling (proofs in Appendix XX). Please also refer to Appendix XX for the technical details.

\paragraph{High-dimensional datasets}
We use 96 real-world HD datasets \cite{jeon25tpami} having diverse characteristics (e.g., number of points, dimensionality). For the datasets exceeding  $3,000$ points, we randomly select a sample of 3,000 points. This is because computing approximated ground truth takes too long (e.g., more than a week) for some datasets.


\paragraph{Baselines}
We use both projection-based and geometric intrinsic dimensionality metrics (\autoref{sec:intdim}) as baselines. We use the PCA-based method for the projection-based metric due to its efficiency (P3) and popularity in literature \cite{espadoto21tvcg, fan2010arxiv, gong19cvpr}. We compute the number of principal components required to explain more than 95\% of data variance, following Espadoto et al. \cite{espadoto21tvcg}. 
For the geometric metric, we use the correlation method due to its robustness in HD compared to the box-counting method \cite{grassberger83pnp}.

\paragraph{Measurement}
We evaluate how well our metrics and baselines correlate with the approximated ground truth structural complexity by computing the capacity of regression models to predict the approximated ground truths from metric scores.
This is done by quantifying the average $R^2$ correlation scores obtained by five-fold cross-validation.
As we only have 96 datasets, the results may depend on how the datasets are split for cross-validation. 
Therefore, for each metric, we repeat the measurement 100 times with different splits and report the maximum score, i.e., the ideal correlation.  

To avoid bias, we use five regression models: linear regression (LR), Polynomial regression (PR), $k$NN regression ($k$NN), Random forest regression (RF), and Gradient boosting regression (GB). Please refer to the Appendix of the original paper for technical details \cite{jeon25tvcg3}.

\paragraph{Hyperparameters}
We set $k=50$ for \MNC.
In terms of \PDSMNC, theoretically, adding more variables always increases the predictive power of regression models. 
However, as we have a small number of datasets, adding more variables can make the models suffer from data sparsity. Our ensemble thus consists of \PDS and three \MNC{}s ($k=25, 50$, and 75).
All other competitors do not have hyperparameters.

\paragraph{Apparatus}
We execute the experiment using a Linux server with 40-core Intel Xeon Silver 4210 CPUs, TITAN RTX, and 224GB RAM. 
We use a single GPU to compute complexity metrics. 
We also use this machine for the following experiments.

\subsection{Results and Discussions}

The following are the findings of our experiment.

\definecolor{red}{RGB}{250, 95, 126}
\definecolor{lightred}{RGB}{247, 163, 180}
\definecolor{lightlightred}{RGB}{245, 208, 216}

\definecolor{myblue}{RGB}{66, 135, 245}
\definecolor{mylightblue}{RGB}{121, 168, 242}
\definecolor{mylightlightblue}{RGB}{197, 213, 240}

\newcommand{\redc}{\cellcolor{red}}
\newcommand{\lredc}{\cellcolor{lightred}}
\newcommand{\llredc}{\cellcolor{lightlightred}}

\begin{table}[t!]
    \centering
    \caption{Results of our analysis on the accuracy of structural complexity metrics in predicting the ground truth (\autoref{sec:acceval}). Each cell depicts the correlation between structural complexity metrics or intrinsic dimensionality metrics (columns) and ground truth structural complexity approximated using five DR evaluation metrics (rows).
    \PDS and \MNC show high correlations with global and local structural complexity, respectively, outperforming intrinsic dimensionality metrics. Their ensemble (\MNCPDS) achieves the best correlation for all cases. 
    }
    \scalebox{0.9}{
    \begin{tabular}{crccccc}
    \toprule
     & & \multicolumn{2}{c}{Local} & Cluster & \multicolumn{2}{c}{Global} \\
     \cmidrule(lr){3-4} \cmidrule(lr){5-5} \cmidrule(lr){6-7} 
     & &  T\&C & MRREs & L-T\&C & S-$\rho$ & P-$r$ \\ 
     \midrule
         \multirow{5}{*}{\makecell{Int. Dim. \\ (Projection)}} & LR  & $.5574$ \llredc& $.5133$\llredc& $.3364$& $.3629$& $.3576$ \\
                           & PR  & $.5334$\llredc& $.4505$\llredc & $.2369$ & $.4231$ \llredc& $.3769$ \\
                           & $k$NN & $.6353$\lredc& $.5537$\llredc& $.3770$ & $.6495$ \lredc&$.6281$ \lredc\\
                           & RF    & $.5815$\llredc&$.4710$ \llredc&$.3013$ &$.6309$ \lredc& $.5996$ \llredc\\
                           & GB   & $.6132$\lredc&$.4021$ \llredc&$.3183$ &$.6278$ \lredc&$.5652$ \llredc \\
    \midrule
    \multirow{5}{*}{\makecell{Int. Dim. \\ (Geometric)}} & LR  & $<0$& $.0142$& $.0065$& $.0689$& $.0478$ \\
                           & PR  & $.0123$& $.2351$& $.0045$& $.0716$ &.$0566$ \\
                           & $k$NN & $.5214$ \llredc& $.3656$& $.3755$& $.5826$\llredc& $.5967$\llredc\\
                           & RF    & $.5054$\llredc& $.3807$& $.3958$& $.6010$ \lredc& $.5881$\llredc\\
                           & GB   & $.5587$\llredc& $.4430$ \llredc&$.3822$ & $.5711$\llredc& $.5730$ \llredc\\
    \midrule
    \midrule
     \multirow{5}{*}{\PDS} & LR  &  $.2781$& $.0673$& $.3404$& $.5152$\llredc& $.5332$\llredc\\
                           & PR  & $.3440$& $.0866$& $.4032$\llredc& $.7304$\lredc& $.7164$\lredc\\
                           & $k$NN & $.4133$\llredc& $.3331$& $.4998$\llredc& $.8003$\redc& $.8180$\redc\\
                           & RF    & $.4119$\llredc& $.3509$ & $.5010$\llredc & $.8282$\redc& $.8217$\redc\\
                           & GB   & $.4523$\llredc& $.3182$ & $.4681$\llredc& $.7959$ \lredc&  $.8075$\redc\\
    \midrule
    \multirow{5}{*}{\MNC} & LR  & $.8454$ \redc& $.6784$ \lredc& $.3692$& $.5677$ \llredc& $.5241$\llredc \\
                           & PR  & $.8807$ \redc& $.7244$ \lredc& $.3174$ & $.5525$ \llredc& $.5140$\llredc \\
                           & $k$NN & $.8780$\redc& $.7007$ \lredc& $.4020$\llredc & $.5962$\llredc& $.5814$\llredc\\
                           & RF    & $.8706$\redc& $.7302$\lredc& $.4189$\llredc& $.6166$\lredc& $.5734$ \llredc\\
                           & GB   & $.8666$\redc& $.7207$\lredc& $.2827$& $.5992$\llredc&  $.5741$\llredc\\
    
    \midrule
    \multirow{5}{*}{\MNCPDS} & LR  & $.8513$\redc& $.7484$\lredc& $.5104$\llredc& $.6772$\lredc& $.7474$\lredc\\
                           & PR  & $\boldsymbol{.8984}$\redc& $\boldsymbol{.8423}$\redc& $.0015$& $.5954$\llredc& $.6572$\lredc\\
                           & $k$NN & $.7472$\lredc& $.6109$\lredc& $.4971$\llredc& $\boldsymbol{.8290}$\redc& $\boldsymbol{.8401}$\redc\\
                           & RF    & $.8881$\redc& $.7506$\lredc& $.5823$\llredc& $.8273$\redc& $.8092$\redc\\
                           & GB   & $.8694$\redc& $.7636$\lredc& $\boldsymbol{.6067}$\lredc& $.8280$\redc&  .8079\redc\\

     \bottomrule
\addlinespace[0.115cm]
\multicolumn{7}{l}{
  1. \textcolor{red}{$\blacksquare$} / \textcolor{lightred}{$\blacksquare$} / \textcolor{lightlightred}{$\blacksquare$}: very strong ($R^2 \geq 0.8)$ / strong ($0.6 \leq R^2 < 0.8)$ / 
} \\
\addlinespace[0.05cm]
\multicolumn{7}{l}{
 \hspace{18mm}moderate ($0.4 \leq R^2 < 0.6)$ correlation \cite{sarjana20jtm}
} \\ 
\addlinespace[0.115cm]
 \multicolumn{7}{l}{
 2. \textbf{Bold} numbers refer to the top score of the column
}
    \end{tabular}
    }
    
    \label{tab:acceval}
\end{table}

\paragraph{Validity of the design of \PDS and \MNC}
The results (\autoref{tab:acceval}) verify that \PDS and \MNC are appropriately designed and work as intended. While \PDS shows at most a very strong or strong correlation with global structural complexity (S-$\rho$ and P-$r$), \MNC achieves at most a very strong correlation with local structural complexity (T\&C and MRREs). 
Meanwhile, both metrics show relatively weak correlations in the opposite cases; showing at most strong or moderate correlations. 
\PDS also shows at most strong correlation with linear and polynomial regression models (LR, PR).
As these models do not work well with non-Gaussian data, such results verify the validity of log transformation (Step 2). 
In contrast, intrinsic dimensionality metrics correlate weakly with the ground truth. For all types of ground truth structural complexity, the best correlation achieved by these metrics is substantially lower than that achieved by $\PDS$, $\MNC$, and $\MNCPDS$. 

\paragraph{Superior predictive power of \PDS and \MNC}
The results also confirm the effectiveness of \PDSMNC.
\PDSMNC achieves the best correlation regardless of DR evaluation metrics used to approximate ground truth structural complexity, making substantial performance gains compared to using \MNC and \PDS individually. For example, \MNCPDS has at most a strong correlation even for cluster-level structural complexity. \MNCPDS is also the sole subject that achieves a very strong correlation for the structural complexity approximated by MRREs. 
Such results underscore the importance of testing alternative strategies to ensemble complexity metrics or identifying the most effective combination. Aligned with this finding, we use \PDSMNC to demonstrate the applicability of our metrics in supporting our dataset-adaptive workflow (\autoref{sec:suitability}).

\paragraph{Weaknesses of our structural complexity metrics}
We also reveal the weaknesses of our complexity metrics. 
\PDS and \MNC show at most a weak correlation with cluster-level structural complexity (L-T\&C). Although \MNCPDS shows at most a strong correlation, the best $R^2$ value obtained is substantially lower compared to the ones from global and local structural complexity. 
Also, \MNC has a relatively weak correlation with local structural complexity approximated by MRREs compared to the one approximated by T\&C. 
We recommend not using the structural complexity metrics for such cases. For example, we recommend using \PDS and \MNCPDS when S-$\rho$ or P-$r$ is used, but not \MNC. The result indicates that we currently lack structural complexity metrics that can be reliably used for cluster-level evaluation metrics (L-T\&C). Such results underscore the need to design new, advanced structural complexity metrics that can complement \PDS, \MNC, and \PDSMNC.

\begin{figure}
    \centering
    \includegraphics[width=0.75\linewidth]{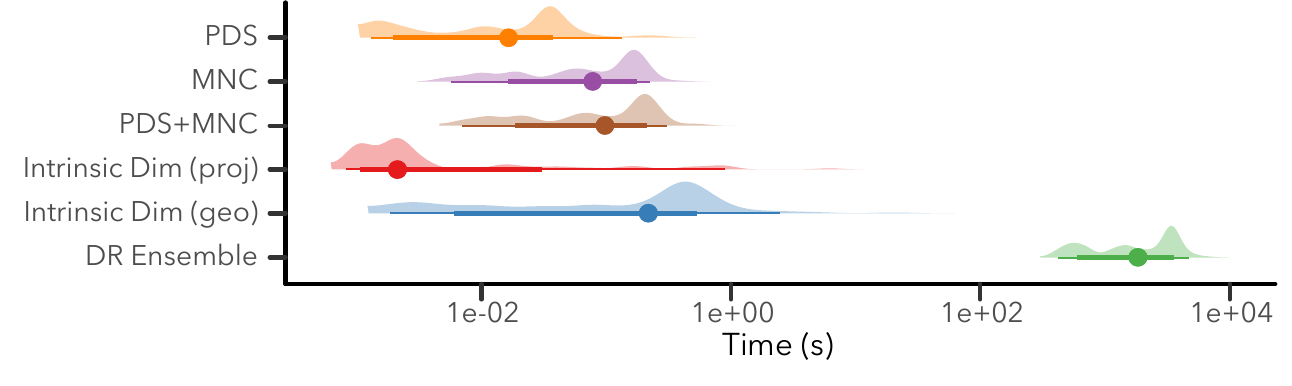}
    \caption{The runtime of our structural complexity metrics, intrinsic dimensionality metrics, and DR ensemble to produce ground truth structural complexity. 
    \PDS, \MNC, and \PDSMNC are faster than the geometric intrinsic dimensionality metric and the ensemble. The thick and thin lines indicate the 66\% and 99\% interval of probability mass.
    }
    \label{fig:efficiency}
\end{figure}

\paragraph{Computational efficiency of the metrics}
We also investigate the execution time of \PDS, \MNC, \PDSMNC, and baselines in being applied to 96 real-world datasets we use. We also examine the execution time for approximating ground truth structural complexity using the ensemble of DR techniques.
Note that we rely on GPU-based parallelization and highly optimized libraries like \texttt{scikit-learn} \cite{pedregosa11jmlr} and PyTorch \cite{paszke19neurips} while implementing intrinsic dimensionality metrics to ensure fairness.

As a result (\autoref{fig:efficiency}), we find that our structural complexity metrics are slower than the projection-based intrinsic dimensionality metric but are faster than other baselines. All three metrics required less than one second for all datasets, substantially scaling up the DR ensemble method. 
Regarding the high accuracy of \PDS and \MNC in predicting the ground truth structural complexity produced by the DR ensemble (\autoref{sec:acceval}), the result indicates that we achieve a favorable balance between efficiency and accuracy.

\section{Experiment 2: Suitability of \PDSMNC for the Dataset-Adaptive Workflow}

\label{sec:suitability}

We want to examine the utility of \PDSMNC, the most advanced structural complexity metric, in properly guiding the dataset-adaptive workflow (\autoref{sec:daworkflow}).
As with a previous experiment (\autoref{sec:acceval}), the experiment also aims to provide guidelines to select structural complexity metrics in practice.
We first evaluate the accuracy of the pretrained regression models in predicting the maximum achievable accuracy of DR techniques from the scores from \PDSMNC (\autoref{sec:maxacc}).
We then evaluate the effectiveness of \PDSMNC in guiding each step of our workflow (\autoref{sec:evalguiding} for Step 1 and \autoref{sec:evalguiding22} for Step 2).

\subsection{Evaluation on Pretraining Regression Models}

We detail our evaluation of \PDSMNC's suitability for pretraining regression models in the dataset-adaptive workflow.

\label{sec:maxacc}

\subsubsection{Objectives and Study Design}

We want to evaluate the utility of \PDSMNC in training regression models that predict the maximum achievable accuracy of DR techniques. 
We assess the performance of regression models that predict the maximum achievable accuracy of DR techniques from metric scores, comparing them with those that use baselines (intrinsic dimensionality metrics).
For 96 datasets that we have, we first obtain maximum scores of DR techniques by optimizing them using Bayesian optimization (detailed settings in Appendix B). 
We then compute the $R^2$ scores of regression models predicting the maximum accuracy from \MNCPDS and baselines.
This is done by (1) splitting 96 datasets into 80 training datasets and 16 test datasets, (2) training the regression model with a training dataset with five-fold cross-validation, and (3) assessing the performance of the model for unseen test datasets.
To make scores comparable with previous experiments (\autoref{sec:acceval}), we repeat the measurement 10 times with different splits and report the ideal performance. 
For the structural complexity metric, we use \MNCPDS as it shows the best correlation with approximated ground truth structural complexity (\autoref{sec:acceval}).

\paragraph{Regression models}
To examine ideal prediction, we leverage AutoML~\cite{he21kbs} by relying on \texttt{auto-sklearn}~\cite{feurer15nips} to train the regression model. 
We train the model for 60 seconds. 
We do not test the five regression models used in our correlation analysis as they hardly achieve ideal predictions.

\paragraph{DR techniques and metrics}
We use the same set of DR techniques and evaluation metrics with the validity analysis (\autoref{sec:acceval}).

\begin{table}[t]
    \centering
    
    \caption{Accuracy of baseline metrics and \PDSMNC on predicting maximum accuracy achievable by different DR techniques (\autoref{sec:maxacc}). \PDSMNC shows strong predictability for the majority of cases, outperforming baselines in the majority of cases. }
    \scalebox{0.90}{
    \begin{tabular}{crccccc}
         \toprule 
                &  &  T\&C & MRREs & L-T\&C & S-$\rho$ & P-$r$ \\
\midrule
               \multirow{6}{*}{\makecell{Int. Dim. \\ (Projection)}}  & UMAP    & $.6540$   \lredc & $.6449$ \lredc  &  $.2656$  &  $.6389$ \lredc &  $.3690$\\
         &$t$-SNE  & $.7615$ \lredc & $.6643$ \lredc & $.0212$ &  $.5746$  \llredc  &  $.1674$\\
        & PCA      & $.8027$ \redc &  $.8430$ \redc& $.6524$ \lredc  &  $.7112$ \lredc & $.6445$\lredc  \\
        & Isomap  & $.7998$ \lredc &    $.7924$ \lredc &  $.6094$ \lredc &  $.7326$   \lredc  &  $.6557$ \lredc \\
        & LLE     & $.7747$ \lredc &  $.7846$  \lredc & $.2987$ &   $.6737$  \lredc &  $.7473$ \lredc \\
        & UMATO   & $.7207$ \lredc & $.6906$ \lredc & $.3222$ & $.7109$ \lredc &  $.2838$ \\
\midrule
               \multirow{6}{*}{\makecell{Int. Dim. \\ (Geometric)}}  & UMAP    & $.7146$ \lredc  & $.6519$  \lredc&  $.4204$  \llredc &  $.6550$ \lredc &  $.0405$\\
         &$t$-SNE  & $.5097$ \llredc & $.5767$ \llredc & $.0029$ &  $.3303$  &  $.1257$\\
        & PCA      & $.8588$ \redc &  $.8417$ \redc & $.1720$ &  $.7989$ \lredc& $.7160$ \lredc\\
        & Isomap  & $.8253$ \redc &    $.7069$ \lredc&  $.5629$ \llredc  &  $.8544$   \redc &  $.8855$ \redc \\
        & LLE     & $.8496$ \redc &  $.4476$  \llredc & $.7058$ \lredc&   $.7123$ \lredc&  $.7539$ \lredc\\
        & UMATO   & $.8438$ \redc & $.7038$ \lredc& $.2275$ & $.5660$ \llredc &  $.1383$ \\ 
        \midrule 
        \midrule
\multirow{6}{*}{\MNCPDS} & UMAP    & $.8955$   \redc   & $.8183$  \redc&  $.3944$  &  $.7978$ \lredc &  $.7578$ \lredc\\
         &$t$-SNE  & $.9471$ \redc& $.7848$ \lredc& $.4303$ \llredc &  $.8600$ \redc   &  $.6569$ \lredc\\
        & PCA      & $.9150$ \redc &  $.8446$ \redc & $.5784$ \llredc &  $.9063$ \redc& $.8836$  \redc\\
        & Isomap  & $.8514$ \redc&    $.9123$ \redc &  $.7962$ \lredc   &  $.9454$ \redc       &  $.8935$ \redc\\
        & LLE     & $.8781$   \redc  &  $.8501$ \redc & $.6986$ \lredc  &   $.7645$ \lredc  &  $.7739$ \lredc\\
        & UMATO   & $.9223$ \redc& $.7823$ \lredc & $.5733$ \llredc & $.8392$ \redc &  $.6880$ \lredc\\
        \bottomrule
        \addlinespace[0.115cm]
         \multicolumn{6}{l}{
         \makecell{
        \textcolor{red}{$\blacksquare$} / \textcolor{lightred}{$\blacksquare$} / \textcolor{lightlightred}{$\blacksquare$}: very strong ($R^2 \geq 0.8)$ / strong ($0.6 \leq R^2 < 0.8)$ / \\ 
        \hspace{13mm}moderate ($0.4 \leq R^2 < 0.6)$ predictive power\cite{sarjana20jtm}
        }
        }
    \end{tabular}
    }
    \label{tab:indmetric}
\end{table}

\subsubsection{Results and discussions}
\autoref{tab:indmetric} depicts the results.
Overall, \MNCPDS has strong predictability with maximum accuracy achievable by DR techniques, outperforming baseline metrics.
For example, while \MNCPDS shows at least a strong predictive power for S-$\rho$ and P-$r$, the baselines fail to do so. 
\MNCPDS also has at least strong predictability for most combinations of DR techniques and evaluation metrics (26 out of 30; 87\%), verifying that the metric can be generally applied to execute the dataset-adaptive workflow in practice. In contrast, projection-based and geometric projection metrics show at least strong predictability for 21 and 17 combinations, respectively. 

However, \MNCPDS relatively works poorly for L-T\&C, showing moderate or weak correlations for four cases. 
Such results align with the results from our previous analysis (\autoref{sec:acceval}), clarifying the need for further development of complexity metrics that complement \PDS and \MNC. \PDSMNC still outperforms baselines also for L-T\&C.

\subsection{Evaluation on Predicting Effective DR Techniques}

\label{sec:evalguiding}

We investigate the effectiveness of \PDSMNC in predicting the accuracy ranking of DR techniques (Step 1; \autoref{fig:workflow} DW1).

\subsubsection{Objectives and Study Design}

We examine whether the pretrained regression models using \PDSMNC can effectively distinguish DR techniques with high and low maximum achievable accuracy in the testing phase.
This capability is important in properly guiding dataset-adaptive workflow to avoid the optimization of suboptimal DR techniques (\autoref{sec:datasetadaptive}, \autoref{fig:workflow} DW1).

We reuse the maximum achievable accuracy of DR techniques computed previously (\autoref{sec:maxacc}). Leveraging this score, we identify the ground truth accuracy ranking of DR techniques. We then train the regression models with 80 training datasets as we do in the previous experiment (\autoref{sec:maxacc}) using \PDSMNC and baselines (intrinsic dimensionality metrics). Based on the models’ approximated accuracy predictions, we derive the predicted accuracy ranking of DR techniques and compute the rank correlation between the ground truth and predicted rankings.
We run the same experiment 10 times with different splits of the datasets and report the average results. 

\paragraph{DR techniques and metrics}
We use the same set of DR techniques and evaluation metrics as the previous experiments (\autoref{sec:acceval}, \ref{sec:maxacc}).

\begin{figure}
    \centering
    \includegraphics[width=0.75\linewidth]{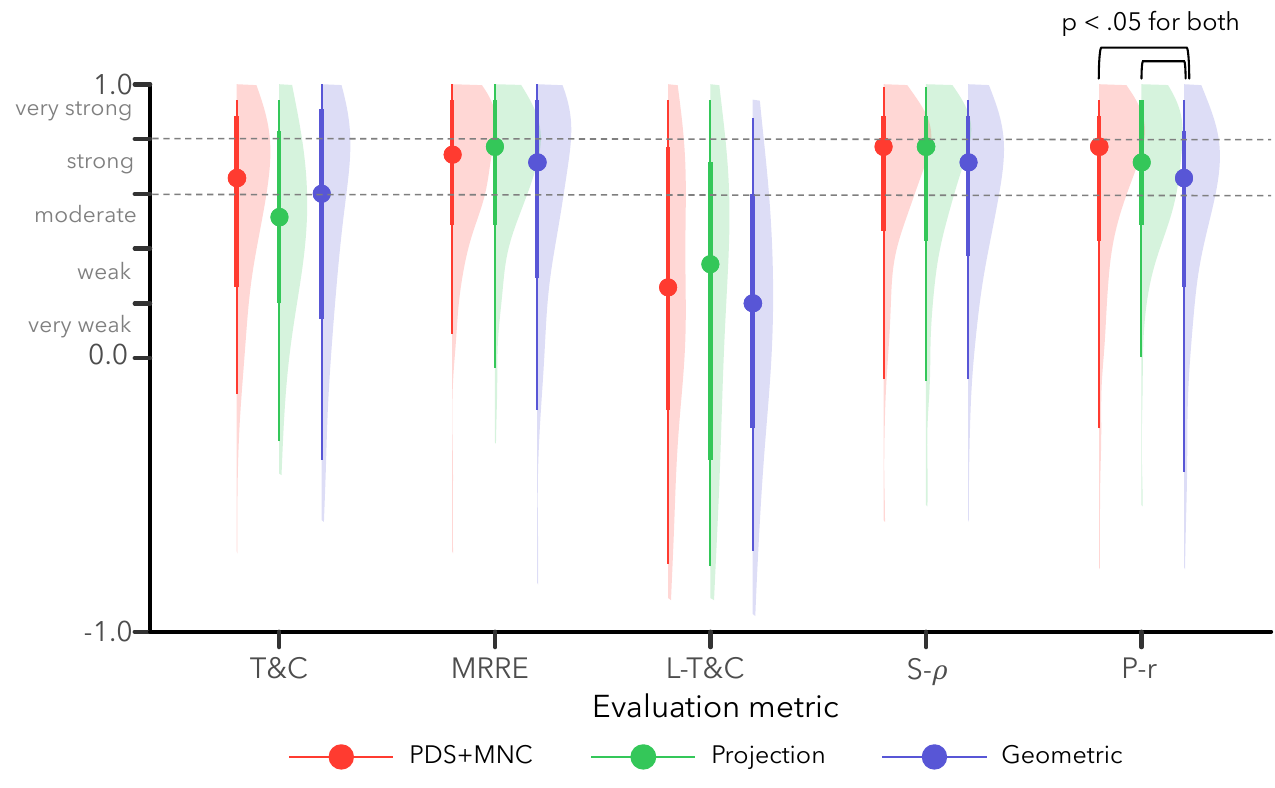}
    \caption{The distribution of the correlation between the true accuracy ranking of DR techniques and the predicted ranking estimated based on \MNCPDS and the baselines (\autoref{sec:evalguiding}). 
    We use the criteria of Prion and Hearing \cite{prion14csn} to explain the strength of the correlation (very strong to very weak).
    Overall, \PDSMNC demonstrates strong predictive power for the majority of cases.
    We also find that all three metrics perform similarly in predicting rankings. The thick and thin lines indicate the 66\% and 99\% probability mass intervals.
    }
    \label{fig:correlation}
\end{figure}

\subsubsection{Results and discussions}
We find that \PDSMNC achieve strong accuracy in predicting the ground truth ranking of DR techniques computed by T\&C, MRRE, S-$\rho$, and P-$r$ (\autoref{fig:correlation} red marks). The result reaffirms the capability of structural complexity metrics in guiding the dataset-adaptive workflow. In contrast, the rankings predicted by \PDSMNC show weak correlation with the ground truth for L-T\&C case, aligning with our previous experiments (\autoref{sec:acceval}, \ref{sec:maxacc}). 

We also find that baselines show good performance in predicting rankings that correlate well with ground truth rankings. We examine the difference in correlations between \PDS and two baselines for each evaluation metric using ANOVA. 
As a result, we find no significant difference for T\&C ($F_{2,477} = 2.26$, $p = .104$), MRRE ($F_{2,477} = 2.71$, $p = .067$), L-T\&C ($F_{2,477} = 1.48$, $p = .226$), and S-$\rho$ ($F_{2,477} = 0.91$, $p = .403$). 
We find a significant difference for P-$r$ case ($F_{2,477} = 5.37$, $p < .01$), and post-hoc test using Tukey's HSD confirms that \PDSMNC and projection-based intrinsic dimensionality metric significantly outperform geometric intrinsic dimensionality metric ($p < .05$ for both). Still, the geometric intrinsic dimensionality metric shows strong predictive power. 
These results indicate that distinguishing between effective and ineffective techniques does not require precise maximum accuracy predictions. We discuss the takeaways of this phenomenon on the practical use of the dataset-adaptive workflow in \autoref{sec:necessity}.

\subsection{Evaluation on Early Terminating Optimization}

\label{sec:evalguiding22}

\renewcommand{\hbar}[3]{%
  \begin{tikzpicture}[baseline=(textnode.base)]
    \pgfmathsetmacro{\barwidth}{0.87} 
    \pgfmathsetmacro{\barheight}{0.25}
    \pgfmathsetmacro{\perc}{#1/#2}
    \draw[blue!07, fill=blue!07] (0,0) rectangle (\barwidth,\barheight);
    \fill[blue!28] (0,0) rectangle (\perc*\barwidth,\barheight);
    \node (textnode) [anchor=mid, inner sep=0, font=\scriptsize] at (\barwidth/2, \barheight/2) {#3};
  \end{tikzpicture}%
}

\begin{table}[t]
    \centering
    \caption{The effectiveness of structural complexity metrics (\PDSMNC) and baselines (projection-based or geometric intrinsic dimensionality metrics) in guiding the early termination of hyperparameter optimization. We report the average error across all trials, the worst-case errors at the 10\%, 5\%, and 1\% percentiles, and the relative time required compared to full optimization. Each blue bar within a table cell represents the cell's relative value compared to the maximum value for each combination of DR evaluation metrics (rows) and measurands (columns). The bold denotes the best performance for each combination.
    The results verify that the approximation using \PDSMNC consistently achieves desirable accuracy while maintaining efficiency. 
    }
    \scalebox{1}{
    \begin{tabular}{clccccc}
         \toprule 
                & & \multicolumn{4}{c}{Error} & \multirow{2.5}{*}{\makecell{Rel. \\Time}}\\
                 \cmidrule{3-6}
                &  & All & 10\% & 5\% & 1\% & \\
                \midrule
        \multirow{3}{*}{\makecell{UMAP \\(w\ T\&C)}} 
        & Projection   & \hbar{.0005}{.0005}{$.0005$} & \hbar{.0112}{.0157}{$\boldsymbol{.0112}$} & \hbar{.0179}{.0192}{$.0179$} & \hbar{.0261}{.0285}{$.0261$} & \hbar{40.6}{40.6}{$40.6$\%} \\
        & Geometric    & \hbar{.0005}{.0005}{$.0005$} & \hbar{.0157}{.0157}{$.0157$} & \hbar{.0192}{.0192}{$.0192$} & \hbar{.0285}{.0285}{$.0285$} & \hbar{29.0}{40.6}{$\boldsymbol{29.0}$\%} \\
        & \PDSMNC      & \hbar{.0004}{.0005}{$\boldsymbol{.0004}$} & \hbar{.0118}{.0157}{$.0118$} & \hbar{.0159}{.0192}{$\boldsymbol{.0159}$} & \hbar{.0246}{.0285}{$\boldsymbol{.0246}$} & \hbar{35.7}{40.6}{$35.7$\%} \\
        \midrule
        \multirow{3}{*}{\makecell{UMAP \\(w\ P-$r$)}} 
        & Projection   & \hbar{.0395}{.0395}{.0395} & \hbar{.1287}{.1473}{.1287} & \hbar{.1560}{.1758}{.1560} & \hbar{.1913}{.2052}{.1913} & \hbar{50.3}{51.0}{50.3\%} \\
        & Geometric    & \hbar{.0380}{.0395}{.0380} & \hbar{.1473}{.1473}{.1473} & \hbar{.1758}{.1758}{.1758} & \hbar{.2052}{.2052}{.2052} & \hbar{44.7}{51.0}{\textbf{44.7\%}} \\
        & \PDSMNC      & \hbar{.0173}{.0395}{\textbf{.0173}} & \hbar{.0639}{.1473}{\textbf{.0639}} & \hbar{.0861}{.1758}{\textbf{.0861}} & \hbar{.1005}{.2052}{\textbf{.1005}} & \hbar{51.0}{51.0}{51.0\%} \\
        \bottomrule
    \end{tabular}
    }
    \label{tab:efferror}
\end{table}

We assess the utility of \PDSMNC and regression models in accelerating hyperparameter optimization by early terminating iterations.

\subsubsection{Objectives and Study Design}
We investigate the effectiveness of \PDSMNC in improving the efficiency of DR hyperparameter optimization by reducing redundant iterations (\autoref{fig:workflow} DW2), comparing with baseline metrics (intrinsic dimensionality metrics). 
We simulate the optimization of hyperparameters for UMAP projections. We evaluate using T\&C and P-$r$ to compare two scenarios: one in which \PDSMNC and baseline metrics exhibit similar performance (T\&C) and one in which \PDSMNC substantially outperforms the baselines in predicting the maximum achievable accuracy of DR techniques (P-$r$). 
This selection is based on our evaluation of pretraining regression models (\autoref{sec:maxacc}).
As with previous experiments, we divide the 96 HD datasets into 80 training datasets and 16 test datasets and train the AutoML regression model to predict the maximum accuracy from the training dataset. We then predict the maximum accuracy of 16 unseen test datasets. 
Note that for T\&C, we use the F1 score of Trustworthiness and Continuity, following the convention of interpreting T\&C as precision and recall of DR \cite{venna10jmlr}.

Finally, we optimize UMAP on 16 datasets with and without interruption based on the predicted maximum accuracy. 
For the former (with), we run Bayesian optimization with 50 iterations and halt the process upon reaching the optimal score. We set the default iteration number as 50, a default value recommended by \texttt{scikit-optimize} \cite{louppe2017bayesian} library. For the latter condition (without), we run the optimization with 50 iterations without halting the process. 
We compare the two settings by evaluating the amount of error introduced and the relative running time compared to the optimizations executed without early termination.
We run the same experiment 10 times with different splits of the datasets.
Note that we report the average error over all trials and specifically examine errors from the worst 10\%, 5\%, and 1\% trials to detail the robustness of the early termination.

\begin{figure}[t]
    \centering
    \includegraphics[width=\linewidth]{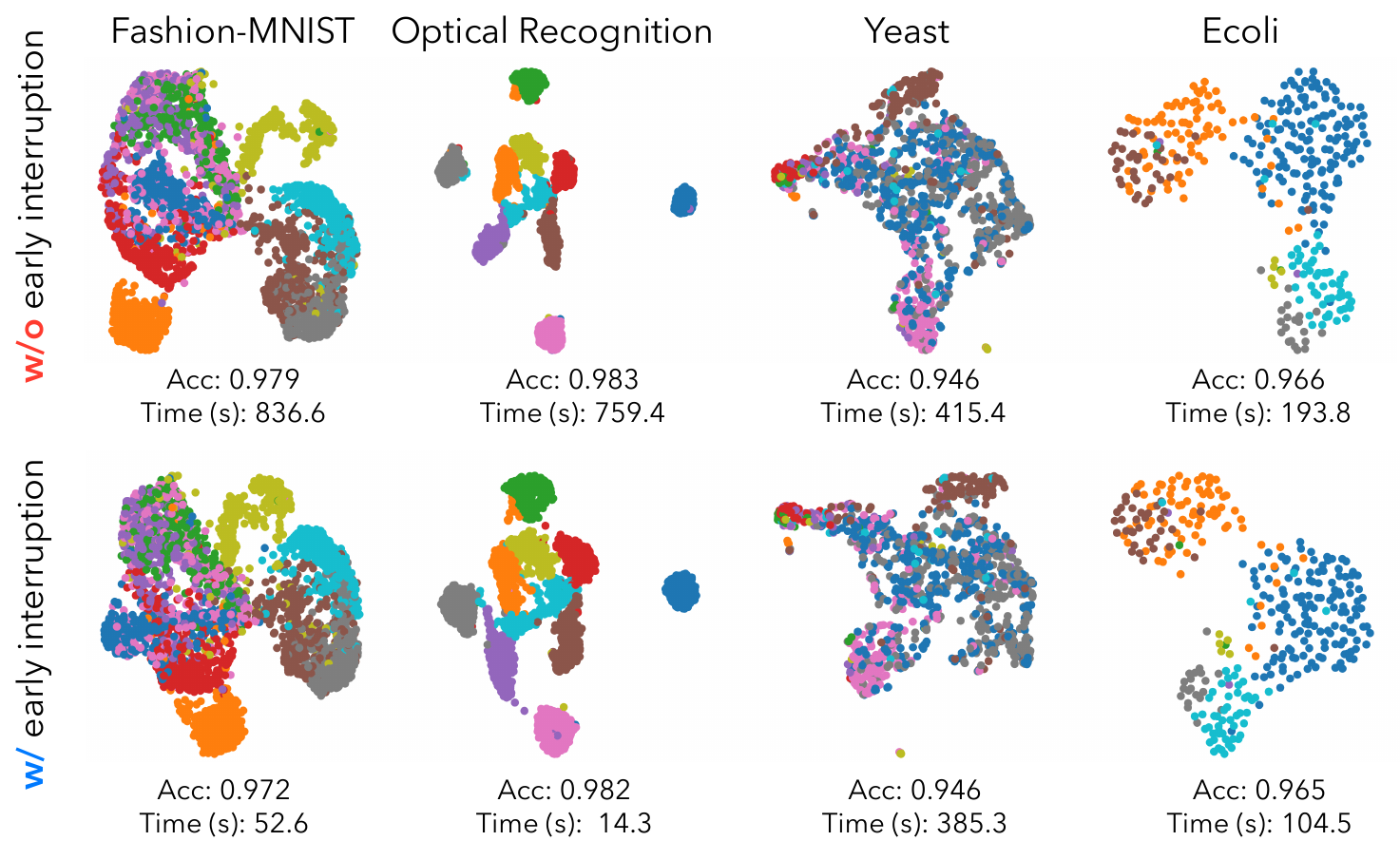}
    \caption{The projections made by optimizing UMAP with and without interruption based on predicted maximum accuracy (\autoref{sec:maxacc}).
    The interruption substantially reduces runtime while maintaining accuracy.}
    \label{fig:projections}
\end{figure}

\subsubsection{Results and discussions}
\autoref{tab:efferror} depicts the results, and \autoref{fig:projections} depicts the subset of projections generated in our experiment.
The results demonstrate the efficacy of \PDSMNC in guiding the early termination of hyperparameter optimization, verifying its usefulness in executing the dataset-adaptive workflow in practice. Specifically, early termination using \PDSMNC substantially reduces runtime while incurring only minimal errors. While reducing runtime by more than half, \PDSMNC outperforms the baselines in terms of error in the majority of cases. 
In the P-$r$ scenario, the error is approximately halved compared to the baselines.

\section{Experiment 3: Effectiveness of the Dataset-Adaptive Workflow}

We evaluate how well our dataset-adaptive workflow  (\autoref{sec:datasetadaptive}) accelerates the process of finding optimal DR projection without compromising accuracy, comparing it with the conventional workflow (\autoref{sec:conventional}).

\subsection{Objectives and Study Design}

We aim to verify two hypotheses: 

\begin{itemize}
    \item[\textbf{H1}] The dataset-adaptive workflow significantly accelerates the DR optimization process compared to the conventional workflow.
    \item[\textbf{H2}] The dataset-adaptive workflow finds DR projections with negligible accuracy loss compared to the conventional workflow.
\end{itemize}
We first split 96 HD datasets into 80 training datasets and 16 test datasets. 
Then, we train regression models for all DR techniques we use previously (\autoref{sec:maxacc}). 
We then execute conventional (\autoref{sec:conventional}) and dataset-adaptive (\autoref{sec:datasetadaptive}) workflows to optimize DR projection on test datasets, setting the default iteration count to 50. 

For our dataset-adaptive workflow, we test two variants: one optimizes the hyperparameters of the top-1 DR technique, and the other optimizes those of the top-3 techniques. This is because we want to examine the tradeoff between the execution time and the accuracy. 
We record the total execution time and the final accuracy obtained by each workflow (conventional, top-1 dataset-adaptive, top-3 dataset-adaptive). We run the same experiment 10 times with diverse dataset splits.

\paragraph{Evaluation metrics}
We want to examine both the full potential of the dataset-adaptive workflow and its effectiveness under worst-case scenarios. We measure the accuracy of DR projections using T\&C and L-T\&C, two metrics that \PDSMNC exhibits the best and worst predictive power in our previous experiment (\autoref{sec:maxacc}), respectively.

\subsubsection{Results and Discussions}

\begin{figure}
    \centering
    \includegraphics[width=0.8\textwidth]{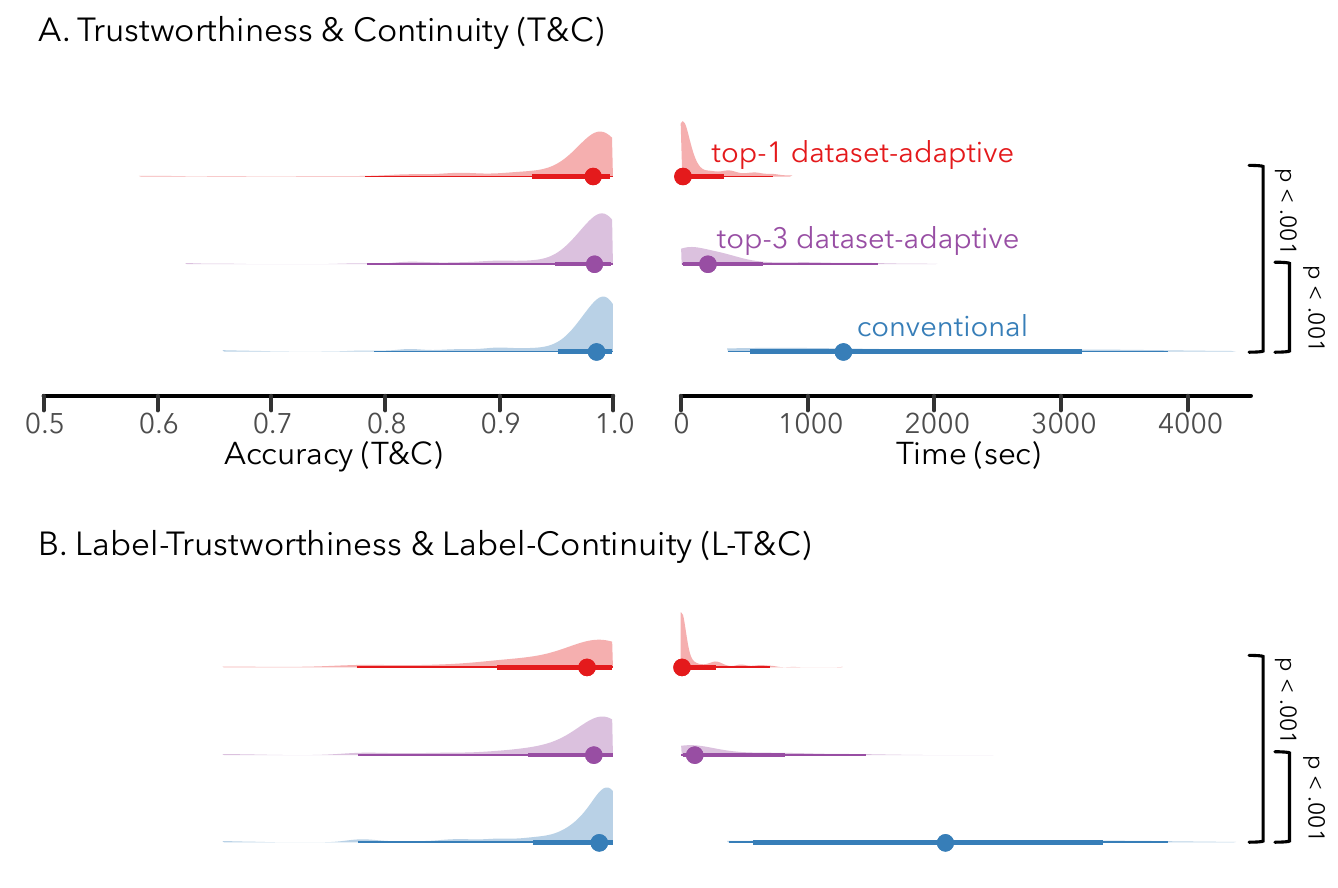}
    \caption{Comparison of the performance of three different workflows in optimizing DR projections. Dataset-adaptive workflows achieve a significant gain in execution time with negligible accuracy loss. The thick and thin lines indicate the 66\% and 99\% probability mass intervals.}
    \label{fig:fullpipe}
\end{figure}

\autoref{fig:fullpipe} depicts the results. 
By running one-way ANOVA on the accuracy scores, we find no significant difference between workflows for both T\&C ($F_{2, 477} = 1.35$, $p = .259$) and L-T\&C ($F_{2, 477} = 2.49$, $p = .084$) cases. The result confirms \textbf{H2}.
In contrast, ANOVA on the execution time indicates the statistically significant difference between the workflows (T\&C: $F_{2, 477} = 194.18$, $p < .001$; L-T\&C: $F_{2, 477} = 190.97$, $p < .001$). 
We find that top-1 and top-3 dataset-adaptive workflows are $\times 5.3$ and $\times 13$ faster than the conventional workflow, respectively.
Post-hoc analysis using Tukey's HSD also confirms that top-1 and top-3 dataset-adaptive workflows require significantly shorter execution time than the conventional workflow ($p < .001$). 
This result confirms \textbf{H1}. 
Meanwhile, we observe no significant difference between the top-1 and top-3 dataset-adaptive workflows (T\&C: $p = .056$, L-T\&C: $p = .077$).
In summary, our experiment verifies both hypotheses, confirming the benefit of using the dataset-adaptive workflow in practice. 

The results also reveal that the dataset-adaptive workflow does not always perform ideally. Although the accuracy differences are not statistically significant, the L-T\&C case yields a relatively low $p$-value (0.084), indicating a slight accuracy difference in the conventional and the dataset-adaptive workflow.
Such results suggest that more efforts should be invested in designing advanced structural complexity metrics and further refining the dataset-adaptive workflow.

\section{Discussions}

We discuss future research directions for dataset-adaptive workflow and structural complexity metrics.

\subsection{Leveraging the Tradeoffs between Predictive Power and Efficiency}

\label{sec:necessity}

We find that \PDSMNC outperforms intrinsic dimensionality metrics in early terminating hyperparameter optimization (\autoref{sec:evalguiding22}) but is similarly effective in identifying optimal DR techniques (\autoref{sec:evalguiding}). This indicates that the predictive power of structural complexity metrics is crucial for the success of the former but not for the latter.

This finding offers a new perspective on designing structural complexity metrics: leveraging the tradeoff between predictive power and efficiency. Although \PDS and \MNC run within a reasonable time frame for our dataset, they may be too slow for enormous datasets with millions of data points. 
In such cases, we can use fast but less accurate structural complexity metrics to select the DR technique in Step 1 and then apply accurate metrics like \PDS and \MNC only to the selected technique in Step 2.
We can also dynamically adjust the optimization procedure for multiple DR techniques. For example, we can run multiple DR techniques in parallel and drop those predicted to produce low values. Applying a progressive visual analytics paradigm \cite{fekete24book} will be effective here. For instance, we can begin optimization with rough estimates of promising DR techniques and eliminate those that prove to be ineffective later in the process.

To pursue this direction, it will also be crucial to investigate the utility of structural complexity metrics in depth. For example, we may examine the gap between the achieved predictive power and the theoretical optimum, and whether this gap is bounded. When metrics are computed progressively, it will also be important to investigate how the error bounds evolve over time.
These endeavors will contribute to making DR-based visual analytics both more responsive and reliable.

\subsection{Complementing Our Structural Complexity Metrics}

Our experiments reveal the necessity of developing new structural complexity metrics that complement \PDS and \MNC, especially the ones that focus on cluster-level structure. This is because \PDS, \MNC, and even \PDSMNC fall short in estimating the ground truth structural complexity and the maximum accuracy of DR techniques computed with L-T\&C, a cluster-level evaluation metric. One idea for developing cluster-level structural complexity metrics is to extend \MNC to the class level. Instead of considering neighbors of ``data points,'' we could consider those of ``classes'' or ``clusters.''

\subsection{\revise{Enhancing Reproducibility}}

\revise{
Dataset-adaptive optimization guarantees to produce projections with near-optimal accuracy; however, it does not guarantee that the resulting visual patterns are reproducible across several executions of optimization. Such limitations can potentially undermine the reliability of visual analytics workflows that rely on DR.
}

\revise{
To address this issue, we propose predicting DR hyperparameters that yield projections with optimal accuracy. The underlying idea is that the dataset-adaptive workflow already estimates which DR technique is optimal for a given dataset, which can be viewed as a coarse-grained prediction of optimal hyperparameters. Extending this approach to predict optimal hyperparameter values that consistently yield accurate projections is, therefore, a natural next step. A key challenge lies in measuring structural complexity at a finer granularity to support such predictions. One approach will be to test a broader range of $k$ values for \MNC.
}

\subsection{Exploring Additional Use Cases}

We hypothesize that structural complexity metrics can improve the \textit{replicability} of benchmarking DR techniques.
DR benchmarks may produce different conclusions about the accuracy of DR techniques depending on the datasets used, i.e., they may have low replicability.
Adding more benchmark datasets may enhance replicability, but this increases the computational burden. 
Here, complexity metrics may contribute to achieving high replicability with fewer datasets.
For example, excluding datasets with simple patterns that are accurately reducible by any DR technique will improve replicability, as the rankings of DR techniques determined based on these datasets may noise the evaluation.
Validating whether this hypothesis confirms or not will be a worthwhile future research avenue to explore.

\section{Conclusion}

It is important to find optimal DR projections to ensure reliable visual analytics. 
However, optimization processes are computationally demanding, leading practitioners to rely on cherry-picking or default hyperparameter values. 
To address this problem, we propose the dataset-adaptive workflow for accelerating DR optimization while maintaining accuracy. 
We introduce two structural complexity metrics, \PDS and \MNC, and verify their effectiveness in terms of precision and efficiency. 
We also demonstrate the utility of the dataset-adaptive workflow guided by these two metrics. 

In summary, we contribute to more reliable visual analytics by motivating practitioners to optimize DR projections rather than relying on arbitrary configurations.
Our proposal opens up broader discussions on how to democratize the benefits of reliable visual analytics.


\chapter{Distortion-aware Brushing for Reliable Cluster Analysis in Dimensionality Reduction Projections}

\label{sec:dabrca}

Brushing is one of the most widely used ways to interact with the scatterplots representing DR projections in visual analytics \cite{kwon18tvcg, chatzimparmpas20tvcg}. 
However, because DR projections are distorted, applying conventional brushing, such as lassoing, as-is can lead to errors while users interact with clusters.

In this chapter, we introduce \textit{\brush}, a new brushing technique that helps analysts to overcome distortions and precisely select high-dimensional clusters that they intended.
\section{Introduction}

Brushing is a process of selecting data points within a continuous region in the 2D space via direct manipulation such as dragging, clicking, or lassoing \cite{heer08chi}. This interaction technique allows users to focus on the selected points by labeling or highlighting them \cite{becker87technometrics, becker87statistical}. 
Since its initial introduction~\cite{fisherkeller75pacific}, brushing has become a common interaction method in visual analytics.
One important use of brushing is the identification and analysis of \textit{clusters} in high-dimensional data via 2D DR projections.
~\cite{jiazhi21tvcg, aupetit14vast, cavallo19tvcg, wenskovitch18tvcg, quadri21tvcg, jeon24tvcg, nonato19tvcg}.

\begin{figure*}[t!]
    \centering
    \includegraphics[width=0.9\textwidth]{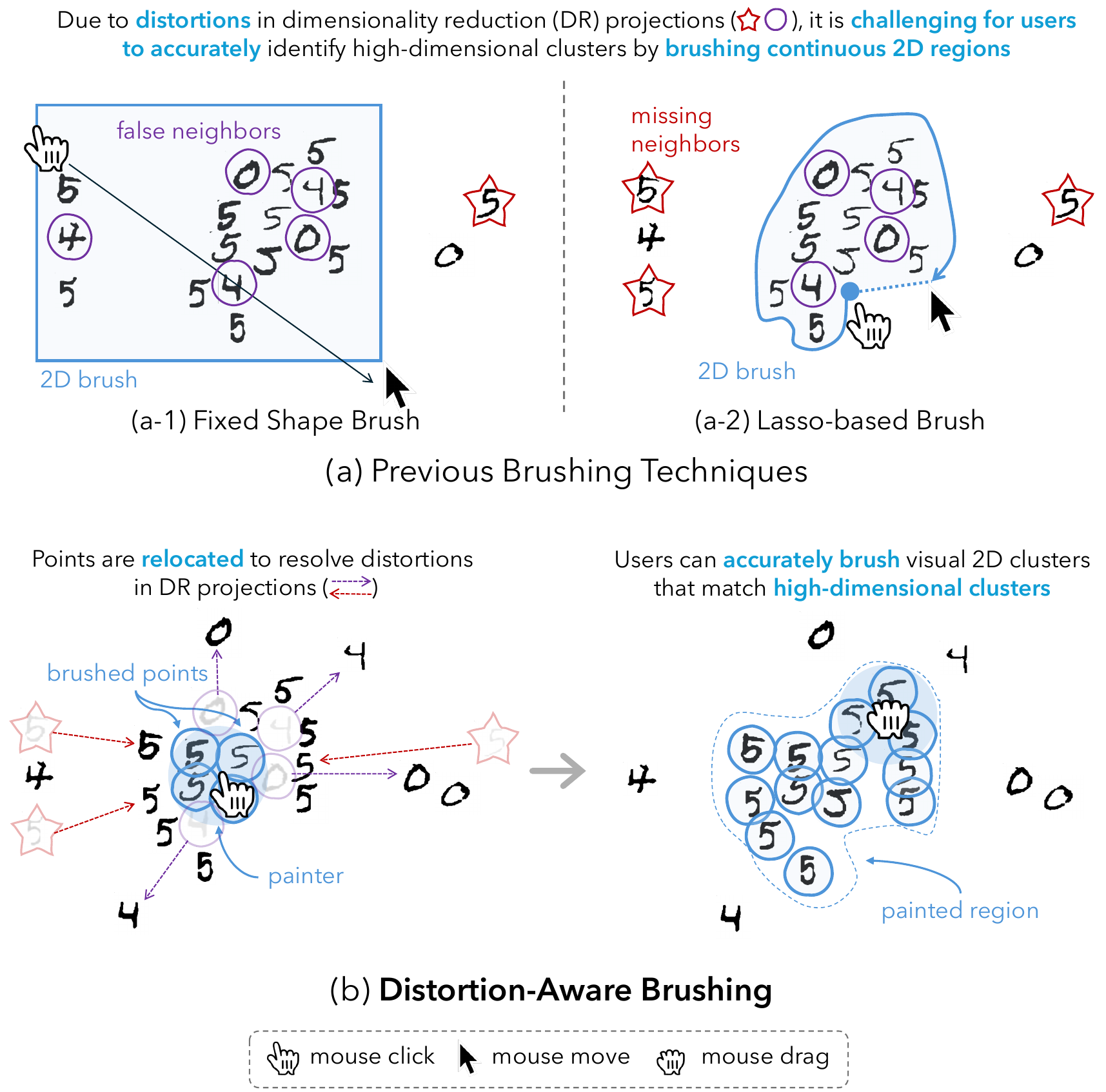}
    \caption{Comparison between existing brushing techniques (\autoref{sec:mdpbrushrel}) and \brush in identifying clusters within high-dimensional data through its 2D projection. (a) Previous brushing techniques work by defining a continuous 2D region via direct manipulation (e.g., lassoing). As projections may not accurately reflect the original high-dimensional data distribution due to distortions, users cannot precisely identify high-dimensional clusters. (b) \brush supports users in precisely extracting clusters by resolving distortions through point relocation.}
    \label{fig:dab:teaser}
\end{figure*}

However, visual analytics of high-dimensional data using conventional brushing interaction on DR projections can be unreliable, i.e., insights from the analysis may not accurately reflect the underlying data. 
Conventional 2D brushing methods typically struggle to detect clusters in the original high-dimensional space because DR projections distort the original data \cite{aupetit07neurocomputing, lespinats11cgf, lespinats07tnn, nonato19tvcg} (\autoref{fig:dab:teaser}a). For instance, data close in the high-dimensional space can be split apart in the 2D layout, forming Missing Neighbors (MN), while nearby points in the layout can come from remote regions in the high-dimensional data space, generating False Neighbors (FN). 
As a result, conventional brushing techniques might capture 2D clusters that are less cohesive or incomplete when mapped back to their original high-dimensional context.

To address this issue, several brushing techniques \cite{novotny06wscg, ward94vis, martin95vis, aupetit14vast} for DR projections have been proposed, yet they still face challenges with distortions.
These techniques generally work by first brushing a specific 2D region and automatically mapping this selection to a high-dimensional region.
This workflow makes the final brushing results vulnerable to distortions as it depends on a continuous 2D region that is subject to these distortions (\autoref{fig:dab:teaser}a). 
These techniques may further constrain data analysis by using fixed shapes for the brushed regions, such as circles and hyperspheres \cite{aupetit14vast}, or rectangles and hypercubes \cite{ward94vis, martin95vis}, which cannot effectively capture clusters with non-trivial shapes in real-world datasets.

We propose \textit{\brush}, a novel brushing technique designed to overcome these issues, enabling users to more accurately identify high-dimensional clusters from their DR projections compared to existing techniques.
Our approach addresses distortions in DR projections by persistently drawing points close in the original data towards the brushed points and repelling those that are farther apart (\autoref{fig:dab:teaser}b).
This relocation ensures that 2D brushes accurately mirror the composition of high-dimensional clusters in 2D space.
Therefore, our technique ensures a more reliable cluster analysis of high-dimensional data even in cases where DR projections suffer from severe distortions.

Through quantitative studies with 24 participants, we
demonstrate that \brush can accurately identify clusters in the original data despite distortions, surpassing previous brushing techniques designed for the projections. 
We also showcase how \brush can be leveraged to support cluster analysis and the interactive labeling of noisy datasets.
We conclude by discussing the benefits and limitations of \brush.

\section{Background and Related Work}
\label{sec:rel}

We discuss two relevant areas: interactive point relocation and brushing techniques for 2D projections of high-dimensional data.

\subsection{Interactive Points Relocation}

Interactive point relocation is widely adopted to explore the underlying structure of high-dimensional data using its projections.
Dust-and-Magnet \cite{yi05dustmagnet} and iPCA \cite{jeong09ipca} allow interactive steering of the projection layout based on the attribute values.
Another approach is to visualize high-dimensional data as snippet images and allow users to arrange projections based on visual similarity between snippets \cite{joia11lamp, xia23tvcg}. Yet another technique \cite{KruigerHSTH17} proposes brushing a cluster of points detected within one projection, freezing it in position, and then visualizing the remaining data in the same layout using another DR technique. 
However, these approaches focus on finding interesting visual cluster patterns rather than preventing distortions; thus, they can be used to identify interesting insights from high-dimensional data but cannot guarantee reliable cluster analysis.

Meanwhile, some previous works aimed to resolve errors locally through relocation. For example, Probing Projections \cite{stahnke16tvcg} transiently relocates points based on their high-dimensional similarity with a user-selected point so that it removes entire MN and FN distortions, but only for the selected point. 
Proxilens \cite{heulot13vamp} focuses on true neighbors of the selected point in the original data by pushing FN to the border of a 2D magic lens centered on that point while highlighting MN with proximity coloring \cite{aupetit07neurocomputing}. 

\paragraphit{Out contribution}
\revise{We leverage point relocation for a new analytical task: selecting data points with brushing. Therefore, }instead of transiently resolving distortions around a single point, our technique maintains the correction of distortions related to a set of points. 
By doing so, we generate a persistent visual pattern (\autoref{fig:dab:teaser}b) that enables the accurate identification and analysis of clusters in the original data.

\subsection{Brushing High-dimensional Data}

\label{sec:mdpbrushrel}

We review brushing techniques for high-dimensional data and categorize them into two groups: \textit{axis-guided brushing} and \textit{data-guided brushing}.
The former works on scatterplot matrices (SPLOMs), aiming to explore how brushed points in one orthogonal projection are distributed across other attribute spaces.
The latter works on a single projection, where the system automatically infers the corresponding high-dimensional region based on the 2D brushed region.

\paragraph{Axis-Guided Brushing in SPLOMs}
The early works on brushing high-dimensional data are designed to brush along the axes. In PRIM-9 \cite{fisherkeller75pacific}, brushing is done by adjusting the range of the 2D rectangular brush region along two axes of an orthogonal projection. Becker et al. \cite{becker87technometrics, becker87statistical} applied the same strategy to SPLOMs consisting of multiple linked orthogonal projections.
However, these techniques allow for the brushing of at most two axes, making it challenging to explore structures that span more dimensions.
To resolve this problem, Ward proposed $N$-dimensional brushing as a feature of XmdvTool \cite{ward94vis}, allowing users to define multiple 2D brushes within different projections of a SPLOM. 
A later version of XmdvTool \cite{martin95vis} enables users to apply logical operators (e.g., \texttt{AND}, \texttt{XOR}) between multiple high-dimensional brushed regions for more flexibility.

As these SPLOM-based techniques are designed to explore high-dimensional data patterns across different data attribute pairs, they are ineffective for identifying clusters with non-linear relationships between data attributes. Moreover, these techniques inherently rely on multiple 2D orthogonal projections, each of which is subject to FN distortions.
Furthermore, SPLOM uses $O(M^2)$ scatterplots to represent $M$-dimensional data, making brushing impractical when $M$ is large. 
Parallel coordinate plots (PCP) \cite{inselberg85vc} provide an alternative axis-based representation of high-dimensional data denser than SPLOM, for which advanced techniques for brushing have been proposed \cite{roberts19pcpbrush, hauser02infovis}. However, PCP shatters the high-dimensional clusters into $M$-linked 1D projections (axes), each generating more FN distortions than 2D projections. 

\paragraph{Data-Guided Brushing in DR Projections}
Data-guided approaches are proposed to identify clusters from a single DR projection. These techniques follow a typical workflow: 
(1) users determine the 2D region through interaction (e.g., painting); (2) a machine automatically constructs the high-dimensional region based on the user-defined 2D region; (3) the brushed points are defined as a union or intersection of the set of points within these two regions. 
Data-driven brushing \cite{martin95vis} allows users to define a 2D region by generating a box that encloses certain areas in the projection, which then generates a high-dimensional region as a minimum-size $M$-cube enclosing all the data corresponding to the painted points. In $M$-ball brushing \cite{aupetit14vast}, users can capture high-dimensional clusters by defining a circular 2D region; the system then automatically formulates an $M$-ball region covering the corresponding data in the original space. Both approaches bound the high-dimensional region to convex shapes, making it hard to discover non-trivial (i.e., any-shaped) clusters. 
Similarity brushing \cite{novotny06wscg} escapes from the problem by allowing users to paint a visual cluster as the 2D region, and defining the high-dimensional region as the area covered by the union of $M$-balls centered on the original data corresponding to the 2D painted points. 

\paragraphit{Our contribution}
All these high-dimensional brushing techniques are vulnerable to distortions.  If the 2D brushed region contains FN, the painted data might belong to more than one cluster in the original space. Moreover, a high-dimensional cluster can be split in the projection due to MN, so users will have to brush each of these 2D clusters separately, or even worse, will ignore them if points are spread apart, not forming clear 2D clusters. 

\brush is a data-guided technique that resolves these issues by continuously relocating points.
Instead of keeping continuous 2D and high-dimensional regions, we only consider the brushed data points. MN and FN are resolved by pulling them toward or away from the currently brushed 2D points, respectively.
In contrast to other techniques, point relocation always generates 2D visual clusters that match high-dimensional ones, faithfully representing users' mental model. The method thus works more reliably for the interactive cluster analysis of high-dimensional data.

\section{Design Objectives}

\label{sec:objectives}

Our design objectives tackle the drawbacks of previous brushing techniques for high-dimensional data (\oone{}, \otwo{}, \othree{}) while maintaining their strengths (\ofour{}).

\subsubsection*{(\oone{}) Guide brushing by visually reflecting high-dimensional clusters}
Previous data-guided brushing techniques work by converting a 2D brushed region into a high-dimensional region.
However, the inconceivability of the high-dimensional space makes it difficult for users to understand this conversion, thus lowering the interpretability and controllability of these techniques.
Instead, \brush performs the conversion in the opposite direction: \textit{2D points are relocated to form a visual cluster that reflects a high-dimensional cluster}.

\subsubsection*{(\otwo{}) Allow brushing to be robust against any kind of distortions}

Previous brushing approaches lead to unreliable cluster analysis as they rely on a compact 2D projection region, which is vulnerable to distortions in DR projections.
In contrast, \brush continuously relocates points to \textit{ensure that 2D neighbors are always true neighbors in the high-dimensional space and that the high-dimensional cluster under focus is not split in the projection}, making the technique robust to distortions regardless of their type or amount.

\subsubsection*{(\othree{}) Allow the brushing of non-trivial-shaped high-dimensional clusters}

In previous brushing techniques \cite{martin95vis, aupetit14vast, ward94vis}, the shape of the high-dimensional region enclosing brushed points is limited to regular, compact domains (hyperspheres or hypercubes).
This limitation makes the techniques hardly support the discovery of clusters with non-trivial shapes typical of real-world data.
For example, fitting such a fixed-shaped brush to a non-trivial-shaped cluster can capture out-of-cluster points. 
On the other hand, reducing the brush size to avoid capturing out-of-cluster points may result in missing in-cluster ones, lowering the clustering accuracy. 
In contrast, \brush manages \textit{a discrete set of brushed points instead of compact 2D and high-dimensional regions}. This enables users to gradually append new points to the 2D brush corresponding to the true high-dimensional neighbors of already brushed points, facilitating the discovery of clusters with arbitrary, non-trivial shapes.

\subsubsection*{\ofour{} Minimize the number of hand-tuned hyperparameters}

Previous brushing techniques have at most one hyperparameter that affects brushing results, making them easy to use and learn. 
Similarly, we design \brush to have \textit{a single hyperparameter that controls the granularity of the brushed clusters} and to automatically optimize its value.

\def\mdtrueneighbors{{\color{mymygreen}{\textit{High-D True Neighbors}}}\xspace}

\def\mdnonneighbors{{\color{mymypurple}{\textit{High-D Non-Neighbors}}}\xspace}
\def\uncertain{{\color{mymyorange}{\textit{Uncertain}}}\xspace}

\section{\brush}

\brush relocates points within and around the current brushed points in the projection by faithfully reflecting the data distribution around the brushed points in the original data (\oone{}, \otwo{}).
The set of brushed points grows progressively to form a visual cluster that accurately reflects the high-dimensional cluster (\othree). By doing so, \brush enables users to conduct a more reliable, detailed analysis of clusters in the high-dimensional data.

In this section, we first describe the design of \brush following the overall workflow of the technique (\autoref{sec:dab:workflow}).
We then describe additional features of the technique developed for its practical usage in visual analytics (\autoref{sec:additional}).

\subsection{Workflow}

\label{sec:dab:workflow}

\begin{figure*}[t!]
  \centering
  \includegraphics[width=\textwidth]{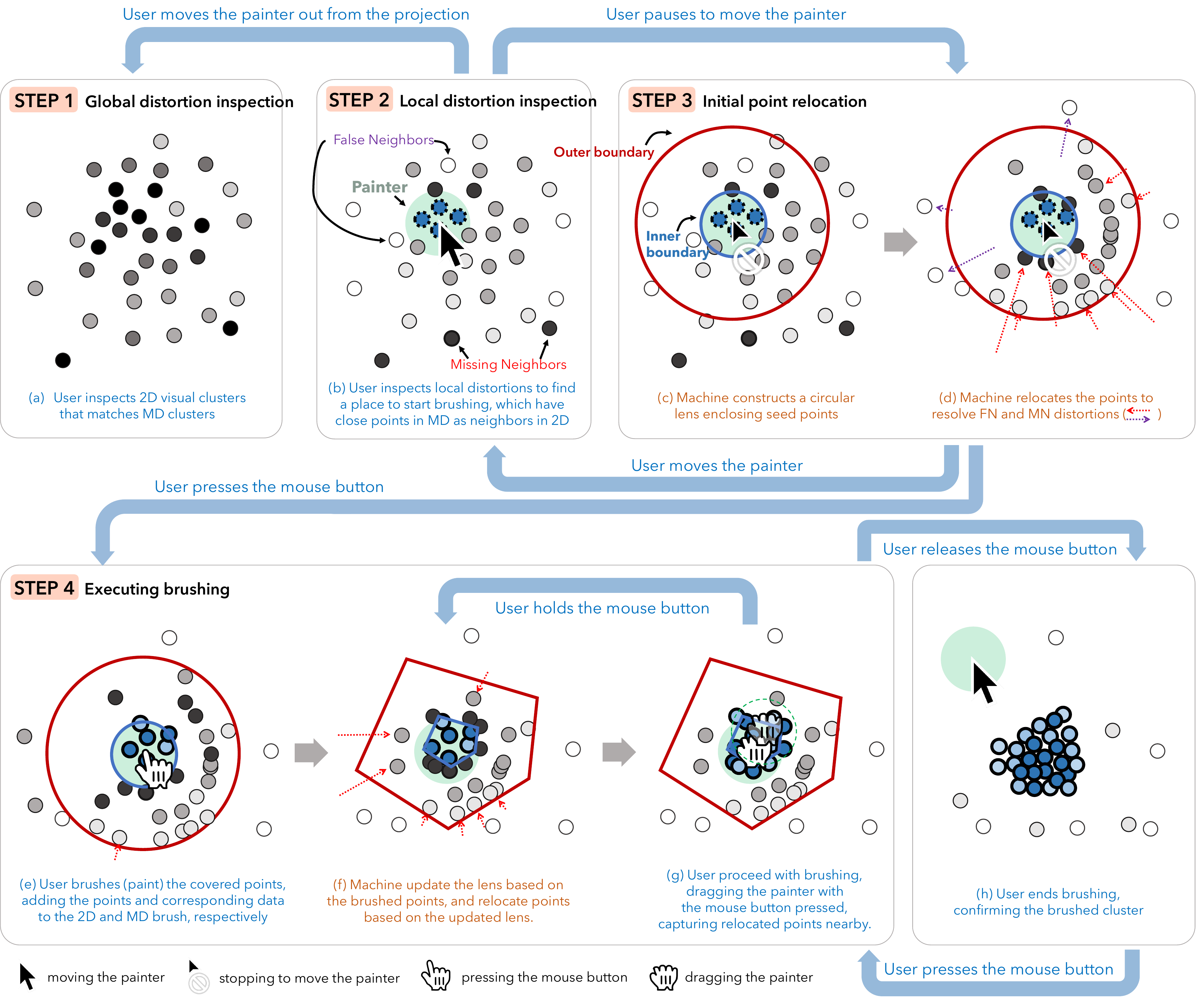}
  \caption{
  Overall workflow of \brush (\ref{sec:dab:workflow}).
  The technique features a lens with inner and outer boundaries depicted as bold blue and red closed lines, respectively.
  Users' actions are explained with blue text and arrows, while the machine's actions are detailed in orange.
  Data points are represented as small circles, i.e., dots, where seed and brushed points are highlighted using thick dotted and solid borders. Seed and brushed points are also highlighted in blue.
  The opacity of data points depicts high-dimensional density in \stepone{} and represents their closeness to the seed or brushed points in the following steps.  
  }
  \label{fig:dab:workflow}
\end{figure*}

\brush follows a four-step workflow (\autoref{fig:dab:workflow}):
\begin{itemize}[leftmargin=0mm]
    
    \item[]  \textbf{(\stepone{})} Users inspect how the high-dimensional data distribution matches the 2D visual clusters to decide the best places to initiate brushing. 

    \item[] \textbf{(\steptwo{})} Users inspect local MN and FN distortions around these candidate places by hovering the painter over the points.

    \item[] \textbf{(\stepthree{})} A pause of mouse move initiates a transient relocation of the points, correcting the local distortions. 

    \item[]  \textbf{(\stepfour{})} Users execute brushing by dragging the painter while the mouse button is pressed, progressively capturing the covered points.
Lens construction and point relocation are performed iteratively based on the current 2D location of brushed points and the high-dimensional distribution in their vicinity.
\end{itemize}

\subsubsection*{(\stepone{}) Inspecting global distortions}

For reliable cluster analysis of high-dimensional data, brushing shall ideally initiate at a place close to the core of an actual high-dimensional cluster. To support the task, we encode each point as a snippet corresponding to the datum in the high-dimensional space (\autoref{fig:dab:teaser}). For example, each point in image datasets can be represented as an image snippet. Also, points in tabular datasets can be represented using glyphs \cite{kammer20tvcg}, e.g., aster plots \cite{kwon17tvcg}. By visualizing snippets, users can spot 2D visual clusters (high mutual proximity) matching with high-dimensional clusters (high similarity between snippets) (\oone{}, \otwo{}), recognizing these locations as good candidates for initiating brushing.

We further aid this step by encoding the high-dimensional density of the data through the opacity of the corresponding points or snippets (\autoref{fig:dab:workflow}a). 
Based on density encoding, users can check the trustworthiness of a 2D visual cluster by comparing whether the 2D density, represented by proximity between points, matches high-dimensional density.
A visual cluster with higher 2D and high-dimensional density than other visual clusters can be considered a good candidate. 
The best location to start the brushing is the high-density central part of the 2D visual cluster, which corresponds to the core region of the high-dimensional cluster.
Other cases showing a density mismatch reflect distortions and should be avoided. More comprehensive distortion visualizations can be used (e.g., visualizing FN and MN distortions \cite{lespinats11cgf, jeon21tvcg}), but we preferred visualizing the density to help users learn the technique more easily

\paragraph{High-dimensional density}
We define the high-dimensional density of a point $p$ as \dens{p} $= \sum_{q \in P}$ \simil{p}{q}, following Density-peak clustering \cite{rodriguez14science, liu18infosci}, where \simil{p}{q} represents the similarity between points $p$ and $q$, and $P$ denotes the entire set of points in the data.

\paragraph{High-dimensional similarity}
We use the Shared-Nearest Neighbors (SNN) similarity \cite{ertoz02siam}, which assigns higher similarity to the pairs of points sharing more $k$-Nearest Neighbors ($k$NN). Formally, the SNN similarity between $p$ and $q$ is defined as \texttt{sim}$_k(p, q) = \sum_{(m,n) \in S_{p, q}} (k + 1 -m) \cdot (k + 1 -n)$; $S_{p,q}$ represents a set containing pairs $(m, n)$ fulfilling $p_m = q_n$ where $p_i$ denotes an $i$-th nearest neighbor of $p$ and $q_i$ denotes an $i$-th nearest neighbor of $q$. 
We select SNN as this metric is shift-invariant \cite{lee11pcs}, making it alleviate the curse of dimensionality \cite{lee11pcs} and thus better represent the cluster structure of high-dimensional spaces compared to other metrics (e.g., $k$NN, Euclidean distance) \cite{ertoz02siam, liu18infosci, jeon21tvcg}. 
We fix $k$ as the square root of the number of points, following the recommendation by Chaudhuri and Dasgupta \cite{chaudhuri14neurips}.

\subsubsection*{(\steptwo{}) Inspecting local distortions}

In this step, users can ``skim'' local distortion of the projection by moving the painter over the points (\autoref{fig:dab:workflow}b). This is done by (1) finding \textit{seed points} within a painter, then (2) visualizing the high-dimensional closeness of any point to the seed points \cite{aupetit07neurocomputing}. 
Compared to \stepone{}, this step provides a more stringent inspection of local distortions that can help users locate the best candidate for initiating brushing. 

The detailed procedure is as follows. First, the machine determines the seed points as condensed points that the painter covers.
This is done by identifying the covered point with the highest high-dimensional density and only using its close neighbors as seed points (see \textit{Finding seed points} below).
It is important to avoid using every point covered by the painter as seed points because they may consist of points coming from two or more distinct high-dimensional clusters due to FN; if this happens, new points brushed from these distinct high-dimensional clusters will erroneously agglomerate into a single visual cluster (\oone{}, \otwo{}). 
The seed points are then highlighted in the color of the current brush.

Then, the remaining points' high-dimensional closeness to the seed points is encoded as their opacity (i.e.,  closer points are darker). This graphical encoding informs users' brushing decisions by indicating more reliable locations to start brushing (\oone{}). Users can identify FN in the projection as points with bright or non-highlighted markers near the seed points and MN as points with dark markers far from the seed points.

\paragraph{Finding seed points}
Constructing a set of seed points starts by identifying the initial seed point $p_{initial}$ with the highest high-dimensional density among the subset of points $C$ covered by the painter: $p_{initial} = \argmax_{p \in C} \texttt{dens}(p)$.
Then, a set of seed points is defined as the $\kappa$ nearest neighbors ($\kappa$NN) of $p_{initial}$ based on SNN similarity in the high-dimensional data,  
where the machine automatically sets $\kappa$ as the maximum value such that all $\kappa$NN are still covered by the painter (\ofour). Users can adjust the size of the painter and so $\kappa$, to make brushing more condensed or relaxed. By this definition, the seed points and the initial brush cannot contain FN. 

\paragraph{High-dimensional closeness}
A closeness between a point $p$ and a set of points $C$ is: \close{p}{C} $=\sum_{q \in (\kappa NN \cap C)} \texttt{sim}_k(q, p) / \sum_{q \in \kappa NN} \texttt{sim}_k(q, p)$.
The more $\kappa$NN of $p$ are members of the cluster $C$, the closer $p$ is to $C$. Moreover, by definition, the seed points of the initial brush have the maximum closeness.
We do not naively average the similarity of $p$ to the points within $C$ because this will make the closeness depend on cluster characteristics like density or size.

\begin{sidewaysfigure}
  \centering
  \includegraphics[width=\linewidth]{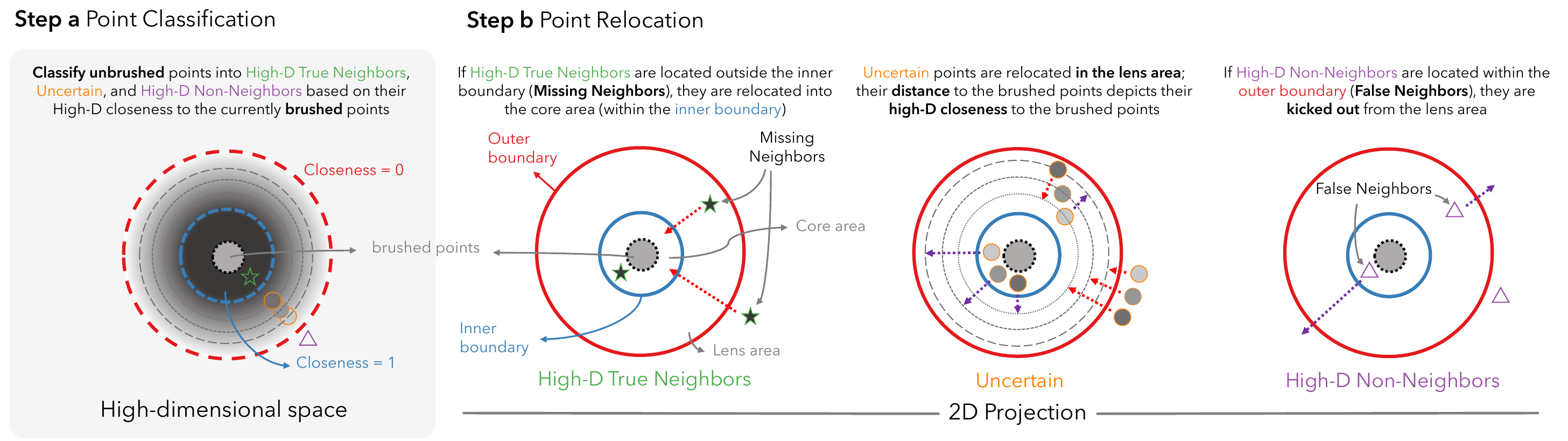}
  \caption{
    Illustration on how \brush relocates points in \stepthree{} and \stepfour{}. Points are relocated based on their initial locations in the projection space and the closeness of their corresponding data to the high-dimensional brush (initially, the seed data).   
    The left view sketches the locations of the high-dimensional data relative to the brush, with one instance of \mdtrueneighbors (star), one instance of \mdnonneighbors (triangle), and three instances of \uncertain data (circle) lying between the other two. The bold dotted circles reflect the plain high-dimensional boundary lines of inner (blue) and outer (red) boundaries in the other views. We assume a circular 2D brush for simplicity, although it can take less regular shapes (Step 4).
  }
  \label{fig:relo_cases}
\end{sidewaysfigure}

\subsubsection*{(\stepthree{}) Initiating point relocation}
Though global (\stepone) and local (\steptwo) distortion inspections help users find a good candidate region, points nearby are likely affected by distortions and do not accurately reflect the local data distribution of the original data. We thus provide a \textit{point relocation} process that corrects the distortions relative to the current brush (initially, the seed points covered by the painter)  (\autoref{fig:dab:workflow}c-d).
Inspired by the Proxilens approach  \cite{heulot13vamp}, users can trigger transient point relocation by halting the painter for a short time, which is determined as 800ms in our implementation through an iterative design process. Correcting for local distortions can be viewed as ``jumping'' into the high-dimensional space, as it makes the 2D distribution around the painter and the currently brushed points better reflect the local high-dimensional data distribution. Users can reverse the current point relocation (jumping back to the 2D space) by moving the painter again.

To perform relocation, the system first constructs a magic lens around seed points in the projection  (\autoref{fig:dab:workflow}c). The lens consists of (1) a core area delimited by an inner boundary that tightly encloses seed points and (2) an annulus lens area around the core, enclosed between the inner boundary and an outer boundary (\autoref{fig:relo_cases}, blue and red solid circular boundaries, respectively).

Afterward, point relocation is performed according to the lenses so that point distribution around seed points in the projection can reflect the distribution around the seed points in the high-dimensional space(\oone{}) (\autoref{fig:dab:workflow}d). 
The relocation of a point thus depends on its high-dimensional closeness to the seed points (\autoref{fig:relo_cases}). 
If the closeness is 1, the point is considered as \mdtrueneighbors, which means that the point belongs to the core high-dimensional cluster formed by the seed points.
If the closeness is 0, the point is categorized with \mdnonneighbors, denoting that it is far apart from the seed points in the high-dimensional space.
If the value is between 0 and 1, the point is considered \uncertain,  forming a ``fuzzy'' neighborhood in the high-dimensional space.
While the machine relocates \mdtrueneighbors (i.e., MNs) into the inner boundary, \mdnonneighbors (i.e., FNs) are repelled from the lens area outside the outer boundary, and \uncertain points are relocated within the lens area, close to the inner boundary, in proportion to their high-dimensional closeness to the core cluster. 
Points are relocated to the correct position with an animated transition. 
Relocation corrects distortions, making points in the inner lens correspond to True Neighbors, as FNs are repelled from the lens and MNs attracted within the lens (\oone{}).

It is worth noting that the \uncertain points and their interpolated relocation are a crucial part of \brush, as it is up to users to decide whether a point will be brushed or not. The \uncertain points falling into the outer lens are natural candidates for brushing; they are also geometrically the next ones that can be captured by the painter (\oone{}, \otwo{}) (see \textit{Dynamical update} in \stepfour{}). 

\paragraph{Initial lens construction}
The initial inner and outer lens boundaries are defined as circular boundaries centered on the painter.
We set $\tau$ as both the inner boundary's radius and the radius of the painter.
By doing so, the painter covers both seed points and \mdtrueneighbors, forcing them to be brushed when users execute brushing (\stepfour{}). 
We also define the radius of the outer boundary as $3\tau$, setting the lens area's width as the painter's diameter (i.e., $2\tau$). 
We justify this decision while describing how we construct the outer lens boundary in \stepfour{} (\textit{Outer boundary construction}).

\subsubsection*{(\stepfour{}) Executing brushing}

Once the transient relocation is settled, users can initiate brushing by pressing the mouse button (\autoref{fig:dab:workflow}e).
The machine appends the points covered by the painter (which naturally contains seed points) to the brushed set of points. The brushed points are highlighted with the color corresponding to the brush. 
Then, the machine updates the lens and relocates the remaining points based on the new set of brushed points (\autoref{fig:dab:workflow}f).
If users drag the painter while keeping the mouse button pressed (\autoref{fig:dab:workflow}g), the brushed points, the painter, and the lens are updated accordingly, and the relocation takes place again in a continuous cycle. 
Such gradual updates enable users to agglomerate new brushed points in an arbitrary direction in the high-dimensional space, thus allowing the brushing of an high-dimensional cluster with an arbitrary shape (O3).

If users decide to end brushing and confirm the brushed high-dimensional cluster, they can end the cycle by releasing the mouse button.
Such decisions can be made when (1) there are no more unbrushed image snippets (i.e., data points) that look similar to the ones inside the set of brushed snippets or (2) newly brushed points have a relatively lower density than the previously added points. Auxiliary visualizations (e.g., parallel coordinates plot or heatmap) can also guide users in deciding the boundary of brushed clusters (\autoref{sec:scenario}).
Here, users can again go back to brushing by pressing the mouse button or erase points that are not intended to be in the brush (\autoref{sec:erasing}).

Note that by allowing users to make the final decision, our technique can effectively handle noisy clusters, i.e., semantic clusters that are not well separated in the data space. For example, \brush can be used to highlight image snippets that are visually distinguishable to humans but indistinguishable by a distance metric. We demonstrate the effectiveness of this feature in \autoref{sec:usecaselabeling}.

During brushing, the brushed points are also relocated to aid interaction. 
First, we uniformize their locations.
The uniformization removes empty space within the inner boundary, making the 2D proximity between each unbrushed point and the brushed points accurately reflect their closeness. 
It also removes overlap between the points, supporting users in visually investigating snippets. 
Then, the points are successively relocated to better reflect the original data distribution of the high-dimensional cluster.
The points in which corresponding high-dimensional data have high closeness to the brushed points move near the center of the core area of the lens, and the ones with low closeness move near the boundary of the inner lens. 
We achieved this by swapping the positions of the points to align their distances to the inner boundary with closeness.
This makes the 2D visual cluster a better match with the high-dimensional cluster (\oone{}) and helps users readily erase points with low closeness (\otwo).

\paragraph{Inner boundary construction}
While brushing, the inner boundary is set as a convex hull enclosing the brushed points. 
We use a convex hull as it is computationally cheap ($O(n\log n)$ for $n$ points) while tightly enclosing the brushed points compared to alternatives (e.g., boundary circle) and has no hyperparameter to tune (\ofour). We do not use alternatives like a concave hull, though it may more tightly enclose the brushed points, as it is relatively expensive to compute and requires hyperparameters that substantially affect the shape of the hull \cite{asaeedi17tcs} (\ofour).
As a result, the brush is always convex, even if the original high-dimensional cluster has a more convoluted shape.

\paragraph{Outer boundary construction}
The outer boundary is constructed by offsetting each corner of the inner boundary to maintain the width of the lens area. Thus, points with the same high-dimensional closeness to brushed points are displayed at the same distance from the core lens. 
This approach also makes the relocation of points isotropic, not biased by the direction from which they originate. It supports our encoding where visual proximity matters, while provenance direction has no specific importance. We detail this justification and design alternatives in Appendix D.

As mentioned in \stepthree{}, we use $2\tau$ offset to match the lens area's width with the painter's diameter. Thus, when an \uncertain point enters the painter, lying on its circular edge, its high-dimensional closeness to the brushed points corresponds roughly to the proportion of the painter area overlapping the core lens, a visual indicator easy to estimate. 
For example, if the painter fully overlaps with the core lens, any point within the painter can belong to the brush but none within the lens area (i.e., the accepted closeness is only $\alpha = 1$).
If the painter $\alpha$-overlap with the core lens, points with high-dimensional closeness above $\alpha$ can belong to the brush (the accepted closeness range is $[\alpha, 1]$).
Finally, if the painter stays entirely within the lens area, outside the core lens, touching the outer boundary,  all the points covered by the painter can belong to the brush (the accepted closeness range is total: $[0,1]$). 

\paragraph{Dynamical update}
As the brush is dynamically updated with the \uncertain points captured by the painter, the brush grows up on the painter's side at a certain speed due to relocation animated transitions and \textit{core lens uniformization} (see below). This increases the overlap area of the core lens with the painter, hence the acceptance threshold $\alpha$, and reduces the painter's covering of the lens area, preventing further \uncertain points from being captured and a positive feedback loop that would lead to a loss of control. Users must keep moving the painter toward the nearby lens area to lower the acceptance threshold back to the desired value, capturing new \uncertain points down to that level. Thus, users intending to brush data carefully will naturally move painter slower, thus maintaining a higher $\alpha$ and a more stringent acceptance of \uncertain points as \mdtrueneighbors.

\paragraph{Core lens uniformization}
We used the centroidal Voronoi tessellation \cite{du99siam} based on Lloyd's algorithm \cite{lloyd82tit} to uniformize the distribution of brushed points. We clip the Voronoi cells to fit the inner boundary so that the points remain within the core lens.

\subsection{Features for Enhanced Usability}

\label{sec:additional}

We discuss features of \brush that improve its usability and applicability in cluster analysis.

\subsubsection{Erasing Mode}

\label{sec:erasing}

Users can erase points that are not intended to be brushed using erase mode. 
The mode is enabled when (1) brushing is paused by releasing the mouse button during \stepfour{} and (2) the keyboard shift button is pressed. When enabled, users can erase points by hovering the painter over the points while keeping the mouse button pressed. The painter's color becomes red.

Natural candidates for erasing are points with low high-dimensional closeness to the brushed points.
As the system pushes such points with low closeness near the inner boundary (\stepfour{}), users can readily erase them by moving the painter in the lens area of the brush slightly overlapping the core lens to capture these outliers.
The erased points are then relocated, as with other unbrushed points, based on their high-dimensional closeness to the updated brush.

\subsubsection{Multiple Brushes}

In previous brushing techniques, multiple brushes are
often provided to compare multiple sets of points, widening the analytic search space \cite{aupetit14vast, ward94vis}.
We also allow users to control multiple brushes while
distinguishing them with different colors. Users can pause the current brush and switch the focus to another brush by pressing a button with the corresponding brush color. We showcase the utility of multiple brushes in our use case (\autoref{sec:scenario}).

\subsubsection{Contextualization within the Original Projection}

\label{sec:contextualization}

The point relocation mechanism corrects distortions around the brushed points (\otwo{}) but can introduce new distortions elsewhere in the projection, making it difficult for users to understand the brushed cluster in the context of the original DR projection.
Moreover, the relocation may reinforce confirmation bias, leading users to preferentially brush points that resemble those already brushed.
To mitigate such side effects, we allow users to restore all points to their original positions through an animated transition while preserving their color to indicate brush identity.
This feature helps users contextualize brushing results within the original projection, reducing selection bias introduced by the current brushing and encouraging the exploration of alternative analytical paths.
Our use cases (\autoref{sec:scenario}, \ref{sec:usecaselabeling}) demonstrate the practical benefits of this contextualization.
Future work may explore history management for clusters \cite{menin21infovis}, which would support hypothesis-driven analysis and enhance navigation across different brushing results---an improvement beneficial to all brushing techniques.

\subsection{Implementation}

\brush is developed as a standalone library, making it available for any web-based application. 
We used \texttt{canvas} for rendering to ensure high scalability while making the library less complicated to extend or maintain. 
The library provides a JavaScript class instance that accepts a \texttt{canvas}, dataset, and projection as parameters and installs the brush on the \texttt{canvas}.

\section{Evaluating the Robustness of Distortion-Aware Brushing}

\label{sec:userstudy}

We conduct two controlled user studies to validate the effectiveness of \brush.
The studies investigate how \brush's performance in extracting high-dimensional clusters (accuracy and task completion time) is affected by the amount of distortions (O2; \autoref{sec:comparisonstudy}) and the non-triviality of the shape of high-dimensional clusters (O3; \autoref{sec:detailstudy}). 
Both studies compare \brush against three existing brushing techniques for high-dimensional data. 
We do not empirically investigate the satisfaction of O1 and O4, as they are fulfilled by design (\autoref{sec:dab:workflow}).

\subsection{User Study 1: Robustness Against Distortions}

\label{sec:comparisonstudy}

We evaluate the effectiveness of \brush in supporting users to brush high-dimensional clusters under various distortion conditions.

\subsubsection{Objectives and Design}

We aim to compare \brush with baselines on accuracy in capturing high-dimensional clusters and their robustness to varying amounts of distortions.
We also want to assess how global  (\stepone) and local (\steptwo) distortion inspections (\autoref{sec:dab:workflow}) affect user performance of the brushing techniques.
To achieve such goals, we design the study to have three independent variables:
\begin{itemize}
    \item Amount and type of distortions (\textsc{DistortionAmount})
    \begin{itemize}
        \item \textit{Low distortion}, \textit{High MN}, and \textit{High FN}
    \end{itemize}
    \item ㅠrushing techniques (\textsc{Techniques})
    \begin{itemize}
        \item Three baselines (\textit{Data-driven brushing} \cite{martin95vis}, \textit{$M$-Ball Brushing} \cite{aupetit14vast}, \textit{Similarity brushing} \cite{novotny06wscg}) and \textit{\brush}.
    \end{itemize}
    \item Availability of global  and local distortion inspections (\textsc{DistortionInspection})
    \begin{itemize}
        \item \textit{No inspection}, \textit{Only global}, and \textit{Both global and local}
    \end{itemize}
\end{itemize}
resulting in a within-subject experiment with 3 \textsc{[DistortionAmount]} $\times$ 4 \textsc{[Techniques]} $\times$ 3 \textsc{[DistortionInspection]} $=36$ trials per participant.
Note that controlling \textsc{DistortionInspection} ensures the fairness of our experiment in comparing \brush and baseline techniques.
This is because baseline techniques' task performance may also benefit from the inspection functionalities.
The detailed study design is presented in the following paragraphs:

\paragraph{Baseline brushing techniques}
We select data-guided brushing techniques as baselines because they share a common analytic goal with \brush: identifying high-dimensional clusters. Axis-guided brushing techniques are not included because they (1) operate on multiple coordinated attribute-based views (i.e., SPLOMs) instead of a single DR projection and (2) serve a different analytic purpose, limiting their capability to identify high-dimensional clusters.

\paragraph{Task}
We ask participants to perform interactive labeling \cite{sacha17tvcg, peltonen13eurovis}. The aim of the task is to label a single designated cluster in the high-dimensional data by brushing it on the 2D DR projection represented by a monochrome scatterplot. 
We pick the task because it is widely used to explore high-dimensional data in visual analytics \cite{sacha17tvcg, peltonen13eurovis, jiazhi21tvcg}, and because it is an important use case for previous brushing techniques \cite{aupetit14vast, novotny06wscg}.

\paragraph{Procedure}
One experimenter manages the experiment for all participants individually and in person.
After a participant signs the consent form, the experimenter explains the concept of high-dimensional data and why 2D DR projections cannot precisely depict them. Then, the experimenter details the tasks and goals of the experiment. During the introduction, the participants are free to ask questions.

Participants are then exposed to 36 trials; each is associated with a single combination of the independent variables.
We divide the trials into four sessions, each assigned to a single \textsc{Technique}.
In each session, after the experimenter demonstrates the technique, participants are given a maximum of five minutes to practice and pose questions.
For the practice, we use a dataset and projections different from those used for the main study (detailed below).
The order of the sessions (i.e., the order of the \textsc{Techniques}) is counterbalanced using a four-level Latin square design (Appendix XX).
Each of these sessions contains nine trials (every combination of \textsc{DistortionAmount} and \textsc{GlobalInspection}), where the order of these trials is again counterbalanced using a Latin rectangular design (Appendix XX). Since each trial takes at most 120 seconds in a pilot study, we set no specific time limit for each trial. 

Finally, we conducted a semi-structured post-study interview investigating the usability of brushing techniques and task difficulty (Questionnaires in Appendix XX). 
The experiment took less than 50 minutes for all participants.

\paragraph{Measurements}
We record labeling results and task completion time of each trial. 
We record the task completion time as the duration from the first mouse hover to the stimuli to the click of the ``finish trial’’ button.
We also convert labeling results into F1 accuracy scores with respect to the ground-truth clusters.

\paragraph{High-dimensional dataset and ground truth clusters}
We use the MNIST dataset as a reference to generate projections that will serve as scatterplot stimuli.
We reduce the dimensionality of the MNIST dataset to 10D using PCA to prevent previous brushing techniques that rely on convex-shaped high-dimensional regions (Data-driven brushing, $M$-ball brushing) 
 to suffer from the curse of dimensionality \cite{bellman66science}, making our experiment fair.
 
We consider each class (i.e., digit) as a potential ground truth cluster. 
We thus render points as snippet images (i.e., digits) so that participants can visually distinguish different digits and readily determine the boundary of brushing.

However, classes may not be well separated in the original high-dimensional space (\autoref{sec:clcl}). 
Therefore, we first identify class pairs with high separability following Aupetit \cite{aupetit14beliv} to create valid ground-truth clusters. 
We first compute the separability of class pairs over 96 labeled datasets available using the adjusted Calinski-Harabasz index ($CH_{A}$) (\autoref{sec:adjusted}), then pick the index value corresponding to the 90th percentile as a separability threshold.
We use $CH_{A}$ as the index is proven to work robustly and fairly regardless of the dimensionality (\autoref{sec:clcl}).
Then, for each stimulus, we select pairs of classes in the MNIST dataset that are mutually separable with a $CH_{A}$ score higher than the separability threshold.

\paragraph{Confounding variables}
We identify and control three confounding variables: the number of points of each cluster, the number of clusters, and the non-triviality of the shape of the cluster designated to be brushed.
To control the non-triviality of a cluster, we first fit the Gaussian Mixture model with a single Gaussian distribution to the cluster in the original high-dimensional space.
We then use the maximum log-likelihood score, quantifying how well the Gaussian fits the data distribution as a proxy for non-triviality. 
We identify the top three classes with high non-triviality and the bottom three classes with low non-triviality of the MNIST dataset as \textit{High} and \textit{Low} non-triviality clusters, respectively, designating the remaining four classes as \textit{Intermediate} clusters. 
We also prepare three different settings for the number of points (\textit{100}, \textit{150}, and \textit{200}) and clusters (\textit{two, three,} and \textit{four}). 
The total number of combinations of the confounding variables is $3 \times 3 \times 3 = 27$, but we reduce it to nine using a three-level Latin square design. 
The number of points thus varies from 200 (100 points $\times$ two clusters) to 800 (200 points $\times$ four clusters).
We maximally equalize the assignment of these nine combinations to all participants (12) and trials (36) using the Latin rectangle design (See Appendix XX).

\begin{figure*}[t]
    \centering
    \includegraphics[width=\linewidth]{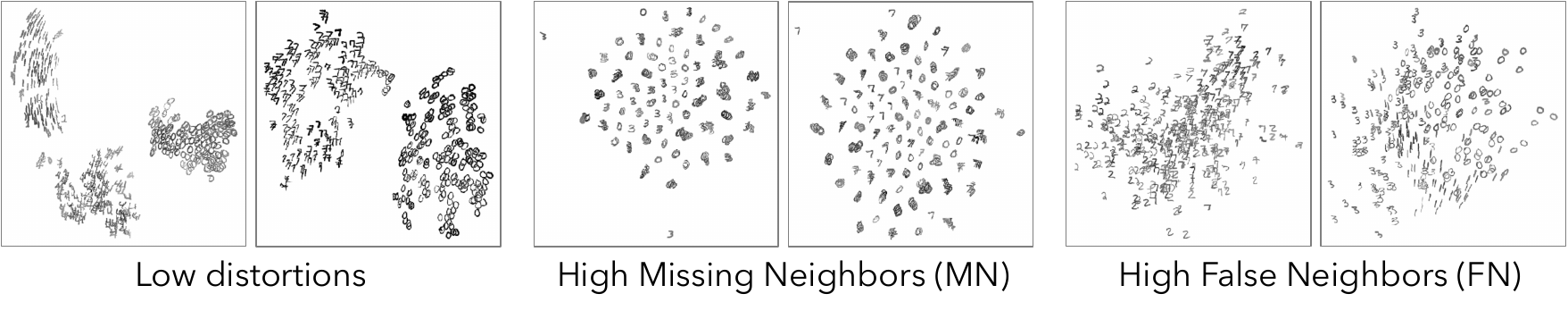}
    \caption{The example projections (i.e., stimuli) used in our experiments with different amounts of distortions. In study 1 (\autoref{sec:comparisonstudy}, we treat the amount of distortions as an independent variable, namely \textsc{DistortionAmount}. In study 2 (\autoref{sec:detailstudy}), it is controlled as a confounding variable. }
    \label{fig:stimuli}
\end{figure*}

\paragraph{Stimuli (i.e., projections) generation}
We have nine combinations of confounding factors and three different \textsc{DistortionAmount} settings; we thus need $9 \times 3 = 27$ types of stimuli. 
When a trial requires a stimulus with $\mathcal{N}$ clusters, $\mathcal{M}$ points per cluster, $\mathcal{L}$ non-triviality, and $\mathcal{O}$ \textsc{DistortionAmount}, we first randomly sample $\mathcal{N}$ clusters while ensuring that at least one of them has $\mathcal{L}$ non-triviality. We then randomly sample data points to make each cluster have $\mathcal{M}$ points and generate a projection having $\mathcal{O}$ \textsc{DistortionAmount} (detailed below) as a final stimulus. One of the clusters with $\mathcal{L}$ non-triviality is randomly designated to be brushed.
As we use the MNIST dataset, we designate the cluster by informing the corresponding class. 
All stimuli are precomputed before the experiment. Please refer to \autoref{fig:stimuli} for the example projections we use.

\begin{figure}
    \centering
    \includegraphics[width=0.6\linewidth]{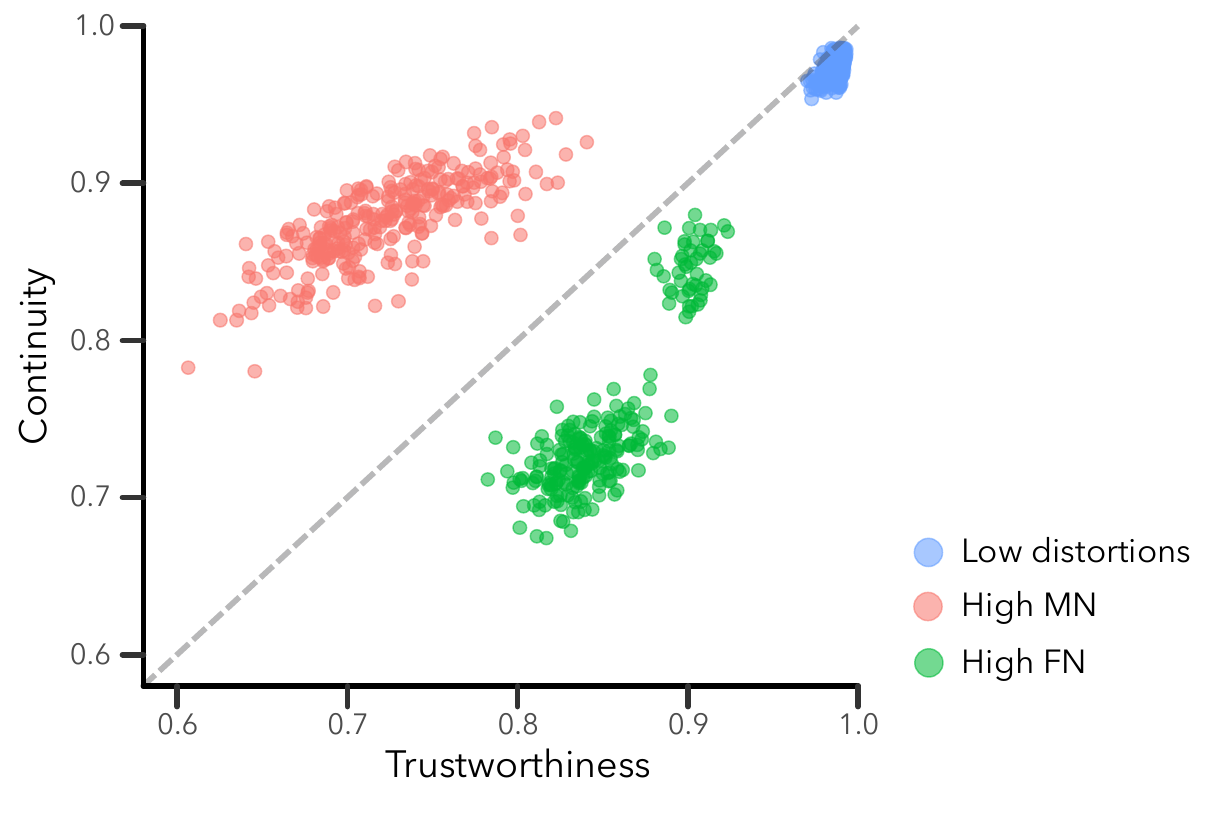}
    \caption{Distribution of T\&C scores over the stimuli projections we use in our user studies. Projections with \textit{low distortions} have high scores in both Trustworthiness and Continuity.
Meanwhile, projections with \textit{High MN} have relatively low continuity scores, and the ones with \textit{High FN} have lower trustworthiness scores. As trustworthiness and continuity aim to quantify the preservation of False and Missing Neighbors \cite{venna06nn}, respectively, the results clearly show the validity of our stimuli.}
    \label{fig:stimuli_tnc}
\end{figure}

\paragraph{Levels of \textsc{DistortionAmount}}
We generate projections with \textit{Low distortion} using $t$-SNE \cite{maaten08jmlr}. We optimize two hyperparameters that significantly impact the projection results, perplexity, and learning rate \cite{gove22vi}. We use Bayesian optimization \cite{snoek12nips} while using the F1 score of Trustworthiness \& Continuity (T\&C) \cite{venna06nn} as the target function.
We use T\&C as they are widely used metrics for detecting FN and MN \cite{jeon21tvcg, jeon24tvcg2, nonato19tvcg, jeon23vis}, respectively.
We also create projections with \textit{High FN} using the random orthogonal projection technique. As the random orthogonal projection always decreases the distances between points, we can expect the resulting projection to have relatively high FN but relatively low MN.
We create \textit{High MN} projections by executing $t$-SNE with an extremely low perplexity setting of 1. Given that low perplexity values cause $t$-SNE to prioritize local structure, setting the perplexity this low can result in the algorithm missing close neighbors and splitting clusters apart, thereby leading to more MN \cite{wattenberg2016tsnetuning}. \autoref{fig:stimuli_tnc} verifies the effectiveness of our generation strategy.

\paragraph{Training session for each brushing technique}
We use PCA projection of a randomly chosen set of three highly separated classes for the training. We use PCA as it is a very standard projection technique that generates FN distortions, but less than random orthogonal projections. Therefore, using PCA, participants can experience each brushing technique's capability to address distortions without being frustrated by the task's difficulty. We turn on both global and local inspection during the training session. We inform participants that these functionalities can be turned off during the main experiment.

\paragraph{Participants}
We recruited 12 participants (identified as E1P1--E1P12)
from a local university (nine males and three females, aged 22--32 [26.4 $\pm$ 3.2]).
All participants had undergraduate-level experience in statistics or linear algebra and thus could readily understand the concept of high-dimensional data and brushing techniques. 
They also had experience in using standard brushing techniques like lasso or box selections.
We compensate each participant with the equivalent of  US \$10 for their participation.

\paragraph{Apparatus}
We execute all brushing techniques in a Chrome web browser running on an Apple MacBook Pro 2021 (Apple M1 Max, 64GB RAM). Participants use techniques through a  27-inch 4K monitor with a regular mouse and keyboard settings. We locate the stimuli in a 1000px $\times$ 1000px box at the center of the screen.

\subsubsection{Quantitative Results}

\begin{sidewaysfigure}
    \centering
    \includegraphics[width=\linewidth]{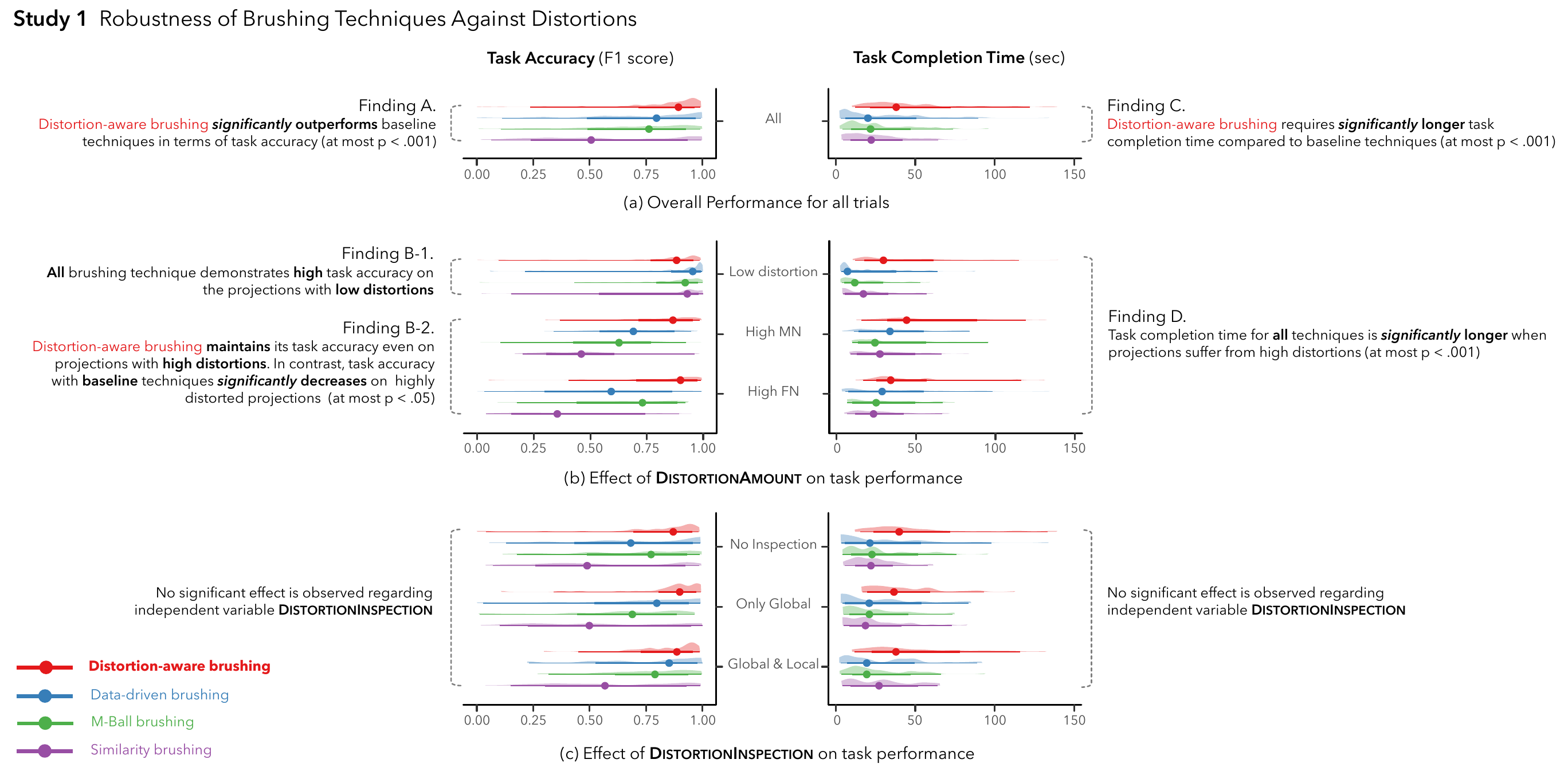}
    \caption{Study 1 results demonstrating the robustness of brushing techniques against the amount of distortions. Overall, \brush significantly outperforms baseline techniques in terms of task completion time, but requires participants to spend more time completing the task.}
    \label{fig:exp1_results}
\end{sidewaysfigure}

We detail our study findings, which are also depicted in \autoref{fig:exp1_results}.

\paragraph{Analysis on overall task accuracy}
Aligned with the main study objective, we investigate how three independent variables (\textsc{DistortionAmount}, \textsc{DistortionInspection}, and \textsc{Techniques}) affect clustering accuracy, i.e., F1 score (\autoref{fig:exp1_results} left).
We execute the three-way repeated measure ANOVA (RM ANOVA) with Bonferroni correction for post-hoc analysis. 

As a result, we find significant main effect on \textsc{Techniques} ($F_{3, 33} = 36.361$, $p < .001$) and \textsc{DistortionAmount} ($F_{2, 22} = 17.102$, $p < .001$). For \textsc{Techniques}, post-hoc analysis reveals that \brush shows significantly higher accuracy compared to baseline techniques ($p <.01$ for Data-driven brushing case, $p < .001$ for other cases). There was no other significant difference.
This leads to our first finding:

\finding{A}{\brush is, overall, the most accurate technique among all competitors for selecting clusters in DR projections.}

\noindent
In terms of \textsc{DistortionAmount}, we find that task accuracy is significantly better on the projections with low distortions compared to the ones with high MN or high FN ($p < .001$ for both cases), which is straightforward.

\paragraph{Interaction effect analysis on task accuracy}
 Using RM ANOVA, we also find a significant interaction effect between \textsc{Techniques} and \textsc{DistortionAmount} ($F_{6, 66} = 7.272$, $p < .001$).  To deep dive into this interaction, we divide the trials into three groups with different \textsc{DistortionAmount} levels (\textit{Low distortion}, \textit{High MN}, and \textit{High FN}). 
We then conduct an additional one-way ANOVA to examine how \textsc{Techniques} affect the F1 accuracy scores within each group. 

As a result, we find that \textsc{Techniques} variable significantly influences the accuracy with the projections having \textit{High MN} ($F_{3, 140} = 20.764$, $p < .001$) or \textit{High FN} ($F_{3, 140} = 19.318$, $p < .001$). However, it has no significant effect for the case with \textit{Low distortions} ($F_{3, 140} = 0.847$, $p = .470$), where all techniques show substantially high performance (over 0.85 in average; \autoref{fig:exp1_results}b left).
 Post-hoc analysis shows that if projections have  \textit{High MN} or \textit{High FN}, \brush significantly outperforms baseline techniques (\textit{High MN}: $p < .01$ for Data-driven brushing case, $p <  .001$ for other cases; \textit{High FN}: $p < .05$ for $M$-Ball brushing case, $p < .001$ for other cases). 
 \autoref{fig:exp1_results}b (right) shows that such a change occurs as the accuracy of \brush stays still while the accuracy of other techniques decreases. The results can be summarized as follows:

\finding{B-1}{Every brushing technique is accurate for the DR projections with low distortions.}

\vspace{-2.5mm}
\finding{B-2}{\brush maintains its high task accuracy on projections with high MN and FN. On the other hand, baseline techniques work poorly on these projections, having significantly lower task accuracy compared to \brush.}
\vspace{-2.5mm}

\noindent
No other significant interaction effect has been found.

\paragraph{Analysis on overall task completion time}
We investigate how three independent variables affect task completion time  (\autoref{fig:exp1_results} right). As with previous analysis on task accuracy, we run the three-way RM ANOVA for that purpose, and use Bonferroni correction for post-hoc analysis.

The analysis reveals significant main effect on \textsc{Techniques} ($F_{3, 33} =8.557$, $p <.001$) and \textsc{DistortionAmount} ($F_{2, 22} =7.162$, $p <.01$). For \textsc{Techniques}, post-hoc analysis shows that \brush shows significantly longer task completion time compared to baseline techniques ($p < .001$ for all cases). This summarizes into:

\finding{C}{\brush requires significantly longer task completion
time than baseline techniques.}

In terms of \textsc{DistortionAmount}, post-hoc analysis reveals that the techniques require longer completion time for the projections with \textit{High MN} and \textit{High FN} compared to the \textit{Low distortions} cases. As no interaction effect coupled with \textsc{DistortionAmount} exists, the results lead us to the following finding:

\finding{D}{Regardless of the type of brushing techniques, the task completion time was significantly
longer when there exist distortions in projections
}

\paragraph{Interaction effect analysis on task completion time}
RM ANOVA identifies significant interaction effect on \textsc{DistortionAmount} and \textsc{DistortionInspection} ($F_{4, 44} = 4.236$, $p < .01$). For follow-up analysis, we divide the trials into three groups based on  \textsc{DistortionAmount} and run one-way ANOVA on each group, investigating the influence of \textsc{DistortionInspection} on task completion time. For all groups, we find no significant main effect found (\textit{Low distortion}: $F_{2, 141} = 1.368$, $p = 0.258$; \textit{High MN}: $F_{2, 141} = 0.340$, $p = 0.712$; \textit{High FN}: $F_{2, 141} = 0.779$, $p = 0.461$).

\paragraph{Visual analysis of baseline techniques' task accuracy}
We visually explored additional factors affecting the interaction between \textsc{Techniques} and \textsc{DistortionAmount} (\autoref{fig:exp1_results}b). 
Although the difference is not statistically significant, we observe that the median task accuracy of $M$-Ball brushing (green) is higher than the one of Data-driven brushing (blue) for the projections with \textit{High FN}, while the opposite happens for the ones with \textit{High MN}.

\subsubsection{Discussions}

\label{sec:study1discuss}

We discuss the takeaways from our main findings. 

\paragraph{\brush is more accurate than competitors when more distortions exist}
Findings A and B verify that \brush significantly outperforms baseline techniques, showing substantial accuracy regardless of distortions. 
These findings clearly imply the benefit of using \brush in high-dimensional data analysis.

\paragraph{Baseline techniques can still be used in low-distortion projections}
Our findings also indicate that baseline techniques remain useful in some situations. 
Finding B-1 indicates that baseline techniques are reliable if we can ensure that the projections exhibit low distortion. In such a case, the analysis will benefit from the faster completion time of baseline techniques (Finding C). 

\paragraph{Local and global inspections are helpful when distortions are high}
The study does not reveal a significant influence on the existence of our inspection functionalities (\stepone{}, \steptwo{}). 
However, in the interaction effect analysis on task completion time, we can observe that $F$-value of ANOVA decreases for \textit{High MN} and \textit{High FN} case, which means that the influence of global and local inspection increases when DR projections suffer from more distortions. Although not supported by statistical evidence, this observation aligns with the qualitative feedback from participants (\autoref{sec:qualinspection}) that our global and local inspection functionalities help users perform brushing tasks faster.

\paragraph{Trade-off between task accuracy and task completion time}
Although \brush shows the best task accuracy, it trades task completion time for such achievement (Finding C, D). 
Our qualitative interview reveals that this tradeoff mainly originates from the enhanced concentration or cautiousness of participants while using \brush (\autoref{sec:qualbenefit}).

\paragraph{Validity of our study design}
The higher task accuracy of M-Ball brushing compared to Data-driven brushing in the \textit{High FN} case, and the lower accuracy in the \textit{High MN} case, reaffirm the validity of our study design and results. Indeed, these findings align with the design of these techniques. M-Ball brushing avoids FN by selecting only true neighbors among points within a 2D circular region, but the corresponding high-dimensional ball is not as extended as the high-dimensional box of the Data-driven brushing in the 10-dimensional data space~\cite{Verleysen2005curseDim}, more likely failing to capture some true neighbors missing in the locally brushed layout. 
In contrast, Data-driven brushing captures more of the missing neighbors but also most of the FN enclosed in its 2D box.

\subsection{User Study 2: Robustness Against the Non-Triviality of High-dimensional Cluster Shape}

\label{sec:detailstudy}

\subsubsection{Objectives and Design}

We aim to investigate the robustness of \brush in maintaining user performance (task accuracy and completion time) across non-trivially shaped clusters.
We also want to check whether the general findings about the performance of \brush we find in Study 1 (\autoref{sec:comparisonstudy}) are maintained. 
To this end, the study has: 
\begin{itemize}
    \item Non-triviality of the shape of a designated high-dimensional cluster to be brushed (\textsc{NonTriviality})
    \begin{itemize}
        \item \textit{High}, \textit{Intermediate}, and \textit{Low},
    \end{itemize}
\end{itemize}
as an independent variable instead of \textsc{DistortionAmount}.
The amount of distortions is controlled for a confounding variable, along with the number of points and clusters.
The study also shares the independent variables \textsc{GlobalInspection} and \textsc{Techniques}, and all other experimental settings with Study 1. 
In summary, this study is identical to the previous one except that we swapped \textsc{DistortionAmount} for \textsc{NonTriviality}, and we work with different sets of participants.

\paragraph{Participants} We recruit 12 participants (E2P1--E2P12) from two local universities (ten males and two females, aged 21--30 [24.6 $\ pm$2.5]) with the same recruiting criteria as Study 1.

\subsubsection{Quantitative Results}

\label{sec:exp2quant}

\begin{sidewaysfigure}

    \centering
    \includegraphics[width=\linewidth]{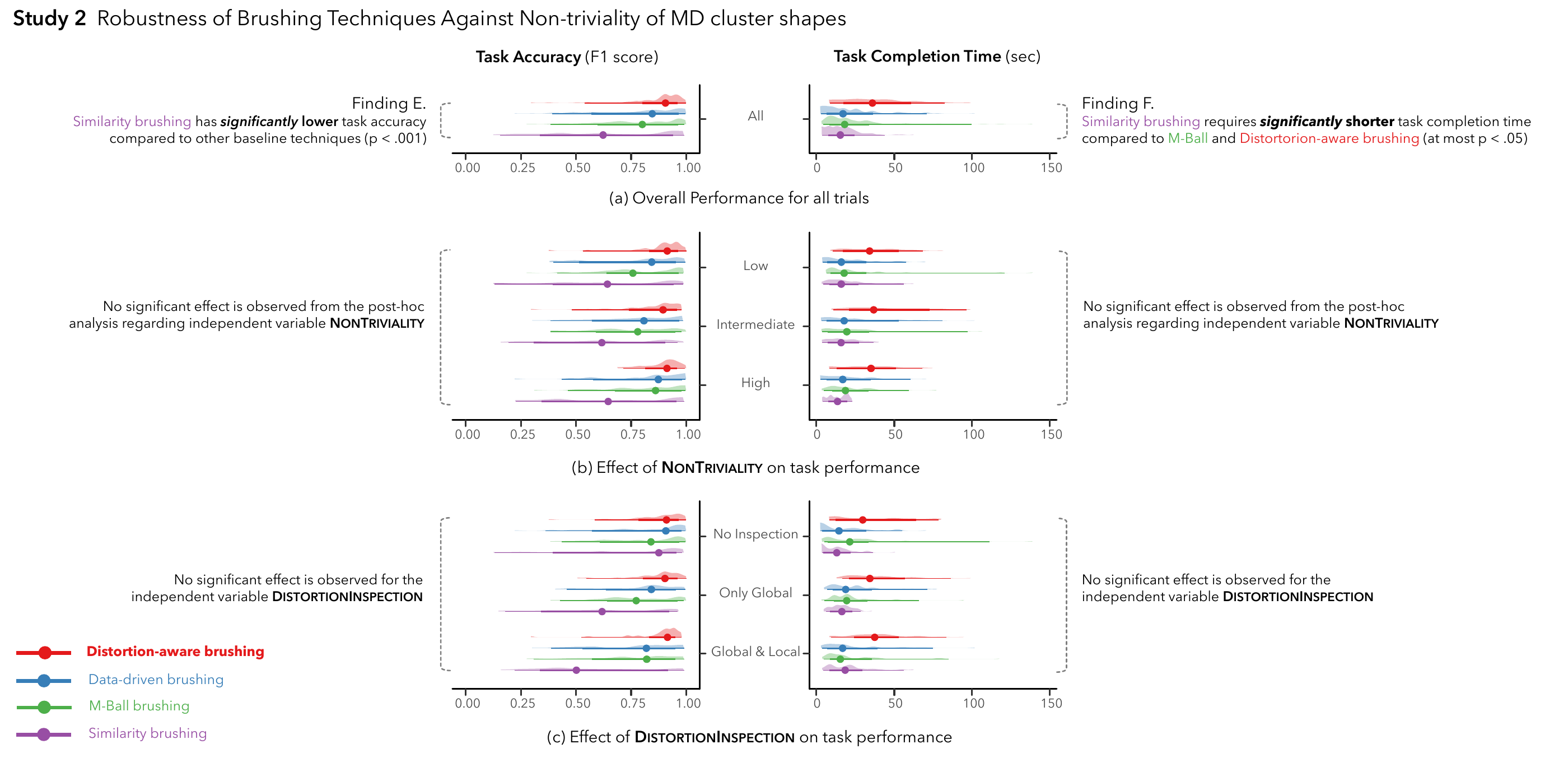}
        \vspace{-6mm}
    \caption{Study 2 results demonstrating the robustness of brushing techniques against the non-triviality of high-dimensional cluster shapes (\textsc{NonTriviality}). The study results reaffirm the superiority of \brush in terms of task accuracy. We find no significant differences between brushing techniques in terms of \textsc{NonTriviality}.}
    \label{fig:exp2_results}
\end{sidewaysfigure}

The following are the findings from the study (\autoref{fig:exp2_results}).
We first check whether our findings from the previous study which are not dependent on \textsc{DistortionAmount} are maintained in this study (Finding A and C). We also discuss new findings from the study.

\paragraph{Analysis on overall task accuracy}
By executing three-way RM ANOVA examining the influence of three independent variables on the F1 score, we find a significant main effect on \textsc{Techniques} ($F_{3, 33} = 34.51$, $p < .001$). Bonferroni post-hoc analysis reveals that \brush significantly outperforms all other techniques ($p < .001$ for all) in terms of accuracy, again confirming Finding A.
Unlike Study 1, post-hoc analysis indicates that Data-driven brushing and $M$-Ball brushing show significantly higher accuracy compared to Similarity brushing ($p<.001$ for both). This leads to a new finding:

\finding{E}{Using similarity brushing leads to lower accuracy compared to not only \brush but also Data-driven brushing and $M$-Ball brushing.}

\noindent
We find a significant main effect for \textsc{NonTriviality} (($F_{2, 22} = 8.55$, $p < .01$), but post-hoc analysis fails to identify significant differences between brushing techniques.

\paragraph{Analysis on overall task completion time}
We run three-way RM ANOVA to investigate how three independent variables affect task completion time (\autoref{fig:exp2_results} right), which is followed by Bonferroni correction for post-hoc analysis.

RM ANOVA reveals a significant main effect on \textsc{Techniques} variable (($F_{3, 33} = 13.84$, $p < .001$)). The post-hoc analysis indicates that \brush requires a significantly longer task completion time compared to all other techniques ($p < .001$ for all). This result again confirms Finding C from the previous study. Moreover, the post-hoc analysis finds that the task completion time with $M$-Ball brushing is significantly longer than Similarity brushing ($p < .05$), hence the finding:

\finding{F}{Using similarity brushing leads to shorter task completion time compared to not only \brush but also $M$-Ball brushing.}

\noindent
We again find a significant main effect for \textsc{NonTriviality} ($F_{2, 22} = 5.22$, $p < .05$), but post-hoc analysis identifies no significant differences between brushing techniques.

\subsubsection{Discussions}

Study 2 again confirms the superiority of \brush regarding task accuracy and the tradeoff between task accuracy and completion time.
Moreover, it also reaffirms that there are no significant effects relevant to \textsc{DistortionInspection}. The following is a takeaway from the new findings of Study 2.

\paragraph{Superiority and inferiority between baseline techniques}
Finding E shows that $M$-Ball Brushing and Data-Driven Brushing have better task accuracy than Similarity brushing. Finding F shows that $M$-Ball brushing requires longer task completion time than Similarity brushing.

Again, these results are consistent with the techniques' design. $M$-Ball brushing progressively and manually expands the brush while avoiding FN distortions, trading time for accuracy. In contrast, Similarity brushing selects all at once the true neighbors of all the points in the painter, which are not necessarily true neighbors of each other (FN), hence trading accuracy for time. Data-driven brushing also captures FN but does not automatically extend the brush to its true neighbors, resulting in fewer distortions than Similarity brushing. 
Our interview results (\autoref{sec:qual}) resonate with the finding that Data-driven brushing achieved better task accuracy than similarity brushing.

\subsection{Post-Study Interview}

\label{sec:qual}

We discuss the post-study interview results and takeaways.

\subsubsection{Benefits in Terms of User Experience}

\label{sec:qualbenefit}

Our interview reveals \brush's positive effects on user confidence, cautiousness, and serendipity. 

\paragraph{\brush makes users more confident}
Participants report that \brush makes them more confident. 
 Specifically, 20 of 24 participants (83\%) report being most confident when using \brush.
 
We find that such benefit mainly originates from the fact that \brush allows users to make final decisions about adding new points to the brush (\stepfour{}). In contrast, baseline techniques automatically formulate the high-dimensional region and only ``notify'' users of the points within the region being brushed. 10 out of 24 participants (42\%) explicitly state they cannot manually fine-tune the brushed points using the baseline techniques, which lowers their confidence. 

We also find that participants feel more confident when using \brush as they can see an high-dimensional cluster as a 2D cluster (\oone{}). Eight out of 24 (33\%) participants mention that they are more confident about their interactions as the visual 2D clusters inform that they are doing well. 
In contrast, participants are not able to visually confirm the correctness of their interaction in baseline techniques, which makes them less confident.

\paragraph{\brush makes users more cautious}
Our interview reveals that the enhanced confidence again stems from participants' heightened cautiousness. E2P5 noted, `\textit{`As I knew that I was doing well, I wanted to perform my best in performing the given task''}. 
The result also aligns well with our quantitative findings. It is intuitive that people will require more time to complete tasks when they are more cautious (Finding C), but this will result in better accuracy (Findings A, B).

\paragraph{Serendipity of \brush}
An unexpected benefit of \brush is that users enjoy using it. 11 out of 24 participants (45\%) note that playing with \brush is fun, and five among them (21\%) especially use the word ``game''. E1P6 noted, \textit{``I feel like playing a game that captures small monsters. I want to achieve a higher score.''}. 
In a similar context, E2P11 said, \textit{"The red trace felt like a laser line, so it was fun for me to use the program like a game."}. Both feedback indicate that they experienced the visual feedback of points relocation as a gamification feature and had fun using the technique.

\subsubsection{Benefits of Global and Local Inspections}

\label{sec:qualinspection}

Although not statistically validated, our studies provide evidence that global and local inspection functionalities (\stepone{} \steptwo{}) provide a positive effect in reducing task completion time 
(\autoref{sec:study1discuss}). Consistent with this result, our interview also provides qualitative evidence of the positive impact of local inspection functionality. Seven out of 24 participants (29\%) reported that the local inspection functionality substantially helps them quickly determine which action to execute next. However, no participants explicitly indicate the benefit of global inspection functionality. We believe this is primarily because image snippets already support easy identification of the correct starting point for brushing. Examining the effect of global inspection without image snippets will be an interesting future avenue to explore.

\subsubsection{Limitations}

\label{sec:techlimit}

The interview also reveals the limitation that \brush requires an initial learning phase.
Participants report that understanding the workflow is not cognitively difficult, but it still takes some time to become proficient with the technique. Sixteen out of 24 participants (67\%) mention that some practice is necessary to fully understand and utilize the functionalities of \brush. This is partly because participants are less familiar with painter-based brushing. 10 of 24 participants (42\%) find the baseline Data-driven brushing more comfortable because they are familiar with box-shaped brushes.
We discuss strategies to make \brush more intuitive for first-time users in \autoref{sec:limitations}.

\section{Use Case 1: Geospatial Cluster Analysis}

\label{sec:scenario}

We demonstrate a use case validating \brush's effectiveness in conducting cluster analysis of high-dimensional geospatial data.

\subsection{Procedure}

\paragraph{Persona and Goals}
We define a persona named Alice, a data analyst hired by a casual dining franchise company. The company wants to open new restaurants in California, but investigating the profitability of every block in the state exceeds its budget. Thus, the company asks Alice to find good candidate areas for detailed examination with three constraints: \textbf{(C1)} report blocks that are geographically similar to each other, which will substantially reduce the investigation cost. \textbf{(C2)} report blocks that are similar to each other for many attributes (i.e., well clustered in the high-dimensional space), also for reducing the investigation cost. 
\textbf{(C3)} focus on the blocks with higher average home prices, as people who have the financial means to pay for the restaurant’s food may live in such areas.

\paragraph{Dataset} 
Alice uses California Housing dataset \cite{pace97statistics}, comprised of nine attributes (\textit{Longitude}, \textit{Latitude}, \textit{Median Income}, \textit{House Age}, \textit{Average Rooms}, \textit{Average Bedrooms}, \textit{Population}, \textit{Average Occupancy}, and \textit{House Price}). Each datum corresponds to an individual block, which is the smallest geographical unit used in the U.S. census. Each attribute value is a statistic that summarizes all households in a corresponding block. Note that we randomly sample 5\% (1031 points) of the original dataset provided by scikit-learn \cite{pedregosa11jmlr} as too many data points may result in visual clutter in projections (\autoref{sec:scalabilitydisc}).

\paragraph{System design}
Alice uses a visual analytics system leveraging \brush consisting of three components: Brushing View, Parallel Coordinate (PC) View, and Attribute View. 

Brushing View provides \brush, where 2D projection is made by mapping longitude and latitude to $x$ and $y$ axes, and the map of California is displayed in the projection background. The high-dimensional space is set as the 7D space formed by all other attributes. The map is hidden when the points are relocated (Steps 3, 4), as the 2D positions of data points have no more direct relation to the map. When users return to the original projection using contextualization (\autoref{sec:contextualization}), they can see the map again. 

PC View helps users to directly monitor how the brushing proceeds and get hints about initiating or terminating brushing. PC dynamically reacts to user interaction in the Brushing View; the lines corresponding to seed points while inspecting local distortions (\steptwo), and the brushed points (\stepfour) are highlighted using the same color as the points. All other points are depicted in black with lower opacity.

Finally, the Attribute (Attr) View displays the distribution of attribute values of (1) brushed clusters, (2) remaining points, and (3) all points using boxplots. Boxplots are dynamically updated, reflecting the brushing status in the Brushing View.

\subsection{Scenario}

We detail our scenario that comprises three stages.

\subsubsection*{(Stage 1) Inspecting local distortions}

\begin{sidewaysfigure}
  \centering
  \includegraphics[width=\linewidth]{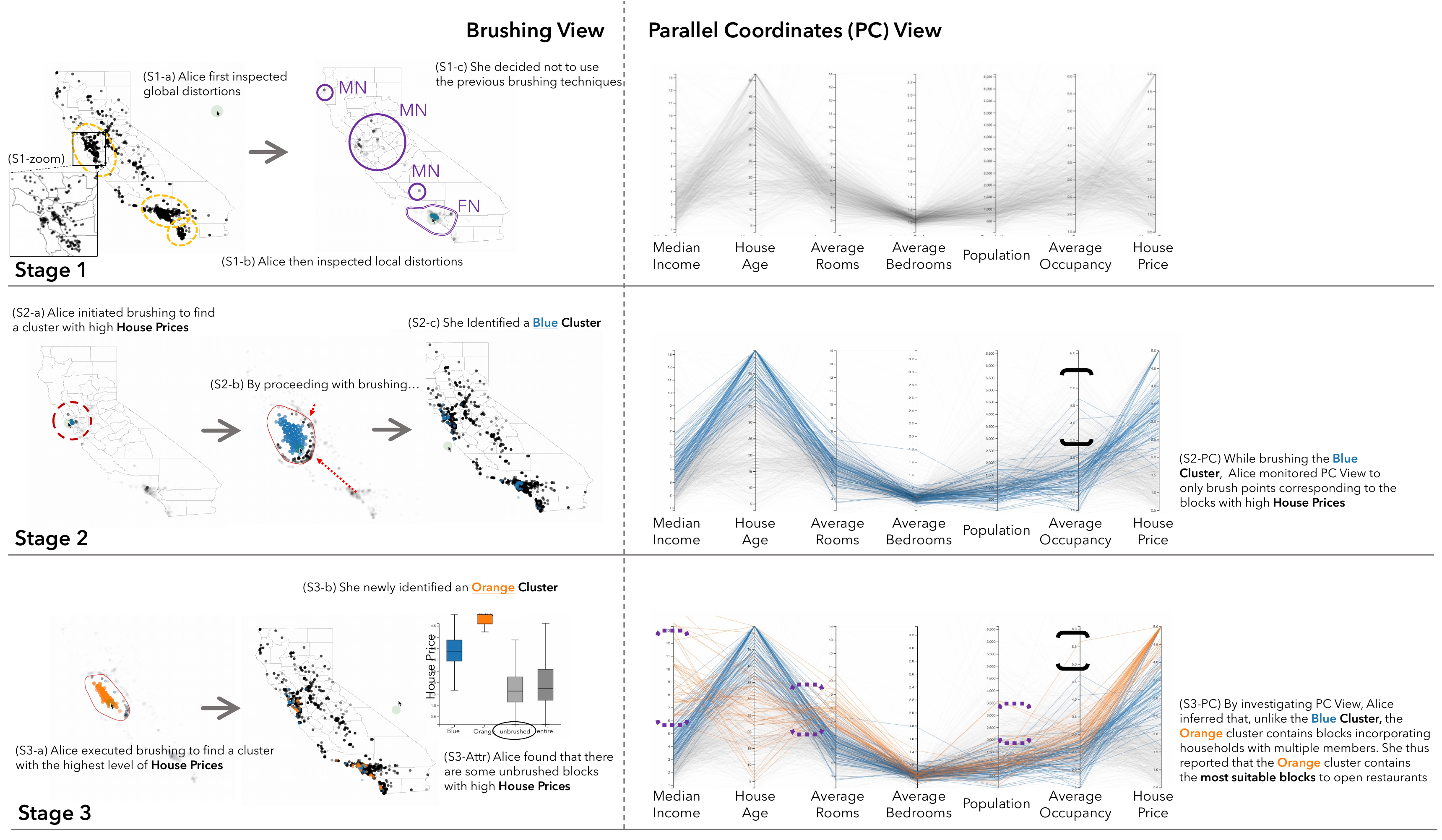}
  \caption{Use Case 1: Alice explores the California Housing Dataset \cite{pace97statistics} (\autoref{sec:scenario}) with \brush to find a good candidate region for opening casual dining restaurants. 
 Alice uses a visual analytics system composed of a brushing view (left) that serves \brush and a Parallel Coordinates (PC) view (right) that shows the attribute values of brushed points. 
 Alice also checks detailed statistics of attribute values in brushed points by checking the Attribute (Attr) view.
 Using a visual analytics tool incorporating \brush, our persona, Alice, extracted clusters of blocks with similar characteristics and successfully conducted the given request.
  }
  \label{fig:scenario}
\end{sidewaysfigure}

Alice first wants to examine whether she could report geographically similar points as clusters (C1). To do so, she checks whether the visual proximity between 2D points matches the high-dimensional similarity. By inspecting the global distortion using density encoding (\stepone), she finds several 2D clusters in the projection (i.e., map) that have high density in the high-dimensional space (yellow dotted ellipses in \autoref{fig:scenario} S1-a). 
Note that she zooms in to remove clutter and verifies there are points with high density on high-dimensional space in such places (\autoref{fig:scenario} S1-zoom).
Thus, she hypothesizes these regions may contain blocks that can also be considered clusters in the high-dimensional space. 

To validate her hypothesis, Alice hovers the painter around three regions to examine their local distortion. As a result, she finds that all regions suffer from both FN (neighboring map locations having low high-dimenisonal similarity, depicted by low opacity; purple doubled line in \autoref{fig:scenario} Stage 1) and MN (map locations far apart having similar attributes; purple solid line in \autoref{fig:scenario} Stage 1), rejecting the hypothesis (\autoref{fig:scenario} S1-b). This means that Alice \textit{cannot use conventional brushing or previous brushing techniques for DR projections} (\autoref{sec:rel}) as they define the 2D brush as a compact, continuous 2D region vulnerable to FN. Therefore, Alice reports that it is difficult to satisfy C1 and decided to keep using \brush for the following analysis (\autoref{fig:scenario} S1-c).

\subsubsection*{(Stage 2) Extracting a cluster with high house prices}

Alice then aims to find points having high \textit{House Prices} (C3) and being well clustered (C2). 
She first moves the painter around the projection while monitoring seed points’ attribute values using PC view. As she expects that such ``wealthy regions'' may be located in urban areas, she especially examined regions around Los Angeles and San Francisco. As a result, she finds an area that contains seed points having \textit{House Prices} ranging from 250,000\$ to 400,000\$, which exceeds the median house price, near San Francisco downtown (red dashed circle in \autoref{fig:scenario} Stage 2; S2-a). She starts brushing from the seed points while monitoring the PC View (Steps 3, 4) (\autoref{fig:scenario} S2-b), and terminates brushing when the brush starts to contain the blocks that have significantly different patterns in PC View. As a result, she brushes the blocks with \textit{House Prices} lower than approximately 200,000\$ (black solid bracket in \autoref{fig:scenario} S2-PC). Following the corresponding brush color, she names the cluster as a \textit{Blue cluster} (\autoref{fig:scenario} S2-c). Note that this \textit{cannot be done by filtering high-price blocks in PC View} as she does not know the threshold of \textit{House Prices} that can accurately discriminate blocks having different patterns in PC View.

\subsubsection*{(Stage 3) Extracting a cluster with the highest house prices}

Through the PC View and the Attribute View, Alice discovers that the portion of blocks with the highest \textit{House Prices} is not incorporated in the Blue cluster.
Therefore, she decides to find data points with similar attribute values (C2) that may comprise these highly-priced blocks (C3). By skimming through the urban area again, she finds seed points with the highest level of \textit{House Prices} and starts brushing again (\autoref{fig:scenario} S3-a). To clearly discriminate the currently brushed cluster from the Blue cluster, she stops brushing when the newly brushed cluster starts to incorporate the blocks with prices lower than 400,000\$ (black solid bracket in \autoref{fig:scenario} S3-PC). She names the cluster the \textit{Orange cluster}, following the same naming convention (\autoref{fig:scenario} S3-b).

Alice finds that the orange cluster not only exceeds the blue cluster in terms of \textit{House Prices} but also in \textit{Median Income}, \textit{Average Room}, and \textit{Average Occupancy} (purple dotted brackets in \autoref{fig:scenario} S3-PC). Higher \textit{Average Occupancy} denotes that the blocks in the orange cluster may contain households consisting of multiple people. Higher \textit{Median Income} and \textit{Average Room} also support this inference, as they will grow proportionally to the number of members in each household. Alice thus concludes that the \textit{Orange cluster} contains blocks that are more appropriate for opening casual dining restaurants, as such restaurants may not target people who eat alone. 

Alice thus reports the blocks within the orange cluster as having top priority for investigation. It is worth noting that blue and orange clusters have no clear segmentation regarding individual attribute values (\autoref{fig:scenario} S3-PC). This means that the \textit{neither brushing axes of the PC nor previous brushing techniques for DR projections can precisely reveal these clusters}. Moreover, points within two clusters are all scattered throughout the projection (\autoref{fig:scenario} Stage 4). This reaffirms that \textit{both naive 2D brushing and previous brushing techniques for DR projections cannot stand out to detect these clusters}, although these clusters have clear differences overall (S3-PC)---yet a clear benefit of \brush in this scenario.

\section{Use Case 2: Visual-Interactive Labeling}

\label{sec:usecaselabeling}

\brush can effectively identify noisy clusters that are semantically distinct but not well-separated in the data space (\stepfour{}).
This enables users to make the final decision on whether to brush certain points.
We present a use case leveraging this utility in supporting the visual-interactive labeling \cite{bernard18tvcg, meng24tvcg} for data with noisy high-dimensional clusters and distorted projections.

\subsection{Procedure}

\begin{figure}
    \centering
    \includegraphics[width=0.7\linewidth]{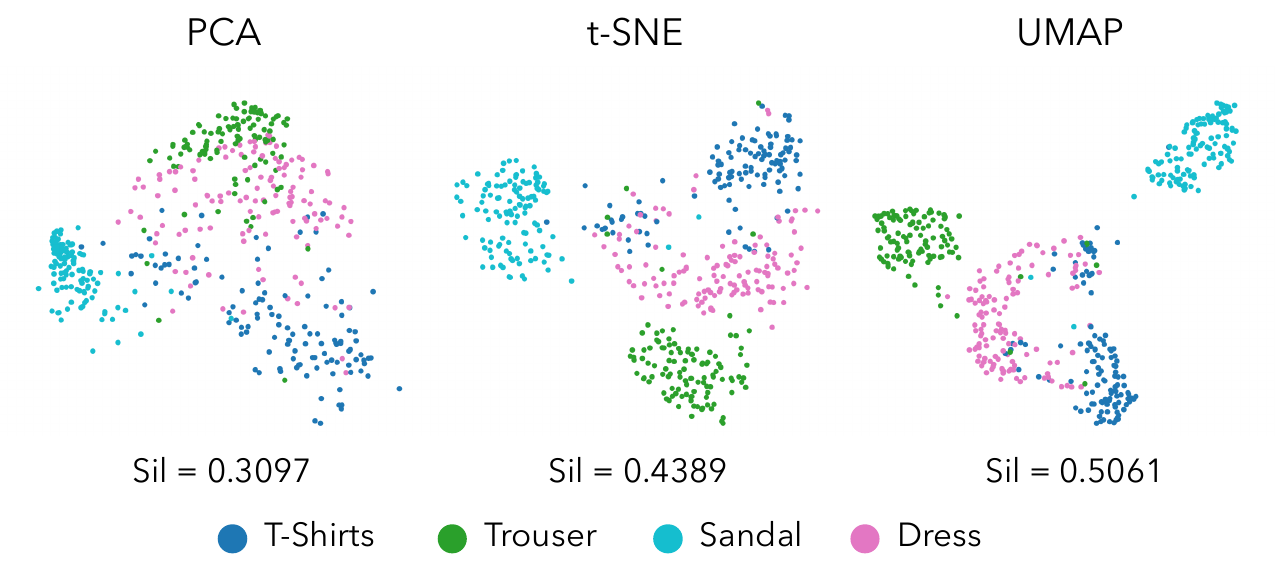}
    \caption{PCA, $t$-SNE, and UMAP projections of the subset of Fashion-MNIST dataset used in our second use case (\autoref{sec:usecaselabeling}). 
    Silhouette (Sil) scores indicate the separability of classes in the projections.
    Data points are initially assumed to be unlabeled, appearing as monochrome scatterplots.}
    \label{fig:usecase2_projection}
\end{figure}

\paragraph{Persona and goals}
Bob, a data engineer, wants to train a classification model on an image dataset. He aims to perform visual-interactive labeling to identify ground-truth classes.

\paragraph{Dataset}
We simulate a dataset with noisy clusters, i.e., clusters that overlap in the high-dimensional data space, making them difficult to detect using automatic clustering algorithms or nonlinear dimensionality reduction techniques. We extract four classes in the Fashion-MNIST dataset \cite{xiao2017arxiv}: \textit{T-Shirts, Trouser, Sandal,} and \textit{Dress} (\autoref{fig:usecase2_projection}). Among them, \textit{T-Shirts} and \textit{Dress} have been verified to have low pairwise separability (\autoref{sec:apptsne}) in the high-dimensional space. We assume that the dataset lacks predefined class labels, requiring Bob to identify these four classes.
As shown in \autoref{fig:usecase2_projection}, these classes are not well-separated in projection spaces. This implies that previous brushing techniques may poorly discriminate these classes (\autoref{sec:userstudy}).
Note that we use the entire 784 dimensions of the original dataset to represent the high-dimensional space.
To avoid visual clutter (\autoref{sec:scalabilitydisc}), we use 2\% (120 points for each cluster) of the original Fashion-MNIST dataset.

\paragraph{System design.}
Bob uses a labeling system equipped with \brush. 
To simulate a scenario where projections distort the representation of original clusters, we use PCA, which shows the worst performance in discriminating classes in terms of Silhouette \cite{rousseuw87silhouette} scores (\autoref{fig:usecase2_projection}).
Each data point is visualized as an image snippet; Bob can thus investigate the semantic difference between data points based on their visual appearance.

\subsection{Scenario}

\subsubsection*{(Stage 1) Brushing a Cluster Separated by Visual Similarity and Proximity}

\begin{sidewaysfigure}
    \centering
    \includegraphics[width=\textwidth]{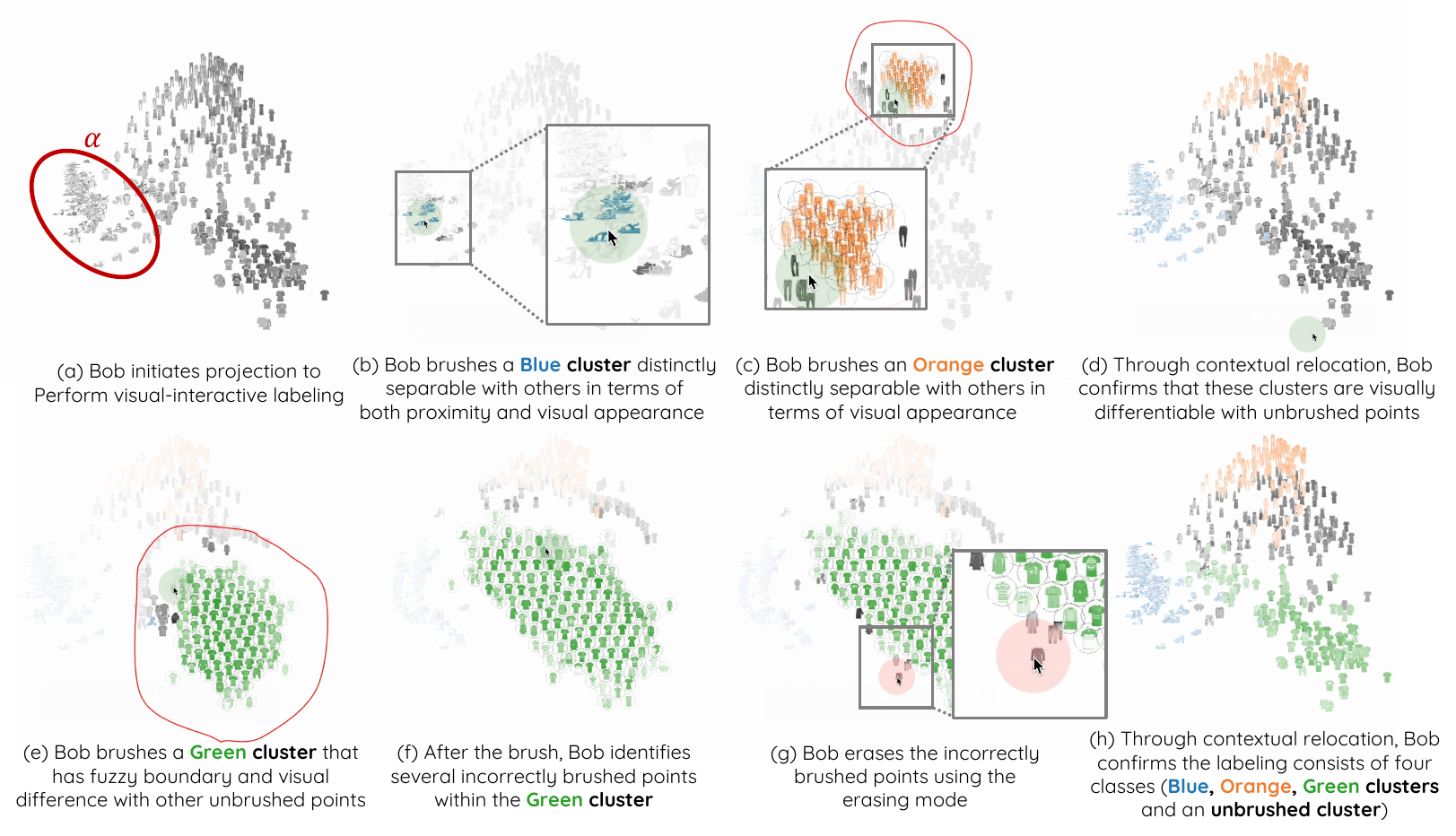}
    \caption{Use Case 2: Bob uses \brush for visual-interactive labeling \cite{bernard18tvcg, meng24tvcg} (\autoref{sec:usecaselabeling}). Bob aims to identify and brush semantically distinct clusters in the dataset to form ground-truth label data for training an automatic classifier. \brush is effective to identify noisy clusters (h) (semantically separated in snippet images but not in the high-dimensional data space) from strongly distorted projections (a).}
    \label{fig:usecase2_scenario}
\end{sidewaysfigure}

Upon initiating the system (\autoref{fig:usecase2_scenario}a), Bob identifies a 2D cluster that is visually and spatially distinct from other data points (\autoref{fig:usecase2_scenario}$\alpha$; corresponds to the \textit{Sandal} class). Bob uses \brush to label this cluster. As the cluster is clearly separated from other points in both the high-dimensional space and the projection, Bob brushes the cluster with no difficulty  (\autoref{fig:usecase2_scenario}b), naming it the \textit{Blue Cluster}.

\subsubsection*{(Stage 2) Brushing a Cluster Separated By Visual Similarity}
Bob identifies that the remaining points consist of two types of fashion items: \textit{tops} (corresponding to \textit{T-Shirts} and \textit{Dress} classes) and \textit{bottoms} (corresponding to \textit{Trouser}). He notices that these two groups are not well-separated in the projection but have semantically distinct visual appearances. To differentiate them, Bob brushes the \textit{bottoms} (\autoref{fig:usecase2_scenario}c).  Although \textit{tops} and \textit{bottoms} overlap in the projection, \brush dynamically relocates the selected points, enabling Bob to discriminate between them with ease. 
He names this cluster the \textit{Orange Cluster}. By contextualizing the brushing results within the original projection (\autoref{sec:contextualization}), Bob confirms that the \textit{Blue Cluster} and \textit{Orange Cluster} represent distinct labels (\autoref{fig:usecase2_scenario}d)

\subsubsection*{(Stage 3) Separating Clusters with Fuzzy Boundaries}
Bob inspects the remaining points that correspond to \textit{tops} and identifies that they consist of two semantically different clusters (\textit{T-shirts} and \textit{Dress} classes). 
However, these two clusters are semantically not clearly separated and partially overlap in the projection. Despite this, Bob brushes one of them, naming it the \textit{Green Cluster} (\autoref{fig:usecase2_scenario}e). As \brush relocates points, it helps Bob consolidate semantically similar points scattered throughout the projection. However, after completing the brushing, he observes that the \textit{Green Cluster} still contains points that do not appear to belong to the correct label (\autoref{fig:usecase2_scenario}f). To refine the selection, he activates the erasing mode (\autoref{sec:erasing}) and removes these points (\autoref{fig:usecase2_scenario}g). Since uncertain points are pushed to the cluster boundary (\stepfour), Bob can easily erase them.

Finally, by once again contextualizing the brushing results within the original projection (\autoref{fig:usecase2_scenario}h), Bob confirms that the labeling now consists of four clusters (\textit{Blue, Orange, Green} and the remaining \textit{unbrushed} clusters).

.

\subsection{Comparison with Existing Systems}

Previous visual-interactive labeling systems \cite{bernard18tvcg, meng24tvcg} provide an initial projection and prompt labelers to label a subset of the data. These labeled points are then incorporated into an underlying classification model, which supports the labeling process by (1) predicting labels for unlabeled points \cite{bernard18tvcg}, or (2) updating projections to better highlight the separation of labeled points \cite{meng24tvcg}. This iterative process continues until labeling is complete.

Our \brush approach offers two main advantages over the previous systems. First, in the earlier systems, the initial labeling can be erroneous because of distortions in initial DR projections, and these errors can propagate through subsequent iterations as the labels are used to train the classification model. Our approach avoids this problem through point relocation. Second, our approach does not require an underlying classification model at all, which simplifies the system by reducing hyperparameters and computational overhead. This also helps users to preserve their mental map throughout the labeling process, as they do not need to wait for model training at each iteration.
However, when DR projections have less distortion and clusters are not noisy (i.e., aligned well with potential classes), \brush may provide little advantages over previous systems. 
Further quantitative comparisons are needed to better evaluate the pros and cons of both approaches.

\section{Discussions}

We revisit the benefit of \brush and discuss limitations and future directions.

\subsection{Comparison to Automatic Clustering Techniques}

\brush and automatic clustering techniques (e.g., $K$-Means) both aim to discover meaningful cluster patterns in high-dimensional data. 
However, by engaging users in the cluster extraction process, \brush makes cluster analysis more insightful. 
First, by allowing users to ``see'' image snippets (\autoref{fig:dab:teaser}) or auxiliary visualizations (\autoref{sec:scenario}), \brush can account for visual similarity between data points that may often be overlooked by clustering techniques that rely on conventional similarity or distance metrics (e.g., Euclidean, $k$NN). 
Our second use case (\autoref{sec:usecaselabeling}) demonstrates the benefits of this approach.
Moreover, as seen in our first use case (\autoref{sec:scenario}), \brush supports users to incorporate their external knowledge into the clustering procedure, specify characteristics of clusters that they want to discover, and set the boundary of clusters themselves. These functionalities enrich data analysis and are not possible in conventional clustering techniques.
Such benefits emphasize both the effectiveness of \brush and the importance of the interactive aspect in cluster analysis.

\subsection{Visual and Computational Scalability}

\label{sec:scalabilitydisc}

The computational time complexity of a single iteration of lens update and point relocation is $O(m\log m + \kappa nm)$, where $n$ denotes the total number of points, $m$ is the size of the convex hull, and $\kappa$ denotes the number of nearest neighbors considered to compute closeness. As $n \gg m$ and $n \gg \kappa$, the running time of the technique is mainly bounded by $n$ and has no relation with the dimensionality, making it highly scalable. 
Note that the preprocessing of our technique requires $O(n^2k^2)$ but is executed only once and does not affect interactivity.

However, the technique may still struggle to deal with large datasets for two reasons. First, the technique may suffer from visual complexity.
When applied to a large dataset, visual clutter can hide image snippets and opacity encoding. 
Enhancing \brush's encoding to mitigate such visual complexity will be an interesting future work. For example, we may adopt density plots \cite{trautner2020sunspotPlots, jo19tvcg} to resolve the clutter. 
Second, the technique requires a substantial amount of memory. We store the similarity matrix in dense format, requiring $O(n^2)$ memory, which may cause an out-of-memory error in the heap space with large datasets. Reducing memory usage, for example, by utilizing sparse matrix formats, will be crucial for enhancing the scalability of \brush when processing large datasets.

\subsection{Reproducibility of Cluster Analysis}

Reproducibility is a fundamental aspect of visual analytics. Since \brush allows analyses to start from different points, potentially leading to varying outcomes (\autoref{sec:contextualization}), it might seem to harm reproducibility. However, our technique mitigates this concern by guiding users in identifying high-dimensional clusters that align with 2D clusters, e.g., through opacity-coding of high-dimensional density (\stepone), which encourages users to initiate clustering in similar regions and reduces confirmation bias. Moreover, \brush is intentionally designed with minimal hyperparameters (\ofour), reducing variability from extensive hyperparameter tuning, which is a common pitfall in automated clustering algorithms. Finally, our user study demonstrates that \brush enables participants to identify high-dimensional clusters that closely align with ground-truth clusters, outperforming conventional brushing approaches. 
\revise{
Nevertheless, given its inherently interactive nature, we recommend using \brush for exploratory data analysis rather than for confirmatory purposes. For example, it would be inappropriate to report findings discovered through \brush in a scientific publication without additional validation.
}

\subsection{\revise{Extension for Subspace Analysis}}

\revise{
In our user studies and use cases, \brush computes high-dimensional similarity between data points by leveraging the full attribute space of the dataset. We believe that extending our technique to allow users to select subspaces for similarity computation would enable a deeper understanding of high-dimensional cluster structures. For example, in our first use case (\autoref{sec:scenario}), the analyst could more easily distinguish a cluster characterized by high house prices (Stage 2) by focusing on relevant attributes (e.g., House Price, Median Income, and Average Number of Rooms), rather than considering the full set of attributes. On the other hand, analysts can interactively examine the extent to which cluster structures differ across subspaces by constructing a projection from one selected subspace while computing high-dimensional similarity using another subspace.}

\revise{
This subspace-based extension of \brush can also be interpreted as a generalization of the query relaxation technique proposed by Heer et al. \cite{heer08chi}, which extends brushing by including data points with similar values along designated attributes. While query relaxation typically focuses on a small number of salient attributes, subspace analysis with \brush enables the examination of topological structures induced by a larger set of attributes and their interactions. Systematically taxonomizing existing approaches that extend brushing---encompassing both Heer et al.’s work \cite{heer08chi} and \brush---represents an interesting avenue for future research.
}

\subsection{Additional Usage Scenarios}

This research contributes \brush to make \textit{cluster analysis of high-dimensional data more reliable}. Our studies and use cases align with this purpose. 

However, the core concept of \brush can be applied to exploring and clustering diverse data entities where discrepancies exist between their 2D spatial positions and semantic meanings. 
For example, \brush can be used to browse computer files within messy directories (e.g., \texttt{Downloads} or \texttt{Desktop}) by interactively clustering the files based on content similarity.
In this case, users would not only rely on the snippet image of each file (i.e., icons) but also benefit from invisible semantic distances between file contents, with distortions computed accordingly to guide the \brush.
Another potential use case of \brush is to compare the hyperparameter settings of machine learning models, where \brush could help users compare different hyperparameter settings, identify a group 
of optimal configurations, and gain insight into how their variations affect model performance.

\subsection{Limitations}

\label{sec:limitations}

\paragraph{Preserving mental map}
While point relocation helps robustly brush high-dimensional clusters regardless of distortions (\otwo), it comes at the cost of the global context provided by the original DR projection. As a result, users may struggle to maintain their mental map, making it harder to perceive the connection between the original projection and the brushed regions. This can potentially hinder tasks that rely on global structure investigation, such as comparing the density of clusters \cite{jiazhi21tvcg}. One possible solution is to juxtapose the interactive projection with the original DR projection, trading visual space for better contextual awareness. 
Our contextualization functionality (\autoref{sec:contextualization} further supports users in recalling the global structure by helping them maintain their mental map throughout the brushing process  

Still, by compromising on global reliability, users can gain substantially high reliability in extracting and examining local clusters, as demonstrated in our user study (\autoref{sec:userstudy}). 
This advantage enables \brush to reliably support the typical data analysis flow of the visual seeking mantra \cite{shneiderman03civ}: \textit{Overview first, zoom and filter, details on demand}. After users inspect the \textit{overview} of data distribution through the original DR projection, they can accurately \textit{zoom into} or \textit{filter} local clusters using the technique, then reliably analyze clusters in \textit{detail} through image snippets or auxiliary visualizations (\autoref{sec:scenario}, \ref{sec:usecaselabeling}).

The strategy to enhance local reliability by sacrificing global reliability is similar to that of Focus-and-Context (FC) (e.g., Fisheye~\cite{gansner05tvcg}). However, while FC approaches only provide transient enhancement of local reliability, \brush continuously maintains the brushed local high-dimensional cluster, which enables more detailed and tangible follow-up analysis (\autoref{sec:scenario}).

\paragraph{Learning curve}
Our user study reveals that \brush takes time for users to become comfortable with (\autoref{sec:techlimit}). 
Post-hoc interviews suggest this is partly due to users' limited familiarity with painter-based brushing systems.
\revise{This issue can potentially introduce uncertainty into early interactions with \brush, thereby affecting the reliability of visual analytics.
To address this problem,}  we plan to redesign \brush to leverage the box-shaped brush regions, which are more familiar to most users. 
Another possible approach is to slow down point relocation, allowing users to control when to perform lens updates and point relocation. While this may slightly increase task completion time, it could help users adapt to \brush more easily.

\paragraph{Limitations in user studies}
We discuss the limitations of our studies. 
First, data points are visualized using image snippets in our experiment, but data points may lack intuitive visual representations, \textit{e.g.} in tabular data or text documents.
We plan to improve the generalizability of our study by testing the situation in which auxiliary visualizations (e.g., \autoref{sec:scenario}) are used to guide the brushing. 
We also want to clarify the need to conduct experiments with more datasets. For instance, \textsc{NonTriviality} may not have a statistically significant effect on the studied brushing techniques because the variation in the non-triviality of cluster shapes between digits in the MNIST dataset was not large enough.
\section{Conclusion}

Although brushing has long been used as an important component to interact with DR projections, previous brushing techniques have struggled to overcome distortions. 
To tackle this problem, we proposed \brush, a brushing technique that relocates points to resolve distortions.  
Our user studies demonstrated the usefulness and usability of \brush in exploring and discovering MD clusters, its robustness to distortions, and its greater accuracy compared to state-of-the-art brushing techniques.

In summary, our work provides a more reliable basis for analysts to communicate with DR projections, advancing the research community one step closer to more reliable cluster analysis of high-dimensional data.

\chapter{Discussions}

We discuss three key questions implied by the contributions and limitations of this thesis.

\section{What will be the Ideal Future Solution?}

We discuss potential future directions that can more fundamentally mitigate the reliability challenges we identified.

\subsection{Automating DR Configuration}

\label{sec:automation}

Although our technical solutions improve the reliability of DR-based visual analytics, they still must be actively leveraged by practitioners.
This, in turn, requires sufficient literacy in DR.
For example, to benefit from \ltc, practitioners must understand that these metrics assess the preservation of cluster structure and know how to execute it.
However, as our investigation on practical usage of DR suggests (\autoref{sec:quantianal}), practitioners often lack such literacy.

One intuitive solution to remedy this problem is to delegate the configuration of DR for visual analytics to machines.
Imagine \textit{VoyagerDR}, a hypothetical programming library with comprehensive knowledge of DR.
This library is equipped with a function that, when given a high-dimensional dataset and a specified analytic task, automatically predicts and recommends ideal DR techniques and hyperparameters that maximize task performance.
In theory, VoyagerDR would always guarantee the proper use of DR and enhance the reliability of visual analytics, even when practitioners lack literacy on DR.

VoyagerDR can be conceptualized to comprise three core components.
First, an \textbf{interpretation module} that interprets user inputs to infer their intended analytic tasks. Advanced LLMs can empower this module to understand and disambiguate vague natural-language descriptions of analytic tasks.
Second, a \textbf{mapping function} that identifies the most suitable DR techniques and evaluation metrics for a given task. The visualization community has already accumulated extensive knowledge on the relationship between DR techniques and analytical tasks through benchmark studies \cite{jiazhi21tvcg, etemadpour15tvcg, atzberger24tvcg} and practical guidelines for proper DR use \cite{coenen19fiar, wattenberg2016tsnetuning}. The challenge lies in formalizing this collective knowledge into a machine-interpretable form---a task for which existing approaches like Draco \cite{moritz19tvcg} provide inspiration. 
The third and most challenging component is \textbf{hyperparameter optimization}. Optimizing DR hyperparameters to find the ideal projections typically requires slow, iterative trial-and-error processes \cite{jeon25arxiv}, which may hinder the usability of VoyagerDR. Nevertheless, predicting optimal model or hyperparameter selection in advance is an active area of research in AutoML \cite{feurer15nips}.
In terms of DR, Jeon et al. \cite{jeon25arxiv} show that dataset complexity can be leveraged to predict the most effective DR technique. We posit that this approach can be extended to encompass a broader range of techniques and the entire hyperparameter space.
Given these existing foundations, we consider the realization of VoyagerDR a feasible goal in the near future.

However, while appealing in concept, adopting VoyagerDR can be controversial because it may harm user agency.
With VoyagerDR, users may heavily rely on this library and their understanding of DR will remain limited. 
Moreover, they may be even less motivated to learn how to use DR properly because machines will do everything for them.
This is problematic because finding an optimal DR projection is not solely about selecting a single DR technique and hyperparameters that best aligns with the target task; it also depends on factors such as available computational resources, time constraints, and the \textit{practitioners' priorities}---such as whether faithfulness, stability, or efficiency takes precedence. 
For example, in time-sensitive scenarios, it might be more effective to bypass hyperparameter optimization in favor of faster system responsiveness.
Consequently, selecting a DR projection is rarely about identifying a single optimal solution; rather, it is often a process of comparing multiple plausible candidates while weighing diverse factors.
In this sense, even with such an idealized DR oracle, we should not abandon efforts to enhance practitioners' literacy as a path toward more reliable visual analytics.

\subsection{Facilitating Discourse}

\label{sec:discourse}

At the opposite end of automation lies a strategy that fosters discourse on the proper use of DR so that practitioners can become more self-motivated to learn how to use DR properly.
To facilitate such discourse, the visualization community should take a more active role.
For example, though papers that providing guidance on proper use of DR exists \cite{jiazhi21tvcg, wattenberg2016tsnetuning, espadoto21tvcg,coenen19fiar}, these papers often require substantial expertise and effort to read and interpret, making them less accessible to many practitioners.
Their impact is thus limited unless accompanied by more approachable forms of engagement.

We thus recommend moving beyond traditional text-based dissemination. 
Organizing workshops, panels, or tutorials at visualization, HCI, and machine learning conferences can help elevate the discourse and emphasize the importance of using DR techniques responsibly. For instance, a tutorial offering hands-on coding exercises for selecting and applying DR techniques properly would provide both practical guidance and increased awareness.
Recent tutorials in EuroVis 2025\footnote{\url{https://hyeonword.com/dr-tutorial/}} \cite{eurovis25evt} and CVPR 2025\footnote{\url{https://cvpr.thecvf.com/virtual/2025/tutorial/35919}} resonates with this path.

Another promising direction involves developing engaging, interactive educational materials. Interactive visualization systems and articles have proven effective for educating users across diverse domains \cite{adar21tvcg}, including machine learning \cite{kahng19tvcg, wang20chiea, wongsuphasawat18tvcg} and genomics \cite{bryan17tvcg, dasu21tvcg}. Similarly, such materials can underscore the risks of misusing DR techniques.
For instance, one might design an interactive visualization that prompts users to estimate dissimilarity between classes or points in \tsne and \umap projections, then reveal discrepancies with the ground truth data to illustrate the limitations of these techniques. 
Distortion visualizations \cite{lespinats11cgf, jeon21tvcg, martins14cg}, which augments projections to show how and where they are distorted from the original data, can also be actively leveraged in this direction.
As these visualizations allow users to assess the faithfulness of projections at a glance, they are likely to be more effective than relying solely on verbal or textual explanations of projection faithfulness and task suitability.
Investigating the effectiveness of other existing visualization and interaction techniques for DR from the visualization community in facilitating the reliable use of DR presents an intriguing avenue for future research.

\subsection{Leveraging Mixed-Initiative Approaches}

\label{sec:mixed}

Between fully automated systems such as VoyagerDR and purely human-driven efforts that promote discourse lies an approach that balances human agency with machine efficiency.
This direction offers a promising path toward both improving the reliability of visual analytics proactively and cultivating practitioners’ DR literacy.
Authoring assistant systems \cite{shin25tvcg, kim25pvis} that guide users in making appropriate configuration choices for DR techniques or issue warnings when improper configurations are detected can be an example.
Another viable approach is to develop web-based interfaces where users can upload their datasets and receive a set of recommended DR projections tailored to different analytic tasks.
Based on the system’s explanations, users can then make the final decisions about which projection to use.
Such approaches preserves users’ agency in decision-making, which in turn contributes to enhancing their literacy \cite{shin23chi}.

A direction that slightly degrades human agency but is more efficient is to make VoyagerDR more explainable. For example, enabling a \texttt{verbose} option by default to demonstrate how the library works internally would be beneficial for people to learn why certain projections are recommended. Furthermore, creating an \textit{Explainer} \cite{lee24vis, kahng19tvcg}, an interactive visualization that explains both VoyagerDR's operational processes and proper DR usage, will also be a plausible direction in terms of motivating people to increase their literacy while improving the reliability of visual analytics using DR instantly.

\subsection{Which Direction Should We Pursue?}

We claim that pursuing all three directions---automation, facilitation of discourse, and mixed-initiative approaches--in parallel will be the ideal approach, as the most effective solution depends on each practitioner’s level of DR literacy and motivation. For instance, a domain researcher new to DR may initially benefit from fully automated solutions (\autoref{sec:automation}) that ensure the safe use of DR while gradually fostering understanding through explanations of why particular projections are recommended (\autoref{sec:mixed}). Conversely, practitioners who are already motivated to deepen their knowledge will find interactive learning systems and educational materials more beneficial (\autoref{sec:discourse}).
Investigating how these three directions can make synergy will be an interesting future avenue to explore.

\section{What are Reliability Challenges Beyond Faithfulness?}

The reliability challenges that we identify and address are all related to the degree to which DR projections faithfully reflect the original data. For example, \ltc (\autoref{sec:clcl}) aims to directly measure the degree to which projections are distorted and unfaithfully represent the original data, and the dataset-adaptive workflow (\autoref{sec:dawadr}) tries to accelerate the process of making DR projections more faithful. 
Finally, Distortion-aware brushing (\autoref{sec:dabrca}) tries to locally resolve distortions to make interactions more accurate. 

However, DR projections can be unreliable not only due to faithfulness but also other factors, such as visual ambiguity of projections or the lack of interpretability \cite{jeon25chi}.
The following sections detail these additional threats, how existing literature addresses them, and the challenges that remain.

\subsection{The Lack of Interpretability}

Visual analytics with DR can also become unreliable due to limited interpretability.
First, the underlying mechanisms of DR techniques and their evaluation metrics are intrinsically complex and often difficult to communicate to analysts \cite{choo10vast, chatzimparmpas20tvcg}.
A second challenge arises from the limitations of scatterplots: they provide little information about the original attribute values \cite{kwon17tvcg, faust19tvcg, dowling19tvcg}, and they primarily represent data items as points, which makes it difficult to convey the structure or format of the underlying data (e.g., text) \cite{kandogan12vast}.
This problem hinders the selection of appropriate techniques that match the analytic goal and contexts, and make sensemaking hardly reflect the original data attribute values. 

These issues are typically addressed by augmenting scatterplots with additional visual encodings \cite{faust19tvcg, heulot13vamp, seifert10eurova} or auxiliary visualizations \cite{chatzimparmpas20tvcg}.
For example, Faust et al. \cite{faust19tvcg} overlay contour lines to depict how attribute values vary across nonlinear projection axes.
Directly encoding data values for each point using glyphs (e.g., aster plots \cite{kwon17tvcg}) or representations that match the underlying data format (e.g., image snippets for image datasets) \cite{bertucci23tvcg, boggust22iui} is also widely used.
A more direct approach allows users to specify visual patterns they expect (or hypothesize) to exist in DR projections and then automatically identifies projections that exhibit these patterns while explaining the influence of each attribute \cite{fujiwara19tvcg, fujiwara22tvcg}.

\paragraphit{Remaining challenges}
Despite the existence of interpretability tools for DR, applying them in analysis remains difficult due to their intricacy \cite{eurovis25evt}.
Furthermore, most interpretability methods are limited to linear projections, whereas contemporary DR-based visual analytics relies heavily on nonlinear techniques such as \tsne and \umap.
This gap also makes existing interpretability tools difficult to apply in practice.
A core challenge here is the computational cost of nonlinear DR.
Recently, Cho et al. \cite{cho25tvcg} proposed an approach that accelerates the generation of nonlinear DR projections via neural network-based approximation, making cluster analysis based on nonlinear projections interpretable. 
Extending this idea to a broader range of analytic tasks and contexts will be an interesting area for future work.

\subsection{Visual Ambiguity}

Visual ambiguity in DR projections, i.e., the extent to which analysts interpret the same projection differently, is another critical factor that can undermine the reliability of visual analytics.
In collaborative settings, such ambiguity can lead analysts to draw conflicting conclusions \cite{xiong20tvcg}, slowing the analysis and creating conflicts in decision-making.
However, most studies on the visual perception of scatterplots, the dominant representation for DR projections, focus on how people perceive patterns in general \cite{abbas19cgf, xia21cgna, wertheimer22pf}.
For example, Gestalt principles \cite{wertheimer22pf} describe how people commonly group visual objects within complex scenes \cite{pinna2010new}.
While valuable, these insights offer limited guidance for addressing reliability issues that arise from visual ambiguity in DR-based visual analytics.

\paragraphit{Remaining challenges}
To the best of our knowledge, the only work that directly addresses this problem is by Jeon et al. \cite{jeon24tvcg2}, who propose a metric that quantifies the visual ambiguity of scatterplots. By incorporating this metric into the DR optimization objective, they produce projections with substantially reduced ambiguity and negligible loss of faithfulness.
However, because this metric is trained on human perception data, it lacks interpretability: we do not know why a particular projection is perceived as ambiguous.
As a result, the metric can help automatically generate projections suitable for collaborative settings but offers little support for human sensemaking. A further obstacle is that the very notion of ambiguity is itself ambiguous \cite{jeon24tvcg2}; people differ in what they perceive as ambiguous.
Developing a concrete definition of ambiguity and an interpretable metric that operationalizes it thus remains a critical avenue for future research.


\subsection{Instability}

One fundamental limitation of DR, especially nonlinear techniques like \tsne and \umap, is that they are unstable: they produce projections with different visual patterns by a slight change in configurations, e.g., hyperparameters or initialization methods \cite{fadel15neurocomp,kobak21nb}. 
The randomness in the optimization procedure can also make DR techniques unstable \cite{wattenberg2016tsnetuning, jung23vis} even with the same configurations.
Such instability harms the reproducibility of DR-based visual analytics, making the resulting insights and decisions less trustworthy.

An intuitive approach is remedy this issue is to make DR techniques more robust to such changes \cite{fadel15neurocomp} or recommend the best configuration \cite{kobak21nb}. 
For example, Kobak and Linderman \cite{kobak21nb} demonstrate that using PCA initialization makes \tsne and \umap projections more stable. 
The researchers also provide similar endeavors to remedy instability in DR evaluation metrics  \cite{johannemann19arxiv, angelini22tvcg, lee10prl}. Recently, Jung et al. \cite{jung24vis, jung25tvcg} proposed alleviating this problem by automatically detecting unstable points and informing the visual analytics process about such instability.

\paragraphit{Remaining challenges}
A core challenge is that existing approaches assess instability solely by measuring how much a projection changes, without considering what kinds of changes occur.
In other words, we lack a notion of task-specific stability: different analytic tasks require sensitivity to different qualitative changes, yet current methods focus only on quantitative magnitude. 
Consequently, analysts receive little guidance on how projection instability affects the accuracy of their tasks.
Establishing such a task-specific notion of instability would also enable systematic comparisons between DR evaluation metrics (which are already linked with tasks) and instability metrics, making it possible to examine trade-offs between projection faithfulness and stability.

\subsection{\revise{Perceptual Misalignment}}

\revise{
Even properly configured and optimized DR projections can yield difficult-to-interpret visual patterns when the similarity computations are misaligned with human perception. In practice, interpretable structures or patterns are often ``hidden'' within the data, as noise can dominate similarity computations and obscure meaningful relationships \cite{fujiwara23pacificvis}. This challenge is further exacerbated in high-dimensional settings, where distances tend to become increasingly similar due to the curse of dimensionality \cite{bellman66science}, making visual patterns highly sensitive to even small perturbations.
}


\revise{
One approach to address this challenge is to automatically identify meaningful structures within datasets. For example, Fujiwara et al. \cite{fujiwara23pacificvis} propose a method that systematically explores diverse subspaces of a dataset and extracts interesting visual patterns. Similarly, Cho et al. \cite{cho25tvcg} introduce techniques that directly search for subspaces that produce visual patterns specified by analysts.
Complementary to these automated approaches, several methods enable analysts to interactively seek interesting structures. For instance, Xia et al. \cite{xia23tvcg}, and Endert et al. \cite{endert11vast} allow users to steer high-dimensional similarity functions by designating examples of similar and dissimilar data points.
}

\paragraphit{\revise{Remaining challenges}}
\revise{Existing approaches primarily aim to find interesting patterns in data processing and visualization stages. 
However, making the data itself focus on meaningful structure prior to DR-based analysis remains an open challenge.
A plausible direction is to leverage advanced deep learning models that learn denoised, semantically meaningful representations. For example, DINO \cite{caron21iccv, simeoni25arxiv} or BYOL \cite{neurips20grill} learn semantic representations from images without supervision, where using such representations as input to DR could yield more interpretable and stable projections. Extending these approaches, originally developed for image data \cite{caron21iccv}, to more general, non-image datasets represents a promising avenue for future research.
}









\section{How Can Visual Analytics Better Reflect Reality?}

In this dissertation, we make DR-based visual analytics more reliable. 
That is, we help analysts to ensure that insights and decisions derived from visual analysis more accurately reflect the underlying data.
However, reliability alone does not guarantee that visual analytics reflects true reality, as datasets themselves are imperfect representations of the real world.
Therefore, to ensure that analysts’ decisions lead to beneficial outcomes in real-world, we need efforts that extend beyond pursuing reliability, more tightly linking visual analytics to the real world.

Our contributions offer insights into how visual analytics might better reflect reality.
This perspective draws on a metaphor: just as a DR projection provides a cross-section of a dataset that reveals only certain structural characteristics, a dataset itself composes a partial cross-section of the real world.
Accordingly, the approaches we develop to make DR-based visual analytics more faithful to the underlying data may also inform how datasets---and visual analytics---can be more faithfully capture the real world. The followings are two directions that we suggest:

\subsection{Dataset Metadata}

We suggest informing the degree to which datasets reflect the real world using metadata.
DR evaluation metrics such as \ltc (\autoref{sec:clcl}) help analysts understand how faithfully a projection represents the underlying high-dimensional data, clarifying the strengths and limitations of different projections for analytic tasks.
Analogously, informing analysts how well and in what ways a dataset reflects the real world may clarify what kinds of insights and decisions that reliably derived from visual analytics.
Appropriate metadata is a key requirement to achieve this goal. 
For example, when a baseball team manager uses player statistics to make scouting decisions, metadata such as the years during which the data were collected or rule changes (e.g., strike-zone adjustments) can be critical for contextual interpretation.

However, collecting and documenting such metadata is labor-intensive and requires deep domain expertise.
Recent work by Shin et al. \cite{shin25arxiv} demonstrates that large language models (LLMs) can act as ``counselors'' that elicit and externalize implicit knowledge through guided questioning.
Building on this idea, we suggest proactively leveraging LLMs' broad knowledge about real-world to support the construction of metadata that clarifies the relationship between datasets and the real-world phenomena they represent. For instance, linking datasets with metadata and the external evidence that supports that metadata (e.g., news articles) can help analysts gain essential contextual knowledge while also enhancing trustworthiness of the metadata \cite{burns24tvcg}.



\subsection{Situated Analytics}

We also claim that making visual analytics situated, that is, contextualizing visual sensemaking in physical spaces, is essential for bringing visual analytics closer to the real world.
For example, when searching for a home, viewing relevant data directly within the house can enable more informed judgments about whether the space is a suitable fit.
This perspective aligns with the spirit of Distortion-aware brushing (\autoref{sec:dabrca}): just as Distortion-aware brushing locally visually restores high-dimensional structure to support more accurate interaction, situated analytics restores the context of physical world while analysts interpret their data. 
Here, a key limitation of situated analytics is its reliance on manual engineering, such as linking relevant databases to specific physical locations, where such costs hinder its widespread adoption.
Addressing this challenge and enabling more people to benefit from the abundance of data embedded in their daily environments represents an important future research direction.













\chapter{Conclusion}

In this thesis, we examine the practical challenges that undermine the reliability of DR-based visual analytics and present technical advances that address these challenges. These contributions open broader discussions on how to democratize reliable visual analytics for wider audiences and extend our findings to visual analytics workflows that do not rely on DR.
In this chapter, we review these contributions and clarify their remaining limitations.


\section{Review of Thesis Contributions}

The contribution of this thesis is to improve the reliability of visual analytics using DR. 
To do so, \textit{we first externalize three prevalent yet previously overlooked problems in practice}.
We conduct extensive literature reviews and an interview study that reveal how practitioners, both within visual analytics and across domains outside visualization, use DR in their analysis. From this investigation, we identify three prevailing challenges: (1) the misuse of popular DR techniques such as \tsne and \umap, (2) cherry-picking of hyperparameters, and (3) erroneous interactions.
\textit{We address each challenge by introducing a corresponding technical solution}, summarized as follows:

\paragraph{Challenge 1: Misuse of famous DR techniques}
We address the prevalent misuse of two popular DR techniques, \tsne and \umap, for tasks for which they are not suitable. This problem undermines the reliability of visual analytics by degrading overall task accuracy. 
We find that this problem stems from practitioners’ perceptual preference for projections that exhibit high apparent class separability: a characteristic that \tsne and \umap deliberately amplify.
To counter this bias, we introduce two new DR evaluation metrics, \textit{\LT} and \textit{\LC} (\ltc), which correct the tendency of existing metrics to reward exaggerated class separability.
\ltc is not only verified to mitigate such bias but also outperforms existing evaluation metrics in terms of accurately quantifying apparent distortions in DR projections.

\paragraph{Challenge 2: Cherry-picking of hyperparameters}
We contribute to resolving the common practice of cherry-picking hyperparameters. 
This practice negatively impacts the reliability of visual analytics by making DR projections less likely to faithfully represent the original data. 
To address this problem, we propose a new \textit{dataset-adaptive workflow} that accelerates the optimization process using dataset characteristics.
This workflow thus  ``motivates'' practitioners to systematically optimize hyperparameters rather than cherry-picking them.
Our experiments verify that this workflow provides a significant gain in optimization time with negligible loss in accuracy.

\paragraph{Challenge 3: Erroneous interactions}
We address the problem that brushing interactions, despite their widespread use in visual analytics, are vulnerable to errors due to distortions.
Such inaccuracies undermine the reliability of visual analytics, as erroneous brushing can propagate to subsequent analyses and lead to misleading conclusions.
To address this weakness, we introduce \textit{Distortion-aware brushing}, a novel brushing technique that resolves distortions while users perform brushing. 
By enabling users to accurately perceive and select high-dimensional clusters in 2D, our approach not only improves task accuracy but also increases users’ confidence in their selections.
We further demonstrate the effectiveness of Distortion-aware Brushing through real-world use cases, including interactive labeling and visual analytics for geospatial data.

\vspace{4pt}
\noindent
\textit{In summary, we advance the reliability of DR-based visual analytics through contributions that span the entire analysis pipeline: from selecting appropriate DR techniques to supporting accruate interaction and sensemaking.}


\section{Summary of Thesis Impact}

The contributions of this thesis advance the data visualization community by clearly establishing and widely disseminating the importance of reliability in DR-based visual analytics.
For example, after \ltc took a seminal step in questioning the validity of DR evaluation metrics as gatekeepers of reliable visual analytics, a series of subsequent works emerged that further examined the validity of evaluation metrics \cite{machado25cgf, smelser24beliv, smelser25arxiv, vanderhoorn25arxiv}. 
Distortion-aware brushing also informed the future development of brushing techniques that are specialized to analyze high-dimensional data \cite{montambault25tvcg}.
Furthermore, the outcomes of this thesis have been recognized through broad adoption within the community. We release \ltc as part of the open-source evaluation-metric library ZADU \cite{jeon23vis}, which has been downloaded more than 20,000 times as of December 2025.
Parts of this thesis were also incorporated into our tutorial on the reliable use of DR for visual analytics, which we delivered at EuroVis 2025.


\section{Final Remarks}

The reliability of visual analytics should be secured to ensure that the decisions and knowledge derived from the analysis genuinely benefit people.
This dissertation contributes to this goal by advancing the reliability through the lens of DR.
We envision that the reliable use of DR in visual analytics---appropriate technique selection, hyperparameter optimization, and the use of tailored interaction techniques---will become a community norm in the near future.
Achieving this vision will require not only technical innovations, but also broader efforts to facilitate relevant discourse and to improve literacy.
We believe this thesis represents a seminal first step toward that broader vision.

\AtNextBibliography{\small\sloppy} 
\printbibliography

\begin{abstractalt} 
\indent 

\vspace{-10mm}
차원 축소(Dimensionality Reduction, DR)는 시각 분석에서 가장 널리 사용되지만 동시에 가장 쉽게 오해되는 도구 중 하나이다.
따라서 DR을 활용한 시각 분석은 신뢰성을 잃기 쉽다. 
다시 말해, 분석에서 얻은 인사이트가 실제 데이터를 정확히 반영하지 못해 잘못된 지식 형성과 의사결정으로 이어질 수 있다.
이러한 문제는 DR 투영이 원래 데이터의 모든 특성을 본질적으로 포착할 수 없으며, 이러한 한계가 분석 과정에서 충분히 고려되지 않는 데에서 비롯된다.

본 학위논문에서는 DR 기반 시각 분석의 신뢰성을 향상시키는 방법을 제안한다.
먼저, 시각 분석에서 DR을 사용할 때 실무자들이 직면하는 신뢰성 문제를 이해한다.
우리는 인터뷰 연구와 문헌 조사를 결합하여 실무자들이 실제로 DR을 어떻게 활용하는지 상세히 분석하였다.
이후 세 가지 주요 문제를 해결하기 위한 기술을 설계하였다.
첫째, \textbf{대표적인 DR 기법인 \tsne와 UMAP의 만연한 오용을 완화한다.}
이 기법들은 군집 및 클래스 분리도를 과도하게 부각시키기 때문에 심미적으로 보기 좋다는 이유로 부적절한 분석 작업에 잘못 사용된다.
기존 DR 평가 지표들은 클래스 레이블을 기반으로 하기 때문에, 클래스를 곧바로 군집의 정답으로 간주하며 이러한 편향을 더욱 증폭시킨다.
이에 우리는 실무자들이 군집 분석을 더 신뢰성 있게 지원하는 투영을 식별할 수 있도록 하는 새로운 평가 지표를 제안한다.
둘째, \textbf{DR 하이퍼파라미터의 체리 피킹 문제를 해결한다.}
적절한 평가 지표가 있더라도, DR 기법 선택과 하이퍼파라미터 최적화는 많은 시행착오를 요구하며, 이는 실무자들이 기본 설정에 의존하거나 임의로 하이퍼파라미터를 골라 쓰게 만든다.
우리는 데이터셋 특성을 자동으로 반영하여 최적의 투영을 효율적으로 탐색하는 적응형 최적화 워크플로우를 제안하여 분석가들이 체계적으로 최적화를 시행하도록 유도한다.
셋째, \textbf{DR 투영에서의 상호작용 오류를 줄인다.}
고차원 공간은 저차원 공간보다 훨씬 높은 자유도를 가지므로, 적절히 최적화된 DR 투영에서도 왜곡을 완전히 피할 수 없다.
이러한 왜곡은 DR 투영 상에서 사용자가 수행하는 브러싱과 같은 상호작용을 부정확하게 만든다.
우리는 사용자가 군집을 탐색할 때 왜곡을 보정해주는 Distortion-aware Brushing 기법을 제안하여, 사용자가 목표한 고차원 군집을 정확하게 포착할 수 있도록 지원한다.

마지막으로, DR 기반 시각 분석의 신뢰성을 근본적으로 강화할 수 있는 미래 연구 방향을 제시한다.
이는 관련 담론의 활성화부터 최적의 DR 투영을 자동으로 선택하는 방향까지를 포괄한다.
본 논문은 신뢰가능한 시각 분석의 더 폭넓은 적용을 위한 기반을 마련하며 마무리된다.
\end{abstractalt}

\begin{acknowledgementalt} 
\textit{I could not have received the Ph.D. degree without the help of my colleagues, advisors, and friends. I dedicate this dissertation to all these people.}

\paragraph{Acknowledgment to my advisor}
I owe my greatest gratitude to my advisor, \textsc{Jinwook Seo}. Your guidance and insightful comments helped me grow as an independent researcher. You were also a wonderful mentor and guiding figure for me, always being kind, patient, understanding, and warm. 
I appreciate your dedication to sharing your time, energy, and knowledge with me over the past six years. I hope to become a professor like you one day.

\paragraph{Acknowledgment to the committee members} I first appreciate \textsc{Jaesik Park} for providing thoughtful feedback that helped me shape my dissertation. The insights drawn from your expertise in computer vision greatly enriched the discussion of future applications and significantly strengthened my work.

I then thank \textsc{Ghulam Jilani Quadri}, not only for providing me with kind advice and chairing my defense, but also for becoming one of my best friends. I am blessed to have met you by chance at CHI 2022 in New Orleans and to collaborate with you.

I also appreciate \textsc{Micha\"el Aupetit}. I recall your kindness in thoughtfully answering the questions from a master's student. Thanks for saying in the email reply that you ``want to provide help in a `tangible' way,'' I was given the opportunity to learn from a researcher who is more thorough than anyone else.

Finally, I express my thanks to \textsc{Kwan-Liu Ma}, for not only advising my dissertation but also hosting me as a visiting Ph.D. student at the University of California, Davis. My time at Davis was invaluable, providing me with new experiences, lasting friendships, and broadened perspectives on research.

\paragraph{Acknowledgment to my closet friends}
I am thankful to \textsc{Sungbok Shin} and \textsc{Hyunwook Lee} for being two of my best colleagues and also closest friends. Though we met only few years ago, you became my emotional pillars and academic mentor with whom I could share even my personal concerns. I sincerely hope our friendship and collaboration will continue.

\paragraph{Acknowledgment to those who made this journey possible}
I provide my thanks to \textsc{Youngtaek Kim} for being one of my first Ph.D. mentors. My academic accomplishment was possible because I learned a lot from you while participating the Githru project.

I am also grateful to \textsc{ Jaemin Jo} for guiding my first Ph.D. project. I still remember your dedication in meeting with me two to three times a week to provide extensive feedback.

My sincere thanks also go to \textsc{Takanori Fujiwara} for guiding me through the journey of dimensionality reduction. You were first my role model as a Ph.D. student, then became my friend, and have now become both mentor and collaborator.

I then appreciate \textsc{Yun-Hsin Kuo} for helping me when I was at Davis and also being my close collaborator. I thank your thoughtful feedback that significantly improved our paper.

My appreciation also goes to \textsc{Keshav Dasu}. Thanks for teaching me not only how to love the visualization community but also why I should embrace it.

I also thank \textsc{Seoyoung Doh} for being my mentee. Through mentoring you, I learned how I should advise students as a senior researcher while realizing that the process is joyful.

Finally, I sincerely thank \textsc{Kwon Ko} for motivating me to initiate my first Ph.D. project. It was not possible for me to provoke \textit{dimensionality reduction considered harmful} if you did not recommend me to read $t$-SNE paper six years ago.

\paragraph{Acknowledgment to the collaborators}
I thank all the collaborators who helped me on my journey in dimensionality reduction.
I would first like to acknowledge 
\textsc{Danielle Albers Szafir}, 
\textsc{Paul Rosen}, 
\textsc{Sungahn Ko}, 
\textsc{Daniel Archambault}, 
\textsc{Bum Chul Kwon}, 
and \textsc{Dae Hyun Kim} 
for their advice, which helped me grow as a researcher. I also thank 
\textsc{DongHwa Shin},
\textsc{Aeri Cho},
\textsc{Seokhyeon Park},
\textsc{Jeongin Park},
\textsc{Soohyun Lee},
\textsc{Jake Hyun},
\textsc{Taehyun Yang},
\textsc{Jinhwa Jang},
\textsc{Gyehun Go},
and \textsc{Lukas Heine}
for providing me with countless supports and help while shaping my work.
A special thanks also goes to \textsc{Rafael M. Martins} for planning and executing the EuroVis tutorial with me. 
Finally, I appreciate 
\textsc{Jean-Daniel Fekete} and
\textsc{Petra Isenberg} 
for hosting and supporting me for my visit to INRIA. 

I also appreciate collaborators who gave me the chance to contribute to their projects. I am grateful to have the opportunity to participate in external collaborations with 
\textsc{Dylan Cashman}, 
\textsc{Mark Keller}, 
\textsc{Qianwen Wang}, 
\textsc{Sanghyun Hong}, 
\textsc{Juhye Ha}, 
\textsc{Changmin Jeon},
\textsc{Naimul Hoque},
\textsc{Tapendra Pandey},
\textsc{Changhoon Oh}, 
\textsc{Amber Assor},
\textsc{Sunghyo Jung},
and \textsc{Niklas Elmqvist}.
I was also fortunate to work with amazing labmates---\textsc{Minsuk Chang},
\textsc{Jaeyoung Kim},
\textsc{Jiyeon Bae},
\textsc{Dong Hun Kim},
\textsc{Youli Chang},
\textsc{Yungun Kim},
\textsc{Guangjing Yan},
\textsc{Sihyeon Lee},
\textsc{Yumin Song},
\textsc{Junhyeong Hwangbo},
\textsc{Jeongmin Rhee},
\textsc{Eugene Choi},
\textsc{Mingyu An},
\textsc{Hyunwoo Kim},
\textsc{Yeongin Kim},
\textsc{Jeanyoon Choi},
\textsc{Jiwon Song},
\textsc{Kiroong Choe},
\textsc{Suyeon Hwang},
\textsc{Min Hyeong Kim},
\textsc{Seokweon Jung},
\textsc{TaeYoung Yeon},
\textsc{Eunhye Kim}, and
\textsc{Sebeom Park}.

\paragraph{Acknowledgment to the friends outside academia}
I appreciate \textsc{Jihoon Lee}, \textsc{Kyusang Jo}, \textsc{Ji-Sahn Kim}, and \textsc{Minhee Park} for being my friends since high school. Our meetings continued throughout the Ph.D. period and became a great source of energy for me. 

\paragraph{Acknowledgment to the community}
The support of the community, especially the HCI and visualization communities, was crucial in my growth as a researcher. I am deeply grateful for this support.
First, I thank the senior researchers who guided me and offered much advice and opportunities, both within and beyond research, 
notably 
\textsc{Yun Jang}, 
\textsc{Michael Sedlmair},
\textsc{Hyunggu Jung}, and
\textsc{Inseok Hwang}.
I also thank my undergraduate advisor, \textsc{Young-Joo Suh}, for his support and advice that helped me to be well-prepared for being a Ph.D. student.
I also appreciate being a co-chair of the IEEE VIS Student Volunteer team, fortunate to work with \textsc{Katy Williams}, \textsc{Jeremy Block}, \textsc{No\"elle Rakotondravony}, \textsc{Yixuan Wang}, \textsc{Zeyang Huang}, \textsc{Magdalena `Momo' Boucher}, and \textsc{Hamza Elhamdadi}.
Finally, I would like to acknowledge my friends and mentors who helped and welcomed me in academia, notably 
\textsc{Keke Wu}, 
\textsc{Jiwon Choi},
\textsc{Leixian Shen}, 
\textsc{Eunkyung Jo}, 
\textsc{Sungwon In},
\textsc{Md Dilshadur Rahman},
\textsc{Sandra Bae},
\textsc{Angelos Chatzimparmpas},
\textsc{John Chung},
\textsc{Linping Yuan},
\textsc{Xiaoyu Zhang},
\textsc{Sangbong Yoo},
\textsc{Myeongwon Jung},
\textsc{Jungeun Lee},
\textsc{Seunghyun Bae},
\textsc{Jacob Miller},
\textsc{Sangwook Lee},
\textsc{Arran Zeyu Wang},
\textsc{Adam Coscia},
\textsc{Joohee Kim}, and
\textsc{David Bauer}.

\paragraph{Acknowledgment to my family}
Finally, and most importantly, I would like to thank my family and relatives. 
I particularly express deep gratitude to my parents, \textsc{Inwon Jeon} and \textsc{Hyeyoung Je}, for their endless and generous love and support. Thanks to your respect and encouragement, I was able to overcome many hardships during my Ph.D. journey.

\vspace{4pt}
\noindent
Due to space constraints, there are many others I have not been able to acknowledge here. I trust there will be opportunities to express my gratitude to them personally.

\paragraphit{Thank you}

\end{acknowledgementalt}

\end{document}